\pdfoutput=1
\documentclass[12pt,letterpaper,UTF8]{report}
\usepackage{natbib}
\usepackage{geometry}
\usepackage{fancyhdr}
\usepackage{afterpage}
\usepackage{graphicx}
\usepackage{amsmath}
\usepackage{amssymb,amsfonts,amsbsy}
\usepackage{mathrsfs}
\usepackage{empheq}
\usepackage{tocloft}
\usepackage{bm}
\usepackage[normalem]{ulem}
\usepackage{xcolor}
\usepackage{booktabs}
\usepackage{CJK}
\usepackage{epstopdf}
\usepackage{etoolbox}
\usepackage[super]{nth}
\usepackage{romannum}
\usepackage{titlecaps}
\Addlcwords{a an the}
\Addlcwords{and but for or nor}
\Addlcwords{aboard about above across %
  after against along amid among anti %
  around as at before behind below %
  beneath beside besides between %
  beyond but by concerning considering %
  despite down during except excepting %
  excluding following for from in %
  inside into like minus near of off %
  on onto opposite outside over past %
  per plus regarding round save since %
  than through to toward towards under %
  underneath unlike until up upon %
  versus via with within without} 
\usepackage[T1]{fontenc} 
\usepackage{ae} 
\usepackage{aecompl}
\usepackage[utf8]{inputenc}
\usepackage[activate={true,nocompatibility},final,tracking=true,kerning=true,spacing=true]{microtype}
\usepackage{asudis}
\usepackage[colorlinks = true, 
            urlcolor   = black, 
            linkcolor  = black, 
            citecolor  = black,
            linktoc    = all,
            bookmarks  = true,
            bookmarksnumbered = true,
            bookmarksdepth = 4,
            hyperindex,breaklinks]{hyperref}
\usepackage{bookmark}

\DeclareRobustCommand*{\APPENDIXcontinue}{APPENDIX \par \afterpage{\lhead{APPENDIX}\rhead{Page}} }

\makeatletter
\AtBeginDocument{
  \hypersetup{
    pdftitle = {\@title},
    pdfauthor = {\@author},
    pdfsubject = {Nuclear Physics},
    pdfkeywords={path integral, quantum, Monte Carlo, light nuclei, chiral, Argonne,
                 ground state, energy, density, radii, response function,
                 High Performance Computing, Arizona State University}
  }
}
\makeatother

\title{Path Integral Quantum Monte Carlo Method for Light Nuclei}
\author{Rong Chen}
\degreeName{Doctor of Philosophy}
\paperType{Dissertation}
\defensemonth{July}
\defenseyear{2020}
\gradmonth{August}
\gradyear{2020}
\chair{Kevin E. Schmidt}
\memberOne{Ricardo O. Alarc\'{o}n}
\memberTwo{Oliver Beckstein}
\memberThree{Joseph R. Comfort}
\memberFour{Igor A. Shovkovy}

\begin{document}

\maketitle

\bookmark[dest=abshyper,startatroot]{ABSTRACT}
\begin{abstract}\hypertarget{abshyper}{}I describe the first continuous space nuclear path integral quantum Monte Carlo method, and calculate the ground state properties of light nuclei including Deuteron, Triton, Helium-3 and Helium-4, using both local chiral interaction up to next-to-next-to-leading-order and the Argonne $v_6'$ interaction.
Compared with diffusion based quantum Monte Carlo methods such as Green's function Monte Carlo and auxiliary field diffusion Monte Carlo, path integral quantum Monte Carlo has the advantage that it can directly calculate the expectation value of operators without tradeoff, whether they commute with the Hamiltonian or not.
For operators that commute with the Hamiltonian, e.g., the Hamiltonian itself, the path integral quantum Monte Carlo light-nuclei results agree with Green's function Monte Carlo and auxiliary field diffusion Monte Carlo results.
For other operator expectations which are important to understand nuclear measurements but do not commute with the Hamiltonian and therefore cannot be accurately calculated by diffusion based quantum Monte Carlo methods without tradeoff, the path integral quantum Monte Carlo method gives reliable results.
I show root-mean-square radii, one-particle number density distributions, and Euclidean response functions for single-nucleon couplings.
I also systematically describe all the sampling algorithms used in this work,
the strategies to make the computation efficient,  the error estimations, and the details of the implementation of the code to perform calculations.
This work can serve as a benchmark test for future calculations of larger nuclei or finite temperature nuclear matter using path integral quantum Monte Carlo.
\end{abstract}

\bookmark[dest=dedhyper,startatroot]{DEDICATION}
\dedicationpage{ \hypertarget{dedhyper}{}
\hspace{0pt}
\vfill
Flowers may bloom over and over again, but people can never be young again.
\par
\begin{CJK}{UTF8}{gkai} 
花有重开日，人无再少年。
\end{CJK}
\\
To my parents Chen Wei and Yuan Mao-Ying.
\par
\begin{CJK}{UTF8}{gkai}
向我的老爸老妈陈伟和袁茂英致意。
\end{CJK}
\vfill
\hspace{0pt}
}

\bookmark[dest=ackhyper,startatroot]{ACKNOWLEDGEMENTS}
\begin{acknowledgements}\hypertarget{ackhyper}{}I am grateful to my advisor prof. Kevin E. Schmidt not only for his transcendental tolerance, extraordinary patience, sagacious guidance, but also for his kindly accepting me at the stage I was at and assigning me an awesome PhD project which enlightened my life.
From him I learned not only delicate and creative physics ideas, but more importantly, the hardcore independent American spirit.
To me he is the high seas Caesar who always tries to train me to be like captain Jack Sparrow who adventures on the seven seas under any whimsical conditions with the Black Pearl.
It was destiny that lead me overseas from Shanghai to Tempe, and eventually brought me the honor to work with prof. Schmidt doing nuclear physics in an \textit{ab initio} way using state of the art quantum Monte Carlo methods.
I thank the United States and prof. Ralph V. Chamberlin for bringing me the Bifr\"{o}st.
I thank prof. Francesco Pederiva, too.

I am honored to have profs. Alarc\'{o}n, Beckstein, Comfort and Shovkovy as my PhD committee.
I also thank profs. Quan Qin, Kong-Thon Tsen, Peter Rez, Cecilia Lunardini, Tanmay Vachaspati, Maulik Parikh, and Damien A. Easson.
I thank Drs. Darya Dolenko and Iwonna Rzanek for teaching assistant instructions.
I thank Lucas Madeira, Cody L. Petrie, and Jie Zhang for discussions.
I thank Zhi-Yu Huang, too.

I am indebted to my MS degree advisor prof. Lie-Wen Chen for laying me a crucial foundation and beyond.
I thank prof. Bao-An Li for collaborations and help.
I thank prof. Dean Lee for beneficial suggestions and kindness.
I thank Bao-Jun Cai, Jing-Jing Yang, Yi-Hua Huang, Bo-Wen Lian, Jie Gu and Jia-Xin Li for long friendship.

Ultimately, I thank my parents for their unmatched love and unconditional support all those years.
These are truly in extra dimensions beyond the Standard Model.

This work was supported by the National Science Foundation Grant No. PHY-1404405.
The High-Performance Computing and storage resources were provided by Research Computing at Arizona State University.
\end{acknowledgements}

\cleardoublepage

\pagestyle{styletoc}
\tableofcontents
\addtocontents{toc}{~\hfill Page \par}

\newpage
\phantomsection
\addcontentsline{toc}{part}{LIST OF TABLES}
\renewcommand{\cftlabel}{Table}
\addtocontents{lot}{Table~\hfill Page \par}
\listoftables
\newpage
\phantomsection
\addcontentsline{toc}{part}{LIST OF FIGURES}
\renewcommand{\cftlabel}{Figure}
\addtocontents{lof}{Figure~\hfill Page \par}
\listoffigures
\newpage
\phantomsection
\renewcommand{\cftlabel}{CHAPTER}
\addtocontents{toc}{CHAPTER \par}

\pagestyle{fancyplain}
\doublespace
\pagenumbering{arabic}
\cleardoublepage

\bookmark[dest=chapterhyper,startatroot]{CHAPTER}
\chapter{INTRODUCTION} \hypertarget{chapterhyper}{}

\section{Background}
\label{secBackground}

In nuclear physics, most of the questions simply cannot be convincingly answered without answering the following two perhaps ultimate questions in the first place:
\begin{enumerate}
  \item What is the interaction between nucleons?
  \item How to solve the many-body problem?
\end{enumerate}

In fact, these two questions are more like big challenges than problems or troubles.
Today, in the beginning of the second decade of the \nth{21} century, which is nearly a century since Erwin Schr\"{o}dinger in 1925 wrote down the equation
$i \hbar \dot{\Psi} = H \Psi$
that rules the quantum world, with discoveries and development in both theory and computation aspects made by several generations of people, the answers to the two questions are on the horizon.

\subsection{Evolution of the Nucleon-Nucleon Interaction}

In regards to the first question, what is the interaction between nucleons? We already know that nucleons are made of quarks and gluons, and quantum chromodynamics (QCD) (\cite{Wilczek73a,Polizer73a}) describes the strong interaction between quarks and gluons within the framework of quantum field theory (QFT) by using the QCD Lagrangian. The interactions between quarks are made by exchanging the gluons which are the `messengers'. Nucleons, i.e. protons and neutrons are actually the bound states of quarks and gluons.
Consequently, in principle we should calculate the interaction between nucleons from QCD, using the degrees of freedom of the quarks and gluons. However, QCD features asymptotic freedom. Namely quarks are weakly interacting at small distance which corresponds to high energies. But the interaction is strong at long distance which corresponds to the low energy region. This causes the confinement of quarks within hadrons like nucleons and mesons.
Therefore, in the high energy region we can perform perturbation calculations but not so easily in the low energy region where hadrons form. This is mainly why it is so difficult to do \textit{ab initio} calculations at the hadronic level directly from QCD. Lattice QCD methods (\cite{davies2002lattice}) are working toward calculating the nucleon many-body problem from quark and gluons degrees of freedom, however the current simulations are not capable of a system beyond two nucleons (\cite{BEANE2008,savage2015nuclear}).

In fact, in the low energy region, we may not have to directly include quark and gluon degrees of freedom, because, the nucleons are in the non-relativistic regime and they are stable; the quarks and gluons hardly ever get excited in the nucleons. In this case, we can treat a nucleon as an elementary particle (nucleons have form factors which can be thought of as the internal structure) with protons and neutrons only distinguished from each other by their isospin. We can treat the interactions between nucleons as instantaneous and therefore we can use a potential to describe the nucleon-nucleon interaction.
Of course, the nucleon-nucleon interaction has to be the residual interactions of the strong interaction described by QCD, just like the van der Waals potential is the residual interaction of the Coulomb interaction at large distance.

\subsubsection{Meson exchange theory}

The earliest version of nucleon-nucleon (NN) interaction is the Yukawa potential (\cite{Yukawa1935}) which is
$V_{\text{Yukawa}}(r)= -g^2 \;\frac{e^{- mr}}{r}$ where $g$ is the coupling constant, $m$ is the mass of the meson and $r$ is the distance between two nucleons. Yukawa theory was pioneering in the 1930s because it first predicted that the interactions between nucleons are made by exchanging a new particle called a meson. The interaction range $r$ between nucleons, which is about 1 fm, can determine the mass of the meson by the uncertainty principle $r \approx 1/m$ so $m\approx 200$ MeV. And such a meson called the pion with mass about 140 MeV was discovered in 1947 (\cite{Meson1947a}). Since then, nuclear and high energy physics has entered a golden age, and there was time where a discovery of a new particle meant a Nobel Prize.

Over the decades, we have more understanding of what the interaction between nucleons looks like.
We at least know that the interactions between nucleons should have momentum dependence. The simplest interaction with momentum dependence is the spin-orbit interaction which is nonlocal, and in fact a realistic nucleon-nucleon interaction cannot be built without it.
Also, there are not only two-nucleon forces (2NF), but also three-nucleon forces (3NF), and even four-nucleon forces (4NF), and even more.
Since the 1990s, several high-precision NN interactions (\cite{Machleidt_2001}) mostly based on meson exchange theory have been developed.
Among them, the Nijmegen potentials (\cite{Stoks1994PRC}), the CD-Bonn potential (\cite{Machleidt1996PRC}), the Argonne series (\cite{Wiringa95PRC,WiringaPieper02a,Veerasamy2011PRC}) of interactions together with Tucson-Melbourne (\cite{Coon1981PRC}) and Urbana and the modern Illinois series of 3NF (\cite{CARLSON198359,WiringaPieper02a}) are widely used.
These meson-exchange based models have achieved great success in explaining the experimental data.
In these potentials, the 2NF includes one-pion exchange (OPE) as well as heavier meson exchange and the 3NF contains terms from two and/or more pion exchange.
However, the parameters in the 3NF models need be adjusted for the chosen 2NF model to fit the binding energy for $A\geq3$ nuclei (\cite{Nogga2000PRL}).
This situation raises a simple question, how do we choose the 2NF and 3NF? Apparently these models are lacking a fundamental theory which can guide us to find or choose the form of 2NF and 3NF within a consistent framework.
From today's point of view, we know that those models are lacking the guidance of the (broken) symmetry from QCD, the chiral symmetry.

\subsubsection{Chiral effective field theory}

Since the 1990s, the importance of NN interactions derived from chiral effective field theory (ChEFT) has been realized by more and more people.
The idea of ChEFT was proposed in 1979 by Steven Weinberg (\cite{WEINBERG1979327}) and after his proposal about how to treat the nuclear interaction at the chiral limit (\cite{WEINBERG1990288,WEINBERG19913}) ChEFT has been rapidly developed in nuclear physics (\cite{Epelbaum2009a,MACHLEIDT20111}). The nuclear interactions derived from ChEFT are connected with the fundamental theory of QCD by the broken chiral symmetry. This is an example of `symmetry dictates interaction' (\cite{Yang1980a}).
To derive the nuclear interaction from ChEFT, one has to identify the relevant symmetries of low-energy QCD first.
Then, one needs to construct the most general Lagrangian including those symmetries and symmetry breakings.
After that, one needs to do a low momentum expansion and design the scheme to collect the important contributions. The scheme is called chiral perturbation theory (ChPT).
Finally, one calculates those Feynman diagrams of the important contributions order by order, in terms of an expansion in the small external momenta over the large scale, $(\frac{Q}{\Lambda_{\chi}})^\nu$. Here Q means the nucleon three-momentum or pion four-momentum or a pion mass which is around 140 MeV, $\Lambda_\chi$ is the chiral symmetry breaking scale (hadronic scale) which is about 1 GeV, $\nu$ is the power of the order.
Determining the power $\nu$ is called power counting.
$\nu=0$ terms are the called the leading order (LO) which contributes the most to the nuclear interaction.
There are no $\nu=1$ terms due to parity and time-reversal invariance, so the next-to-leading order (NLO) terms are for $\nu=2$.
The next-to-next-to-leading order (N$^2$LO) terms are for $\nu=3$, and this is where a nonvanishing 3NF occurs (\cite{Klock1994a,Epelbaum2002PRC}). The next-to-next-to-next-to-leading order (N$^3$LO) are for $\nu=3$ and that is where a 4NF occurs.
Note that $(\frac{Q}{\Lambda_{\chi}})<1$, and the 2NF occurs at $\nu=0$, the 3NF occurs at $\nu=3$, and the 4NF occurs at $\nu=4$. $(\frac{Q}{\Lambda_{\chi}})^0 \gg (\frac{Q}{\Lambda_{\chi}})^3 \gg (\frac{Q}{\Lambda_{\chi}})^4$, therefore ChPT explains the hierarchy of nuclear force $2NF \gg 3NF \gg 4NF$ which we knew empirically before.
Because the Chiral interaction are based on momentum expansion $(\frac{Q}{\Lambda_{\chi}})^\nu$, it naturally carries the momentum dependence.
It is worth mentioning that (\cite{Gerzerlis13a}) up to N$^2$LO (beginning from N$^3$LO, non-local operators cannot be avoided however their effects should be negligible and can be treated perturbatively), one can write down a mostly local version of the chiral interaction\footnote{That is, the momentum dependence is predominantly determined by the momentum transfer which Fourier transforms to a local in space potential}.
The local chiral interaction will be particularly useful for any quantum Monte Carlo (QMC) method in continuous space which includes the method used in this work.

\subsection{Methods of Solving the Many-Body Problem}
\subsubsection{Phenomenological approaches}
For the second main question: how do we solve for the many-body problem given the interactions between the particles?
There are some approaches developed in the past several decades.
Some of them are based on mean field theory and the phenomenological NN interactions. For example, models based on non-relativistic mean field like Skyrme-Hartree-Fock (\cite{SKYRME1958615}), Gogny-Hartree-Fock (\cite{Gogny1980PRC})  models\footnote{
Those are the methods I was mainly using in my master degree career (\cite{ChenRong2012PRC}).
},
relativistic mean field theory (RMF) (\cite{Reinhard_1989,RING1996193}), etc.
In those methods, for bulk matter people assume that the each particle is in the mean field of the others and so the wave function of the particles is approximately plane wave and it obeys a Fermi-Dirac distribution in phase space. The energy per nucleon or energy density can be calculated by integrating over phase space and by using a phenomenological nuclear interaction such as Skyrme or Gogny interaction specially designed for this mean field approximation.
Those methods have been successful in many aspects, in particular, they are computationally cheap so it is easy to reach a high density region for either isospin symmetric or asymmetric nuclear matter and to do finite temperature calculations. Also, they can give analytic expression for the equation of state (EOS) of nuclear matter. Therefore a lot of quantities, like the symmetry energy, single nucleon potential, pressure, incompressibility coefficient, etc.,  can be conveniently calculated from the EOS.
However, the mean field assumption and phenomenological interaction are not given from first principles, and different sets of parameters in the models can give different results for physical quantities especially at high nucleon density (\cite{LWChen2008a}). So, one may argue that those methods may not be among the most reliable methods in predicting the properties of nuclear matter or nuclei at high density and/or finite temperature.
Therefore, people may prefer to seek the answers from first-principle calculations.

\subsubsection{First principle approaches}
First principle, or \textit{ab initio} calculations are referring to those methods which are aiming at directly solving the Schr\"{o}dinger equation of the system. The no-core shell-model (NCSM) (\cite{Navr_til_2009,BARRETT2013131}), coupled cluster theory (\cite{Hagen_2014}), and quantum Monte Carlo method (QMC) (\cite{CarlsonRMP15a}) are by far the three most successful and most mentioned methods. Each of them have their own advantages in their ability to approach specific problems.

NCSM is basically a method for solving for the eigenvalues of the Hamiltonian matrix in the time-independent Schr\"{o}dinger equation, by adding the truncated Jacobi-coordinate and/or the single-nucleon Slater determinant Harmonic oscillator (HO) Hamiltonian to the original NCSM Hamiltonian, and therefore use a truncated HO basis to calculate the modified Hamiltonian matrix. The effects from HO Hamiltonian are subtracted in order to get correct results.
The number of HO basis has to be truncated to make NCSM calculable.
However that raises a problem. If the potential is hard instead of soft\footnote{By `soft' we mean the potential remain relatively small when the distance $r$ between the nucleons $r \rightarrow 0$ and it varies not too abruptly with $r$. `Hard' is the opposite.},
then the truncated momentum for the HO basis may not be high enough for the hard potential and the calculation will not converge. Usually NCSM works best with a soft potential.

Coupled cluster method is mostly used to calculate the properties of closed-shell or nearly closed-shell nuclei. It defines a similarity-transformed Hamiltonian $\bar{H}=e^{-T} H_N e^{T}$ where $H_N$ is a normal ordered Hamiltonian derived from the original Hamiltonian $H$, and $T$ is the cluster operator.
It finds the ground state by using a variational method to minimize the energy functional $E(L,T) \equiv \langle \phi | L e^{-T} H_N e^{T} | \phi \rangle     =  \langle \phi | L \bar{H} | \phi \rangle $, where $L$ is the de-excitation operator and $| \phi \rangle$ is a trial state.
The variation of $E(L,T)$ with respect to the truncated operator $L$ yields a set of so called coupled clusters with singles and doubles (excitations) approximation (CCSD) equations.
Solving for the cluster amplitudes from $T$ in those CCSD equations, one can show that they give the ground state energy $E_0= \langle \phi |\bar{H} | \phi \rangle $.
We can also do variations with respect to the cluster amplitudes in $T$, that will yield other sets of equations.
The solved coefficients from $L$ in those equations can be used to solve for the excited energies or other physics quantities.
In coupled cluster calculations, $\bar{H}$ needs to be expanded by using the Baker-Campbell-Hausdorff (BCH) expansion
$\overline{H} = H_N + \left[H_N,T\right] + {1\over
2!}\left[\left[H_N,T\right],T\right]+ \ldots .$
and this can be represented by diagrams just like the Feynman diagrams in high energy physics.
For ground state calculations, the diagrams with three or more body nature will vanish so one only needs to calculate those with two-body nature. This feature makes the coupled cluster method computationally attractive. The computational cost in coupled cluster are mainly in solving those non-linear CCSD equations, and the number of the equations depends on the size of the basis. So, with the current technique, it must truncate the basis (usually HO basis) in its model space, especially for medium and large nuclei calculations. Consequently, like NCSM, it works best with soft potentials, otherwise one needs to modify the potential in some way to make it soft.

Unlike NCSM which aims at solving for the eigenvalues of the effective Hamiltonian matrix and the coupled-cluster method which requires solving for many non-linear equations, QMC methods seeks the solution of many-body system from the stochastic process called the Monte Carlo (MC) method (\cite{Kalos1986a}), and so QMC codes can naturally be made highly parallel and therefore particularly suited for modern high performance computing.
QMC methods are among the most trusted methods since approximations, if any, can be well characterized.
In the last 30 years, significant progress has been made either in continuous space time (\cite{CarlsonRMP15a}) or on the lattice (\cite{LEE2009117,Lee2012PRL}). Nuclear QMC methods have achieved wide success in accurately calculating the properties of light nuclei and neutron matter using realistic nuclear interactions such as the Argonne potentials (\cite{Wiringa95PRC,WiringaPieper02a}) and the ChEFT potentials (\cite{Lynn14a,Gerzerlis14a,Lynn2017PRC,Lonardoni2018PRL,Lonardoni2018PRC}).

QMC methods in continuous space are called continuum QMC. In this work, we focus on continuum QMC.
These methods are mainly variational Monte Carlo\footnote{Brief introductions to VMC is in Appendix \ref{secVMC}.} (VMC), Diffusion Monte Carlo (DMC)\footnote{Brief introductions to DMC is in Appendix \ref{secDMC}.}  (\cite{HammondMCbook}) and path integral Monte Carlo (\cite{Ceperley95a}).
DMC and PIMC both project out the ground state from the trial wave function by applying the short imaginary time propagator $e^{-H\Delta \tau}$ repeatedly until all the excited states are removed, but in a different way which will be introduced in chapter \ref{ChapPIMC}.
Usually one begins from VMC to find some optimized parameters in the trial wave function that minimize the expectation value of the energy. One starts the DMC or PIMC from the optimized trial wave function to project out the ground state.

Green's function Monte Carlo (GFMC) (\cite{Carlson87a}) and auxiliary field diffusion Monte Carlo\footnote{Brief introductions to AFDMC is in Appendix \ref{secAFDMC}.} (AFDMC) (\cite{Schmidt99a}) are both DMC methods, and they are among the most successful ones in nuclear physics.
GFMC for bosons was introduced by Malvin Kalos (\cite{Kalos1974PRA}),
nowadays the GFMC for nuclear physics is the one introduced in late 1980s by Joseph Carlson (\cite{Carlson87a}) who included the spin and isospin degrees of freedom for nucleons.

In GFMC, for a system of $A$ nucleons with $Z$ protons, charge conservation gives $C_A^Z \equiv  \frac{A!}{Z!(Z-A)!}$ isospin states, and $2^A$ spin states, so the trial wave function is composed of $2^A C_A^Z =\frac{2^A \times A!}{Z!(Z-A)!}$ spin/isospin basis states. GFMC yields very accurate results because it sums over all the spin isospin states, but the number of basis states for the wave function will be too big for heavy nuclei to be calculable. For this reason, at the current stage, GFMC calculations are not able to go far beyond $^{12}$C because the computational cost is too high\footnote{For $^{12}$C, GFMC cost about 130 million core hours (\cite{lovato2020ab}) on the MIRA supercomputer at Argonne National Laboratory located in Lemont, Illinois, USA.}.

AFDMC in nuclear physics was invented in late 1990s by Kevin E. Schmidt and Stefano Fantoni (\cite{Schmidt99a}). The idea was to  use a trial wave function that scales better than GFMC's, and sample the spin isospin states by introducing auxiliary variables (fields) using a Hubbard-Stratonovich (HS) transformation (\cite{CarlsonRMP15a}) of the operators in the potential.
 A convenient form of wave functions in AFDMC is a Slater determinant which has $A$ single-particle orbitals, so the wave function is a determinant of an $A \times A$ matrix and the computational cost is the order of $A^3$ operations (\cite{wikiCompcost}) which is significantly cheaper than GFMC's.
 Another choice can be the Pfaffian pairing wave function (\cite{Bajdich2006PRL}) which also scales as $\mathcal{O}(A^3)$.
Currently, including 3NF from ChEFT, AFDMC was able to accurately calculate nuclei up to $^{16}\textrm{O}$ (\cite{Lonardoni2018PRL}), and it has calculated systems with up to 100 nucleons interacting with a 2NF.

\section{Motivation}

All the current continuous space nuclear QMC calculations (such as GFMC and AFDMC) are performed within the framework of DMC.
Due to the nature of DMC, when we perform ground state calculations,
the operators will typically be sandwiched between the trial wave function $\Psi_T$ and the ground state wave function $\Phi_0$. Therefore, DMC based methods can accurately calculate the ground state expectation values of the operators which commute with the Hamiltonian, e.g., the ground state energy $E_0$ itself. However, if the operators do not commute with the Hamiltonian, the ground state expectation values cannot be obtained without tradeoff\footnote{
Diffusion based QMC methods typically use a technique called `future walking' (\cite{HammondMCbook}) or `forward walking' (\cite{Runge92PRB,Casulleras1995PRB,Samaras1999ForwardwalkingGF}) to deal with operators which do not commute with $H$. This technique is sensitive to the quality of the trial wave function $\Psi_T$.
}.
In fact, there are a lot of operators which do not commute with the Hamiltonian, such as the root-mean-square (rms) radius, the particle number density, response functions, etc. Therefore, if we want to accurately calculate the ground state properties beyond the energy itself, we may want to use methods other than DMC.

Path integral Monte Carlo (PIMC) method (\cite{Ceperley95a}) can be a natural choice for such calculations. Unlike DMC methods, in PIMC, the operator can indeed be sandwiched easily between the ground states. So, whether the operator commutes with the Hamiltonian or not, we can always naturally calculate its ground state expectation value.
This is the most attractive feature of PIMC.

There were important applications of PIMC in calculating the properties of condensed Helium atoms (\cite{Ceperley95a}) before.
There are also PIMC calculations of nuclei on the lattice (\cite{LEE2009117,Lee2012PRL}).
However, there are no PIMC with the realistic NN potentials such as Argonne potential or ChEFT potentials to calculate the properties of any nuclei in continuous space time before this work.

\begin{figure} [htbp!]
\centering
\includegraphics[scale=0.47]{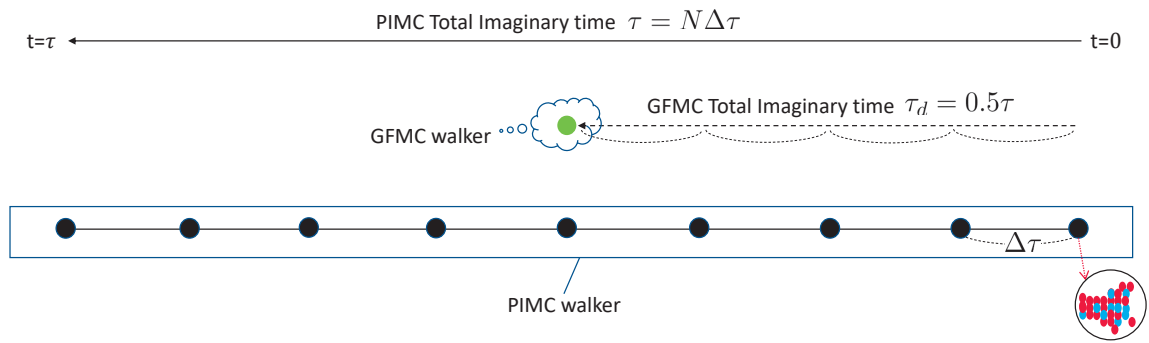} 
\caption[Illustration of a GFMC Walker and a PIMC Walker]{Illustration of a GFMC walker and a PIMC walker.}
\label{PIMCwalker}
\end{figure}

One of the main reasons is that continuous space PIMC calculations can be computationally expensive (even more expensive than GFMC).
As illustrated in Fig. \ref{PIMCwalker}, for a PIMC walker,  just like in the path integral, we need to separate the total imaginary time of the path into $N$ slices, insert a complete basis at each slice, and form a bead which contains the spatial configuration of the nucleus. Computing the beads one by one will be more and more time consuming as $N$ increases,
especially considering that the realistic nuclear interactions contain spin and isospin structure.
As a rough estimation, the amount of work a PIMC walker does is about $N$ times more than a GFMC walker does.
However, due to the nature of DMC, a GFMC walker needs to evolve $N/2$ steps to reach $\tau_d$ which is half of the total imaginary time $\tau$ a PIMC walk does, in order to project out the ground state.
Besides, for a given amount of computation time, the GFMC walkers can be more uncorrelated with each other than PIMC walkers. That is to say, PIMC actually needs more than twice the computation time to have equally uncorrelated walkers as GFMC walkers.
So, overall the cost of a PIMC walker can be more than twice the cost of a GFMC walker.
Therefore PIMC can be more expensive than GFMC.
The storage space a PIMC walker requires is also $N$ times bigger than a GFMC walker, because the configurations of all the nucleons in the nucleus at each of $N$ beads need to be stored.

Nonetheless, ever since the early nuclear QMC simulation such as GFMC performed in 1987 (\cite{Carlson87a}), in the last more than three decades, human beings have witnessed an impressive increase in supercomputers' power. A typical CRAY system in the 1980s could do about $10^9$ floating point operations per second (FLOPS) (\cite{ChodoshEVO1946}),
and today's most powerful supercomputer Fugaku\footnote{As of June 2020 (\cite{wiki:Fugaku2020}), the most powerful supercomputer in the world is Fugaku (peak performance is around 0.54 exaFLOPS) which is installed at RIKEN Center for Computational Science in Kobe, Japan. It is preceded by the Summit (peak performance is around 0.2 exaFLOPS) which is located at the Oak Ridge National Laboratory in Tennessee, USA.}
can do more than $10^{17}$ FLOPS.
With a magnitude of $10^8$ increase in computing power as well as the better understanding of NN interaction such as those from ChEFT, we have seen more and more ChEFT based GFMC and AFDMC calculations (\cite{Lynn14a,Gerzerlis14a,Lynn2017PRC,Lonardoni2018PRL,Lonardoni2018PRC}).
Therefore, we will naturally ask, what can PIMC do in nuclear physics?

The motivation of this work is to show that (as a benchmark test to begin with),
it is now possible to perform a reliable continuous space PIMC calculation of light nuclei, by using modern realistic NN interactions (particularly the chiral interaction).

\section{Outline}

The dissertation is organized as follows.

In chapter \ref{ChapPIMC}, I introduce the framework of PIMC first,
then I introduce the detailed form of the Hamiltonian we used and its model (Hilbert) space, the trial wave function, the short-time approximation of the propagator and the time step error structure of our calculation.
After writing down the details of all the calculations and equations which are suitable for PIMC simulations first, I then describe how the path is calculated and efficiently updated. Particularly, I describe how the path is efficiently calculated by introducing the path diagrams, which visualize the path and make the calculation more accessible than merely the equations alone.
The sign problem is discussed, too.

In chapter \ref{compalgorithm}, I describe the Monte Carlo sampling techniques used in the PIMC.
I begin by introducing the Markov chain Monte Carlo and the Metropolis algorithm. I introduce how they are connected to several sampling methods used in PIMC calculations, particularly the multi-level sampling which is efficient and is mostly used. Other sampling methods are introduced, too. I describe how the path is efficiently updated and the sampling strategies implemented in the code.

In chapter \ref{RsDs}, I show the results of the ground state PIMC calculations of the light nuclei up to $^4$He.
I introduce how I choose the total imaginary time $\tau$, time step $\Delta \tau$ and the cost of the computation. I show the PIMC results of ground state energy which is in agreement with GFMC and AFDMC results. I also show how to verify if the ground state has been really reached.
After presenting the results of the ground state energy, I show results of the operators which do not commute with the Hamiltonian but are also important, such as root-mean-square radii, particle number density and angle averaged Euclidean response functions for single-nucleon couplings.

The summary and outlook of this work are finally given in chapter \ref{secSumOutlook}.

\chapter{Path Integral Quantum Monte Carlo}
\label{ChapPIMC}

\section{Theoretical Framework}
\label{TheoFrame}
PIMC directly calculates the expectation value of an operator $\langle \hat{O} \rangle$ as
\begin{equation}
\langle \hat{O} \rangle= \frac{  \langle \Psi_T| e^{-H \tau_1}  \hat{O} e^{-H \tau_2 } | \Psi_T \rangle  }
{\langle \Psi_T| e^{-H \tau} | \Psi_T \rangle}, \label{PIMCopexp}
\end{equation}
where $\tau_i$ is an imaginary time and the total imaginary time $\tau=\tau_1+\tau_2$. $\Psi_T$ is a trial wave function and it can serve as the initial state of the system. $H$ is the Hamiltonian of the system,
\begin{equation}
H=T+V, \label{HTV}
\end{equation}
where $T$ and $V$ are kinetic and potential energy operators.
Since the operator $\hat{O}$ is Hermitian, the numerator and denominator in Eq.(\ref{PIMCopexp}) are real, so in PIMC it is actually calculated as,
\begin{equation}
\langle \hat{O} \rangle
= \frac{ Re  \langle \Psi_T| e^{-H \tau_1}  \hat{O} e^{-H \tau_2 } | \Psi_T \rangle  }
{ Re \langle \Psi_T| e^{-H \tau} | \Psi_T \rangle} .
\label{PIMCopexpdetail}
\end{equation}

On the one hand, if $\tau_i=0$, Eq.(\ref{PIMCopexp}) becomes,
 \begin{equation}
\langle \hat{O} \rangle  = \frac{ \langle \Psi_T | \hat{O}   | \Psi_T  \rangle }   { \langle \Psi_T | \Psi_T  \rangle  }
=\frac{ Re \langle \Psi_T | \hat{O}   | \Psi_T  \rangle }   { Re \langle \Psi_T | \Psi_T  \rangle  } , \label{VMCoexp}
\end{equation}
so PIMC becomes VMC which gives an upper bound to the true ground state energy.

On the other hand, the trial state $| \Psi_T \rangle$ can be written as the linear combination of the eigenstates $|\Phi_n\rangle$ of the Hamiltonian whose eigenvalues are $E_n$, and with corresponding amplitude $a_n$ such that,
\begin{equation}
|\Psi_T \rangle = \sum_{n=0}^{\infty} a_n |\Phi_n\rangle .
\label{PsiTexpansion}
\end{equation}
After operating with the propagator $e^{-H \tau_1}$, $| \Psi_T \rangle$ becomes,
\begin{equation}
 e^{-H \tau_1} |\Psi_T \rangle =
a_0e^{-E_0 \tau_1}
\left[   | \Phi_0 \rangle  + \sum_{i=1}^{\infty}\frac{a_i}{a_0}e^{-(E_i-E_0) \tau_1} | \Phi_i \rangle \right].
\label{eHPsiT}
\end{equation}

Therefore, if $\tau_i$ is big enough, the factor $e^{-(E_i-E_0) \tau_1}$ will make all the excited states negligible compared with the ground state $| \Phi_0 \rangle$, and we find that,
\begin{equation}
\lim_{\tau_i \rightarrow \infty } e^{-H \tau_i} | \Psi_T  \rangle \propto | \Phi_0 \rangle.
\end{equation}

So, $e^{-H \tau_i} | \psi_T \rangle$ project out the ground state $| \Phi_0 \rangle$.
In this case Eq.(\ref{PIMCopexp}) will become the ground state expectation value of $\hat{O}$,
\begin{equation}
\langle \hat{O} \rangle  = \frac{ \langle \Phi_0 | \hat{O}   | \Phi_0  \rangle }   { \langle \Phi_0 | \Phi_0  \rangle  }
=\frac{ Re \langle \Phi_0 | \hat{O}   | \Phi_0  \rangle }   { Re \langle \Phi_0 | \Phi_0  \rangle  }. \label{PIMCopgndexp}
\end{equation}

In all kinds of calculations, however, it is usually not easy to directly sample the whole exact propagator $e^{-H \tau_i}$.
Instead, we break $\tau_i$ into $N_i$ pieces of time step $\Delta \tau $ such that $\tau_i=N_i \Delta \tau$, and the whole propagator can be written as a product of $N_i$ short-time propagator $e^{-H \Delta \tau}$,
\begin{equation}
e^{-H \tau_i}=  (e^{-H \Delta \tau})^{N_i} . \label{wholepropagator}
\end{equation}

Typically we approximate $e^{-H \Delta \tau}$ using a Trotter breakup (\cite{SchmidtLee95a}),
\begin{equation}
e^{-H \Delta \tau}  \approx e^{-\frac{V}{2} \Delta \tau}e^{- T \Delta \tau}e^{- \frac{V}{2} \Delta \tau} ,
\label{Trotter}
\end{equation}
where each of the terms $e^{-\frac{V}{2} \Delta \tau}$ and $e^{- T \Delta \tau}$ can be calculated exactly or well approximated, as shown later.

If we want to calculate ground state properties, we need to use a sufficiently large $\tau_i$ first such that $e^{-H \tau_i}$ can project out the ground state $|\Phi_0\rangle$, then we choose a proper $N_i$ to make sure $\Delta \tau$ is small enough such that time step error are negligible for the quantity we want to calculate.
However, if $\Delta \tau$ is too small it can cost an unnecessary large amount of computational resources, because then we need a very big $N_i$ which means many short-time propagators $e^{-H \Delta \tau}$.
Therefore, time step $\Delta \tau$ should be set at a reasonable point between accuracy and computational cost.

Once $\tau_i$, $\Delta \tau$ and $N_i$ are chosen, then we apply Eq.(\ref{wholepropagator}) in Eq.(\ref{PIMCopexpdetail}) such that we accurately multiply the $N_i$ short-time propagators $e^{-H \Delta \tau}$ together to replace $e^{-H \tau_i}$, and therefore the sampling of Eq.(\ref{PIMCopexpdetail}) will be equivalent with Eq.(\ref{PIMCopgndexp}) the ground state calculation.
So we calculate Eq.(\ref{PIMCopexpdetail}) by choosing a sufficiently large $\tau_1$ and $\tau_2$ first (usually we set $\tau_1=\tau_2$), then replacing $e^{-H \tau_1}$ and $e^{-H \tau_2}$ with corresponding $(e^{-H \Delta \tau})^{N_1}$ and $(e^{-H \Delta \tau})^{N_2}$ such that,
\begin{equation}
\langle \hat{O} \rangle
= \frac{ Re  \langle \Psi_T| e^{-H \Delta \tau}...e^{-H \Delta \tau} \hat{O} e^{-H \Delta \tau} ...e^{-H \Delta \tau}| \Psi_T \rangle  }
{ Re \langle \Psi_T| e^{-H \Delta \tau}...e^{-H \Delta \tau} e^{-H \Delta \tau} ...e^{-H \Delta \tau} | \Psi_T \rangle} .
\label{PIMCopexpdetailshorttime}
\end{equation}

Eq.(\ref{PIMCopexpdetailshorttime}) is basically how PIMC works.
As we can see from Eq.(\ref{PIMCopexpdetailshorttime}), a PIMC calculation not only depends on the operator $\hat{O}$, but also depends on the Hamiltonian $\hat{H}$, the trial wave function $\Psi_T$ and the short-time propagator $e^{-H \Delta \tau}$. All of them will be introduced one by one in the following sections.

As to the operator $\hat{O}$, on the one hand, it may commute with the Hamiltonian. In this case, $\hat{O}$ can be moved to the leftmost or the rightmost position, and Eq.(\ref{PIMCopexpdetail}) can then be written as the average of the two,
\begin{equation}
\langle \hat{O} \rangle
= \frac{
\frac{1}{2}  Re \! \left( \!   \langle \Psi_T| \hat{O} e^{-H \tau}  | \Psi_T \rangle
\! + \!
 \langle \Psi_T|  e^{-H \tau} \hat{O} | \Psi_T \rangle
\! \right)
 }
{  Re \langle \Psi_T| e^{-H \tau} | \Psi_T \rangle}
.
\label{PIMCopexpdetailOH}
\end{equation}
This is useful when the application of the operator on the trial wave function is simpler than its application directly on the propagator $e^{-H \tau}$, or has lower variance.

On the other hand, however, the operator $\hat{O}$ can also be other operators which do not commute with $H$.
In fact, a big advantage of PIMC is that it can directly deal with operators which do not commute with $H$,
while diffusion based QMC methods like GFMC and AFDMC cannot easily do so without tradeoff.

Because, fundamentally speaking, unlike Eq.(\ref{PIMCopexpdetail}), all the diffusion based QMC are trying to use a large enough total imaginary time $\tau_d$ in $e^{-H \tau_d }$ to project out the ground state and calculate the following quantity,
\begin{equation}
\langle \hat{O} \rangle
= \frac{ Re  \langle \Psi_T| \hat{O} e^{-H \tau_d } | \Psi_T \rangle  }
{ Re \langle \Psi_T| e^{-H \tau_d} | \Psi_T \rangle}
=\frac{ Re  \langle \Psi_T| \hat{O} | \Phi_0 \rangle  }
{ Re \langle \Psi_T | \Phi_0 \rangle},
\label{DMCopexpdetail}
\end{equation}
which is called mixed estimator\footnote{It can be seen from Eq.(\ref{DMCopexpdetail}) that the DMC total imaginary time $\tau_d$ can be just half of the PIMC total imaginary time $\tau$, $\tau_d=0.5 \tau$, because DMC project out the ground state only from one side while PIMC does so from both sides.}.
It works fine as long as $\hat{O}$ commutes with $H$.
However if $\hat{O}$ does not commute with $H$, diffusion based QMC cannot directly calculate the ground state of $\langle \hat{O} \rangle$, and often the mixed estimator along with extrapolation of the errors (\cite{HammondMCbook}) are used instead.
However, the extrapolated mixed estimator is not as accurate as the true ground state estimator Eq.(\ref{PIMCopexpdetail}).
And in fact, a lot of operators do not commute with $H$, and they can only be accurately and directly calculated by PIMC (or higher variance forward walking additions to diffusion methods).
That is one of the main reasons to choose PIMC.

\begin{figure} [hptb!]
\centering
\includegraphics[scale=0.46]{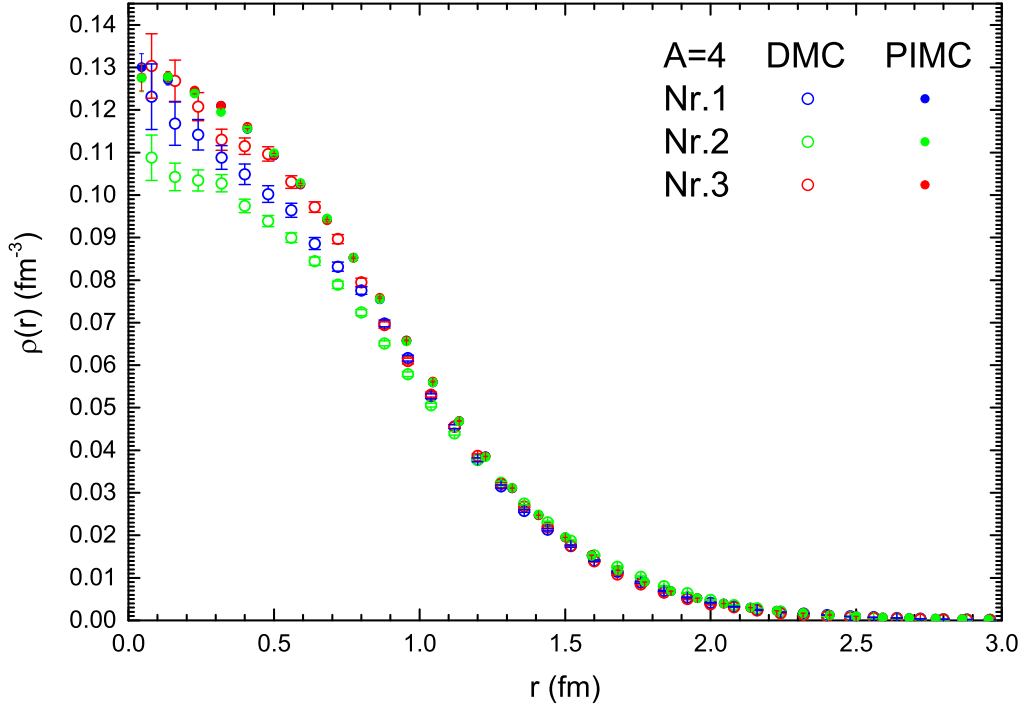}
\caption[Comparison Between PIMC and DMC 4-Particle Calculations for One-particle Number Density $\rho(r)$ Using 3 Different Trial Wave Function and with Malfliet-Tjon Interaction]{Comparison between PIMC and DMC 4-particle calculations for one-particle number density $\rho(r)$ using 3 different trial wave function (\cite{ZABOLITZKY1981114}) and with Malfliet-Tjon interaction (\cite{MALFLIET1969161}).}
\label{rhopimcdmc}
\end{figure}

Fig. \ref{rhopimcdmc} is an example to illustrate the advantage of PIMC.
The one-particle number density operator $\hat{O}=\hat{\rho}(r)$ whose definition is Eq.(\ref{rhordef}) does not commute with $H$, therefore straightforward DMC can only calculate the mixed estimator of $\rho(r)$ which is different from the exact $\rho(r)$.
As one can see from Fig. \ref{rhopimcdmc}, in the 4-particle system, using three different trial wave function $\Psi_T$ marked as Nr.1, Nr.2 and Nr.3, the extrapolated estimator $\rho(r)$ from DMC are different with each other and none of them is the true ground state expectation value\footnote{the DMC results are obtained by using the extrapolated estimator Eq.(\ref{Apest}).}.
However, no matter which of the 3 trial wave functions $\Psi_T$ are used, the $\rho(r)$ projected out by PIMC are the same within error bar, which means they are the true ground state $\rho(r)$. Obviously the PIMC results are more reliable than the DMC in this case.

It needs to be pointed out that, due to the fact that $T$ and $V$ do not commute with each other as well as $V$ is usually non-trivial, we use approximated short-time propagator $U(\Delta \tau) \approx e^{-H \Delta \tau}$ in the calculation instead of $e^{-H \Delta \tau}$. And $U(\Delta \tau)$ and $e^{-H \Delta \tau}$ are different by a time step $\Delta \tau$ error,
\begin{equation}
U(\Delta \tau) = e^{-H \Delta \tau} + \mathcal{O}(\Delta \tau^l),
\end{equation}
where different choices of $U(\Delta \tau)$ will give different integer $l$, and the error between the approximate short-time propagator $U(\Delta \tau)$ and exact short-time propagator $e^{-H \Delta \tau}$ is of $\Delta \tau^l$ order.

Therefore, what we really calculate in PIMC is $\langle O(\Delta \tau) \rangle$, which is an approximation of the true expectation value $\langle O \rangle$ in Eq.(\ref{PIMCopexpdetailshorttime}),
\begin{equation}
\langle \hat{O} (\Delta \tau) \rangle
= \frac{ Re  \langle \Psi_T| [U(\Delta \tau)]^{N_1} \hat{O} [U(\Delta \tau)]^{N_2}| \Psi_T \rangle  }
{ Re \langle \Psi_T| [U(\Delta \tau)]^{N_1+N_2} | \Psi_T \rangle} .
\label{PIMCopexpdetailshorttimeapprox}
\end{equation}
In the limit of the time step $\Delta \tau$ becomes zero, $\langle O(\Delta \tau) \rangle$ will be the same as $\langle O \rangle$,
\begin{equation}
\lim_{\Delta \tau \rightarrow 0}\langle \hat{O} (\Delta \tau) \rangle
= \langle O \rangle .
\end{equation}
So we need to calculate $\langle O(\Delta \tau) \rangle$ for different time step $\Delta \tau$, then extrapolate the $\langle O(\Delta \tau) \rangle$ at $\Delta \tau=0$ in order to find the true expectation value $\langle O \rangle$.
If the operator $\hat{O}$ in Eq.(\ref{PIMCopexpdetailshorttimeapprox}) is placed exactly at the middle of the path such that $N_1=N_2$, the PIMC will essentially be equivalent with VMC whose trial function is $[U(\Delta \tau)]^{\frac{N}{2}}| \Psi_T \rangle$, in this case $\langle O(\Delta \tau) \rangle$ will be the upper bound of the true expectation $\langle O \rangle$.

\section{Hamiltonian}
\label{secHamilton}

In this work, the Hamiltonian $H$ of our light nuclei calculations is non-relativistic. It can be written as,
\begin{equation}
H = \sum_{i=1}^A \frac{\bm{p}_i^2}{2m}  + \sum_{i<j}V_{ij}, \label{PIMCHamilton}
\end{equation}
where $A$ is total number of nucleons in the system, $\bm{p}_i=-i\hbar \nabla_i$ is the momentum of particle $i$ in real space.
The complete two body interaction for a given $ij$ pair of particles, $V_{ij}$, is composed of the nucleon-nucleon interaction $V^{NN}_{ij}$ and the electric magnetic force $V^{EM}_{ij}$,
\begin{equation}
V_{ij}=V_{ij}^{NN}+V^{EM}_{ij}.\label{Vij}
\end{equation}

By using both the local chiral interaction with N$^2$LO (\cite{Gerzerlis14a,Lynn2017PRC,Lonardoni2018PRC}) and the Argonne $v'_6$ (AV6') interaction  (\cite{WiringaPieper02a}), we use PIMC to calculate the ground state properties of light nuclei such as $^2$H, $^3$H, $^3$He and $^4$He.
The nucleon-nucleon interaction is written as,
\begin{equation}
V^{NN}_{ij}=\sum_{p=1}^7 v_p(r_{ij}) O_{ij}^p , \label{AV6'}
\end{equation}
where $r_{ij}$ is the length of $\bm{r}_{ij}$, $v_p(r_{ij})$ is the radial function for the $p^{\text{th}}$ operator, and they are different for chiral and AV6' interactions. The first 6 of the operators $O_{ij}^p$ are 1, $\bm{\tau}_i \cdot \bm{\tau}_j$, $\bm{\sigma}_i \cdot \bm{\sigma}_j$, $\bm{\sigma}_i \cdot \bm{\sigma}_j \bm{\tau}_i \cdot \bm{\tau}_j$, $S_{ij}$ and $S_{ij} \bm{\tau}_i \cdot \bm{\tau}_j$, where $\bm{\sigma}$ and $\bm{\tau}$ are spin and isospin operators.
The tensor force $S_{ij}$ which comes from one-pion-exchange (OPE) is ,
\begin{equation}
S_{ij}=3 \bm{\sigma}_i \cdot \hat{r} \bm{\sigma}_j \cdot \hat{r} - \bm{\sigma}_i \cdot \bm{\sigma}_j , \label{AV6'O}
\end{equation}
where $\hat{r}$ is the unit vector of $\bm{r}_{ij}$. Note that the spin term $\bm{\sigma}_i \cdot \bm{\sigma}_j = 2P_{ij}^{\sigma} -1 $ and isospin term $\bm{\tau}_i \cdot \bm{\tau}_j = 2P_{ij}^{\tau} -1 $ where $P_{ij}^{\sigma}$ and $P_{ij}^{\tau}$ are spin and isospin exchange operators. The tensor force flips the spins of nucleons.

The AV6' is a local interaction, which is a six operator truncation of the full Argonne AV18 NN interaction (\cite{Wiringa95PRC}) which consists of 18 operators.
The prime symbol means that the parameters in AV6' are refitted and therefore they are different from those in AV18.
The AV6' only contain the first 6 operators, it does not have the \nth{7} operator $O_{ij}^7$, so its $v_7(r_{ij})=0$.

When we say a potential is local, the word `local' means the operators can be determined once the positions of particles are given. If an interaction contains mostly local operators, it is a local interaction. If an interaction contains some non-local operators, usually like momentum operator $\bm{p}$ and/or powers of $\bm{p}$, it is `non-local'. This is because momentum operator in real space is a differential operator, if it operates on a function $f(x)$ then we need to calculate the derivative of the function $f'(x)$. The value of $f(x)$ is not enough to determine $f'(x)$, we need to know the value of $f(x+dx)$ at $x+dx$ so that we can calculate $f'(x)$, so the information at the local point $x$ is not enough, information at $x+dx$ is needed, so it is called `non-local'.
If an interaction contains $\bm{p}^n$, then in order to compute $f^{(n)}(x)$ at $x$, we need to know all the non-local information of $f(x+dx)$, $f(x+2dx)$, ..., $f(x+ndx)$, so as the integer $n$ increase, the interaction will become more and more `non-local'.

Chiral interactions from ChEFT are obtained based on a momentum expansion so they are naturally non-local, so computing them is usually not easy for real space QMC methods.
However, it was shown that up to N$^2$LO, we are able to obtain a mostly local version of chiral interaction (\cite{Gerzerlis13a}) which is suitable for QMC calculations (\cite{Lynn14a,Gerzerlis14a,Lynn2017PRC,Lonardoni2018PRL,Lonardoni2018PRC}).
For chiral local N$^2$LO interaction, it has the same 6 local operators as AV6', just with different radial functions.
However it has the \nth{7} operator in the $V^{NN}_{ij}$ in Eq.(\ref{AV6'}), and it is the spin-orbit term (\cite{Lonardoni2018PRC}) which is
\begin{equation}
O_{ij}^7 = \bm{L} \cdot \bm{S},
\end{equation}
where $\bm{L}$ is the relative angular momentum operator and $\bm{S}$ is the total spin operator,
\begin{align}
\bm{L} & = \frac{1}{2\hbar}(\bm{r}_i-\bm{r}_j) \times (\bm{p}_i-\bm{p}_j) , \\
\bm{S} & = \frac{1}{2} (\bm{\sigma}_i+\bm{\sigma}_j) .
\end{align}
With the radial function $v_7(r_{ij})$, the spin-orbit interaction $v_{LS}(r_{ij})$ is,
\begin{equation}
v_{LS}(r_{ij})=v_7(r_{ij}) \bm{L} \cdot \bm{S},
\label{vlsrij}
\end{equation}

The complete chiral N$^2$LO interaction also has a three body interaction (\cite{Lonardoni2018PRC}), $V_{3b}=\sum_{i<j<k}V_{ijk}$,
which to the linear order approximation in the propagator can be straightforwardly added in light nuclei PIMC calculations. Since our PIMC calculation is mainly a benchmark test, we did not include three body interaction in this work.

For the electromagnetic (EM) force $V^{EM}_{ij}$, here we only consider the Coulomb force $v^{C}(r_{ij})$ between proton and proton (pp) which contains the proton charge form factor,
\begin{equation}
V^{EM}_{ij}=v^{C}(r_{ij}) P^p_i P^p_j . \label{VEMpp}
\end{equation}
where $P^p_i= (1+\tau_{iz}  )/2$ is the proton projection operator and $\tau_{iz} $ is the isospin z operator of $i^{\text{th}}$ particle.
Coulomb interaction is long range and the NN interaction is short range (typically less than 2 fm).

The total potential energy operator $V$ is,
\begin{equation}
V = \sum_{i<j} V_{ij} = \sum_{i<j} (V_{ij}^{NN}+V^{EM}_{ij}) . \label{PE}
\end{equation}

\begin{figure}[tb!]
\centering
\includegraphics[scale=0.5]{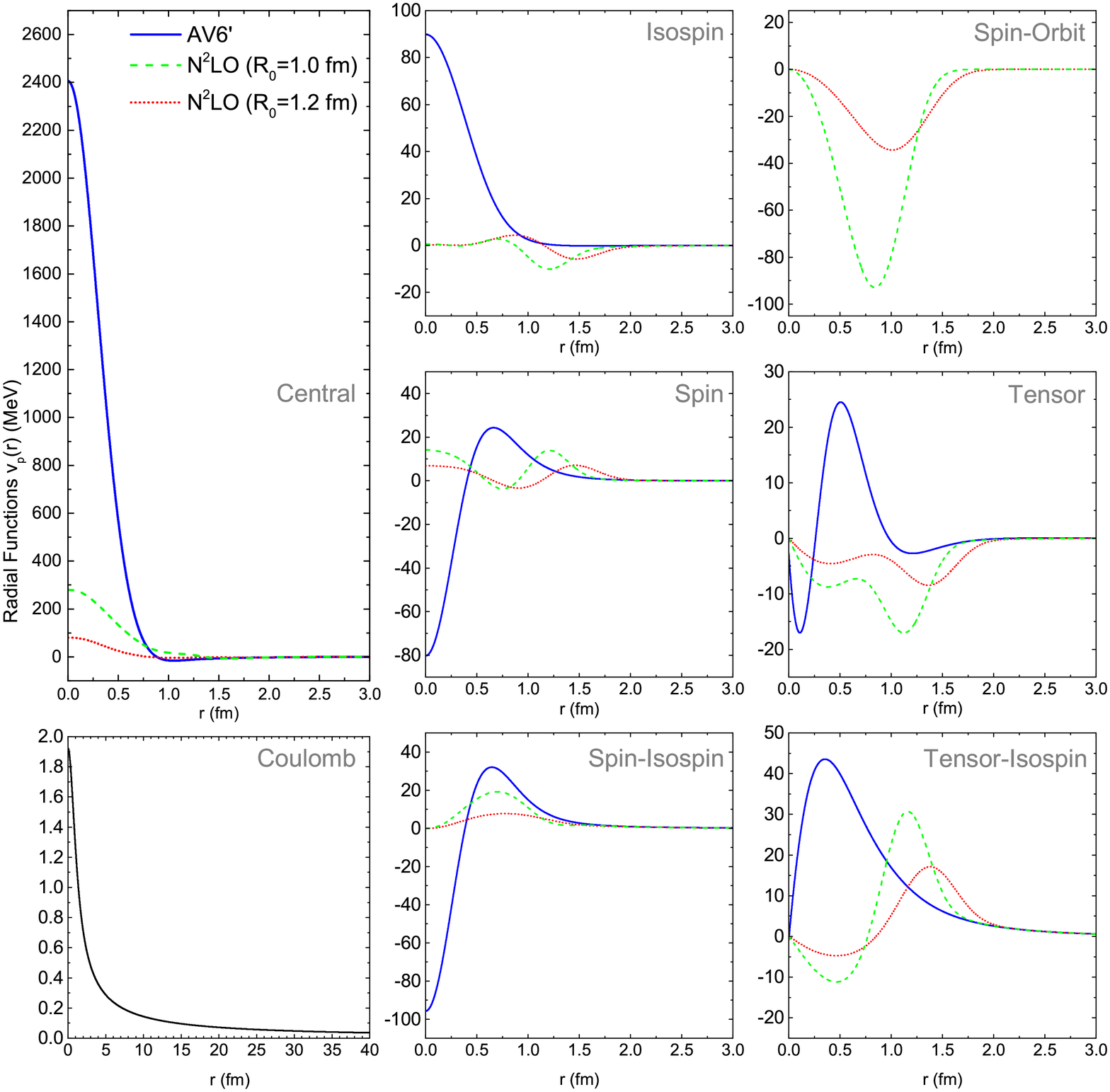}
\caption[Radial Functions of the AV6' and the Chiral N$^2$LO Interaction]{Radial functions of the AV6' and the chiral N$^2$LO interaction. }
\label{AV6N2LOradial}
\end{figure}

Fig. \ref{AV6N2LOradial} shows the radial function of each operator in AV6' and chiral N$^2$LO interaction. For the chiral N$^2$LO interactions, their spectral-function-regularization (SFR) cutoff $\tilde{\Lambda}$ is chosen to be 1000 MeV, and
we list the radial functions for coordinate-space cutoff $R_0=1.0$ fm and $R_0=1.2$ fm cases (\cite{Gerzerlis14a}).
We find that the chiral N$^2$LO interactions are much softer than AV6' interaction. Usually a soft potential is computationally more easy to simulate than hard ones. Because for soft potentials the chance that we meet with very big or very small values of potentials are much lower than hard ones, and this may lead to less variance (smaller error bar) in the results.

\section{Model Space}
\label{secModSpace}

The interaction used in this work will not change the total number of $Z$ protons and $A-Z$ neutrons\footnote{The AV6' and local chiral N$^2$LO we use in this work conserve the number of isospin states. For example, for $^4$He we begin with $\binom {4}{2}=6$ different isospin states, the interaction we use always generate isospin states within these 6 states, they do not generate isospin states other than these 6 states.}, so the number of possible isospin states is $\binom {A}{Z}$. However the tensor part $S_{ij}$ can flip spin, so the possible number of spin states is $2^A$. Therefore, as stated before,  the total number of spin-isospin basis states is,
\begin{equation}
N_{\rm{tot}}= \binom {A}{Z} \times 2^A = \frac{A!}{Z!(Z-A)!} \times 2^A.
\label{Nbasistotal}
\end{equation}

Take $^4$He as an example, it has $N_{\rm{tot}}=6 \times 16=96$ spin-isospin basis states.
In other words, our Hilbert space is constructed by the 96 basis states and therefore a general state is the linear combination of these basis states. We mark these basis states as $| S \rangle$ and $S$ can be labeled from 1 to 96,
\begin{equation}
| S \rangle \equiv  | s_1\rangle \otimes | s_2 \rangle \otimes | s_3 \rangle \otimes | s_4 \rangle \equiv | s_1 s_2 s_3 s_4 \rangle, \label{basisS}
\end{equation}
where $s_i$ means the spin isospin state of particle i, it can be any state from neutron spin up $|n \uparrow \rangle$, neutron spin down $|n \downarrow\rangle$, proton spin up $|p \uparrow\rangle$, proton spin down $|p \downarrow\rangle$.
In Eq.(\ref{basisS}), we implicitly mean particle 1 is at $s_1$ state, particle 2 is at $s_2$ state, particle 3 is at $s_3$ state and particle 4 is at $s_4$ state. To be clear, it is really,
\begin{equation}
| S \rangle \equiv  | s^1_1\rangle \otimes | s^2_2 \rangle \otimes | s^3_3 \rangle \otimes | s^4_4 \rangle
\equiv | s^1_1 s^2_2 s^3_3 s^4_4 \rangle, \label{basisSmoreclear}
\end{equation}
where the subscript (the state label) 1 to 4 represents state 1 to state 4, and the superscript (particle label) from 1 to 4 represents particle 1 to particle 4.

We denote the spatial configuration of a system with $A$ particles as $R$,
\begin{equation}
R \equiv ( \bm{r}_1, \bm{r}_2, ..., \bm{r}_A  ) , \label{Rdef}
\end{equation}
where $\bm{r}_i=(x_i,y_i,z_i)$ is the coordinates of particle i.
The coordinate space configuration $R$ of $^4$He can also be written as a state $| R \rangle$,
\begin{equation}
| R \rangle \equiv  | \bm{r}_1\rangle \otimes | \bm{r}_2 \rangle \otimes | \bm{r}_3 \rangle \otimes | \bm{r}_4 \rangle \equiv | \bm{r}_1 \bm{r}_2 \bm{r}_3 \bm{r}_4 \rangle. \label{basisR}
\end{equation}
Similar with Eq.(\ref{basisSmoreclear}), Eq.(\ref{basisR}) really means,
\begin{equation}
| R \rangle \equiv  | \bm{r}^1_1\rangle \otimes | \bm{r}^2_2 \rangle \otimes | \bm{r}^3_3 \rangle \otimes | \bm{r}^4_4 \rangle
\equiv | \bm{r}^1_1 \bm{r}^2_2 \bm{r}^3_3 \bm{r}^4_4 \rangle. \label{basisRmoreclear}
\end{equation}

Taking $R$ and the spin isospin state $S$ in to account, our basis state $| RS \rangle$ can finally be written as,
\begin{equation}
| RS \rangle = | R \rangle \otimes | S \rangle
\equiv | \bm{r}_1 \bm{r}_2 \bm{r}_3 \bm{r}_4 \rangle \otimes | s_1 s_2 s_3 s_4 \rangle , \label{basisRS}
\end{equation}

The correspondingly  identity operators for our basis are,
\begin{flalign}
 \int dR | R \rangle  \langle R | &=  1 , \label{identityR} \\
 \sum_S | S \rangle  \langle S | &=  1 , \label{identityS} \\
 \sum_S  \int dR | RS \rangle  \langle RS | & =
\int dR | R \rangle  \langle R |  \otimes  \sum_S  | S \rangle  \langle S | = 1 .  \label{identityRS}
\end{flalign}

In fact,
without the Coulomb interaction, for $^4$He it is possible to use two $|T=0\rangle$ states (T is the total isospin quantum number) to couple with 16 spin states to form 32 basis. And due to the fact that the Hamiltonian has time-reversal symmetry, only 16 of the 32 basis states are independent, therefore we can use 16 states to form the Hilbert space instead of 96.
However the Coulomb interaction does not commute with the total isospin operator, it does not conserve total isospin and can generate $T\neq 0$ states, therefore the 16 spin/isospin states with $T=0$ cannot be used when the Coulomb interaction was present.
And this is why we use the 96 states in this work.

\section{Wave Function}
From Eq.(\ref{PIMCopexp}) we know that an initial trial wave function $\Psi_T$ is needed. Like all QMC calculations, the closer the trial wave function $\Psi_T$ is to the ground state, the faster and the lower variance the calculation will become. In fact, obtaining a high quality trial wave function is always an important task for QMC.

The state of the trial wave function $| \Psi_T \rangle$ in this work takes the following form (\cite{CarlsonRMP15a}),
\begin{equation}
|\Psi_T \rangle = \mathcal{F} |\Phi \rangle, \label{PsiTstateF}
\end{equation}
where $| \Phi \rangle$ is the model state, and $\mathcal{F}$ is the correlation operator.

Since nucleons are fermions, the model state $|\Phi \rangle$ has to be antisymmetric under the exchange of the labels between any of the two particles. For $A\leq 4$ it can be decomposed as a spatial part and a spin isospin part,
\begin{equation}
|\Phi \rangle =  |\Phi_\textrm{R} \rangle \otimes | \Phi_\textrm{S} \rangle.\label{Phi}
\end{equation}
The spatial part $|\Phi_\textrm{R} \rangle$ is symmetrized which can be chosen as,
\begin{equation}
|\Phi_\textrm{R} \rangle
= \int dR  | R \rangle, \label{PhiR}
\end{equation}
such that
\begin{equation}
\langle R'|\Phi_\textrm{R} \rangle=\int dR \langle R'  | R \rangle = \int dR \delta(R'-R)=1. \label{normPhiR}
\end{equation}

Since the spatial part is symmetric, the spin-isospin part $|\Phi_\textrm{S} \rangle$ has to be antisymmetric under the exchanges of labels between any of the two particles.
For light nuclei with $A\leqq 4$, we can pick a basis state
$|S' \rangle \equiv  | {s'_1}^1\rangle \otimes | {s'_2}^2 \rangle \otimes .. \otimes | {s'_A}^A \rangle
\equiv | {s'_1}^1 {s'_2}^2 ...{s'_A}^A \rangle$ from any one of the $N_S=\binom {A}{Z}\times 2^A$ spin-isospin basis $| S \rangle$, then construct $|\Phi_\textrm{S} \rangle$ using $A$ by $A$ Slater determinant of single particle states in such a form,
\begin{eqnarray}
|\Phi_\textrm{S} \rangle \!=\!
\mathcal{A}   | {s'_1}^1 {s'_2}^2 .. {s'_A}^A \rangle
= \sum_{N=1}^{N_S} \phi_{N} | N \rangle,
\label{PhiS}
\end{eqnarray}
where the number $\phi_{N}$ is either -1, 1 or 0 depends on whether a certain spin-isospin state $| N \rangle$ from the $N_S$ states $| S \rangle$ has odd permutation (exchanging the particle labels between two particles), even permutation or any overlap with the chosen $| S' \rangle$ under any permutation or not.

The projection of $|\Phi_\textrm{S} \rangle $ on state $| S \rangle \equiv | s^1_1 s^2_2 s^3_3 s^4_4 \rangle$ is the amplitude $\langle S | \Phi_\textrm{S} \rangle$ which is calculated by,
\begin{align}
\langle S |\Phi_\textrm{S} \rangle =
\left |
\begin{array}{ccccc}
\langle s^1_1 | {s'}^1_1  \rangle & \langle s^2_2 | {s'}^2_1 \rangle & ... &  \langle s^{A-1}_{A-1} | {s'}^{A-1}_1 \rangle & \langle s^A_A| {s'}^A_1 \rangle  \\
\langle s^1_1 | {s'}_2^1  \rangle & \langle s^2_2 | {s'}^2_2 \rangle & ... & \langle s^{A-1}_{A-1} | s^{A-1}_2 \rangle & \langle s^A_A | s^A_2 \rangle  \\
\vdots & \vdots &  \ddots & \vdots & \vdots  \\
\langle s^1_1 | s^1_{A-1}  \rangle & \langle s^2_2 | {s'}^2_{A-1} \rangle & ... & \langle s^{A-1}_{A-1} | s^{A-1}_{A-1} \rangle & \langle s^A_A | s^A_{A-1} \rangle  \\
\langle s^1_1 | s^1_A  \rangle &\langle s^2_2 | {s'}^2_{A} \rangle & ... & \langle s^{A-1}_{A-1} | s^{A-1}_A \rangle & \langle s^A_A | s^A_A \rangle
\end{array}
\right |
= \phi_S,
\label{SPhiS}
\end{align}
where the overlap between the spin-isospin state $| {s'}^N_j  \rangle$ of particle $N$ and  $| {s'}^M_j  \rangle$ of particle $M$ is,
\begin{equation}
\langle s^M_i | {s'}^N_j  \rangle = \langle s_i | {s'}_j  \rangle \delta_{MN} = \delta_{s_i,{s'}_j} \delta_{MN} .
\end{equation}

Note that $\langle S | \Phi_\textrm{S} \rangle$ is a fermion wave function and it is antisymmetric under the particle label exchange between two particles. If $| S''  \rangle$ is obtained from $| S \rangle$ by exchanging the $i$ and $j$ particle labels, then $\langle S'' | \Phi_\textrm{S} \rangle$ is the determinant in Eq.(\ref{SPhiS}) with $i$ and $j$ columns exchanged, which obviously becomes $-\langle S | \Phi_\textrm{S} \rangle$.

By using Eqs.(\ref{normPhiR}--\ref{SPhiS}), the wave function $\langle RS | \Phi\rangle$ is calculated as,
\begin{equation}
\langle RS | \Phi\rangle
= \langle R | \Phi_\textrm{R} \rangle
\langle S | \Phi_\textrm{S} \rangle
=\sum_{N=1}^{N_S} \phi_N \langle S | N \rangle
=\phi_S. \label{RSPhi}
\end{equation}

Again, take $^4$He as an example, we can pick any one of the states from the 96 basis, and then antisymmetrize it to form the spin isospin part $| \Phi_\textrm{S} \rangle$ which will then be a linear combination of $4!=24$ non-zero basis states. We can pick a basis state $ |S' \rangle= | {n \uparrow}^1 {n \downarrow}^2 {p \uparrow}^3 {p \downarrow}^4  \rangle$ from the basis, then antisymmetrize it to form $| \Phi_\textrm{S} \rangle$ of $^4$He,
\begin{equation}
|\Phi_\textrm{S} \rangle
= \sum_{N=1}^{96}  | N \rangle \langle N | \Phi_\textrm{S} \rangle
= \sum_{N=1}^{96} \phi_N | N \rangle  , \label{PhiSHe4}
\end{equation}

where $\phi_N=\langle N | \Phi_N\rangle$ is either $-1$ or $1$ depending on the antisymmetrization for each of the 24  basis states, and 0 for the rest 72 basis states. For example, using Eq.(\ref{SPhiS}) one can find  if $ | N\rangle =| {n \uparrow}^1 {n \downarrow}^2 {p \uparrow}^3 {p \downarrow}^4  \rangle$ then $\phi_N$ is 1, if $ | N \rangle =| {n \uparrow}^2 {n \downarrow}^1 {p \uparrow}^3 {p \downarrow}^4  \rangle$ then  $\phi_N$ is -1, if $ | N\rangle =| {n \uparrow}^2 {n \downarrow}^1 {p \uparrow}^3 {p \uparrow}^4  \rangle$ then  $\phi_N$ is 0, etc.

The correlation operator $\mathcal{F}$ is represented by the products of two-body correlation operators $F_{ij}$,
\begin{equation}
{\mathcal F} =  {\mathcal S} \prod_{i<j} F_{ij} \, , \label{coropF}
\end{equation}
where $\mathcal{S}$ is the symmetrization operator, and $F_{ij}$ is,
\begin{equation}
F_{ij} =  \sum_{p=1}^6 f_{ij}^p O_{ij}^p \, , \label{coropFij}
\end{equation}
where $O_{ij}^p$ are the same AV6' operators, and $f_{ij}^p $ is the corresponding correlation function operator (\cite{Lomnitz1981NPA}). After operating on $| R \rangle$, $f_{ij}^p $ will become a function $f_{ij}^p (r_{ij}) $ which only depends on the distance $r_{ij}$ between particle $i$ and $j$.

By using Eqs.(\ref{basisRS}--\ref{coropFij}), the trial wave function $ \langle R S | \Psi_T  \rangle $ is calculated as,
\begin{equation}
\langle R S | \Psi_T  \rangle \! = \!
\langle R  | \otimes  \langle S | \mathcal{F} |\Phi_{\textrm{R}} \rangle \otimes | \Phi_{\textrm{S}} \rangle
\!=\!
\left\langle S \left|
{\mathcal S}  \prod_{i<j}
\left[
\sum_{p=1}^6   f_{ij}^p (r_{ij})  O_{ij}^p
\right]
\sum_{N=1}^{N_S} \phi_N \right| N \right\rangle
 . \label{RSPsiT}
\end{equation}

The trial wave function $\langle R S | \Psi_T  \rangle$ with correlation operators Eq.(\ref{coropFij}) enriches the structure of the wave function, especially, it has the important tensor structure (\cite{CarlsonRMP15a}), and thus is an improvement over those without $O_{ij}^p$ operators.

However, such a trial wave function is usually expensive to calculate.
Even if $O^p$ are just 1, it will take at least $96 \times 6= 576$ operations for a given pair. And we have 4 particles so 6 pairs, so $576\times 6=3456$ operations. Then there is a symmetrizer $\mathcal{S}$ which contribute another factor of $6!=720$. So totally at least $3456*720=2488320$ operations for computing just one $\langle R S | \Psi_T  \rangle$. In general, for a nuclei, without further simplifying $\langle R S | \Psi_T  \rangle$, brute force calculation for AV6' can take at least $ \binom{A}{Z} \times 2^{A} \times 6 \times [A(A-1)/2] \times [A(A-1)/2]!$ operations. This is computationally too expensive.
Because of that, in this work, as well as in most other QMC methods like AFDMC (\cite{Schmidt99a}),  it is sampled instead of summing over all the $[A(A-1)/2]!$ symmetrized terms.

In our calculation, we use `$l$' to denote a sampled order $\langle \Psi^l_T |$ from $\langle \Psi_T |$, and use `$r$' to denote a sampled order $| \Psi^r_T \rangle$ from $| \Psi_T \rangle$. So, it is,
\begin{align}
\langle \Psi_T | & = \sum  _l\langle \Psi^l_T |,  \\
| \Psi_T \rangle &= \sum_r | \Psi^r_T \rangle.
\end{align}
We can write $\langle \Psi_T | RS\rangle$ and $\langle RS | \Psi_T \rangle$ as,
\begin{align}
\langle \Psi_T | RS \rangle &= \sum  _l\langle \Psi^l_T |  RS \rangle , \label{lorderdef}\\
\langle RS | \Psi_T \rangle &= \sum_r \langle RS | \Psi^r_T \rangle, \label{rorderdef}
\end{align}
and $\langle R S | \Psi_T^r  \rangle$ and $\langle \Psi_T^l | R S \rangle$  are,
\begin{align}
\langle R S | \Psi_T^r  \rangle &= \!
\left\langle S \left|
\sideset{}{^r}
 \prod_{i<j}
\left[
\sum_{p=1}^6   f_{ij}^p (r_{ij})  O_{ij}^p
\right]
\sum_{N=1}^{96} \phi_N
\right| N \right\rangle
, \label{RSPsiTr} \\
\! \! \! \! \! \!  \langle \Psi_T^l | R S \rangle &= \!
 \sum_{N=1}^{96} \left\langle N \left|\phi^*_N
\sideset{}{^l} \prod_{i<j}
\left[
\sum_{p=1}^6  {f_{ij}^p}^* (r_{ij})   O_{ij}^p
\right]
\right| S \right\rangle ,\label{RSPsiTl}
\end{align}
where we use symbols $\sideset{}{^r} \prod_{i<j}$ and $\sideset{}{^l} \prod_{i<j}$ to denote a particular chosen `$r$' and `$l$' order from $ {\mathcal S} \prod_{i<j}$.
This $l$ and $r$ order sampling totally reduces the cost by the factor of $\left\{[A(A-1)/2]!\right\}^2$. For $^4$He that is $(6!)^2=518400$.
Since the operator commutators are relatively small, so all the $[A(A-1)/2]!$ terms from either $\langle \Psi^l_T |$ or $| \Psi^r_T \rangle$ are almost the same, therefore the variance increase from sampling this operator order is also small.

It needs to be pointed that, $|\Phi_\textrm{S} \rangle $ (and therefore $|\Phi \rangle $) does not have to be just one Slater determinant, it can be (sometime it better be) a linear combination of several different Slater determinants formed from different antisymmetrized basis states. Overall, that is to say, using a linear combination of Slater determinant to form $| \Phi \rangle$ may be required to obtain the correct quantum numbers for a nucleus. To say the least, the deuteron ground state must have $J=S=1$ (\cite{SiemensJensenbook}).

For the $A> 4$ system, merely the four spin-isospin states $|n\uparrow \rangle$, $|n\downarrow \rangle$, $|p\uparrow \rangle$ and $|p\downarrow \rangle$ are not enough to form the antisymmetric state $| \Psi \rangle$ in Eq.(\ref{PsiTstateF}).
If we were to use single-particle states to form $| \Psi \rangle$ by Slater determinants, they have to contain some orbital quantum number information. For example we can use single particle orbital (\cite{zengjinyanQMbook,Xuzizong2009book}) inspired from shell model (\cite{mayer1964shell}).

\begin{figure}[hptb!]
\centering
\includegraphics[scale=0.5]{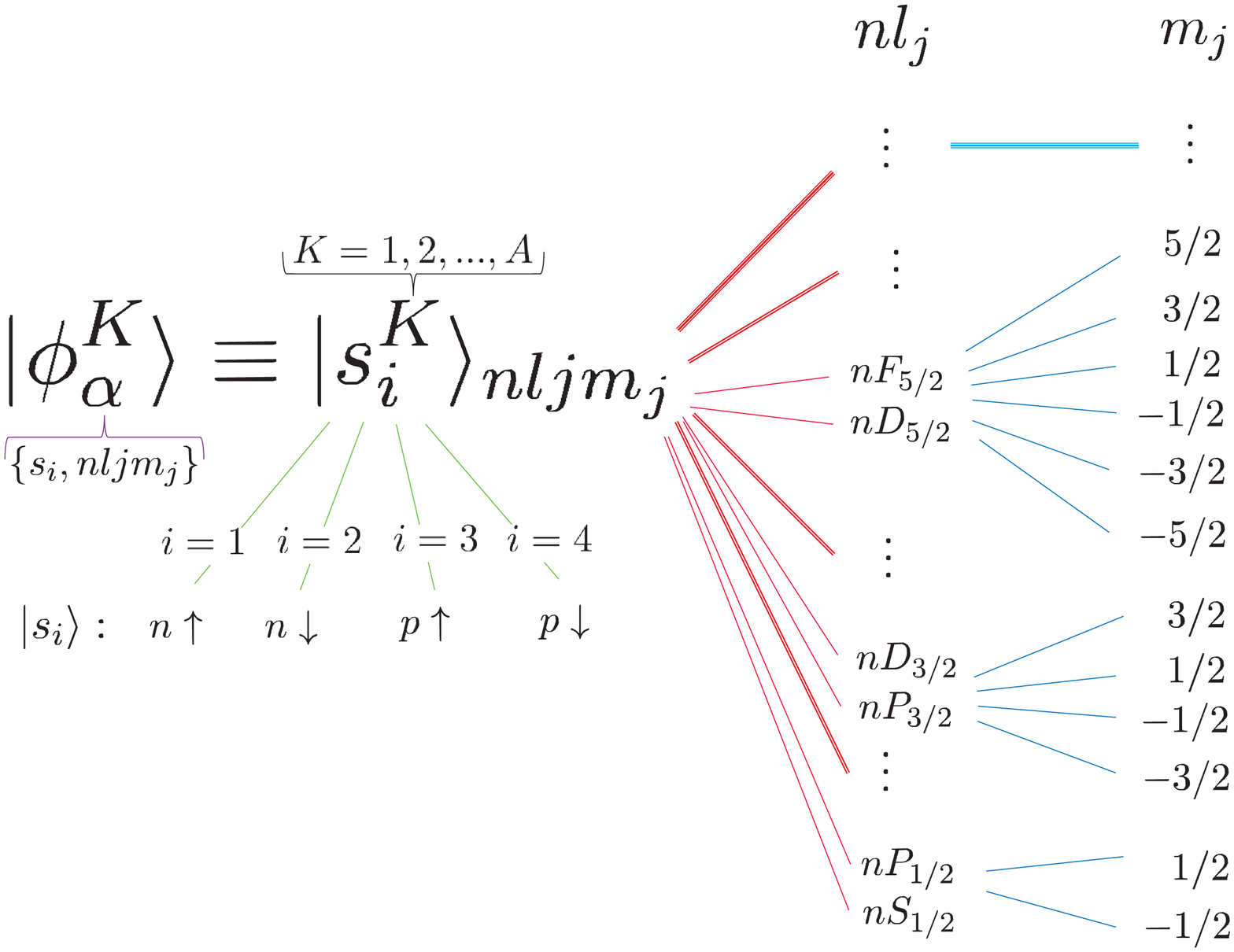}
\caption[Single-particle Orbital States $| s^K_i \rangle_{nljm_j}$]{Single-particle orbital states $| s^K_i \rangle_{nljm_j}$.  }
\label{singleobitalstates}
\end{figure}

We can mark such a shell-model-like single-particle state as $| s^K_i \rangle_{nljm_j}$ as illustrated in Fig. \ref{singleobitalstates}. The symbol $K$ means it is the state of particle $K$. Letter $i$ can be from 0 to 4 and they denote spin-isospin state $n\uparrow$, $n\downarrow$, $p\uparrow$, $n\downarrow$ accordingly. Quantum number $n$ is the radial quantum number, $l$ is the orbit angular momentum quantum number, $j$ is the total angular momentum quantum number and $m_j$ is the $j$ projection quantum number.
For simplicity, we can use $|\phi^K_\alpha\rangle$ to denote $| s^K_i \rangle_{nljm_j}$,
\begin{equation}
  |\phi^K_\alpha\rangle \equiv | s^K_i \rangle_{nljm_j} ,
\end{equation}
where symbol $\alpha$ contains all the spin-isospin information $s_i$ and orbital information $nljm_j$,
\begin{equation}
  \alpha=\{s_i,nljm_j\}.
\end{equation}
and the amplitude between $ |\phi^K_\alpha\rangle$ and particle $M$'s basis $| \bm{r}^M_j s_j^M \rangle$ is  $\langle \bm{r}^M_j s_j^M|\phi^K_\alpha\rangle $ which can be written as,
\begin{equation}
\langle \bm{r}^M_j s_j^M|\phi^K_\alpha\rangle= \phi_\alpha(\bm{r}_j,s_j) \delta_{s_j,s_i} \delta_{MK}.
\label{rsphialpha}
\end{equation}
The function $\phi_\alpha(\bm{r}_j,s_j)$ is usually called an orbital and it can be written as,
\begin{equation}
\phi_\alpha(\bm{r}_j,s_j) = R_{nl}(r_j) \mathscr{Y} _l^{jm_j},
\end{equation}
where $R_{nl}(r)$ is a radial function which can be found from the solution of some central potential such as Woods-Saxon potential (\cite{Lonardoni2018PRC}),
and $\mathscr{Y} _l^{jm_j}$ is a simultaneous eigenfunction of $\bm{l}^2$, $\bm{s}^2$ and $\bm{j}^2$ called spin-angular function (\cite{sakurai1993}).
$\mathscr{Y} _l^{jm}$ is a spin spherical harmonic $Y_l^m$ multiplied by some coefficients which are functions of $l$, $j$ and $m_j$. ,

Assuming there are $n_\phi$ ways to pick $A$ suitable single-particle states $|\phi^K_\alpha\rangle$ which can produce the correct total angular momentum $J$, parity $\pi$ and total isospin quantum number $T$ of a give A-nucleon nucleus, and the state $|\Phi\rangle$ can be written as a linear combination of $n_\phi$ Slater determinants $| \mathcal{D}_{\Phi}^q \rangle$,
\begin{equation}
|\Phi\rangle = \sum_{q=1}^{n_\phi} c_q | \mathcal{D}_{\Phi}^q \rangle,
\label{PhiSlaterDsum}
\end{equation}
where $c_q $ is some coefficient and the amplitude $\langle RS | \mathcal{D}_{\Phi}^q \rangle  $ can be calculated by using Eq.(\ref{rsphialpha}) as,
\begin{align}
\langle RS | \mathcal{D}_{\Phi}^q \rangle
= \left |
\begin{array}{ccc}
\langle \bm{r}^1_1 s_1^1 | {\phi}^1_1  \rangle  & ...  & \langle \bm{r}^{A}_{A} s_{A}^{A} | {\phi}^A_1 \rangle  \\
\vdots  &  \ddots  & \vdots  \\
\langle \bm{r}^1_1 s_1^1 | {\phi}^1_A  \rangle &  ...  & \langle \bm{r}^{A}_{A} s_{A}^{A} | {\phi}^A_A \rangle
\end{array}
\right |
=
\left |
\begin{array}{ccc}
\phi_1(\bm{r}_1 s_1)  & ...  & \phi_1(\bm{r}_A s_A)   \\
\vdots  &  \ddots  & \vdots  \\
\phi_A(\bm{r}_1 s_1)   \rangle &  ...  & \phi_A(\bm{r}_A s_A)
\end{array}
\right |.
\end{align}

\begin{figure}[hptb!]
\centering
\includegraphics[scale=0.5]{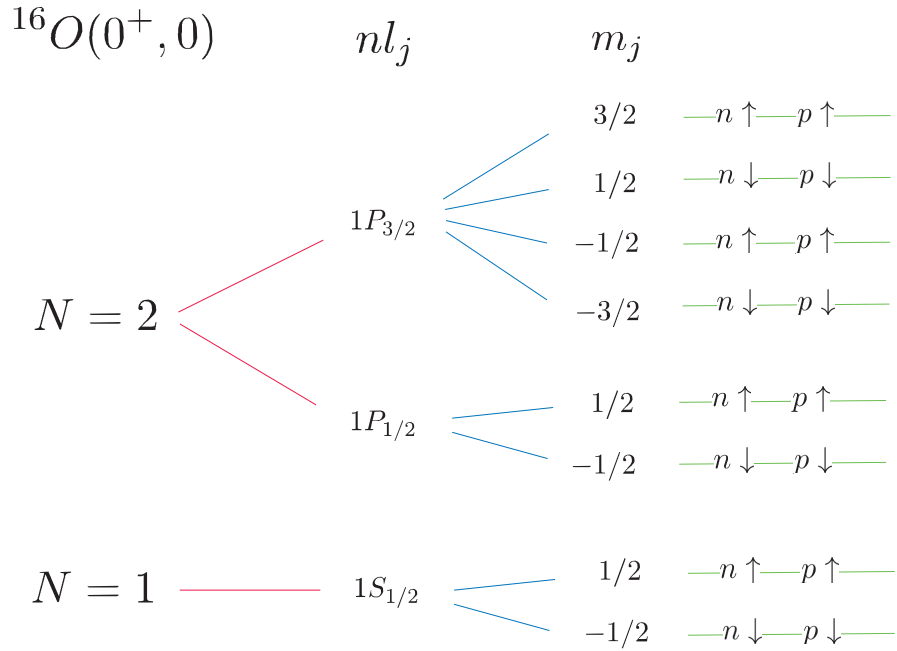}
\caption[Single-particle Orbital States of $^{16}$O$(0^+,0)$]{Single-particle orbital states of $^{16}$O$(0^+,0)$ }
\label{O16shell}
\end{figure}

For example, consider $^{16}$O$(0^+,0)$ whose total angular momentum quantum number and parity is $J^\pi=0^+$ and isospin quantum number is $T=0$. From the knowledge of shell model, as illustrated in Fig. \ref{O16shell}, we know that $^{16}$O$(0^+,0)$ is a closed shell nucleus.
All the 16 particles fully occupy the $N=1$ and $N=2$ shells. Among them, 4 occupy the $1S_{1/2}$ states, 4 occupy the $1P_{1/2}$ states and 8 occupy the $1P_{3/2}$ states. So totally 16 particles occupy 16 different orbital. So we can form one 16 by 16 Slater determinant to describe the wave function of $^{16}$O.

However, for open-shell nuclei, there are more single-particle orbital states than the number of nucleons $A$. Therefore in order to form the $|\Phi \rangle$ in Eq.(\ref{PhiSlaterDsum}),  all the possible orbital states may need to be picked to form the accordingly total number of $n_\phi$ Slater determinants. E.g., $^6$He$(0^+,1)$ needs 9 Slater determinants, $^6$Li$(1^+,0)$ needs 32 and $^12$C$(0^+,0)$ needs 119 (\cite{Lonardoni2018PRC}).
Also, considering each Slater determinants needs to be operated by the correlation operator $\mathcal{F}$ in Eq.(\ref{coropF}) in order to form $|\Psi_T \rangle$ in Eq.(\ref{PsiTstateF}), the computational cost of calculating open-shell nuclei is much higher than closed-shell nuclei.

Last but not least, it needs to be pointed out that, the wave function does not have to be formed by the Slater determinants using single-particle orbital. We can also use paired states to form the antisymmetrized fermion wave functions in the Pfaffian  form (\cite{Bouchaud1988a,Bajdich2006PRL}). In fact, making the wave function relatively cheap to calculate while making it containing as much of ground state information as possible is one of the most important tasks in QMC calculations.

\section{Propagator}
\label{propagator}
Efficiently sampling the imaginary time propagator $e^{-H \Delta \tau}$ by using a short-time approximated propagator $U(\Delta \tau)$ is important for PIMC.
Due to the fact that the local chiral N$^2$LO interaction contains the spin-orbit term $v_{LS}(r_{ij})$ in Eq.(\ref{vlsrij}), the approximated  short-time propagator $U(\Delta \tau)$ for $e^{-H \Delta \tau}$ is calculated as,
\begin{align}
U(\Delta \tau) & \equiv
e^{-\frac{\mathcal{V}^\dagger}{2} \Delta \tau }
\mathcal{G}
e^{-T \Delta \tau} e^{-\frac{\mathcal{V}}{2} \Delta \tau }
= e^{-H \Delta \tau} + \mathcal{O}(\Delta \tau^2),
\label{propagatorN2LOVTV}
\end{align}
where $\mathcal{V}$ is the NN potential only with the first 6 operators in Eq.(\ref{AV6'}) plus the EM force in Eq.(\ref{VEMpp}),
\begin{equation}
\mathcal{V}  =  \sum_{i<j} \left[\sum_{p=1}^6 v_p(r_{ij}) O_{ij}^p + V^{EM}_{ij}(r_{ij}) \right]
\equiv \sum_{i<j} \mathcal{V}_{ij} .
\label{PER}
\end{equation}

For $e^{-\frac{\mathcal{V}}{2}\Delta \tau }$, we pick an order of each $ij$ pair and calculate it using the following approximation,
\begin{equation}
e^{-\frac{\mathcal{V}}{2} \Delta \tau }
= e^{\frac{-\mathcal{V}_{A-1 A} \Delta t}{2}}
e^{\frac{-\mathcal{V}_{A-2 A} \Delta t}{2}}
...
e^{\frac{-\mathcal{V}_{13}  \Delta t}{2}}
e^{\frac{-\mathcal{V}_{12} \Delta t}{2}}
+ \mathcal{O}(\Delta \tau)^2 ,
\label{propagatorv6half}
\end{equation}
where the $\mathcal{O}(\Delta \tau)^2$ terms are due to $\mathcal{V}_{ij}$ do not commute with each other for different $ij$ pairs, and $e^{-\frac{\mathcal{V}^\dagger}{2} \Delta \tau}$ is the Hermitian conjugate of the $e^{-\frac{\mathcal{V}}{2}\Delta \tau }$ approximation which is the reverse order of Eq.(\ref{propagatorv6half}).
Note that the first 6 operators in $V^{NN}_{ij}$ in Eq.(\ref{AV6'}), $V^{NN}_{ij}$ commutes with the $V_{ij}^{EM}$, so we have,
\begin{equation}
e^{ - \frac{\mathcal{V}_{ij }}{2}\Delta \tau}
=  e^{ - \frac{\sum_{p=1}^6 v_p(r_{ij}) O_{ij}^p }{2}\Delta \tau}
 e^{ - \frac{V^{EM}_{ij} }{2}\Delta \tau} .
 \label{propeV6eVem}
\end{equation}

The first 6 operators in  $V^{NN}_{ij}$ are the same AV6' operators and they form a group.
$e^{ - \sum_{p=1}^6 v_p(r_{ij}) O_{ij}^p  \Delta \tau}$ in Eq.(\ref{propeV6eVem}) can be linearized by the original 6 operators $O^p_{ij}$ in the following way,
\begin{equation}
e^{ - \sum_{p=1}^6 v_p(r_{ij}) O_{ij}^p \Delta \tau }  =
 \sum_{p=1}^6 u^p_{ij} (r_{ij}) O^p_{ij} ,
\label{eVlinear}
\end{equation}
where the six coefficients $u^p_{ij} (r_{ij}) $ can be solved once $r_{ij}$ is given\footnote{See Appendix \ref{secAV6'linear} for details about linearizing the AV6' operators in the propagator. In order to save time, in the code all the $u^p(r)$ are tabulated for the relevant range of $r$ before performing the Monte Carlo simulations. And the interpolations of the table are done by Lagrange 4-point interpolation polynomials.}.

The symbol $\mathcal{G}$ is the approximated short-time propagator for the spin-orbit term $e^{-V_{LS} \Delta \tau }$ to the linear order\footnote{
The three body interaction $V_{3b}=\sum_{i<j<k}V_{ijk}$ may be added in the same way as the spin-orbit term in Eq.(\ref{propagatorN2LOVTV}) to the linear order, such that
\begin{equation}
\mathcal{G}\equiv 1 -\sum_{i<j} v_{LS}(\hat{r}_{ij}) \Delta \tau -\sum_{i<j<k}V_{ijk} \Delta \tau .
\end{equation}
For light nuclei this linear approximation may be sufficient because the three-body force and the spin-orbit force usually have opposite sign and similar strength, so $\mathcal{G}$ is close to 1 and the fluctuation in the propagator will not be too big.
Better approximations for the 3-body propagator are straightforward since they do not contain derivatives.
},
\begin{equation}
\mathcal{G} \equiv 1 -\sum_{i<j} v_{LS}(\hat{r}_{ij}) \Delta \tau
= e^{-V_{LS} \Delta \tau } + \mathcal{O}(\Delta \tau^2).
\label{defdownarrow}
\end{equation}

The approximated short-time propagator $U(\Delta \tau)$ in our basis is calculated as,
\begin{flalign}
& \langle R' S' |  U(\Delta \tau)   | R S \rangle \nonumber \\
&= \left\langle R'  \left| \otimes \left\langle S' \left|
e^{-\frac{\mathcal{V}^\dagger}{2} \Delta \tau }
\mathcal{G}
e^{-T \Delta \tau} e^{-\frac{\mathcal{V}}{2} \Delta \tau }
\right| R \right\rangle \otimes \right| S \right\rangle \nonumber \\
&= \left\langle S' \left| e^{-\frac{\mathcal{V}^\dagger(R')}{2} \Delta \tau }
\left\{ \left[1 -\sum_{i<j} v_7(r'_{ij}) \bm{L}_{ij} \cdot \bm{S}_{ij} \Delta \tau\right]
 \left\langle R' \left|  e^{-T \Delta \tau}  \right| R \right\rangle \right\}
e^{-\frac{\mathcal{V}(R)}{2} \Delta \tau } \right| S \right\rangle \nonumber \\
&=
\left\langle S' \left| e^{-\frac{\mathcal{V}^\dagger(R')}{2} \Delta \tau }
\bm{\mathcal{G}}
e^{-\frac{\mathcal{V} (R)}{2} \Delta \tau } \right| S \right\rangle
\times G^f_{R'R}.
\label{propagatorN2LORS}
\end{flalign}

The symbol $G^f_{R'R}$ in Eq.(\ref{propagatorN2LORS}) means the products of the free particle propagator of each particle in the system,
\begin{equation}
G^f_{R'R}  \equiv
\langle R' | e^{-T \Delta \tau}  | R \rangle
= \prod_{i=1}^{A}  \langle \bm{r}'_i | e^{-\frac{p_i^2 \Delta \tau}{2m}    } | \bm{r}_i \rangle
\equiv  \left( \frac{m}{2\pi \Delta \tau \hbar^2} \right)^{\frac{3A}{2}}  e^{  -\frac{ (R'-R)^2 }{2 \Delta \tau \frac{\hbar^2}{m}}  } ,
\label{propagatorFree}
\end{equation}
and $(R'-R)^2$ is calculated as,
\begin{eqnarray}
\!\!\!\!\! (R'\!-\!R)^2 \!  = \!
\sum_{i=1}^A  (\bm{r}_i'-\bm{r}_i)^2
=\sum_{i=1}^A \left[ (x'_i\!-\!x_i)^2 \!+\! (y'_i\!-\!y_i)^2 \!+\! (z'_i\!-\!z_i)^2 \right] . \label{Rdiffdef}
\end{eqnarray}

The symbol $\bm{\mathcal{G}}$ in Eq.(\ref{propagatorN2LORS}) is the effective $\mathcal{G}$ which is written as,
\begin{equation}
\bm{\mathcal{G}} =
1 - \left(\frac{-m}{4i  \hbar^2 } \right)\sum_{i<j} v_7(r'_{ij}) \left[ \bm{r}'_{ij} \times \frac{\Delta \bm{r}^{R',R}_{ij}}{\Delta \tau}
\cdot (\bm{\sigma}_i+\bm{\sigma}_j) \right] \Delta \tau  ,
\label{defdownarroweff}
\end{equation}
where $r'_{ij}$ is the distance between particle $i$ and $j$ in configuration $R'$,
$\bm{r}'_{ij}\equiv \bm{r}_i-\bm{r}_j$ in $R'$,
and
$\Delta \bm{r}^{R',R}_{ij} \equiv \Delta \bm{r}^{R',R}_i-\Delta \bm{r}^{R',R}_j $
where $\Delta \bm{r}^{R',R}_i \equiv \bm{r}^{R'}_i - \bm{r}^R_i $
and
$\Delta \bm{r}^{R',R}_j \equiv \bm{r}^{R'}_j - \bm{r}^R_j $,
and the symbol $\bm{r}^{R'(R)}_{i(j)}$ simply means the $\bm{r}_{i(j)}$ in configuration $R'(R)$.
Note that Eq.(\ref{defdownarroweff}) does not explicitly contain $\Delta \tau$ anymore.
However, as will be shown in chapter \ref{compalgorithm}, $\Delta \bm{r}^{R',R}_{ij}$ will essentially have the order of $\Delta \tau$ because $\bm{r}^{R'}_j$ will be effectively sampled around $\bm{r}^{R}_j$ from a gaussian function whose variance is the order of $\Delta \tau$.

\section{Time Step Error Structure}
\label{secErrEst}
For the AV6' interaction, there are only the first 6 operators in Eq.(\ref{AV6'}), so $\mathcal{G}$ is 1, and Eq.(\ref{propagatorN2LOVTV}) becomes the high-accuracy Trotter formula\footnote{All the error structures of the propagator can be found by repeatedly using Campbell-Baker-Hausdorff (CBH) formula (\cite{Reinsch2000JMP}),
\begin{eqnarray}
e^A e^B = e^{A+B+\frac{1}{2}[A,B]+\frac{1}{12}[A-B,[A,B]]+\frac{1}{24}[B,[A,[B,A]]] + ...},
\label{CBHformula}
 \end{eqnarray}
where A and B are operators or matrices. From Eq.(\ref{CBHformula}) we can further find that,
\begin{equation}
e^{\frac{B}{2} }  e^{A}  e^{\frac{B}{2}}
= e^{A+B
+ \frac{1}{24}[2A+B, [A,B]] + ... }.
\end{equation}
and that is how we can write down (\cite{VesaPDF}) Eq.(\ref{HighTrotter}).} (\cite{Feynman1948RMP,SchmidtLee95a}) which has $\mathcal{O}(\Delta \tau)^3$ accuracy and can be denoted as $U_S(\Delta \tau)$,
\begin{equation}
U_S(\Delta \tau) \equiv
e^{-\frac{\mathcal{V}^\dagger}{2} \Delta \tau }
e^{-T \Delta \tau} e^{-\frac{\mathcal{V}}{2} \Delta \tau }
= e^{- H \Delta \tau
- \frac{ {\Delta \tau}^3}{24}[2T+\mathcal{V}, [T,\mathcal{V}]] + ... }
= e^{- H \Delta \tau} + \mathcal{O}(\Delta \tau)^3,
\label{HighTrotter}
\end{equation}
where the subscript letter $S$ means symmetric because $U_S(\Delta \tau)=U^\dagger_S(\Delta \tau)$.
Note that although the approximation of $e^{-\frac{\mathcal{V}}{2} \Delta \tau }$ which is Eq.(\ref{propagatorv6half}) alone has $\mathcal{O}(\Delta \tau^2)$ error, when written the form of Eq.(\ref{HighTrotter}), the $\mathcal{O}(\Delta \tau^2)$ cancels and one is left with only $\mathcal{O}(\Delta \tau^3)$ error.
Furthermore, we find
\begin{equation}
U_S(\Delta \tau) U_S(-\Delta \tau)=1,
\label{USuni}
\end{equation}
this ensures the error in the exponential of $U_S(\Delta \tau)$ only contains odd orders of $\Delta \tau$,
\begin{equation}
U_S(\Delta \tau) = e^{-H \Delta \tau
+ \hat{f}_3 \Delta \tau^3
+ \hat{f}_5 \Delta \tau^5
+ \hat{f}_7 \Delta \tau^7 + ... } ,
\label{propagatorerrstruc}
\end{equation}
where $\hat{f}_i$ are functions of operators which do not depend on $\Delta \tau$.

For the AV6' interaction, in the PIMC calculation, we use $N$ short-time propagator $[U_S(\Delta \tau)]^N$ instead of $(e^{-H \Delta \tau})^N$ in calculating Eq.(\ref{PIMCopexpdetailshorttime}),
and the total imaginary time $\tau=N\Delta \tau$, so $N\mathcal{O}(\Delta \tau^n)=\tau \mathcal{O}(\Delta \tau^{n-1})$.
Considering Eq.(\ref{propagatorerrstruc}), this means that for a given $\tau$, the error of $[U_S(\Delta \tau)]^N$ only contain the even orders of $\Delta \tau$,
\begin{equation}
\left[U_S(\Delta \tau)\right]^N =
\left(e^{ -H \Delta \tau
+ \hat{f}_3 \Delta \tau^3
+ \hat{f}_5 \Delta \tau^5 + ... }\right)^N
=e^{-H \tau} + \mathcal{O}(\Delta \tau^2) + \mathcal{O}(\Delta \tau^4) + ... ,
\label{propagatorNerrstruc}
\end{equation}
and this means
$\langle \hat{O} (\Delta \tau) \rangle$ in Eq.(\ref{PIMCopexpdetailshorttimeapprox}) and the true expectation value $\langle \hat{O} \rangle$ of Eq.(\ref{PIMCopexpdetailshorttime}) are related by the following equation,
\begin{align}
\langle \hat{O} (\Delta \tau) \rangle
& = \frac{ Re  \langle \Psi_T| [U_S(\Delta \tau)]^{N_1} \hat{O} [U_S(\Delta \tau)]^{N_2}| \Psi_T \rangle  }
{ Re \langle \Psi_T| [U_S(\Delta \tau)]^{N} | \Psi_T \rangle} \nonumber \\
& = \langle \hat{O} \rangle \times
\frac{1 + \mathcal{O}(\Delta \tau^2) + \mathcal{O}(\Delta \tau^4) + ... }
{1+ \mathcal{O}(\Delta \tau^2) + \mathcal{O}(\Delta \tau^4) + ... } .
\label{PIMCUSdetail}
\end{align}
For small $\Delta\tau$, in order to extrapolate the expectation value $\langle \hat{O} \rangle$, we need to use,
\begin{equation}
\langle \hat{O} (\Delta \tau) \rangle
= \langle \hat{O} \rangle
+ \sum_{n=1}^{n_c} C_{2n} (\Delta \tau^{2n}),
\label{Oextrapolation}
\end{equation}
where $C_{2n}$ are the fitting parameters for even orders of $\Delta \tau$, and $n_c$ is the cutoff integer for $n$. The theoretical value of $C_{2n}$ depends on the trial wave function and the total imaginary time $\tau$.
In principle, a general extrapolation formula which suits a wide range of $\Delta \tau$ should have the same form as Eq.(\ref{PIMCUSdetail}).

For the local chiral N$^2$LO interaction, we use the $U(\Delta \tau)$ in Eq.(\ref{propagatorN2LOVTV}) as the short-time propagator and it has errors $\mathcal{O}(\Delta \tau^2)$, so when the total imaginary time $\tau=N\Delta \tau$, we multiply $U(\Delta \tau)$ the short-time propagator $N$ times we have error of $N\mathcal{O}(\Delta \tau^2)=\tau \mathcal{O}(\Delta \tau)$ which is $\mathcal{O}(\Delta \tau)$,
\begin{equation}
\left[U(\Delta \tau)\right]^N = e^{-H \tau} + \mathcal{O}(\Delta \tau),
\label{propagatorNerrstruc}
\end{equation}
this means that the error of $\langle \hat{O} (\Delta \tau) \rangle$ in Eq.(\ref{PIMCopexpdetailshorttimeapprox}) has $\mathcal{O}(\Delta \tau)$ error,
\begin{align}
\langle \hat{O} (\Delta \tau) \rangle
& = \frac{ Re  \langle \Psi_T| [U(\Delta \tau)]^{N_1} \hat{O} [U(\Delta \tau)]^{N_2}| \Psi_T \rangle  }
{ Re \langle \Psi_T| [U(\Delta \tau)]^N | \Psi_T \rangle} \nonumber \\
& = \langle \hat{O} \rangle \times
\frac{1 + \mathcal{O}(\Delta \tau) + \mathcal{O}(\Delta \tau^2) + \mathcal{O}(\Delta \tau^3) + ... }
{1+ \mathcal{O}(\Delta \tau) + \mathcal{O}(\Delta \tau^2) + \mathcal{O}(\Delta \tau^3) + ... } .
\label{PIMCUN2LOdetail}
\end{align}
Similar with the Taylor expansion of $\frac{1}{1+x}=1-x+...$ requires $|x|<1$ to converge,
when the error in the denominator in Eq.(\ref{PIMCUN2LOdetail}) is less than one,  Eq.(\ref{PIMCUN2LOdetail}) can be written as,
\begin{align}
\langle \hat{O} (\Delta \tau) \rangle
= \langle \hat{O} \rangle
+ \mathcal{O}(\Delta \tau) + \mathcal{O}(\Delta \tau^2) + ... .
\end{align}
So, usually, for a given total time $\tau$ and small $\Delta \tau$, in order to get the true expectation value $\langle \hat{O} \rangle$ of Eq.(\ref{PIMCopexpdetailshorttime}), we need to calculate $\langle \hat{O} (\Delta \tau) \rangle$ for difference time step $\Delta \tau$ and error bar, and then use the following polynomial equation to extrapolate $\langle \hat{O} \rangle$ and its error bar,
\begin{equation}
\langle \hat{O} (\Delta \tau) \rangle
= \langle \hat{O} \rangle
+ \sum_{n=1}^{n_c} C_{n} (\Delta \tau^{n}),
\label{ON2LOextrapolation}
\end{equation}
where $C_n$ are fitting parameters and the theoretical value of $C_{n}$ depends on the trial wave function and the total imaginary time $\tau$. In principle a general extrapolation expression should be Eq.(\ref{PIMCUN2LOdetail}).

From the error structure Eq.(\ref{PIMCUSdetail}) or Eq.(\ref{PIMCUN2LOdetail}) we can see that, even if $\Delta \tau$ itself is not very small, as long as the $\Delta \tau$ errors (especially the leading order errors) in the numerator and the denominator of Eq.(\ref{PIMCUSdetail}) or Eq.(\ref{PIMCUN2LOdetail}) are comparable, $\langle \hat{O} (\Delta \tau) \rangle$ can still be close to the true expectation value $\langle \hat{O} \rangle$. That is to say the results of PIMC is robust because they may not be very sensitive to the time step $\Delta \tau$.

\section{Computational Details}
\label{comdetails}

\subsection{Recast}
\label{Recast}

We need to recast Eq.(\ref{PIMCopexpdetailshorttime}) into a form that is suitable for PIMC calculations.
We set the total time $\tau$ as $\tau=\tau_1+\tau_2=N \Delta \tau$, and we insert $N$ short-time propagators $U(\Delta \tau)$ in Eq.(\ref{PIMCopexpdetailshorttime}) as well as inserting corresponding identity operators in Eqs.(\ref{identityRS}) at each of them, then Eq.(\ref{PIMCopexpdetailshorttime}) with short-time propagator becomes,
\begin{eqnarray}
\resizebox{.95\hsize}{!}{
$
\!\! \langle  \hat{O} (\Delta \tau) \rangle  \!=\!
\frac{ Re \! \sum\limits_{\mathcal{S}'} \! \int \! \mathcal{D} \mathcal{R}'   \langle \Psi_T| R_0 S_0\rangle
\langle R_0 S_0 |U(\Delta \tau) | R_1 S_1 \rangle...\langle R_I S_I |\hat{O} | R_{I\!+\!1} S_{I\!+\!1} \rangle ...
\langle R_{N\!+\!1} S_{N\!+\!1}  | \Psi_T \rangle  }
{ Re \! \sum\limits_{\mathcal{S}} \! \int \! \mathcal{D} \mathcal{R}  \langle \Psi_T| R_0 S_0\rangle
\langle R_0 S_0 | U(\Delta \tau) | R_1 S_1 \rangle ... \langle R_{N\!-\!1} S_{N\!-\!1} | U(\Delta \tau) | R_N S_N \rangle
\langle R_{N} S_{N}  | \Psi_T \rangle }, ~~
$}
\label{Oraw}
\end{eqnarray}
where $\sum\limits_{\mathcal{S}'}$ and $\int \mathcal{D} \mathcal{R}' $ in the numerator means $\sum\limits_{S_0,...,S_{N+1}}$ and $\int d R_0 ... dR_{N+1} $, and $\sum\limits_{\mathcal{S}}$ and $\int \mathcal{D} \mathcal{R}$ in the denominator means $\sum\limits_{S_0,...,S_{N}}$  and  $\int d R_0 ... dR_{N} $.

For any local operator $\hat{O}$ in Eq.(\ref{Oraw}), we can write down,
\begin{eqnarray}
\langle R_I S_I |\hat{O} | R_{I+1} S_{I+1} \rangle
= \langle S_I | O_I | S_{I+1} \rangle \langle R_I | R_{I+1} \rangle
= \langle S_I | O_I | S_{I+1} \rangle \delta (R_I-R_{I+1}) ,
\label{Orawreduce}
\end{eqnarray}
where $O_I$ means operator $\hat{O}$ with spatial configuration $R_I$. Note that Eq.(\ref{Orawreduce}) allows us to remove $R_{I+1}$ from the integral in the numerator of Eq.(\ref{Oraw}). So the $N+2$ configurations $R_0$ to $R_{N+1}$ in the numerator can finally be reduced to $N+1$ configurations from $R_0$ to $R_{N}$ after relabeling. So the actual number of beads in the numerator is the same as in the denominator.

For the propagator in Eq.(\ref{Oraw}) we insert $\sum_{S_I}  |S_I \rangle \langle S_{I} | =1$ in Eq.(\ref{identityS}) and we can write down,
\begin{flalign}
& \sum_{S_I}  \langle R_{I-1} S_{I-1} | U_S(\Delta \tau) | R_I S_I \rangle
\langle R_I S_I | U_S(\Delta \tau) | R_{I+1} S_{I+1} \rangle \nonumber \\
& = \sum_{S_I} \bigg\{ \langle S_{I-1} | \otimes \langle R_{I-1}   |  e^{-\frac{\mathcal{V}^\dagger}{2} \Delta \tau } \mathcal{G} e^{-T \Delta \tau} e^{-\frac{\mathcal{V}}{2} \Delta \tau }| R_I \rangle \otimes | S_I \rangle \nonumber \\
&  ~~~~~~~ \times
\langle S_I | \otimes \langle R_I |  e^{-\frac{\mathcal{V}^\dagger}{2} \Delta \tau } \mathcal{G}  e^{-T \Delta \tau} e^{-\frac{\mathcal{V}}{2} \Delta \tau }| R_{I+1} \rangle \otimes | S_{I+1} \rangle  \bigg\} \nonumber \\
&=
\langle S_{I-1} | e^{- \frac{\mathcal{V}^\dagger_{I-1} \Delta \tau }{2} } \bm{\mathcal{G}}  e^{- \frac{\mathcal{V}_{I} \Delta \tau }{2}}
 e^{- \frac{\mathcal{V}^\dagger_{I} \Delta \tau }{2} } \bm{\mathcal{G}} e^{- \frac{\mathcal{V}_{I+1} \Delta \tau }{2} }
| S_{I+1} \rangle \times G^f_{R_{I-1} R_I}  G^f_{R_{I} R_{I+1}} ,
\label{eHrawreduce}
\end{flalign}
where $\mathcal{V}_I$ means the potential operator $\mathcal{V}$ with spatial configuration $R_I$, and $e^{- \frac{\mathcal{V}_{I} \Delta \tau }{2}}$ is calculated by Eq.(\ref{propagatorv6half}).
The free particle propagator $G^f_{R_I R_J}=\langle R_{I} | e^{-T \Delta \tau} | R_J \rangle$ can be calculated by Eq.(\ref{propagatorFree}).
By repeatedly using Eq.(\ref{eHrawreduce}) in Eq.(\ref{Oraw}) we can remove any $S_I$ sandwiched between two short-time propagators $U(\Delta \tau)$. Finally we are left with 2 spin-isospin symbols in the trial wave functions to be summed over, we label them $S_0$ and $S_N$, where $S_0$ is for the trial wave function on the left $\langle \Psi_T | R_0 S_0 \rangle$, and $S_N$ is for the trial wave function on the right $\langle R_N S_N   | \Psi_T \rangle$.

Furthermore, consider the $l$ and $r$ order picked from the trial wave function $\Psi_T$ as shown in Eqs.(\ref{lorderdef},\ref{rorderdef}), the approximated expectation value $\langle \hat{O}(\Delta \tau) \rangle$ in Eq.(\ref{Oraw}) for PIMC calculations can finally be written as a general form,
\begin{flalign}
\langle \hat{O} (\Delta \tau)\rangle &=
\frac{ \overbrace{Re \! \sum\limits_{S_0,S_N} \! \int \! \mathcal{D} \mathcal{R}   \langle \Psi_T| R_0 S_0\rangle
\langle R_0 S_0 | [U(\Delta \tau)]^{N_1} \hat{O} [U(\Delta \tau)]^{N_2}| R_{N} S_{N} \rangle
\langle R_{N} S_{N}  | \Psi_T \rangle}^{=\sum_l\sum_r \int \mathcal{D} \mathcal{R} Re[g(\mathcal{R})_{lr}]}  }
{ \underbrace{Re \! \sum\limits_{S_0,S_N} \! \int \! \mathcal{D} \mathcal{R}  \langle \Psi_T| R_0 S_0\rangle
\langle R_0 S_0 | [U(\Delta \tau)]^{N} | R_N S_N \rangle
\langle R_{N} S_{N}  | \Psi_T \rangle}_{=\sum_l\sum_r \int \mathcal{D} \mathcal{R} Re[f(\mathcal{R})_{lr}]  } } \nonumber \\
&= \frac{\sum_l\sum_r \int \mathcal{D} \mathcal{R}
\left\{\frac{Re[ g(\mathcal{R})_{lr}]}{|Re[ f(\mathcal{R})_{lr}]|  } \right\}
\left|Re[ f(\mathcal{R})_{lr}] \right|  }
{ \sum_l\sum_r \int \mathcal{DR}
\left\{\frac{Re[ f(\mathcal{R})_{lr}]}{|Re[ f(\mathcal{R})_{lr}]|  } \right\}
\left|Re[ f(\mathcal{R})_{lr}] \right|  } \nonumber \\
&=
\frac{\sum_l\sum_r \int \mathcal{D} \mathcal{R} A_{lr} (\mathcal{R}) P_{lr} (\mathcal{R}) }
{ \sum_l\sum_r \int \mathcal{D} \mathcal{R} B_{lr} (\mathcal{R}) P_{lr} (\mathcal{R})  }  \nonumber \\
&=\frac{\langle A_{lr}(\mathcal{R}) \rangle }{\langle B_{lr}(\mathcal{R}) \rangle}
\bigg|_{ \{l,r,\mathcal{R}\} \in P_{lr}(\mathcal{R})  } ,
\label{OABP}
\end{flalign}
where we use symbol $\mathcal{R}$ to denote all the spatial configurations $\{R_0,R_1,...R_N \}$,
\begin{equation}
\mathcal{R} \equiv \{R_0,R_1,...R_N \} ,
\label{defmathcalR}
\end{equation}
and its integral is the spatial integral over all the configurations $R_I$,
\begin{equation}
\int \mathcal{D} \mathcal{R} \equiv  \prod_{I=0}^N \int d R_I . \label{defmathcalRint}
\end{equation}

In Eq.(\ref{OABP}), $A_{lr} (\mathcal{R})$ and $B_{lr} (\mathcal{R})$ are real functions,
\begin{align}
A_{lr} (\mathcal{R}) &= \frac{Re[ g(\mathcal{R})_{lr}]}{|Re[ f(\mathcal{R})_{lr}]|  } ,
\label{Alrform}
\\
B_{lr} (\mathcal{R}) &= \frac{Re[ f(\mathcal{R})_{lr}]}{|Re[ f(\mathcal{R})_{lr}]|  } ,
\label{Blrform}
\end{align}
and we can see that $B_{lr} (\mathcal{R})$ is either 1 or -1.

In Eq.(\ref{OABP}), $P_{lr} (\mathcal{R})$ is the probability distribution,
\begin{equation}
P_{lr} (\mathcal{R}) = \frac{|Re[ f(\mathcal{R})_{lr}]|}{ \mathcal{N}  },  \label{PlrR}
\end{equation}
which is normalized to one in the following way,
\begin{equation}
\sum_{l}\sum_{r} \int \mathcal{D} \mathcal{R} P_{lr}(\mathcal{R}) =1,
\end{equation}
and its normalization factor is
$\mathcal{N}=\sum_l \sum_r \int \mathcal{D} \mathcal{R} |Re[ f(\mathcal{R})_{lr}]|$.

\subsection{Path Diagrams}
\label{secPathDiagrams}

Function $f(\mathcal{R})_{lr}$ corresponds to the denominator of Eq.(\ref{Oraw}) and it does not depend on where the operator $\hat{O}$ is placed,
\begin{flalign}
f(\mathcal{R})_{lr} &=
{ \sum\limits_{S_0,S_N}  \langle \Psi^l_T| R_0 S_0\rangle
\langle R_0 S_0 | [U(\Delta \tau)]^{N} | R_N S_N \rangle
\langle R_{N} S_{N}  | \Psi^r_T \rangle } \nonumber \\
& =
f^V_{lr}(\mathcal{R})
g^F(\mathcal{R}) . \label{fRlr}
\end{flalign}

Function $g(\mathcal{R})_{lr}$ corresponds to the numerator of Eq.(\ref{Oraw}), and it depends on $\hat{O}_M$ defined in Eq.(\ref{Orawreduce}) which means the operator $\hat{O}$ is placed at bead $M$ (its spatial configuration is $R_M$),
\begin{flalign}
g(\mathcal{R})_{lr} &=
\sum\limits_{S_0,S_N} \langle \Psi^l_T| R_0 S_0\rangle
\langle R_0 S_0 | [U(\Delta \tau)]^{N_1} \hat{O}_M [U(\Delta \tau)]^{N_2}| R_{N} S_{N} \rangle
\langle R_{N} S_{N}  | \Psi^r_T \rangle
\nonumber \\
&=
g^V_{lr,M}(\mathcal{R})
 g^F(\mathcal{R}). \label{gRlr}
\end{flalign}

\begin{figure}[hptb!]
\centering
\includegraphics[scale=0.45]{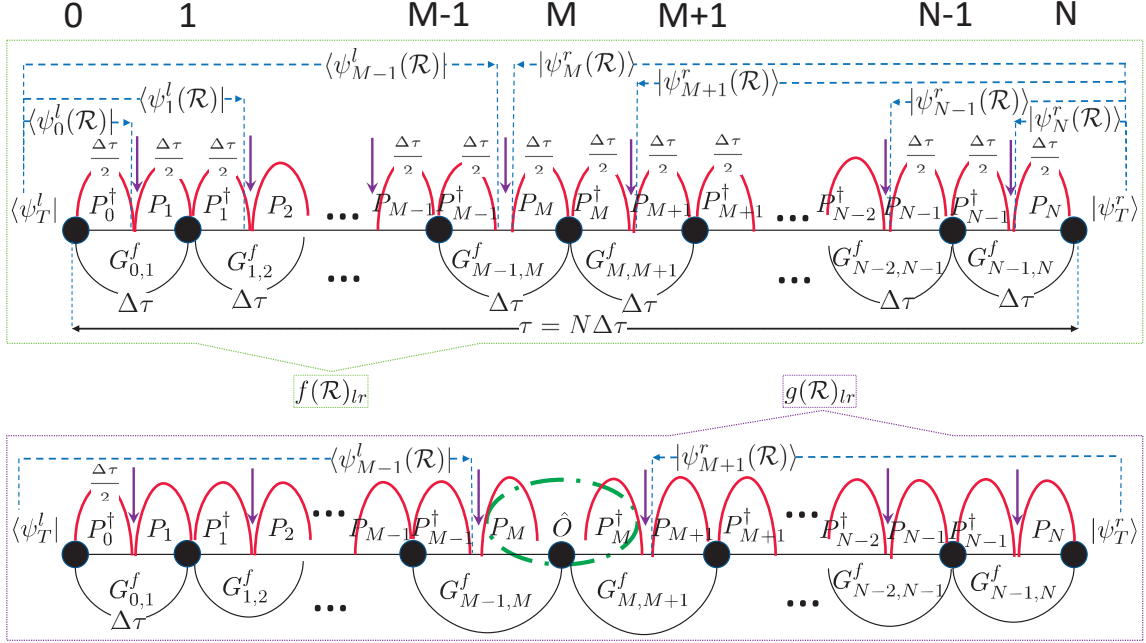}
\caption[Path Diagrams Which Illustrate the PIMC Beads, $ f(\mathcal{R})_{lr}$, and $ g(\mathcal{R})_{lr}$]{Path diagrams which illustrate the PIMC beads, $ f(\mathcal{R})_{lr}$, and $ g(\mathcal{R})_{lr}$. }
\label{PIMCbeads}
\end{figure}

Function $g^F(\mathcal{R})$ in Eqs.(\ref{fRlr},\ref{gRlr}) is the total free particle propagator,
\begin{equation}
g^F(\mathcal{R}) = \prod_{I=0}^{N-1} G^f_{I, I+1} ,
\label{gF}
\end{equation}
where the free particle propagator $G^f_{I, I+1}$ in Eq.(\ref{gF}) is defined as\footnote{
For any $R_I$ and $R_J$ on the time line which is represented by the long horizontal lines in Fig. \ref{PIMCbeads} on which there are black points (beads), the general form of the free particle propagator can be written as,
\begin{equation}
G^f_{I, J} \equiv \langle R_I | e^{-T \Delta \tau |J-I| }  | R_{J} \rangle
= \left( \frac{m}{2\pi |J-I| \Delta \tau \hbar^2} \right)^{\frac{3A}{2}}
e^{  -\frac{ (R_I-R_{J})^2 }{2 |J-I| \Delta \tau \frac{\hbar^2}{m}}  },
\label{Gfij}
\end{equation}},
\begin{equation}
G^f_{I, I+1} \equiv \langle R_I | e^{-T \Delta \tau}  | R_{I+1} \rangle
= \left( \frac{m}{2\pi \Delta \tau \hbar^2} \right)^{\frac{3A}{2}}   e^{  -\frac{ (R_I-R_{I+1})^2 }{2 \Delta \tau \frac{\hbar^2}{m}}  },
\label{Gfii1}
\end{equation}
and each of these free particle propagator is represented by the black curved line from below in Fig. \ref{PIMCbeads}, it carries $\Delta \tau$ and it connects configuration $R_I$ and $R_{I+1}$ (or we can say it connects beads $I$ and $I+1$).
Actually, We can see that the $\prod_{I=0}^{N-1}  G^f_{I, I+1}$ in functions $A_{lr} (\mathcal{R})$ and $B_{lr} (\mathcal{R})$ actually canceled out because they exist in both the denominators and the numerators.

Note that in $A_{lr} (\mathcal{R})$ and $B_{lr} (\mathcal{R})$ in Eqs.(\ref{Alrform},\ref{Blrform}), both $f(\mathcal{R})_{lr} $ and $g(\mathcal{R})_{lr} $ contain $g^F(\mathcal{R})$,
therefore $A_{lr} (\mathcal{R})$ and $B_{lr} (\mathcal{R})$ can be simplified as,
\begin{align}
A_{lr} (\mathcal{R}) &= \frac{Re[ g^V_{lr,M}(\mathcal{R})]}{|Re[ f^V_{lr}(\mathcal{R})]|  } ,
\label{Alrsimpform} \\
B_{lr} (\mathcal{R}) &= \frac{Re[ f^V_{lr}(\mathcal{R})]}{|Re[ f^V_{lr}(\mathcal{R})]|  } .
\label{Blrsimpform}
\end{align}
The superscript $V$ in $g^V_{lr,M}(\mathcal{R})$ and $f^V_{lr}(\mathcal{R})$ means the potential propagators are involved.

Fig. \ref{PIMCbeads} is a visualization of $ f(\mathcal{R})_{lr}$ in Eq.(\ref{fRlr}) and $g(\mathcal{R})_{lr}$ in Eq.(\ref{gRlr}).
The calculations of functions $g^V_{lr,M}(\mathcal{R})$ in Eq.(\ref{Alrsimpform}) and $f^V_{lr}(\mathcal{R})$ in Eq.(\ref{Blrsimpform}) can be more clear with the help of Fig. \ref{PIMCbeads} which contains diagrams that visualize the path.
By inserting $N+1$ complete basis marked by the black dots and labeling them from 0 to $N$, the total time $\tau$ is split into $N$ small time step $\Delta \tau$ so $\tau=N \Delta \tau$.
We call each black dot a bead.
For a given whole spatial configuration $\mathcal{R}$ defined by Eq.(\ref{defmathcalR}), each of the beads carries its own spatial configuration. For example, the $I^{\text{th}}$ bead has spatial configuration $R_I$.
The long black horizontal line represents the total imaginary time of the path $\tau$ and we can call it time line, and it has been split into $N$ pieces and occupied by $N+1$ beads.
Each of the black curved lines from the bottom represents the corresponding free particle propagator, e.g., if the black curved line connects beads $I$ and $I+1$, the black curved line just represents the free particle propagator $G^f_{I, I+1}$ in Eq.(\ref{gF}).
The red curved line from the upper, on the other hand, represents the accordingly potential part of the propagator,
e.g., we use $P_I$ to denote the red curved line as the corresponding $e^{\frac{- \mathcal{V}_I \Delta \tau }{2}}$ where $\mathcal{V}_I$ is defined in Eq.(\ref{PER}) and it has spatial configuration $R_I$ while $P^\dagger_I$ is the Hermitian conjugate of $P_I$ which gives the reverse order of Eq.(\ref{propagatorv6half}).
The symbol $\downarrow$ means the additional propagator inserted when needed. For the AV6' interaction, $\downarrow$ means 1, for the local chiral N$^2$LO interaction it means the effective $\bm{\mathcal{G}}$ which is Eq.(\ref{defdownarroweff}).

For a given $l$ and $r$ order in the trial wave function $\Psi_T$, and a given $\mathcal{R}$ of all the beads, the calculations of $f^V_{lr}(\mathcal{R})$ and $g^V_{lr,M}(\mathcal{R})$ are straightforward. Here we describe the computation rule which is basically a two-step process.
\begin{description}
  \item[Step one] First of all, we just need to follow the order of beads in Fig. \ref{PIMCbeads} and replace all the red curves and $\downarrow$ symbols above the black time line with corresponding $e^{\frac{- \mathcal{V}_I \Delta \tau }{2}}$ or $e^{\frac{- \mathcal{V}^\dagger_I \Delta \tau }{2}}$ and the effective $\bm{\mathcal{G}}$.
       We next replace all of the black curve under the black time line with corresponding $g^F_{IJ}$ which are gaussian functions.
  \item[Step two] Then, for $f^V_{lr}(\mathcal{R})$ in the upper part in Fig. \ref{PIMCbeads}, we just need to sandwich all the ordered operators above the black time line between $\langle \psi_T^l |$ and $| \psi_T^r \rangle$ to give a (complex) number. This number is the corresponding $f^V_{lr}(\mathcal{R})$. We multiply $f^V_{lr}(\mathcal{R})$ by $g^F(\mathcal{R})$ which is the product of all the black curved lines under the time line to find $f(\mathcal{R})_{lr}$.
\end{description}

The calculations of $g^V_{lr,M}(\mathcal{R})$ and $g(\mathcal{R})_{lr}$ are similar. The difference is that when calculating $g^V_{lr,M}(\mathcal{R})$, the additional operator $\hat{O}$ needs to be placed at the appropriate imaginary time as illustrated in the lower panel in Fig. \ref{PIMCbeads}.

\subsection{Calculating the Path}
\label{secCalcPath}

In order to write down the expressions for $f^V_{lr}(\mathcal{R})$ and $g^V_{lr,M}(\mathcal{R})$, as illustrated in Fig. \ref{PIMCbeads}, we define states $| \psi^r_M (\mathcal{R}) \rangle $ and $| \psi^l_M (\mathcal{R})\rangle$ along the path for computational convenience.
Following the computation rule described in chapter. \ref{secPathDiagrams},
the states $| \psi^r_M (\mathcal{R}) \rangle $ and $| \psi^l_M (\mathcal{R})\rangle$ are defined as
\begin{eqnarray}
&& | \psi^r_M (\mathcal{R}) \rangle \nonumber \\
&  \!\!\!\! \! \equiv & \! \!\!
\begin{cases}
\underbrace{  \sum\limits_{S}
   e^{\frac{- \mathcal{V}_M  \Delta \tau }{2}}  e^{\frac{- \mathcal{V}^\dagger_M  \Delta \tau }{2}} \bm{\mathcal{G}}
e^{\frac{- \mathcal{V}_{M\!+\!1}  \Delta \tau }{2}}  e^{\frac{- \mathcal{V}^\dagger_{M\!+1\!}  \Delta \tau }{2}} \bm{\mathcal{G}}
...
e^{\frac{- \mathcal{V}^\dagger_{N\!-\!1}  \Delta \tau }{2}} \bm{\mathcal{G}}
e^{\frac{- \mathcal{V}_N  \Delta \tau }{2}}
| S \rangle \langle R_N  S | \psi^r_T \rangle,
}_{ M \neq  N,  0.} \\
\underbrace{ \sum\limits_{S}
e^{\frac{- \mathcal{V}_N  \Delta \tau }{2}}
| S \rangle \langle R_N  S | \psi^r_T \rangle,
}_{M = N.} \\
\underbrace{  \sum\limits_{S} \! e^{\frac{- \mathcal{V}^\dagger_0  \Delta \tau }{2}} \bm{\mathcal{G}}
   e^{\frac{- \mathcal{V}_1  \Delta \tau }{2}}  e^{\frac{- \mathcal{V}^\dagger_1  \Delta \tau }{2}} \bm{\mathcal{G}}
e^{\frac{- \mathcal{V}_{2}  \Delta \tau }{2}}  e^{\frac{- \mathcal{V}^\dagger_{2}  \Delta \tau }{2}} \bm{\mathcal{G}}
 ...
e^{\frac{- \mathcal{V}^\dagger_{N\!-\!1}  \Delta \tau }{2}} \bm{\mathcal{G}}
e^{\frac{- \mathcal{V}_N  \Delta \tau }{2}}
| S \rangle \langle R_N  S | \psi^r_T \rangle.
}_{M =0.}
\end{cases} \label{psirmRdef}
\end{eqnarray}
\begin{eqnarray}
&&| \psi^l_M (\mathcal{R}) \rangle \nonumber \\
&  \!\!\!\! \! \equiv & \! \!\!
\begin{cases}
\underbrace{ \sum\limits_{S}
\left( \bm{\mathcal{G}}  e^{\frac{- \mathcal{V}_1  \Delta \tau }{2}}  e^{\frac{- \mathcal{V}^\dagger_1  \Delta \tau }{2}}
\bm{\mathcal{G}}
...
\bm{\mathcal{G}}
 e^{\frac{- \mathcal{V}_{M}  \Delta \tau }{2}}  e^{\frac{- \mathcal{V}^\dagger_{M}  \Delta \tau }{2}}
 \right) ^\dagger
e^{\frac{- \mathcal{V}_{0}  \Delta \tau }{2}}
| S \rangle \langle R_0 S | \psi^l_T \rangle,
}_{M\neq 0,N.} \\
\underbrace{ \sum\limits_{S}
e^{\frac{- \mathcal{V}_{0}  \Delta \tau }{2}}   | S \rangle \langle R_0  S | \psi^l_T \rangle,
}_{M=0.} \\
\underbrace{ \sum\limits_{S}\!   e^{\frac{- \mathcal{V}^\dagger_{N}  \Delta \tau }{2}}
\left( \bm{\mathcal{G}}  e^{\frac{- \mathcal{V}_1  \Delta \tau }{2}}  e^{\frac{- \mathcal{V}^\dagger_1  \Delta \tau }{2}}
\bm{\mathcal{G}}
...
\bm{\mathcal{G}}
 e^{\frac{- \mathcal{V}_{N-1}  \Delta \tau }{2}}  e^{\frac{- \mathcal{V}^\dagger_{N-1}  \Delta \tau }{2}} \bm{\mathcal{G}}
 \right) ^\dagger
e^{\frac{- \mathcal{V}_{0}  \Delta \tau }{2}}
| S \rangle \langle R_0 S | \psi^l_T \rangle.
}_{M=N.}
\end{cases} \label{psilmRdef}
\end{eqnarray}

Both $| \psi^l_M (\!\mathcal{R}\!)\rangle$ and $| \psi^r_M (\!\mathcal{R}\!)\rangle$ can be expanded by the $N_{\rm{tot}}$ spin isospin basis states $| S \rangle$ with accordingly probability amplitude $\phi^r_{M}(\mathcal{R},S)$ and $\phi^l_{M}(\mathcal{R},S)$,
\begin{align}
\phi^r_{M}(\mathcal{R},S) &= \langle S | \psi^r_M (\!\mathcal{R}\!) \rangle ,
\label{phirMR}\\
\phi^l_{M}(\mathcal{R},S)  &= \langle S | \psi^l_M (\!\mathcal{R}\!)\rangle .
\label{philMR}
\end{align}
So that,
\begin{flalign}
| \psi^r_M (\!\mathcal{R}\!) \rangle &= \sum_S | S \rangle \langle S | \psi^r_M (\!\mathcal{R}\!) \rangle  = \sum_S \phi^r_{M}(\mathcal{R},S) | S \rangle , \label{phirMexpansion} \\
| \psi^l_M (\!\mathcal{R}\!)\rangle &= \sum_S | S \rangle \langle S | \psi^l_M (\!\mathcal{R}\!)\rangle = \sum_S  \phi^l_{M}(\mathcal{R},S) | S \rangle.
\label{philMexpansion}
\end{flalign}

Function $g^V_{lr,M}(\mathcal{R})$ depends on which bead $M$ the operator $\hat{O}$ is placed at.
For a given $M$, it is calculated as
\begin{equation}
g^V_{lr,M}(\mathcal{R}) =
\begin{cases}
\langle \psi^l_{M\!-\!1}  (\mathcal{R}) | \bm{\mathcal{G}}
   e^{\frac{- \mathcal{V}_M  \Delta \tau }{2}} \hat{O}_M   e^{\frac{- \mathcal{V}^\dagger_M  \Delta \tau }{2}} \bm{\mathcal{G}}
| \psi^r_{M\!+\!1} (\mathcal{R}) \rangle,
 &  M \neq 0, N . \\
\sum\limits_{S} \langle \psi^l_T| \hat{O} | R_0 S\rangle \langle S | \psi^r_0 (\mathcal{R}) \rangle,
 &  M=0. \\
\sum\limits_{S} \langle \psi^l_N (\mathcal{R}) | S \rangle \langle R_N S| \hat{O} | \psi^r_T \rangle,
 & M=N .
\end{cases}
\label{gVRlrdef2}
\end{equation}
In particular, if the operator $\hat{O}$ commutes with Hamiltonian, then $\hat{O}$ can be moved to leftmost or the rightmost position, and the result Eq.(\ref{PIMCopexpdetail}) can then be written as Eq.(\ref{PIMCopexpdetailOH}), and the corresponding $g(\mathcal{R})_{lr}$ becomes,
\begin{equation}
  g(\mathcal{R})_{lr} =  \frac{1}{2}
\left[ g^V_{lr,M}(\mathcal{R}) |_{M=0} + g^V_{lr,M}(\mathcal{R}) |_{M=N}  \right]
g^F(\mathcal{R}) .
\label{gRlrOH}
\end{equation}
This is for example how we calculate the ground state energy $\hat{O}=H$.

For the AV6' interaction which does not have a spin-orbit term, the $\bm{\mathcal{G}}$ in Eq.(\ref{gVRlrdef2}) is just 1.
Note that the integer $M$ in $g^V_{lr,M}(\mathcal{R})$ is usually set at the middle between 0 and N, such that $e^{\frac{- \mathcal{V}_M  \Delta \tau }{2}} \hat{O}  e^{\frac{- \mathcal{V}^\dagger_M  \Delta \tau }{2}} $ is exactly placed in the middle of all the beads, so we make sure operator $\hat{O}$ is truly sandwiched between the same state which is the ground state if total imaginary time $\tau$ is big enough.

The function $f^V_{lr}(\mathcal{R})$ does not depend on $M$. Its value is the same as long as the left and right order $l$, $r$, and $\mathcal{R}$ are the same.
For any integer $M$ from 0 to $N-1$, it can be calculated as,
\begin{align}
f^V_{lr}(\mathcal{R})
= \langle \psi^l_M (\mathcal{R}) |
\bm{\mathcal{G}}
| \psi^r_{M+1} (\mathcal{R}) \rangle
= \sum_S \langle \psi^l_M (\mathcal{R}) | S \rangle
\langle S| \bm{\mathcal{G}} | \psi^r_{M+1} (\mathcal{R}) \rangle,
\label{fVRlrN2LOdef2}
\end{align}
where
\begin{flalign}
& \langle S| \bm{\mathcal{G}} | \psi^r_{M+1} (\mathcal{R}) \rangle \nonumber \\
&=
\left\langle S \left|
\left\{1 + \frac{m}{4i  \hbar^2} \sum_{i<j} v_7(r^M_{ij}) \left[ \bm{r}^M_{ij} \times \Delta \bm{r}^{M,M+1}_{ij} \cdot (\bm{\sigma}_i+\bm{\sigma}_j) \right] \right\}
\right|\psi^r_{M+1} (\mathcal{R})
\right\rangle ,
\label{propagatorN2LORSf}
\end{flalign}
and the definitions of $r^M_{ij}$, $\bm{r}^M_{ij}$ and $\Delta \bm{r}^{M,M+1}_{ij} $ are the same form as for $r'_{ij}$, $\bm{r}'_{ij}$ and $\Delta \bm{r}^{R',R}_{ij} $ used in Eq.(\ref{propagatorN2LORS}).
The $M$ and $M+1$ simply mean the configurations $R_M$ and $R_{M+1}$.
For the AV6' interaction we have $\mathcal{G}=1$ so $f^V_{lr}(\mathcal{R})$ is simplified to the following form,
\begin{align}
f^V_{lr}(\mathcal{R})
&= \langle \psi^l_M (\mathcal{R}) | \psi^r_{M+1} (\mathcal{R}) \rangle \nonumber \\
&= \sum_S \langle \psi^l_M (\mathcal{R}) | S \rangle \langle S |\psi^r_{M+1} (\mathcal{R}) \rangle \nonumber \\
&= \sum_{S} {\phi^{l}_{M}}^*(\mathcal{R},S) \phi^r_{M+1}(\mathcal{R},S).
 \label{fVRlrdef2}
\end{align}

In our PIMC code, the spin, isospin states are written in binary representation (\cite{Koonin91}) and we label the $A$ particles from 0 to $A-1$.
In the code, we use 0 for neutron and spin down, 1 for proton and spin up.
We use a 2 dimensional complex array $a(l_i,l_s)$ to represent a state.
For example, in Eq.(\ref{philMexpansion}),
a certain basis state is $\phi^r_{M}(\mathcal{R},S) | S \rangle$, if $|S\rangle=| n \uparrow n \downarrow p \uparrow p \uparrow  \rangle$,
then this basis state can be stored as $a(3,11)=\phi^r_{M}(\mathcal{R},S) $.
Integer $l_i=3$ for the isospin in binary is 0011 which,
from right to left, means \nth{0} and \nth{1} particles are protons, \nth{2} and \nth{3} particles are neutrons, so it is $| nnpp \rangle$. Integer $l_s=11$ for the spin in binary is 1011, means that the \nth{0} and \nth{1} particles are spin up, \nth{2} particles is spin down, \nth{3} one is spin up, so it is $| \uparrow \downarrow \uparrow \uparrow \rangle$. All those $| \psi^l_M (\!\mathcal{R}) \rangle$ and $| \psi^r_M (\!\mathcal{R}) \rangle$ states defined in Eqs.(\ref{phirMexpansion},\ref{philMexpansion}) for each bead, are represented by their coefficients $\phi^r_{M}(\mathcal{R},S)$ and $\phi^l_{M}(\mathcal{R},S)$ defined in Eqs.(\ref{phirMexpansion},\ref{philMexpansion}), and therefore stored in those 2 dimensional complex arrays like $a(l_i,l_s)$. All the quantities can be calculated from $| \psi^r_M (\!\mathcal{R}) \rangle$ and $| \psi^l_M (\!\mathcal{R})\rangle$.

It will be slow if we calculate $f^V_{lr}(\mathcal{R})$ and $g^V_{lr,M}(\mathcal{R})$ by first calculating all the $N_{\rm{tot}}$ by $N_{\rm{tot}}$ matrices of $ \langle S' |e^{\frac{- \mathcal{V}_I  \Delta \tau }{2}} | S \rangle $ and $\langle S' |\bm{\mathcal{G}} | S \rangle $ along all the beads in Fig. \ref{PIMCbeads}, then multiplying all of them.
This is because these matrices are sparse.
Any 2-body charge conserving operator will only have either 4 or 8 nonzero entries in a row or column.

We can instead from right to left directly calculate $| \psi^r_{N-1} (\mathcal{R}) \rangle $ from $| \psi^r_N (\mathcal{R}) \rangle $, then $| \psi^r_{N-2} (\mathcal{R}) \rangle $ from $| \psi^r_{N-1} (\mathcal{R}) \rangle $, ..., until $| \psi^r_{M+1} (\mathcal{R}) \rangle $ from $| \psi^r_{M+2} (\mathcal{R}) \rangle $.
Also, from left to right, we can calculate $| \psi^l_{1} (\mathcal{R}) \rangle $ from $| \psi^l_0 (\mathcal{R}) \rangle $, then $| \psi^l_{2} (\mathcal{R}) \rangle $ from $| \psi^l_{1} (\mathcal{R}) \rangle $, ..., until $| \psi^l_{M} (\mathcal{R}) \rangle $ from $| \psi^l_{M-1} (\mathcal{R}) \rangle $.
The recursion relations from Eqs.(\ref{psirmRdef}) and (\ref{psilmRdef}), can be written as,
\begin{eqnarray}
 | \psi^r_M (\mathcal{R}) \rangle  =
\begin{cases}
 \sum\limits_{S}
 \left(  e^{\frac{- \mathcal{V}_M  \Delta \tau }{2}}  e^{\frac{- \mathcal{V}^\dagger_M  \Delta \tau }{2}}
\bm{\mathcal{G}} \right) | S \rangle
\langle S| \psi^r_{M+1} (\mathcal{R}) \rangle,
 & M \neq  N,  0. \\
 \sum\limits_{S}
e^{\frac{- \mathcal{V}_N  \Delta \tau }{2}}
| S \rangle \langle R_N  S | \psi^r_T \rangle,
 &  M = N. \\
  \sum\limits_{S} \! e^{\frac{- \mathcal{V}^\dagger_0  \Delta \tau }{2}} \bm{\mathcal{G}}
  | S \rangle \langle S | \psi^r_1 (\mathcal{R}) \rangle,
 &  M =0.
\end{cases} \label{phirMRrecur}
\end{eqnarray}
\begin{eqnarray}
| \psi^l_M (\mathcal{R})\rangle  =
\begin{cases}
 \sum\limits_{S}
\left( \bm{\mathcal{G}}   e^{\frac{- \mathcal{V}_M  \Delta \tau }{2}}  e^{\frac{- \mathcal{V}^\dagger_M  \Delta \tau }{2}}
 \right)^\dagger | S \rangle
\langle S | \psi^l_{M-1} (\mathcal{R}) \rangle,
 & M\neq 0,N. \\
 \sum\limits_{S}
e^{\frac{- \mathcal{V}_{0}  \Delta \tau }{2}}   | S \rangle \langle R_0  S | \psi^l_T \rangle,
 &  M=0. \\
 \sum\limits_{S}\!   e^{\frac{- \mathcal{V}^\dagger_{N}  \Delta \tau }{2}} \bm{\mathcal{G}}^\dagger | S \rangle
\langle S | \psi^l_{N-1} (\mathcal{R}) \rangle,
 &  M=N.
\end{cases} \label{philMRrecur}
\end{eqnarray}

\begin{figure}[hptb!]
\centering
\includegraphics[width=\textwidth]{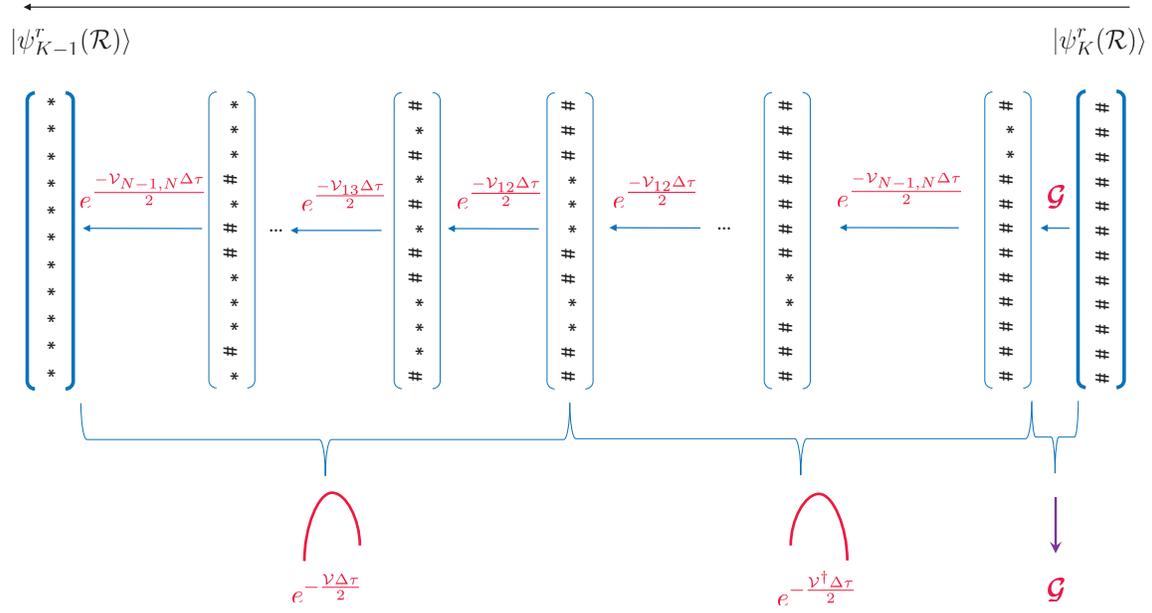}
\caption[Illustration of Calculating $| \psi^r_{K} (\mathcal{R}) \rangle $ and $| \psi^r_{K-1} (\mathcal{R}) \rangle $ from Right to Left Directly.]{Illustration of calculating $| \psi^r_{K-1} (\mathcal{R}) \rangle $ from $| \psi^r_K (\mathcal{R}) \rangle $ from right to left directly.}
\label{FigdirectCalcStates2}
\end{figure}

Once we have the corresponding $| \psi^r_{M} (\mathcal{R}) \rangle $ and $| \psi^l_{M} (\mathcal{R}) \rangle $ calculated, we can use Eqs.(\ref{fVRlrN2LOdef2}) and (\ref{gVRlrdef2}) to calculate  $f^V_{lr}(\mathcal{R})$ and $ g^V_{lr,M}(\mathcal{R}) $ and then $A_{lr}(\mathcal{R})$ and $B_{lr}(\mathcal{R})$ and thus $\langle \hat{O} (\Delta \tau)  \rangle$.
This updating strategy for $| \psi^r_{M} (\mathcal{R}) \rangle $ and $| \psi^l_{M} (\mathcal{R}) \rangle $ is much faster than direct $N_{\rm{tot}}$ by $N_{\rm{tot}}$ sparse matrix multiplication method.
This is why we define states $| \psi^r_M (\!\mathcal{R}\!) \rangle $ and $| \psi^l_M (\!\mathcal{R}\!)\rangle$ and use Eqs. (\ref{phirMRrecur}) and (\ref{philMRrecur}) to calculate them and then finally calculate $f^V_{lr}(\mathcal{R})$ and $ g^V_{lr,M}(\mathcal{R})$ by using Eqs.(\ref{fVRlrN2LOdef2}) and (\ref{gVRlrdef2}).

Fig. \ref{FigdirectCalcStates2} illustrates the right to left direct calculation of $| \psi^r_{K-1} (\mathcal{R}) \rangle $ from $| \psi^r_K (\mathcal{R}) \rangle $. In the direct calculation $| \psi^r_{K-1} (\mathcal{R}) \rangle $ and  $| \psi^r_K (\mathcal{R}) \rangle $ are always treated as column vectors, we directly operate the $\bm{\mathcal{G}}$, $e^{\frac{-\mathcal{V}^\dagger \Delta t}{2}}$ and $e^{\frac{-\mathcal{V} \Delta t}{2}}$ on the column vector $| \psi^r_K (\mathcal{R}) \rangle $ and get new column vectors each time until we get the next column vector $| \psi^r_{K-1} (\mathcal{R}) \rangle $.
By doing that we prevent writing each of the operators as $N_{\rm{tot}}$ by $N_{\rm{tot}}$ sparse matrix and doing the slow matrix multiplications.

\subsection{Calculating the Samples }

The reason to recast Eq.(\ref{PIMCopexpdetailshorttimeapprox}) into the form of Eq.(\ref{OABP}) is because once we have the detailed form of Eq.(\ref{PlrR}), the normalized probability distribution $P_{lr}(\mathcal{R})$, we can sample it by using suitable algorithms.
And then after we take enough uncorrelated samples of $ A_{lr} (\mathcal{R}) $ and $ B_{lr}(\mathcal{R})$ in Eq.(\ref{OABP}) from the probability distribution $P_{lr} (\mathcal{R})$ in Eq.(\ref{PlrR}), we have the corresponding average values $\langle A \rangle$ for  $ A_{lr} (\mathcal{R}) $ and $\langle B \rangle $ for $ B_{lr}(\mathcal{R})$ and their errors $\sigma_A$ and $\sigma_B$, finally $\langle \hat{O} (\Delta \tau) \rangle$ and its error bar $\sigma$ can be estimated by,
\begin{align}
\langle \hat{O} (\Delta \tau)\rangle =
\frac{\sum_l\sum_r \int d \mathcal{R} A_{lr} (\mathcal{R}) P_{lr} (\mathcal{R}) }{ \sum_l\sum_r \int d \mathcal{R} B_{lr} (\mathcal{R}) P_{lr} (\mathcal{R})  }
=\frac{\langle A_{lr}(\mathcal{R}) \rangle }{\langle B_{lr}(\mathcal{R}) \rangle}
\bigg|_{ \{l,r,\mathcal{R}\} \in P_{lr}(\mathcal{R})  }
= \frac{\langle A \rangle}{\langle B \rangle}
\label{OABPdtave}
\end{align}
\begin{align}
\sigma =
\sqrt{ \left[ \frac{\partial (\langle A \rangle / \langle B \rangle )}{ \partial \langle A \rangle } \sigma_A \right]^2
+ \left[ \frac{\partial (\langle A \rangle / \langle B \rangle )}{ \partial \langle B \rangle } \sigma_B \right]^2  }
=
\sqrt{ \left( \frac{\sigma_A}{\langle B\rangle} \right)^2 + \left(\frac{ \langle A \rangle \sigma_B }{ \langle B \rangle^2 } \right)^2    }.
\label{OABerror}
\end{align}

Alternatively, we can also find $\langle \hat{O} \rangle$ and its error bar $\sigma$ from the block average of $A$ and $B$.
For example we can arrange all the $N$ samples of $A$ and $B$ into much smaller blocks. Each block contain $n$ samples of $A$ and $B$. So there are $n_b=N/n$ blocks. We can mark each of the block average as $\bar{A}_n$ and $\bar{B}_n$.
And the average of block average is the average value, namely
\begin{align}
\langle \bar{A}_n \rangle &= \langle A \rangle, \\
\langle \bar{B}_n \rangle &= \langle B \rangle.
\end{align}
The standard deviation of the $n_b$ samples of $\bar{A}_n$ and $\bar{B}_n$ are $\sigma_{A_n}$ and $\sigma_{B_n}$.
According to central limit theorem (\cite{Negele1998Quantum,VesaPDF}), the independent block averages of $\bar{A}_n$ and $\bar{B}_n$ each forms a gaussian distribution $P(\bar{A}_n)$ and $P(\bar{B}_n)$ with average value $\langle A \rangle$ and $\langle B \rangle$ and their corresponding standard deviations are
\begin{align}
\sigma_{P_{\bar{A}_n}} &= \frac{\sigma_{\bar{A}_n}}{\sqrt{n_b}}, \\
\sigma_{P_{\bar{B}_n}} &= \frac{\sigma_{\bar{B}_n}}{\sqrt{n_b}}.
\end{align}
Then we find that,
\begin{align}
\left\langle\frac{\bar{A}_n}{ \bar{B}_n }\right\rangle &=
\int d \bar{A}_n d \bar{B}_n \frac{\bar{A}_n}{\bar{B}_n}
P(\bar{A}_n) P(\bar{B}_n) \nonumber \\
&=
\int d \bar{A}_n \bar{A}_n P(\bar{A}_n)
\int d \bar{B}_n \frac{1}{\bar{B}_n} P(\bar{B}_n) \nonumber \\
&=\langle A   \rangle \int d \bar{B}_n \frac{1}{\bar{B}_n} P(\bar{B}_n).
\label{OABclt1}
\end{align}
Note that when $\sigma_{P_{B_n}}$ is small, the gaussian distribution $P(\bar{B}_n)$ becomes a $\delta$ function $\delta( \bar{B}_n -\langle B \rangle  )$,
so Eq.(\ref{OABclt1}) in this case becomes,
\begin{align}
\left\langle\frac{\bar{A}_n}{ \bar{B}_n }\right\rangle &=
\langle A   \rangle \int d \bar{B}_n \frac{1}{\bar{B}_n} \delta \left( \bar{B}_n -\langle B \rangle  \right)=
\frac{\langle A \rangle}{\langle B \rangle}
=\langle \hat{O} \rangle ,
\label{OABclt2}
\end{align}
and we can take the standard deviation of the mean of ${\bar{A}_n}/{ \bar{B}_n }$ as the error bar $\sigma$. In all cases we ensure that the number of samples is large enough that any bias from dividing two stochastic quantities is smaller than the statistical error.

\section{Sign Problem}
\label{secSignProblem}

Since wave functions for fermions are antisymmetric under interchange of
the particles, they necessarily can change sign. Except for a few special
cases where the Monte Carlo sampling can be arranged so
that sign changes for two types of fermions occur together, fermion quantum
Monte Carlo has a sign or phase problem (\cite{Umrigar2007PRL,Alhassid1994PRL,LiZiXiang2019a}).
This means that for a long
enough path it is possible to propagate from a region where the wave function
is positive to one where it is negative (or has a different phase).
In such a case as the total imaginary time for the path increases,
these sign changes make the signal to noise ratio decay exponentially,
or, for a fixed amount of computer time, the variance increases exponentially
with the total imaginary time.

For the light nuclei here, there would be no sign problem if the potential
were spin-isospin independent. In that case, the four spin-isospin
states proton up, proton down, neutron up, neutron down would not be
mixed by the interaction, and can be interpreted as different kinds of
particles. The spatial part of the wave function would be positive definite.

Realistic interactions change the spin and isospin, but the fact that
the wave function is still predominantly in this ``s-shell'' means that
the fermion sign problem is weak. As shown in this work, for imaginary
times that converge the system to the ground state, the fraction of
negative paths is small and we simply ignore the sign problem. That is,
our calculations need no approximations to deal with the fermion sign problem\footnote{
In PIMC calculation, the sign problem is equivalent to say that the $B_{lr} (\mathcal{R})$ in Eq.(\ref{Blrsimpform}) we sampled may frequently oscillate between 1 and -1, which can make the $\langle B_{lr} (\mathcal{R}) \rangle$ in Eq.(\ref{OABP}) close to zero and therefore lead to a big error bar.
However, In our light nuclei PIMC calculation, e.g. the $^4$He calculation, we find $\langle B_{lr} (\mathcal{R}) \rangle \approx 0.9$ which means there are only about $5\%$ of the $B_{lr} (\mathcal{R})$ samples flip their signs from 1 to -1. So we do not have any serious sign problems in this work, partially because our total imaginary time is not big, and also we are summing over all the spin isospin states and this makes the sign flipping less likely to happen.}.

Extensions of this method to larger systems will necessary require dealing
with a stronger sign or phase problem. We expect that the approximate
methods developed for GFMC and AFDMC can be applied.
Those methods include fixed node approximation (\cite{Moskowitz1982a,JBAnderson2008}), constrained-path approximation (\cite{ZhangShiwei2003PRL,Gandolfi2009PRCa,Motta2018}), fixed-phase approximation (\cite{Carlson87a}), eigenvector continuation method (\cite{DLee2018PRL}), etc.

\chapter{Monte Carlo Algorithms}
\label{compalgorithm}

In this chapter, we introduce how we sample the probability distribution $P_{lr}(\mathcal{R})$ in Eq.(\ref{PlrR}) and also our algorithm for performing the other parts of the PIMC calculations, the sampling is based on a Markov chain process (\cite{Ceperley95a,HammondMCbook}) along with the Metropolis algorithm (\cite{Metropolis53a,Kalos1986a}).
\\

\section{Metropolis Algorithm}
\label{MarkovChian}

In general, for a normalized probability distribution $\pi(s)$ such that $\sum_s \pi(s)=1$, $\langle F \rangle$ the expectation value of function $F(s)$ is,
\begin{equation}
\langle F \rangle = \sum_s F(s) \pi(s),
\label{Fave}
\end{equation}
where the letter $s$ means a set of variables which can describe a state. In the Monte Carlo method, if we can sample $n$ states which effectively explore the state space of $\pi(s)$ and are distributed according to $\pi(s)$, denoting those states as $\{s_0,s_1,s_2...,s_n\} \in s $, then the expectation value $\langle F \rangle$ can be calculated by,
\begin{equation}
\langle F \rangle = \frac{\sum_{i=0}^n F(s_i) }{n} .
\label{FaveMC}
\end{equation}

So it is clear that we need to make a process which can continuously generate a series of states \{$s_0,s_1,s_2,s_3$...\} such that they distributed according to a target distribution $\pi(s)$. In order to do so, we need a transition probability $\mathcal{P}(s\rightarrow s')$ which is the probability for state $s$ to transit to $s'$. Once this process reaches its equilibrium status, it must satisfy,
\begin{equation}
\sum_s \pi(s) \mathcal{P}(s\rightarrow s')=\pi(s').
\label{equilibriumeq}
\end{equation}

We can require the number of states that make transition from any $s$ to $s'$ equals that transit from $s'$ back to $s$,
\begin{equation}
\pi(s) \mathcal{P}(s\rightarrow s')=\pi(s') \mathcal{P}(s'\rightarrow s).
\label{detailbalance}
\end{equation}
Eq.(\ref{detailbalance}) is the detailed balance condition (\cite{Ceperley95a}). Note that $\sum_{s'} \mathcal{P}(s\rightarrow s')=\sum_{s} \mathcal{P}(s'\rightarrow s)=1$, so once we reach the detailed balance, Eq.(\ref{equilibriumeq}) the equilibrium status will automatically become valid and therefore the states we generated will indeed be distributed according to $\pi(s)$.

The transition probability can be written as,
\begin{equation}
\mathcal{P}(s\rightarrow s')=T(s\rightarrow s') A(s\rightarrow s'),
\label{PeqTA}
\end{equation}
where $T(s\rightarrow s')$ is a proposed transition probability and $A(s\rightarrow s')$ is the acceptance probability which will be used to judge if the proposed transition is accepted or not. With Eq.(\ref{PeqTA}), the detailed balance equation Eq.(\ref{detailbalance}) can be further written as,
\begin{equation}
\frac{A(s\rightarrow s')}{A(s'\rightarrow s)}=\frac{\pi(s')T(s'\rightarrow s)}{\pi(s) T(s\rightarrow s') } .
\label{detailbalanceAoA}
\end{equation}

A common choice of the acceptance probability $A(s\rightarrow s')$ which can satisfy Eq.(\ref{detailbalanceAoA}) and thus the detailed balance condition is (\cite{Kalos1986a}),
\begin{equation}
A(s\rightarrow s')=
\min \left[  1, \frac{\pi(s')T(s'\rightarrow s)}{\pi(s) T(s\rightarrow s')} \right] .
\label{Accptequation}
\end{equation}

As we can see, the normalization factors in $\pi(s)$ and $\pi(s')$ and that in $T(s \rightarrow s')$ and $T(s'\rightarrow s)$ will be canceled out in Eqs.(\ref{detailbalanceAoA},\ref{Accptequation}) so those factors are irrelevant in calculating $A$. In order to make the value of $A(s\rightarrow s')$ larger, we can try to make $T(s'\rightarrow s)$ similar with $\pi(s)$, and $T(s'\rightarrow s)$ similar with $\pi(s')$, such that the numerator and denominator in Eq.(\ref{Accptequation}) can largely cancel. That will usually make the calculation efficient. In the ideal case, if we can make $T(s'\rightarrow s)=\pi(s)$ and $T(s'\rightarrow s')=\pi(s')$, then $A(s\rightarrow s')=1$ so we just need to always accept the new proposed states, but this is normally not possible, and if possible the Metropolis method would be unnecessary.

The basic procedure for the Monte Carlo method is described as follows.
\begin{description}
  \item[Step 1] For the current state $s$, propose a new state $s'$ according to the proposed transition probability $T(s\rightarrow s')$.
  \item[Step 2] For the proposed new state $s'$, use Eq.(\ref{Accptequation}) to calculate the acceptance rate $A(s\rightarrow s')$.
  \item[Step 3] Generate a uniform random number $a$ which ranges from 0 to 1.
  If $a \leq A(s\rightarrow s')$, the new state $s'$ is accepted, so set the accepted state $s'$ as the current state $s$, namely $s = s'$, and keep $s'$ as a sample, then go to step 1 and continue. Else, the new state $s'$ is rejected which means the state $s$ is accepted and kept as the sample, so set the accepted state $s$ as our current state $s$, then go to step 1 and continue.
\end{description}

After looping over the procedure for enough times, all those `current state' $s$ in step 1 will eventually be distributed according to the desired distribution $\pi(s)$.

Note that the choice of $T$ and therefore $A$ are not unique.
In order to make the algorithm more robust,
we usually include different types of moves with correspondingly proposed transition probability $T$, and acceptance rate $A$,
and this will still lead to the target distribution $\pi(s)$,
as long as the detailed balance condition Eq.(\ref{detailbalanceAoA}) holds for each type of move.
Any such choice that has transitions between any two allowed states by a series of moves and has some rejections will converge to samples of $\pi(s)$.
Also, the choice of $T$ determines how efficient the Monte Carlo procedure is. A poorly chosen $T$ can make the code run for a very long time but still not able to form $\pi(s)$. So $T$ is crucial in the algorithm, we will discuss our choice of $T$ in the next section.

\section{Methods of Sampling}
\label{secSamplingMethod}

In this section, we apply the Monte Carlo method described in section \ref{MarkovChian} to our PIMC algorithm. The $P_{lr} (\mathcal{R})$ in Eq.(\ref{PlrR}) is exactly the $\pi(s)$ in section \ref{MarkovChian}. State $s$ in $\pi(s)$ means a particular choice of \{$l,r,\mathcal{R}$\}, and $\sum_s$ in Eq.(\ref{Fave}) is the same as $\sum_l \sum_r \int d \mathcal{R}$ in Eq.(\ref{OABP}). We have
\begin{align}
s &= \{ l,r,\mathcal{R} \},
\label{state_s}\\
\pi(s)  &= \mathcal{P}_{lr}(\mathcal{R}), \\
\sum_s \pi(s) &= \sum_l \sum_r \int \mathcal{DR} \mathcal{P}_{lr}(\mathcal{R}) =1 .
\end{align}

The proposed transition probability $T(s \rightarrow s')$ can be decomposed into two parts. First we propose a transition for the left and right order of the trial wave function $\{ l,r \}$ to transit to a new pair $\{ l',r' \}$, mark it $T_{lr \rightarrow l'r'}$.
Then, based on the new $\{ l',r' \}$, we choose a new configuration $\mathcal{R}'$ based on the new proposed transition probability $T_{l'r'}(\mathcal{R} \rightarrow \mathcal{R}')$. In such a way we have
\begin{equation}
T(s \rightarrow s') = T_{lr \rightarrow l'r'} T_{l'r'}(\mathcal{R} \rightarrow \mathcal{R}').
\end{equation}
We can just randomly pick $\{l',r'\}$ so that $T_{lr \rightarrow l'r'}$ equal to its inverse transition probability $T_{l'r' \rightarrow lr} $, so they are canceled in Eq.(\ref{Accptequation}). Considering Eqs.(\ref{PlrR},\ref{fRlr},\ref{gRlr}), we can write down,
\begin{equation}
\frac{\pi(s')T(s'\rightarrow s)}{\pi(s) T(s\rightarrow s')} = \frac{ f^V_{l'r'}(\mathcal{R}') g^F(\mathcal{R}')  T_{lr}(\mathcal{R}' \rightarrow \mathcal{R})  }
{ f^V_{lr}(\mathcal{R}) g^F(\mathcal{R}) T_{l'r'}(\mathcal{R} \rightarrow \mathcal{R}') }, \label{Acc2}
\end{equation}
and we can judge if we accept the new state $s'$ by Eq.(\ref{Accptequation}) which now becomes,
\begin{equation}
A (s\rightarrow s') \! = \!
\min \! \left[  1,
\frac{ f^V_{l'r'}(\mathcal{R}') g^F(\mathcal{R}')  T_{lr}(\mathcal{R}' \! \rightarrow \! \mathcal{R})  }
{ f^V_{lr}(\mathcal{R}) g^F(\mathcal{R}) T_{l'r'}(\mathcal{R} \! \rightarrow \! \mathcal{R}') }
\right] .
\label{Accptequationdetail1}
\end{equation}

In the program, we do an equivalent two-stage judgement.
At stage one, we randomly pick $\{l',r'\}$, and judge if we accept the new order by
\begin{equation}
A (s\rightarrow s') \! = \!
\min \! \left[  1,
\frac{ f^V_{l'r'}(\mathcal{R}) g^F(\mathcal{R})  }
{ f^V_{lr}(\mathcal{R}) g^F(\mathcal{R})  }
\right] .
\end{equation}
No matter $\{l',r'\}$  is accepted or not (the acceptance rate for $\{l,r\} \rightarrow \{l',r'\} \approx 99\%$), we mark the left and right order after the judgement as $\{l,r\}$.
At the stage two, we propose the new configuration $\mathcal{R}'$ and judge if it is accepted or not by
\begin{equation}
A (s\rightarrow s') \! = \!
\min \! \left[  1,
\frac{ f^V_{lr}(\mathcal{R}') g^F(\mathcal{R}')  T_{lr}(\mathcal{R}' \! \rightarrow \! \mathcal{R})  }
{ f^V_{lr}(\mathcal{R}) g^F(\mathcal{R}) T_{lr}(\mathcal{R} \! \rightarrow \! \mathcal{R}') }
\right] .
\end{equation}

Since $f^V_{l'r'}(\mathcal{R}')$ and $f^V_{lr}(\mathcal{R})$ are sandwiched between the trial wave functions, there is no obvious ways to cancel them out in the numerator and denominator, so we leave them alone.
However, the proposed probability $T_{l'r'}(\mathcal{R} \rightarrow \mathcal{R}')$ is something we can deal with.
There can usually be two choices.
\begin{description}
  \item[Choice 1] Make the proposed probability as,
\begin{equation}
T_{l'r'}(\mathcal{R} \rightarrow \mathcal{R}') = T_{lr}(\mathcal{R}' \rightarrow \mathcal{R})  ,
\label{tlrnewtlrold}
\end{equation}
and by doing this, Eq.(\ref{Acc2}) can be simplified as,
\begin{equation}
\frac{\pi(s')T(s'\rightarrow s)}{\pi(s) T(s\rightarrow s')} =
\frac{ f^V_{l'r'}(\mathcal{R}') g^F(\mathcal{R}')  }{ f^V_{lr}(\mathcal{R}) g^F(\mathcal{R})  }, \label{acc2simp2}
\end{equation}
and then we can judge if we accept the new state $s'$ by,
\begin{equation}
A(s\rightarrow s')=
\min \left[  1, \frac{ f^V_{l'r'}(\mathcal{R}') g^F(\mathcal{R}')  }{ f^V_{lr}(\mathcal{R}) g^F(\mathcal{R})  } \right] .
\label{Accptequationsimp2}
\end{equation}
  \item[Choice 2] Make the proposed transition probability $T_{l'r'}(\mathcal{R} \rightarrow \mathcal{R}')$ and $T_{lr}(\mathcal{R}' \rightarrow \mathcal{R})$ in a way such that
\begin{equation}
g^F(\mathcal{R}') T_{lr}(\mathcal{R}' \rightarrow \mathcal{R}) = g^F(\mathcal{R}) T_{l'r'}(\mathcal{R} \rightarrow \mathcal{R}')   ,
\label{gtnewgtold}
\end{equation}
and by doing this, Eq.(\ref{Acc2}) can be simplified as,
\begin{equation}
\frac{\pi(s')T(s'\rightarrow s)}{\pi(s) T(s\rightarrow s')} = \frac{ f^V_{l'r'}(\mathcal{R}') }{ f^V_{lr}(\mathcal{R})  }, \label{acc2simp}
\end{equation}
and then judge if we accept the new state $s'$ by the simplified version of Eq.(\ref{Accptequation}),
\begin{equation}
A(s\rightarrow s')=
\min \left[  1, \frac{ f^V_{l'r'}(\mathcal{R}') }{ f^V_{lr}(\mathcal{R})  }\right] .
\label{Accptequationsimp}
\end{equation}
\end{description}

In our work, different methods of sampling means different choices of proposed transition probability $T_{l'r'}(\mathcal{R} \rightarrow \mathcal{R}')$ and $T_{lr}(\mathcal{R}' \rightarrow \mathcal{R}) $. No matter what the choice is, they all correspond to the same target distribution $\pi(s)$.  However, different choices of proposed transition probability will lead to different convergence speed. So in order to achieve the target distribution $\pi(s)$ efficiently and quickly, we need to mainly use efficient proposed transition probabilities.

\subsection{Single-Bead Sampling}
\label{Sample1bead}

From $\mathcal{R}=\{ R_0, ..., R_K, ..., R_N \}$, we can propose to move just one bead, bead $K$, we move $R_K$ to $R_K'$, then $\mathcal{R}'=\{ R_0, ..., R_K', ..., R_N \}$. This is single bead sampling.

For the $K = 0$ or $N$ case, the $g^F(\mathcal{R}')$ in Eq.(\ref{gF}) becomes,
\begin{align}
g^F(\mathcal{R}')|_{R_0 \rightarrow R_0'} &= G^f_{0', 1} \prod_{I=1}^{N-1} G^f_{I, I+1} , \label{gFRnew1bead0} \\
g^F(\mathcal{R}')|_{R_N \rightarrow R_N'} &= G^f_{N-1, N'} \prod_{I=0}^{N-2} G^f_{I, I+1} , \label{gFRnew1beadN}
\end{align}
where $G^f_{0' 1 }$ and $G^f_{N-1, N'}$ are as defined in Eq.(\ref{Gfii1}) and can be calculated by Eq.(\ref{propagatorFree}),
\begin{align}
G^f_{0', 1}  &=  \langle R'_{0} | e^{-T \Delta \tau}  | R_{1} \rangle =  \left( \frac{m}{2\pi \Delta \tau \hbar^2} \right)^{\frac{3A}{2}}
 e^{  -\frac{ ({R'_{0}}-{R_1})^2 }{2 \Delta \tau  \frac{\hbar^2}{m}  }}   ,
\label{gFRnew1bead0a} \\
G^f_{N-1, N'}  &=  \langle R_{N-1} | e^{-T \Delta \tau}  | R'_{N} \rangle
=  \left( \frac{m}{2\pi \Delta \tau \hbar^2} \right)^{\frac{3A}{2}}
 e^{  -\frac{ (R_{N-1}-R'_{N})^2 }{2 \Delta \tau \frac{\hbar^2}{m} }  }  .
\label{gFRnew1beadKa}
\end{align}

We can just set the proposed transition probability as,
\begin{align}
T_{l'r'}(\mathcal{R} \rightarrow \mathcal{R}') |_{R_0 \rightarrow R_0'} &=
G^f_{0', 1},
 \label{Tlrnew0} \\
T_{l'r'}(\mathcal{R} \rightarrow \mathcal{R}') |_{R_N \rightarrow R_N'} &=
G^f_{N-1, N'},
 \label{TlrnewN}
\end{align}
and then Eq.(\ref{Acc2}) becomes Eq.(\ref{acc2simp}) and we use Eq.(\ref{Accptequationsimp}) to judge if we accept $R_K'$ and thus $\mathcal{R}'$ or not.

For the $K \neq 0$ or $N$ case, the corresponding $g^F(\mathcal{R}')$ becomes,
\begin{equation}
g^F(\mathcal{R}') = G^f_{K-1, K'} G^f_{K', K+1} \prod_{I=0}^{K-2} G^f_{I, I+1} \prod_{I=K+2}^{N-1} G^f_{I, I+1},
\label{gFRnew1bead}
\end{equation}
where $G^f_{K-1, K'}$ and $G^f_{K', K+1}$ are the free propagators calculated by Eq.(\ref{Gfii1}) or Eq.(\ref{propagatorFree}).

In order to proceed, we introduce a useful relation in our calculation.
For any coefficients $a$ and $b$ and any configuration $R_M$, $R_N$ and $R_L$,
\begin{eqnarray}
\!\! \!\! \! a(R_M-R_N)^2\!+\!b(R_N-R_L)^2
\! = \! (a+b)
\!  \left(  R_N \!-\! \frac{aR_M\!+\!bR_L}{a\!+\!b} \right)^2
\!\! \! + \! \frac{ab}{a\!+\!b} \! \left( R_M \!-\! R_L  \right)^2, \label{Rdiffadd}
\end{eqnarray}
where each square can be calculated by using Eq.(\ref{Rdiffdef}). In particular, if $a=b=1$, Eq.(\ref{Rdiffadd}) just becomes,
\begin{eqnarray}
(R_M-R_N)^2+(R_N-R_L)^2
= 2\left(  R_N -\frac{R_M+R_L}{2} \right)^2
+ \frac{\left( R_M-R_L  \right)^2}{2} . \label{Rdiffadd1}
\end{eqnarray}

By using Eq.(\ref{Rdiffadd1}), the product of the two free particle propagators $G^f_{K-1, K'} G^f_{K', K+1}$ in Eq.(\ref{gFRnew1bead}) becomes,
\begin{equation}
G^f_{K-1, K'} G^f_{K', K+1}  =
\left( \frac{m}{2\pi \Delta \tau \hbar^2} \right)^{3A}
 e^{ -\frac{ \left({R_{K}'}-\frac{R_{K\!-\!1}+R_{K\!+\!1}}{2} \right)^2 }{ \Delta \tau  \frac{\hbar^2}{m}  } }
e^{ - \frac{ \left({R_{K\!-\!1}}-{R_{K\!+\!1}}\right)^2 }{ 4 \Delta \tau  \frac{\hbar^2}{m}  } } .
 \label{GKm1KpKpkp1}
\end{equation}

We set the proposed transition probability $T_{l'r'}(\mathcal{R} \rightarrow \mathcal{R}')$ from Eq.(\ref{GKm1KpKpkp1}) as,
\begin{equation}
T_{l'r'}(\mathcal{R} \rightarrow \mathcal{R}') =
\left( \frac{m}{\pi \Delta \tau \hbar^2} \right)^{\frac{3A}{2}}
 e^{  -\frac{ \left({R_{K}'}-\frac{R_{K\!-\!1}+R_{K\!+\!1}}{2} \right)^2 }{ \Delta \tau  \frac{\hbar^2}{m}  }} .
 \label{Tlrnew}
\end{equation}
The corresponding reverse transition probability $T_{lr}(\mathcal{R}' \rightarrow \mathcal{R})$ can be written as,
\begin{equation}
T_{lr}(\mathcal{R}' \rightarrow \mathcal{R}) =
\left( \frac{m}{\pi \Delta \tau \hbar^2} \right)^{\frac{3A}{2}}
 e^{  -\frac{ \left({R_{K}}-\frac{R_{K\!-\!1}+R_{K\!+\!1}}{2} \right)^2 }{ \Delta \tau  \frac{\hbar^2}{m}  }} .
 \label{Tlrold}
\end{equation}

Now we can see that $g^F(\mathcal{R}')  T_{lr}(\mathcal{R}' \! \rightarrow \! \mathcal{R}) = g^F(\mathcal{R}) T_{l'r'}(\mathcal{R} \! \rightarrow \! \mathcal{R}') $, so Eq.(\ref{Acc2}) becomes Eq.(\ref{acc2simp}) and we use Eq.(\ref{Accptequationsimp}) to judge if we accept $R_K'$ and thus $\mathcal{R}'$ or not.

For particle i at bead $K$, we define its spatial coordinates as $\bm{r}^K_i \equiv (x^K_i,y^K_i,z^K_i)$, and its new proposed coordinates as $ \bm{r'}^K_i \equiv ({x'}^K_i,{y'}^K_i,{z'}^K_i)$. So for the A-nucleon nuclei, $R_K=(\bm{r}^K_1, \bm{r}^K_2, ..., \bm{r}^K_A)$ and $R_K'=(\bm{r'}^K_1, \bm{r'}^K_2, ..., \bm{r'}^K_A)$. We also define a 3 dimensional vector $\bm{g}_i$ for particle $i$,
\begin{equation}
\bm{g}_i=(g_{ix},g_{iy},g_{iz}), \label{gaussgi}
\end{equation}
where $g_{ix},g_{iy},g_{iz}$ are all independent gaussian random numbers with average value 0 and variance 1.

The basic procedures of the algorithm is described as follows.
\begin{description}
  \item[Step 1] Randomly pick an integer $K$ from 0 to $N$.
  \item[Step 2] For particle $i$ at bead $K$,
  \begin{enumerate}
    \item if $K=0$, according to Eq.(\ref{Tlrnew0}), propose its new coordinates ${\bm{r}'}^0_i$ according to ${\bm{r}}^{1}_i$ at bead $1$,
\begin{equation}
{\bm{r}'}^0_i = \bm{r}^{1}_i + \bm{g}_i \sqrt{  \Delta \tau  \frac{\hbar^2}{m} } ,
\end{equation}
    \item If $K=N$, according to Eq.(\ref{TlrnewN}), propose its new coordinates ${\bm{r}'}^N_i$ according to ${\bm{r}}^{N-1}_i$ at bead $N-1$,
\begin{equation}
{\bm{r}'}^N_i = \bm{r}^{N-1}_i + \bm{g}_i \sqrt{  \Delta \tau  \frac{\hbar^2}{m} } ,
\end{equation}
    \item
If $K\neq 0,N$, according to Eq.(\ref{Tlrnew}), propose its new coordinates ${\bm{r}'}^K_i$ according to ${\bm{r}}^{K-1}_i$ at bead $K-1$ and ${\bm{r}}^{K+1}_i$ at bead $K+1$,
\begin{equation}
{\bm{r}'}^K_i = \frac{\bm{r}^{K-1}_i+\bm{r}^{K+1}_i}{2} + \bm{g}_i \sqrt{ \Delta \tau  \frac{\hbar^2}{2 m} } ,
\end{equation}
  \end{enumerate}
  \item[Step 3] Loop over step 2 for all the particles, $i$ from 1 to A. Then we have the new configuration $R_K'$.
  \item[Step 4] Use Eq.(\ref{Accptequationsimp}) to judge if we accept the proposed configurations $\mathcal{R}'$ or not. Then go to step 1 and continue.
\end{description}

In the program, we store all the probability amplitudes $\phi^r_{M}(\mathcal{R},S)$ in Eq.(\ref{phirMR}) and $\phi^l_{M}(\mathcal{R},S)$ in Eq.(\ref{philMR}) for $M$ from 0 to $N$. Once any $R'_K$ is accepted, we need to update all the $\phi^r_{M}(\mathcal{R}',S)$ for $K \leq M \leq N$ and $\phi^l_{M}(\mathcal{R}',S)$ for $0\leq M \leq K$.
This means we need to do about $N$ updates for just one accepted $R'_K$ (which is not economical), and therefore totally about $N^2$ updates for all the beads if they are accepted.
Besides, single bead sampling usually has a very high acceptance rate, this indicates that the beads may not be moved far enough in the configuration space.
Therefore it should not be used often in the program.
It is however simple, and can be used as a check for more complicated algorithms.

\subsection{Multi-level Sampling}
\label{secMultibeadSample}

Due to the disadvantage of single bead sampling, for a given amount of computation time, we have to sample multiple beads at a time instead of just one in order to make the sampling efficient.
For example, if we propose to move $u$ portion of the $N$ beads, so about $uN$ beads will be proposed to move at a time, and the acceptance rate is $a$. Then in order for all the $N$ beads to be moved, we need at least $1/(ua)$ trials. For each trial, in order to calculate acceptance rate $A(s\rightarrow s')$, we need to do $uN$ updates for either $\phi^r_{M}(\mathcal{R},S)$ in Eq.(\ref{phirMR}) or $\phi^l_{M}(\mathcal{R},S)$ in Eq.(\ref{philMR})  with accordingly range of $M$. So for the proposed moves there are $uN/(ua)=N/a$ updates. Besides, among all the  $1/(ua)$ trials, only about $a/(ua)=1/u$ are accepted, for those accepted moves, we need to update some beads for the $\phi^r_{M}(\mathcal{R},S)$ and some for the $\phi^l_{M}(\mathcal{R},S)$. This will actually results in updating about $N$ beads. So for the accepted moved, there will be $N/u$ updates. So totally, there will be $N/u+N/a=N(1/u+1/a)$ updates. And the factor $(1/u+1/a)$ is usually much smaller than $N$, for example, N=500, when $u=1/3$, $a$ could be around 0.1 so $10\%$, so $(1/u+1/a) \approx 13$ which is much smaller than $N=500$. So $N(1/u+1/a)$ updates much more cheaper than $N^2$ updates needed for the single beads sampling.

In order to find some ways to sample multiple beads at a time, we need to make full use of Eq.(\ref{Rdiffadd1}).
From the total $N+1$ beads on the time line from bead $0$ to bead $N$, we randomly pick $2^n+1$ contingent beads from bead $I$ to bead $I+2^n$, where $n$ is a positive integer. To make the notations easier, here we temporarily mark the configurations from bead $I$ to bead $I+2^n$ as $R_0$ to $R_{2^n}$, and we mark their configurations as,
\begin{equation}
\mathcal{R}=\{R_0, R_1, R_2, ..., R_{2^n-1}, R_{2^n} \}. \label{mathcalRmulbeads}
\end{equation}
The configurations $R_0$ to $R_{2^n}$ in Eq.(\ref{mathcalRmulbeads}) are just the relabeling of the original $R_I$ to $R_{I+2^n}$ among the total $N+1$ beads on the path.

The product of the free particle propagators $g^F(\mathcal{R})$ can be written as,
\begin{equation}
g^F(\mathcal{R}) = \prod_{I=0}^{2^n-1} G^f_{I, I+1}.  \label{gF2n}
\end{equation}
We can group the $2^n+1$ beads $\{R_0, R_1, ..., R_{2^n} \}$ in $\mathcal{R}$ into $n$ levels as the following,
\begin{empheq}[box=\fbox]{align}
\begin{array}{cccc}
\textrm{Level } 0 :&  R_0, R_{2^n} .  & ({\textrm{2 beads}})  & L^n_0(\mathcal{R})  \nonumber \\
\textrm{Level } 1 :&  R_{\frac{1}{2}2^n}  .  & ({\textrm{1 bead}}) & L^n_1(\mathcal{R})  \nonumber \\
\textrm{Level } 2 :& R_{\frac{1}{4} 2^n} , R_{\frac{3}{4} 2^n} . & ({\textrm{2 beads}})  & L^n_2(\mathcal{R})   \nonumber \\
 \vdots  & \vdots & \vdots & \vdots   \nonumber \\
\textrm{Level } k :&  R_{\frac{1}{2^k} 2^n} , R_{\frac{3}{2^k} 2^n}, R_{\frac{5}{2^k} 2^n}, ..., R_{\frac{2^k-1}{2^k} 2^n} .   & ({\textrm{$2^{k-1}$ beads}}) & L^n_k(\mathcal{R}) \nonumber \\
 \vdots   &  \vdots &  \vdots &  \vdots \nonumber \\
\textrm{Level } n :& R_{1}, R_{3}, R_{5}, R_{7}, ..., R_{2^n-1} . & ({\textrm{$2^{n-1}$ beads}}) & L^n_n(\mathcal{R})
\end{array}   \nonumber
\end{empheq}

We can then use Eq.(\ref{GKm1KpKpkp1}) to rearrange Eq.(\ref{gF2n}) level by level and Eq.(\ref{gF2n}) becomes,
\begin{equation}
g^F(\mathcal{R}) = \prod_{k=0}^{n} L^n_k(\mathcal{R}) ,
\label{gFRplevels}
\end{equation}
where $L^n_k(\mathcal{R})$ is defined as the product of gaussian functions at level $k$,

\begin{equation}
 L^n_k(\mathcal{R}) =
\begin{cases}
G^f_{0, 2^n},  &  k = 0. \\
 \prod\limits_{I=1}^{2^{k-1}}  G^f_{\frac{2I-1}{2^k}2^{n}; \frac{2I-2}{2^k}2^{n}, \frac{2I}{2^k}2^{n}}  , & k \neq 0
.\end{cases} \label{Lfkn}
\end{equation}
For any non negative integer $J$, $K$, $L$, $M$, $N$, the $G^f_{J, K}$ is calculated by Eq.(\ref{Gfij}), and the $G^f_{M;L,N}$ in Eq.(\ref{Lfkn}) is,
\begin{align}
& G^f_{M;L,N} =
\left( \frac{2m}{\pi |N-L| \Delta \tau \hbar^2} \right)^{\frac{3A}{2}}
 e^{  -   \frac{ \left(  R_{M}  - \frac{ R_{L} + R_{N}   }{2}   \right)^2   }   {2\times \frac{|N-L|}{2}\Delta \tau \frac{\hbar^2}{2m}   }    }.
\label{gFRplevels}
\end{align}
Given Eqs.(\ref{Lfkn}) and (\ref{gFRplevels}), we can propose to move multiple beads, from bead $1$ to $2^n-1$ such that the new configuration is
\begin{equation}
\mathcal{R}'=\{R'_0, R'_1, R'_2, ..., R'_{2^n-1}, R'_{2^n} \},  \label{mathcalRpmulbeads}
\end{equation}
where $R_0$ and $R_{2^n}$ remain the same although they are marked as $R'_0$ and $R'_{2^n}$ in Eq.(\ref{mathcalRpmulbeads}). The product of the new free particle propagators $g^F(\mathcal{R}')$ can be written as,
\begin{equation}
g^F(\mathcal{R}') = \prod_{k=0}^{n} L^n_k(\mathcal{R}') . \label{gFRp}
\end{equation}

We can set the proposed transition probability $T_{l'r'}(\mathcal{R} \rightarrow \mathcal{R}')$ as,
\begin{equation}
T_{l'r'}(\mathcal{R} \rightarrow \mathcal{R}') = g^F(\mathcal{R}') = \prod_{k=0}^{n} L^n_k(\mathcal{R}').
 \label{Tlrnewmbeads}
\end{equation}
Eq.(\ref{Tlrnewmbeads}) means that we can set the proposed transition probability at each level $k$ as $T^n_k(\mathcal{R}')$ such that
\begin{equation}
T^n_k(\mathcal{R}') =  L^n_k(\mathcal{R}'),
 \label{Tlrnewmbeadslevelk}
\end{equation}
and $T_{l'r'}(\mathcal{R} \rightarrow \mathcal{R}')$ can be just written as
\begin{equation}
T_{l'r'}(\mathcal{R} \rightarrow \mathcal{R}') = \prod_{k=0}^{n} T^n_k(\mathcal{R}').
 \label{Tlrnewmbeadslbl}
\end{equation}
The reverse transition probability $T_{lr}(\mathcal{R}' \rightarrow \mathcal{R})$ can be written as,
\begin{equation}
T_{lr}(\mathcal{R}' \rightarrow \mathcal{R}) =
\prod_{k=0}^{n} T^n_k(\mathcal{R}) .
 \label{Tlroldmbeadslbl}
\end{equation}

\begin{figure}[hptb!]
\centering
\includegraphics[scale=0.455]{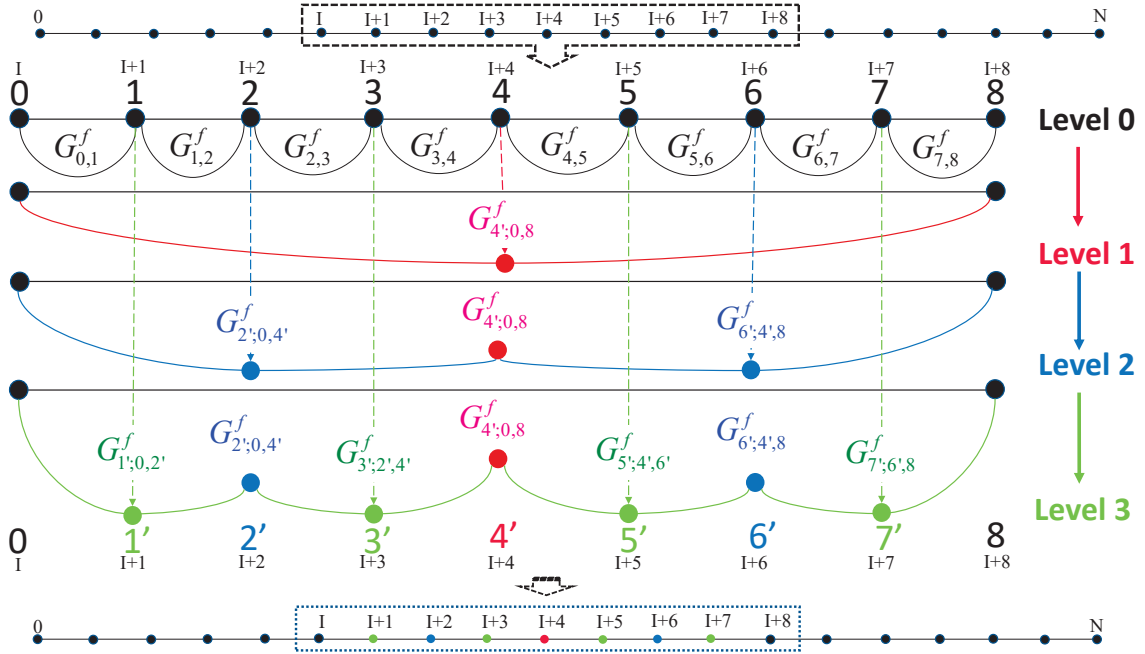}
\caption[Illustration of a 3-Level Sampling Algorithm]{Illustration of a 3-level sampling algorithm. }
\label{PIMCBisection}
\end{figure}

We can call this multi-level or n-level sampling, because we can propose to move the beads level by level from level 0 to level $n$, by using $T^n_k(\mathcal{R}')$ at each level.
By doing proposed moves in this way, we implement Eq.(\ref{Tlrnewmbeadslbl}). This makes $g^F(\mathcal{R}')  T_{lr}(\mathcal{R}' \! \rightarrow \! \mathcal{R}) = g^F(\mathcal{R}) T_{l'r'}(\mathcal{R} \! \rightarrow \! \mathcal{R}') $ and it enables Eq.(\ref{Acc2}) to become Eq.(\ref{acc2simp}), and we can use Eq.(\ref{Accptequationsimp}) to judge if accept $\mathcal{R}'$ or not. Fig. \ref{PIMCBisection} is the illustration of 3-level sampling.

The algorithm can be described as follows.
\begin{description}
  \item[Step 1] For the total $N+1$ beads on the time line as shown in Fig. \ref{PIMCbeads}, select a proper integer $n$, randomly pick $2^n+1$ contingent beads from $R_I$ to $R_{I+2^n}$ (note that in the program, $n$ does not need to be a fixed number, we can pick different $n$ each time when we are at step 1).
  For notational convenience, temporarily call them $R_0$ to $R_{2^n}$ instead of $R_I$ to $R_{I+2^n}$, and we use $\mathcal{R}$ to denote the group, $\mathcal{R}=\{R_0, R_1, R_2, ..., R_{2^n-1}, R_{2^n} \}$.

  For example, in Fig. \ref{PIMCBisection}, it is $n=3$ so $2^n=8$. We randomly pick $R_I$ to $R_{I+8}$ beads from the total $N+1$ beads, we temporarily mark them as $R_0$ to $R_8$.

  \item[Step 2]
  \begin{enumerate}
  \item If we keep $R_0$ and $R_{2^n}$ unchanged, this can be taken as a `fake' level 0 move, so $R'_0=R_0$ and $R'_{2^n}=R_{2^n}$. In this case, $L^n_0(\mathcal{R}') = L^n_0(\mathcal{R}) $, we set
    \begin{equation}
T_{l'r'}(\mathcal{R} \rightarrow \mathcal{R}') = \prod_{k=1}^{n} L^n_k(\mathcal{R}') ,
 \label{Tlrnewmbeadsf1}
\end{equation}
    and it can still make Eq.(\ref{Acc2}) become Eq.(\ref{acc2simp}). And we go to step 3.
  \item  Else, if we keep $R_{2^n}$ unchanged and just propose $R'_0$ (so $R'_{2^n}=R_{2^n}$), then according to Eq.(\ref{Lfkn}), $L^n_0(\mathcal{R}')=\exp \left[ - \frac{(R'_0-R_{2^n})^2}{2\times 2^n \Delta \tau \frac{\hbar^2}{m}  } \right]$, we propose
      \begin{equation}
  {\bm{r}'}_i^0 =
   {\bm{r}}_i^{2^n}
  + \bm{g}_i \sqrt{ 2^n\Delta \tau \frac{\hbar^2}{m}} ,
 \label{}
\end{equation}
    for $i$ from 1 to $A$, and we have $R_0'$ and go to step 3 for the rest $L^n_k(\mathcal{R}')$.
  \item     Else, if we keep $R_{0}$ unchanged and just propose $R'_{2^n}$ (so $R'_0=R_0$), then according to Eq.(\ref{Lfkn}), $L^n_0(\mathcal{R}')=\exp\ \Big[ - \frac{(R_0-R'_{2^n})^2}{2\times 2^n \Delta \tau \frac{\hbar^2}{m}  } \Big]$, we propose
      \begin{equation}
  {\bm{r}'}_i^{2^n} =
   {\bm{r}}_i^{0}
  + \bm{g}_i \sqrt{ 2^n\Delta \tau \frac{\hbar^2}{m}} ,
 \label{}
\end{equation}
   for $i$ from 1 to $A$, and we have $R_{2^n}'$ and go to step 3 for the rest $L^n_k(\mathcal{R}')$.
   \item
    Else, we can also use the one bead sampling algorithm described in section \ref{Sample1bead}, propose $R'_0$ according to $R_1$ and $R'_{2^n}$ according to $R_{2^n-1}$ and then judge and go to step 1.
\end{enumerate}

For example, in Fig. \ref{PIMCBisection},  we keep $R_0$ and $R_8$ unchanged, so $R_0'=R_0$ and $R_8'=R_8$.

  \item[Step 3] According to the product of gaussian functions of $L^n_k(\mathcal{R})$ in Eq.(\ref{Lfkn}), we propose the moves level by level, so $k$ from $1$ to $n$. At level $k$, $2^{k-1}$ beads need be to moved, which means from $I=1$ to $2^{k-1}$, each $R_{\frac{2I-1}{2^k}{2^{n}}}$ need to be proposed to a new configuration $R'_{\frac{2I-1}{2^k}{2^{n}}}$. And we propose $R'_{\frac{2I-1}{2^k}{2^{n}}}$ according to the
      $ G^f_{\left(\frac{2I-1}{2^k}2^{n}\right)'; \left(\frac{2I-2}{2^k}2^{n}\right)', \left(\frac{2I}{2^k}2^{n}\right)'} $ in $ L^n_k(\mathcal{R}') $ in Eq.(\ref{Tlrnewmbeads}), such that ${\bm{r}'}_i^{{(2I-1){2^{n-k}}}}$ (the new coordinates of particle $i$) in $R_{\frac{2I-1}{2^k}{2^{n}}}$ is proposed according to,
  \begin{equation}
  {\bm{r}'}_i^{{\frac{2I-1}{2^{k}}{2^{n}}}} \!\!=
  \frac{ {\bm{r}'}_i^{\frac{2I-2}{2^{k}}{2^{n}}} \!\! + {\bm{r}'}_i^{\frac{2I}{2^{k}}{2^{n}}}      }{2}
  + \bm{g}_i \sqrt{ \frac{2^n}{2^k}\Delta \tau \frac{\hbar^2}{2m}} .
 \label{Lfknp}
\hspace{-1em}
\end{equation}
For $i$ from 1 to $A$, we propose to move particle $i$ according to Eq.(\ref{Lfknp}) in each configuration in each level.

Note that the level $k$ beads are only proposed based on the beads of previous levels, so this allow us to propose the new beads level by level.
After we propose the new configurations level by level, we have $\mathcal{R}'=\{R'_0, R'_1, R'_2, ..., R'_{2^n-1}, R'_{2^n} \}$.
For example, in Fig. \ref{PIMCBisection}, level 1, 2 and 3 are,
\begin{description}
    \item[Level 1] $k=1$, $2^{1-1}=1$, $I=1$, each $[{(2I-1){2^{3-1}}}]^{\text{th}}$ bead needs to be moved, so $R_4$ is the only bead that needs to be moved, and we propose $R_4'$ based on $R'_0$ and $R'_8$ according to Eq.(\ref{Lfknp}), and each particle is moved as
          \begin{equation}
  {\bm{r}'}_i^{4} =
  \frac{ {\bm{r}'}_i^{0}  + {\bm{r}'}_i^{8}      }{2}
  + \bm{g}_i \sqrt{ 4 \Delta \tau \frac{\hbar^2}{2m}} .
\end{equation}
       After looping over all the particles, go to level 2.
    \item[Level 2] $k=2$, $2^{2-1}=2$, so $I=1,2$, each $[(2I-1){2^{3-2}}]^{\text{th}}$ bead needs to be moved, so $R_2$ and $R_6$ are the two beads that need to be moved, and according to Eq.(\ref{Lfknp}), we propose $R'_2$ based on $R'_0$ $R'_4$, and $R'_6$ based on $R'_4$ and $R'_8$. E.g., each particle in $R'_2$ is moved as
          \begin{equation}
  {\bm{r}'}_i^{2} =
  \frac{ {\bm{r}'}_i^{0}  + {\bm{r}'}_i^{4}      }{2}
  + \bm{g}_i \sqrt{ 2 \Delta \tau \frac{\hbar^2}{2m}} ,
\end{equation}
and that in  $R'_6$ is moved as
          \begin{equation}
  {\bm{r}'}_i^{6} =
  \frac{ {\bm{r}'}_i^{4}  + {\bm{r}'}_i^{8}      }{2}
  + \bm{g}_i \sqrt{ 2 \Delta \tau \frac{\hbar^2}{2m}}.
\end{equation}
 After looping over all the particles, go to level 3.
    \item[Level 3] $k=3$, $2^{3-1}=4$, so $I=1,2,3,4$, each $[(2I-1){2^{3-3}}]^{\text{th}}$ beads need to be moved, and so $R'_1$, $R'_3$,$R'_5$, $R'_7$ need to be proposed according to Eq.(\ref{Lfknp}) similarly as described before.
  \end{description}

  \item[Step 4] Now we need to look at the whole set of $N+1$ beads on the time line, and use consistent notations in the work. The $\{R'_0, R'_1, R'_2, ..., R'_{2^n-1}, R'_{2^n} \}$ in step 3 are in fact as mentioned in step 1, really $\{R'_I, R'_{I+1}, R'_{I+2}, ..., R'_{I+2^n-1}, R'_{I+2^n} \}$ on the time line. So overall, the proposed configuration $\mathcal{R}'$ is,
  \begin{equation}
\mathcal{R}'=\{R_0,R_1,...,R'_I, R'_{I+1}, ..., R'_{I+2^n}, ..., R_N \}.
\end{equation}
    The original configuration $\mathcal{R}$ is just Eq.(\ref{defmathcalR}) for the whole $N+1$ original beads.

    For example, in Fig. \ref{PIMCBisection}, once we have the new $\{R'_0, R'_1, ..., R'_{8} \}$, we change their labels to the original ones $\{R'_I, R'_{I+1}, ..., R'_{I+8} \}$. So finally $\mathcal{R}'$ is the same as $\mathcal{R}$ except that $R_I$ to $R_{I+8}$ are replaced by $R'_I$ to $R'_{I+8}$.

  \item[Step 5] Use Eq.(\ref{Accptequationsimp}) to judge if we accept the proposed configurations $\mathcal{R}'$ or not. Then go to step 1 and continue.

      For example, in Fig. \ref{PIMCBisection}, we use Eq.(\ref{Accptequationsimp}) to judge if we accept the proposed configurations $\mathcal{R}'$ or not.

\end{description}

It should be pointed out that, the proposed trial moves at each level in the multi-level sampling are the same as in the bisection method (\cite{Ceperley95a}).
But the two are not the same.
We propose all the n-level new configurations and then do an overall acceptance or rejection.
However in the bisection method, we accept or reject at each level, and if the proposed configuration is rejected at level $k$, then the whole proposed configurations from level 1 to $k$ will be completely rejected and discarded.

In the bisection method, the lower the level, the less time consuming it is to do the rejection.
In other words, `if you want to reject, reject early.'
It works best if the first level's acceptance rate is highly distinct from the rest levels'.

For example, consider a 3-level bisection, if the first level acceptance rate is $30\%$, the second and third levels' are $80\%$ and $90\%$, then the total acceptance rate is $21.6\%$.
But since the first level acceptance rate is significantly lower than the rest,
actually most trial moves are quickly rejected at the first level and there is no need to spend time to go to \nth{2} and \nth{3} level,
and those trial moves which get to the \nth{2} and \nth{3} level are mostly be accepted.
Therefore, not much time is wasted at the \nth{2} and \nth{3} levels.
In this case, bisection method will be faster than multi-level sampling simply because it can reject early.
Bisection works well for systems with potentials with highly repulsive cores where core overlap can be deleted early and rejected.

However, if the three levels' acceptance rate are similar,
then the bisection method does not have an advantage over multi-level sampling.
If the \nth{1}, \nth{2} and \nth{3} level's acceptance rate are $50\%$, $60\%$ and $70\%$, then the total acceptance is still similarly $21\%$.
Nonetheless, more moves will go to level 2 and 3 and be judged and then be rejected.
This slows down the bisection method.
We have tested the bisection method and find it lack of efficiency in our calculations.
In the bisection method we need to accept or reject using an effective Eq.(\ref{Accptequationsimp}) at each level which involves the calculation of an effective $f^V_{l'r'}(\mathcal{R}')$ and $f^V_{lr}(\mathcal{R})$.
They contain the sum over all the spin and isospin states.
This is different from the situations in most bisection method calculations which mostly deal with central interaction which does not have spin isospin structure.
This spin and isospin sum makes each level's acceptance rate not too much different from each other, just like the $50\%$, $60\%$ and $70\%$ example above.

Therefore, we find the bisection method does not have a computational advantage over multi-level sampling.
As expected, multi-level sampling has a slightly higher acceptance rate than bisection method.
This is because the bisection method can reject early.
For example, say the total acceptance rate is $a=20\%$ for a 2-level sampling. In bisection method it decomposes the acceptance rate $a$ in to $a_1$ for level 1 and $a_2$ for level 2, and theoretically $a=a_1a_2$.
If $a_1$ and $a_2$ are both less than $100\%$ then the acceptance rate of bisection and 2-level sampling are the same, both are $a=20\%$.
But, in bisection it sometimes happens that $a_1=10\%$ and $a_2=200\%$
(when it is bigger than $100\%$ it simply means always accepting the moves),
so theoretically $a_1a_2$ is still $20\%$, but actually since $a_1=10\%$, $90\%$ of the moves have already been rejected at the \nth{1} level.
The real total acceptance is around $10\%$ instead of the $20\%$ for 2-level sampling.
We use multi-level sampling in our calculations, because it allows us to take more samples of new configurations per unit time.

\subsection{Regular Sampling}
\label{secregularsampling}

In section \ref{Sample1bead} and section \ref{secMultibeadSample}, we showed how to sample the free particle propagator for single bead and multiple beads. We proposed the transition probabilities $T_{l'r'}(\mathcal{R} \rightarrow \mathcal{R}') $ according to the free particle propagator part $g^F(\mathcal{R}')$, so we have Eq.(\ref{gtnewgtold}) which is $g^F(\mathcal{R}') T_{lr}(\mathcal{R}' \rightarrow \mathcal{R}) = g^F(\mathcal{R}) T_{l'r'}(\mathcal{R} \rightarrow \mathcal{R}')  $, therefore we can use a simple version of acceptance rate $A(s\rightarrow s')$ which is Eq.(\ref{Accptequationsimp}), to judge if we accept the new state $s'$ or not.
Those moves can simplify the acceptance rate $A(s\rightarrow s')$, but they can make the spatial configurations of involved beads more or less correlated.

So, In order to make the algorithm more robust,
we can add some moves such that the new configuration of each bead does not depend on other beads.
For example, as is used in most Monte Carlo simulations, we can add moves with proposed transition probabilities obeying Eq.(\ref{tlrnewtlrold}) which is $T_{l'r'}(\mathcal{R} \rightarrow \mathcal{R}') = T_{lr}(\mathcal{R}' \rightarrow \mathcal{R}) $, and use Eq.(\ref{Accptequationsimp2}) to judge if we accept the new state $s'$ or not.
In these cases, the common choices for $T_{l'r'}(\mathcal{R} \rightarrow \mathcal{R}')$ can be a constant over an interval or independent gaussian functions. We can randomly select $m$ beads from bead $0$ to bead $N$, $m$ can be at least $1$ and at most $N+1$, we put the label of the $m$ beads into a group call it $\mathcal{M}$, e.g., if we select bead $0$, $2$, $4$, then $\mathcal{M}=\{ 0,2,4 \}$, etc.

The proposed transition probability $T_{l'r'}(\mathcal{R} \rightarrow \mathcal{R}')$ can be written as,
\begin{equation}
T_{l'r'}(\mathcal{R} \rightarrow \mathcal{R}') = \prod_{I  \in \mathcal{M} } \left( \prod_{i=1}^A  T_{I}(\bm{r}_i \rightarrow \bm{r}_i') \right) ,
 \label{}
\end{equation}
where $T_{I}(\bm{r}_i \rightarrow \bm{r}_i') $ is the proposed transition probability for particle $i$ in bead $I$, and it is normalized to 1,
\begin{equation}
\int d \bm{r}_i'  T_{I}(\bm{r}_i \rightarrow \bm{r}_i') =1 .
 \label{}
\end{equation}
There can usually be two choices.

\begin{description}
  \item[Choice 1] For constant $T_{l'r'}(\mathcal{R} \rightarrow \mathcal{R}')$, we can choose $T_{I}(\bm{r}_i \rightarrow \bm{r}_i') $ as,
\begin{equation}
T_{I}(\bm{r}_i \rightarrow \bm{r}_i') = \frac{1}{ (2 \alpha)^3}.
 \label{}
\end{equation}
In this case the new position ${\bm{r}'}_i$ of particle $i$ in a bead can be proposed from its current position ${\bm{r}}_i$ by,
\begin{equation}
  {\bm{r}}'_i =
   {\bm{r}}_i
  + \alpha \bm{u}_i  ,
 \label{uniformmove}
\end{equation}
where $\bm{u}_i=(u_x,u_y,u_z)$ is a 3 dimensional random number and each of its components $u_x$, $u_y$, $u_z$ ranges from $-1$ to $1$ uniformly.
$\alpha$ is a number which controls the step size of proposed move, it can be chosen to make the correspondingly acceptance rate $A(s\rightarrow s')$ around $20 \sim 80\%$.
  \item[Choice 2] For independent gaussian function $T_{l'r'}(\mathcal{R} \rightarrow \mathcal{R}')$, we can choose $T_{I}(\bm{r}_i \rightarrow \bm{r}_i') $ as,
\begin{equation}
T_{I}(\bm{r}_i \rightarrow \bm{r}_i') = \left(\frac{1}{\sqrt{2\pi \sigma^2}}\right)^3 e^{ - \frac{( \bm{r}_i' - \bm{r}_i )^2 }{2 \sigma^2} }.
 \label{}
\end{equation}
In this case the new position ${\bm{r}'}_i$ of particle $i$ in a bead can be proposed from its current position ${\bm{r}}_i$ by,
\begin{equation}
  {\bm{r}}'_i =
   {\bm{r}}_i
  + \sigma \bm{g}_i  ,
 \label{gaussmove}
\end{equation}
where $\bm{g}_i$ is the gaussian random number defined in Eq.(\ref{gaussgi}), and  $\sigma$ is the standard deviation of the gaussian function, which can control the corresponding acceptance rate $A(s\rightarrow s')$.
\end{description}

In the above two choices,
$T_{l'r'}(\mathcal{R} \rightarrow \mathcal{R}') = T_{lr}(\mathcal{R}' \rightarrow \mathcal{R}) $,
and the new configuration $R'_I$ is proposed from the current $R_I$, with a $3A$ dimensional vector $\Lambda_I$ which can be either composed of $\bm{u}_i$ or $\bm{g}_i$ for each particle,
\begin{equation}
 R'_I=R_I+\Lambda_I.
 \label{regsampling1}
\end{equation}

If we propose new configurations for all the beads with the same $3A$ dimensional random vector $\Lambda$ at the same time, namely,
\begin{equation}
 R'_I=R_I+\Lambda,
 \label{shiftmove}
\end{equation}
it will still obey Eq.(\ref{tlrnewtlrold}), $T_{l'r'}(\mathcal{R} \rightarrow \mathcal{R}') = T_{lr}(\mathcal{R}' \rightarrow \mathcal{R}) $.
We can this kind of move as a `shift'.

The regular sampling methods mentioned in this section are useful to guarantee uncorrelated samples but we do not perform them very often.
This is because they are not as efficient as the multi-level sampling. If we just move few beads a time, it is not computationally efficient since we need to update the whole path anyway.
If we move the same number of or more beads than the multi-level sampling method,
we need to use very small $\alpha$, $\sigma$ or $\Lambda$ values in order to achieve similar acceptance rates as the multi-level sampling.
This means they are not able to explore the configuration space as efficiently as multi-level sampling does.

\subsection{Directional Importance Sampling}
\label{secimpsampling}

In this section, we discuss a special case of the importance sampling method.
Sometimes, the proposed transition probability $T(s\rightarrow s')$ has a certain symmetry along $n$ directions in the state space, so there are always $n$ unique states $s'_1,s'_2,...,s'_n$ which share the same proposed transition probability $T(s\rightarrow s'_i)$,
\begin{equation}
T(s\rightarrow s'_1) = T(s\rightarrow s'_2) = ... =T(s\rightarrow s'_n).
 \label{Timpnew}
\end{equation}
So, from the current state $s$, there is the same chance of proposing any of the states from $s'_1$ to $s'_n$.
Similarly, for a given state $s'$, there can be several states $s_1,...,s_n$ which have the same proposed transition probability $T(s'\rightarrow s_i)$. For the reverse proposed transition probability we have,
\begin{equation}
T(s'\rightarrow s_1) = T(s'\rightarrow s_2) = ... = T(s'\rightarrow s_n).
 \label{Timprev}
\end{equation}
The summation of the proposed transition probability alone each direction is the same,
\begin{equation}
\sum_{s_i'}T(s\rightarrow s'_i) = \sum_{s_i}T(s'\rightarrow s_i)
= \frac{1}{n}.
 \label{T1on}
\end{equation}
The total summation of the proposed transition probability is,
\begin{equation}
\!\!\!
\sum_{s'}T(s\rightarrow s') = \sum_i \sum_{s_i'}T(s\rightarrow s'_i) =\sum_{s}T(s'\rightarrow s) = \sum_i \sum_{s_i}T(s'\rightarrow s_i)= 1.
 \label{sumT1on}
\end{equation}

\begin{figure}[hptb!]
\centering
\includegraphics[scale=0.52]{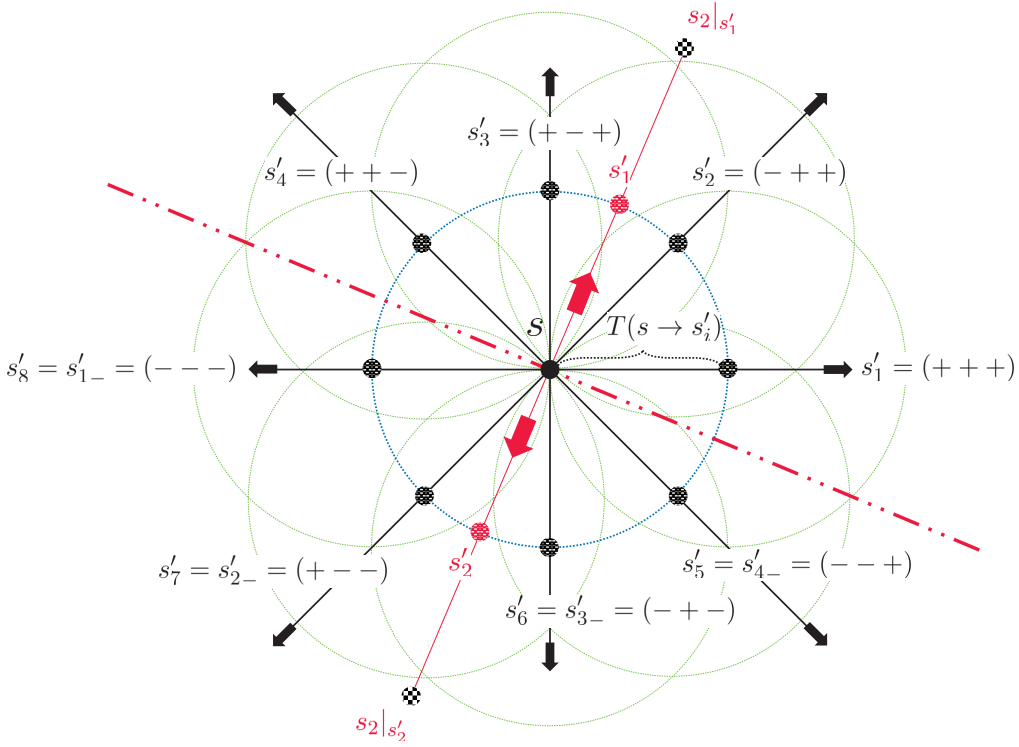}
\caption[Illustration of a State Space with 8-Direction Symmetry]{Illustration of a state space with 8-direction symmetry. }
\label{FigImpnsample}
\end{figure}

For example, Fig. \ref{FigImpnsample} is an illustration for $n=8$ case.
In this case, in the configuration space, a particle located at $\bm r=(x,y,z)$ can be described as in state $s$.
We can propose to move it to a new state $s'$ with $r'=(x',y',z')$ according to,
\begin{equation}
T(s \rightarrow s') = \left(\frac{1}{\sqrt{2\pi \sigma^2}}\right)^3
e^{ - \frac{( x' - x )^2 }{2 \sigma^2} }
e^{ - \frac{( y' - y )^2 }{2 \sigma^2} }
e^{ - \frac{( z' - z )^2 }{2 \sigma^2} }.
 \label{}
\end{equation}
On each axis there are 2 choices which correspond to the same value of $T(s \rightarrow s')$.
They are $x'$ and $x''=2x-x'$, $y'$ and $y''=2y-y'$, $z'$ and $z''=2z-z'$.
That means there are in total $n=2^3=8$ directions and unique states correspond to the same $T(s \rightarrow s')$.
We can label these 8 states from $s'_1$ to $s'_8$.
E.g., set $s'_1$ as the $(x',y',z')$ state, and if $x'>x$, $y'>y$ and $z'>z$, we can simply denote $s'_1 \equiv (+++) $. Set $s'_2$ as the $(x'',y',z')$ state, we can denote $s'_2 \equiv (-++) $, etc. Finally set $s'_8$ as the $(x'',y'',z'')$ state, we can denote $s'_8 \equiv (---) $. So now, $s'_1$ to $s'_8$ simply mean the $n=8$ unique directions in the state space.
Similarly, for each of the 8 possible $s'$ with $r'=(x',y',z')$, there are 8 possible $s$ (denoted by the light circles for each $s'$). Their coordinates are $x$ and $x_-=2x'-x$, $y$ and $y_-=2y'-y$, $z$ and $z_-=2z'-z$.

For the summations over $s'_i$, we have $\sum_{s_i'}T(s\rightarrow s'_i)=1/8$.
For example, the summation over all the $s'_1=(+++)$ states is,
\begin{flalign}
\sum_{s_1'}T(s\rightarrow s'_1) =
\frac{\int_{x}^{+\infty} dx'
\int_{y}^{+\infty} dy'
\int_{z}^{+\infty} dz'
e^{ - \frac{( x' - x )^2 }{2 \sigma^2} }
e^{ - \frac{( y' - y )^2 }{2 \sigma^2} }
e^{ - \frac{( z' - z )^2 }{2 \sigma^2} } }
{\left(\sqrt{2\pi \sigma^2}\right)^3}
= \frac{1}{8}.
\label{sumpppimp}
\end{flalign}
The summation over all the $s'_2=(-++)$ states is $\sum_{s_2'}T(s\rightarrow s'_2)$ which has the same integrand as Eq.(\ref{sumpppimp}) but with a different integral range $\int_{-\infty}^{x} dx
\int_{y}^{+\infty} dy
\int_{z}^{+\infty} dz$, and $\sum_{s_2'}T(s\rightarrow s'_2)$ is also $1/8$.
The summation over all the $s'_8=(---)$ states is $\sum_{s_8'}T(s\rightarrow s'_8)$ which has the same integrand as Eq.(\ref{sumpppimp}) and again with a different integral range $\int_{-\infty}^{x} dx
\int_{-\infty}^{y} dy
\int_{-\infty}^{z} dz$, and $\sum_{s_8'}T(s\rightarrow s'_8)$ is $1/8$, too.
The summation for all the other states are similar and $\sum_{s_i'}T(s\rightarrow s'_i)$ are all $1/8$. Therefore Eq.(\ref{T1on}) and Eq.(\ref{sumT1on}) are valid.

If $T(s\rightarrow s')$ has directional symmetry, we can mark the current state $s$ as $s_i$ because it implicitly carries the directional notation $i$.
When we proposed a new state $s'$ from such a $T(s\rightarrow s')$, then we can temporarily mark $s'$ as $s'_1$.
However, we do not immediately pick this state $s'_1$ as the new trial state. Because from $T(s\rightarrow s')$, each of the $s'_1$ to $s'_n$ states has the same chance to be picked, we want to further pick the `best' state from $s'_1$ to $s'_n$.
We can pick a state $s'_j$ according to $\tilde{\pi}(s'_j)$ such that,
\begin{equation}
\tilde{\pi}(s'_j)=\frac{\pi(s'_j)}{\sum_{k=1}^n \pi(s'_k)} .
\label{imppinewdist}
\end{equation}

This can be called directional importance sampling.
In this way, the correspondingly proposed transition probability for directional importance sampling $T^{\textrm{imp}}(s_i\rightarrow s'_j) $  can be written as,
\begin{equation}
T^{\textrm{imp}}(s_i\rightarrow s'_j)
= \left[ \sum_{j=1}^n T(s_i\rightarrow s'_j)  \right] \tilde{\pi}(s'_j)
= \frac{ n T(s_i\rightarrow s'_j) \pi(s'_j)}{\sum_{k=1}^n \pi(s'_k)}.
 \label{Timpnewsi}
\end{equation}
The reverse directional importance sampling transition probability $T^{\textrm{imp}}(s'_j\rightarrow s_i) $  can be similarly written as,
\begin{equation}
T^{\textrm{imp}}(s'_j\rightarrow s_i)
= \frac{ n T(s'_j\rightarrow s_i) \pi(s_i)}{\sum_{i=1}^n \pi(s_i)},
\label{Timpreverse}
\end{equation}
where states $s_1$ to $s_n$ all have the same proposed transition probability $T(s'_j\rightarrow s_i)$\footnote{
Note that $T^{\textrm{imp}}(s\rightarrow s'_i) $ is a valid proposed transition probability. Because for Eq.(\ref{Timpreverse}), with the help of Eq.(\ref{Timpnew}) and Eq.(\ref{T1on}), we can see that,
\begin{align}
\sum_{s'} T^{\textrm{imp}}(s\rightarrow s') &=  \sum_{i=1}^n \sum_{s'_i} T^{\textrm{imp}}(s\rightarrow s'_i) \nonumber \\
&= n \sum_{i=1}^n \sum_{s'_i}  \frac{T(s\rightarrow s'_i) \pi(s'_i)}{\sum_{j=1}^n \pi(s'_j)} \nonumber \\
&= n  \sum_{s'_i} T(s\rightarrow s'_i) \frac{\sum_{i=1}^n \pi(s'_i)}{\sum_{j=1}^n \pi(s'_j)} \nonumber \\
&= n \sum_{s'_i} T(s \rightarrow s'_i) \nonumber \\
&= 1 . \label{Timpnewnorm}
\end{align}
So $T^{\textrm{imp}}(s\rightarrow s'_i) $ based on the importance sampling algorithm is a valid proposed transition probability because it is non-negative and its summation over all the possible states is 1.
Actually Eq.(\ref{Timpnewnorm}) may not really need to be required. As long as the normalization factors of $T^{\textrm{imp}}$ can be canceled in acceptance rate $A(s \rightarrow s')$ it will be correct.
}.

Therefore after picking $s'_1$ from $T(s\rightarrow s')$ we further pick $s'_j$ by Eq.(\ref{imppinewdist}).
Finally we judge if $s'_j$ is accepted or not by the acceptance probability $A(s_i\rightarrow s'_j)$,
\begin{equation}
\!\!\!\!\!\!\!\!
A(s_i \rightarrow s'_j)=
\min \left[  1, \frac{\pi(s'_j)T^{\textrm{imp}}(s'_j\rightarrow s_i)}{\pi(s_i) T^{\textrm{imp}}(s_i \rightarrow s'_j)} \right]
=  \min \left[  1, \frac{ T(s'_j\rightarrow s_i) \sum \limits_{k=1}^n \pi(s'_k)  }{ T(s_i\rightarrow s'_j) \sum \limits_{j=1}^n \pi(s_j)  } \right] .
\label{Accptequationimp}
\end{equation}
If $T(s_i\rightarrow s'_j) =T(s'_j\rightarrow s_i)  $, Eq.(\ref{Accptequationimp}) simply becomes,
\begin{equation}
A(s_i\rightarrow s'_j)
=  \min \left[  1, \frac{\sum_{j=1}^n \pi(s'_j)}{\sum_{i=1}^n \pi(s_i)} \right] .
\label{Accptequationimp2}
\end{equation}

If there are $s'_j$ with bigger $\pi(s'_j)$ than the $\pi(s'_1)$ from the initially chosen $s'_1$  by $T(s_i\rightarrow s'_1)$, then the directional importance sampling acceptance rate of Eq.(\ref{Accptequationimp2}) can be higher than acceptance rate from gaussian sampling of Eq.(\ref{Accptequationsimp}) or regular sample of Eq.(\ref{Accptequationsimp2}).
In other words, with this additional importance sampling, we can move more beads in gaussian sampling with a larger step size in regular sampling than those without importance sampling, with the same acceptance rate as without importance sampling.
Adding importance sampling can enable us to explore the configuration space more efficiently.
The tradeoff is that we need to evaluate all the relevant $\pi(s'_j)$ and  $\pi(s_i)$ which can be time-consuming.

As long as the number $n$ is an even number,
and each state $s'_i$ can find its opposite direction among the $n$ directions,
it can be reduced to a simple 2-directional case. We can pick $n/2$ directions and combine them into one generalized direction, and combine the rest $n/2$ opposite directions into the other generalized direction.
Once we picked $s'_1$ from one direction, there is always a state $s'_2$ in exactly the opposite direction.
For example, in the $n=8$ case illustrated by Fig. \ref{FigImpnsample}, as indicated by the red arrows and lines, we can denote $s$ as $s_1$, and we group $s'_1$ to $s'_4$ into one generalized direction.
And, since $s'_5$ is at the opposite direction of $s'_4$, $s'_6$ is the opposite of $s'_3$, $s'_7$ is the opposite of $s'_2$, and $s'_8$ is the opposite of $s'_1$, so we group $s'_5$ to $s'_8$ into the other generalized direction.
By doing this, we reduced the $n=8$ case into an $n=2$ case.
Every time we proposed $s'_1$, we take $s'_2$ in the opposite direction of $s'_1$ as the only alternative state.
We can just mark $s'_2$ as ${s'_1}_{-}$.
For the chosen $s'_1$ or $s'_2$, there is a corresponding $s_2$ such that $T(s'_i\rightarrow s_1)=T(s'_i\rightarrow s_2)$.
In our PIMC, usually one can find an `opposite direction' of $\mathcal{R}'$ which can be marked as $\mathcal{R}''$ such that $T_{l'r'}(\mathcal{R} \rightarrow \mathcal{R}')=T_{l'r'}(\mathcal{R} \rightarrow \mathcal{R}'')$, and then we can use directional importance sampling to further pick either $\mathcal{R}'$ or $\mathcal{R}''$.

\begin{figure}[hptb!]
\centering
\includegraphics[scale=0.6]{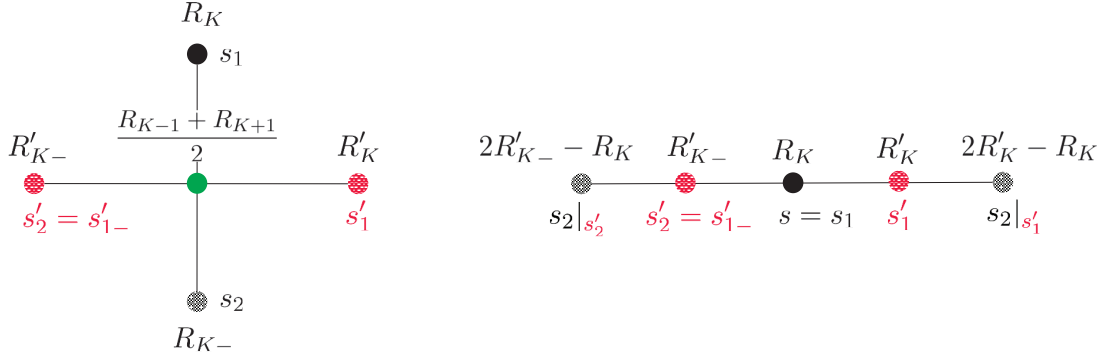}
\caption[Illustration of 2-Direction Importance Samplings]{Illustration of 2-direction importance samplings. The left one is for single bead sampling plus 2-direction importance sampling. The right one is for regular sampling plus 2-direction importance sampling.}
\label{FigImpnsample2D}
\end{figure}

As a concrete example, we can apply the $n=2$ importance sampling to the single bead sampling in Sections \ref{Sample1bead}.
The algorithm is described as follows.
\begin{description}
  \item[Step 1] For the current state $s$, randomly pick bead $K$, we propose to move its configuration from $R_K$ to $R'_K$ according to Eq.(\ref{Tlrnew}).
  \item[Step 2] It is 3 dimensional space and the particle number is $A$, each particle along one dimension has 2 different choices of coordinates, so there are $2^{3A}$ different $R'_K$ which give the same value of Eq.(\ref{Tlrnew}).
      For any proposed $R'_K=\frac{R_{K-1}+R_{K+1}}{2}+\Delta R$, we only pick ${R'_K}_{-}=\frac{R_{K-1}+R_{K+1}}{2}-\Delta R=R_{K-1}+R_{K+1}-R'_K$,
      because it not only corresponds to the same value of Eq.(\ref{Tlrnew}) but also the opposite direction of $R'_K$. Call the state with total configuration $\mathcal{R'}=\{R_1,..,R'_K,...,R_N\}$ as $s'_1$, and call the state with  $\mathcal{R'}_{-}=\{R_1,..,{R'_K}_{-},...,R_N\}$ as $s'_2$ or ${s'_1}_{-}$. We now calculate,
\begin{flalign}
\frac{\pi(s'_1)}{\pi(s'_1)+\pi({s'_1}_{-})} & = \frac{ f^V_{l'r'}(\mathcal{R}') g^F(\mathcal{R}')   }{ f^V_{l'r'}(\mathcal{R}') g^F(\mathcal{R}')
+ f^V_{l'r'}(\mathcal{R}'_{-}) g^F(\mathcal{R}'_{-})   } \nonumber \\
& = \frac{ f^V_{l'r'}(\mathcal{R}')   }{ f^V_{l'r'}(\mathcal{R}') + f^V_{l'r'}(\mathcal{R}'_{-})  } .
\label{imp1beadexample}
\end{flalign}
Generate a random number $a$ ranging uniformly from 0 to 1. If $a \leq $ Eq.(\ref{imp1beadexample}) we choose $R'_K$ which means $s'_1$, otherwise we choose ${R'_K}_{-}$ which means $s'_2$.

\item[Step 3]  Denote the original state $s$ as $s_1$. According to Eq.(\ref{Tlrold}), no matter which $s'_i$ is selected, there are only two choices for $s_1$ and $s_2$. Since we denote the original state $s$ with $R_K$ as $s_1$, $s_2$ must include a another choice of $R_K$ call it ${R_K}_{-}$ such that $\left({R_{K}}-\frac{R_{K\!-\!1}+R_{K\!+\!1}}{2} \right)^2=\left({R_{K-}}-\frac{R_{K\!-\!1}+R_{K\!+\!1}}{2} \right)^2$.
Similar with ${R'_K}_{-}$, we have $R_K-\frac{R_{K-1}+R_{K+1}}{2}$ = $\frac{R_{K-1}+R_{K+1}}{2}-{R_K}_{-}$ for ${R_K}_{-}$.
We denote ${R_K}_{-}=R_{K-1}+R_{K+1}-R_K $ as the opposite position of $R_K$, and we denote $\mathcal{R}_{-}=\{R_1,..,{R_K}_{-},...,R_N\}$ as $s_2$ or ${s_1}_{-}$.
The left part of Fig. \ref{FigImpnsample2D} illustrates the $R_K$, $R_{K-}$, $R'_{K-}$, and $R'_{K-}$.
For the picked new state $s'_i$, use Eq.(\ref{Accptequationimp}) to calculate the acceptance probability $A(s_i\rightarrow s'_i)$ and we find that
\begin{flalign}
A(s_1\rightarrow s'_i) \! &= \! \min \! \left[
 1,
 \frac{ T(s'_i\rightarrow s_1) \sum \limits_{j=1}^2 \pi(s'_j)  }{ T(s_1\rightarrow s'_i) \sum \limits_{j=1}^2 \pi(s_j)  }
 \right]  \nonumber \\
\! &= \! \min \! \left \{ \!
 1,
 \frac{ e^{  -\frac{ \left({R_{K}}-\frac{R_{K\!-\!1}+R_{K\!+\!1}}{2} \right)^2 }{ \Delta \tau  \frac{\hbar^2}{m}  }} \! \! \left[f^V_{l'r'}(\mathcal{R}') g^F(\mathcal{R}') + f^V_{l'r'}(\mathcal{R}'_{-}) g^F(\mathcal{R}'_{-}) \right]  }
 { e^{  -\frac{ \left({R_{K}'}-\frac{R_{K\!-\!1}+R_{K\!+\!1}}{2} \right)^2 }{ \Delta \tau  \frac{\hbar^2}{m}  }} \! \! \left[f^V_{lr}(\mathcal{R}) g^F(\mathcal{R}) + f^V_{lr}(\mathcal{R}_{-}) g^F(\mathcal{R}_{-}) \right]  } \!
 \right \} \nonumber \\
\! &= \! \min \! \left \{
 1,
 \frac{f^V_{l'r'}(\mathcal{R}')+ f^V_{l'r'}(\mathcal{R}'_{-}) }{f^V_{lr}(\mathcal{R})+f^V_{lr}(\mathcal{R}_{-})}
 \right \} .
\end{flalign}

  \item[Step 4] Generate a uniform random number $b$ which ranges from 0 to 1. If $b \leq A(s_1\rightarrow s'_i)$, the new state $s'_i$ is accepted, so set it as the current state $s$, namely $s = s'_i$, then go to step 1 and continue. Else, the new state $s'_i$ is rejected (in other words the new state $s'_i$ is just the current state $s$), then go to step 1 and continue.
\end{description}

The importance sampling can also be applied on the regular single bead sampling in sections \ref{secregularsampling}.
The algorithm is very similar.
The difference is that, in step 1, we propose a move for the selected bead $K$ according to Eq.(\ref{regsampling1}).
In step 2,
$R'_K-R_K=R_K-{R'_K}_{-}$ so ${R'_K}_{-}=2R_{K}-R'_K$, and denote the new state with $R'_K$ as $s'_1$, with ${R'_K}_{-}$ as $s'_2$.
In step 3, denote the current state as $s_1$, if $s'_1$ is chosen,
$R_K-R'_K=R'_K-{R_K}_{-}$ so ${R_K}_{-}=2R'_{K}-R_K$.
If $s'_2$ is chosen,
$R_K-{R'_K}_{-}={R'_K}_{-}-{R_K}_{-}$ so ${R_K}_{-}=2{R'_{K}}_{-}-R_K$.
Based on the chosen $s'_1$ or $s'_2$, denote the corresponding ${R_K}_{-}$ as $s_2$.
The right part of Fig. \ref{FigImpnsample2D} illustrates the relevant states and the coordinates.
Since $T(s_1\rightarrow s'_i) =T(s'_i\rightarrow s_1)  $, the $A(s_1\rightarrow s'_i)$ is simply Eq.(\ref{Accptequationimp2}) and we have,
\begin{eqnarray}
\!\!\!\!\!\!\!\! A(s_1\rightarrow s'_i) \!\!\!\! &=& \!\!\!\! \min \left[
 1,
 \frac{ \sum \limits_{j=1}^2 \pi(s'_j)  }{  \sum \limits_{j=1}^2 \pi(s_j)  }
 \right]  \nonumber \\
 &=& \!\!\!\! \min \left \{
 1,
 \frac{ f^V_{l'r'}(\mathcal{R}') g^F(\mathcal{R}') + f^V_{l'r'}(\mathcal{R}'_{-}) g^F(\mathcal{R}'_{-}) }
 { f^V_{lr}(\mathcal{R}) g^F(\mathcal{R}) + f^V_{lr}(\mathcal{R}_{-}) g^F(\mathcal{R}_{-})  }
 \right \} \nonumber \\
&=& \!\!\!\! \min \left \{
 1,
 \frac{f^V_{l'r'}(\mathcal{R}') G^f_{K-1, K'} G^f_{K', K+1}
 + f^V_{l'r'}(\mathcal{R}'_{-}) G^f_{K-1, K'_{-}} G^f_{K'_{-}, K+1}  }
 {f^V_{lr}(\mathcal{R}) G^f_{K-1, K} G^f_{K, K+1}
 +f^V_{lr}(\mathcal{R}_{-}) G^f_{K-1, K_{-}} G^f_{K_{-}, K+1}  }
 \right \} . ~~~
\end{eqnarray}
The $K_{-}$ or $K'_{-}$ appear at the $I$ or $J$ subscript in $G^f_{I,J}$ means ${R_K}_{-}$ or ${R_K}'_{-}$ in the corresponding place in Eq.(\ref{Gfij}).

\section{Path Updating Strategy}
\label{secUpdateStrat}

\begin{figure}[hptb!]
\centering
\includegraphics[width=\textwidth]{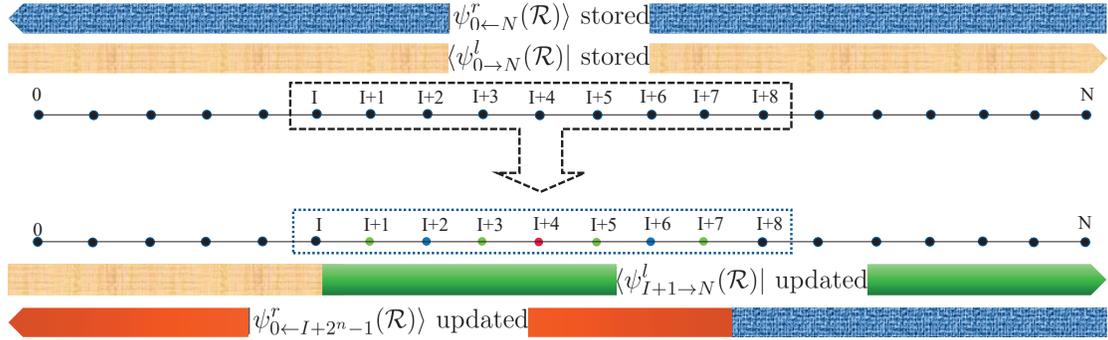}
\caption[Illustration of the Path Updating]{Illustration of the path updating.}
\label{PIMCpathstore}
\end{figure}

In each type of sampling, after we judge whether to accept the new $l$ and $r$ order in the trial wave function or not, we mark the order after judgement as $l$ and $r$.
In order to make the computation fast, as illustrated in Fig. \ref{PIMCpathstore} by the blue and yellow bands with texture, we store all the states from $| \psi^r_0 (\mathcal{R}) \rangle$ to $| \psi^r_N (\mathcal{R}) \rangle$ and $| \psi^l_0 (\mathcal{R}) \rangle$ to $| \psi^l_N (\mathcal{R}) \rangle$, by storing all the coefficients from $\phi^r_{0}(\mathcal{R},S)$ to $\phi^r_{N}(\mathcal{R},S) $ and from $\phi^l_{0}(\mathcal{R},S)$ to $\phi^l_{N}(\mathcal{R},S) | S \rangle$ in the arrays.
We also store $f^V_{lr}(\mathcal{R})$ which is calculated by Eq.(\ref{fVRlrN2LOdef2}).

Then, we propose new moves as $\mathcal{R}'$. In order to judge if the new spatial configuration $\mathcal{R}'$ is accepted or not, we need to calculate ${ f^V_{lr}(\mathcal{R}')}/{ f^V_{lr}(\mathcal{R})}$.
For the $\mathcal{R}'$, it is usually the same as $\mathcal{R}$ except for several contiguous beads.
For example in the multi-level sampling with level $n$, for a chosen integer $I$, we only propose new configurations from bead $I+1$ to bead $I+2^n-1$ in $\mathcal{R}'$, and the rest of the beads remain the same as those in $\mathcal{R}$.
When we propose new moves from bead $I+1$ to bead $I+2^n-1$,
we only need to calculate from right to left, from $\phi^r_{I+2^n-1}(\mathcal{R}',S)$ until $\phi^r_{I+1}(\mathcal{R}',S)$,
by using the recursion relations Eq.(\ref{phirMRrecur}) starting from $\phi^r_{I+2^n}(\mathcal{R},S)$ which is already stored.
Then we calculate
$f^V_{lr}(\mathcal{R}')= \left\langle \psi^{l'}_I (\mathcal{R}) \left| \bm{\mathcal{G}}\right| \psi^{r'}_{I+1} (\mathcal{R}') \right\rangle $,
and since $ f^V_{lr}(\mathcal{R})$ is also already stored, we can readily calculate ${ f^V_{l'r'}(\mathcal{R}')}/{ f^V_{lr}(\mathcal{R})}$.
If the proposed new moves are accepted, we update all the rest beads.
From right to left we update from $\phi^r_{I}(\mathcal{R}',S)$ until $\phi^r_{0}(\mathcal{R}',S)$, and from left to right we update from $\phi^r_{I+1}(\mathcal{R}',S)$ until $\phi^r_{N}(\mathcal{R}',S)$.
In Fig. \ref{PIMCpathstore}, the corresponding updated parts of the path from left to right are marked as green band, and those updated from right to left are marked as the red band.

The path updating process is the most time-consuming part, especially if the number of beads $N$ is large.
So we find a proper $n$ in multi-level sampling to make sure $n$ is big enough to ensure the acceptance rate is relatively low,
in order to optimize the decorrelation of the path within the computer time.

\section{Sampling Strategies}
\label{secSampStrat}

We have tested all the algorithms described in this chapter.
We have also tested the bisection method (\cite{Ceperley95a}), as well as the reptation Monte Carlo method (\cite{Baroni99a,Baroni10a}) for our calculation.
For our light nuclei PIMC calculation, as pointed out in section  \ref{secMultibeadSample}, we did not find any advantage for bisection over the multi-level sampling method.
The reptation method basically samples the spatial configuration of the path by adding some beads at one end of the path,
and removing some beads to the other end of the path.
However, since we have spin and isospin sums in the path, the removing and adding of beads, even if just one bead is involved, requires the recalculation of the whole path $f^V_{l'r'}(\mathcal{R}')$.
Therefore it is not economical to perform reptation moves.
After initial testing we did not pursue reptation moves further.

The multi-level sampling is efficient so we do not do regular sampling described in section \ref{secregularsampling} very often.
The importance sampling also does not make much difference.
Considering that the importance sampling at least needs to calculate the two possibilities $\pi(s'_1)$ and $\pi(s'_2)$ and an additional $\pi(s_2)$, we did not find it particularly increase the efficiency. Again we turned off the importance sampling in our calculations.
So when performing our calculations, about $90\%$ moves are proposed by multi-level sampling described in section \ref{secMultibeadSample}, and the other $10\%$ are proposed by different kinds of regular sampling described in section \ref{secregularsampling}.

For multi-level sampling, for the total $N+1$ beads along the path, we pick two lengths, $n_1 \approx N/3$ and $n_2 \approx N/6$.
And, $80\%$ of the moves are proposed for $n_1$ beads and the others are for $n_2$ beads.
The acceptance rate for sampling $n_1$ beads is about $20\%$, for $n_2$ beads is about $40\%$.
The reason for sampling $n_2$ beads is mainly to include the \nth{0} and the $N^{\text{th}}$ beads.
For example, if we have a total of 400 beads, we can pick $n_1=2^7=128$ and $n_2=2^6=64$. $80\%$ moves are proposed for the $n_1=128$ beads with acceptance rate around $20\%$,
and the other $20\%$ of the moves among the total multi-level samplings are for $n_2=64$ beads with an acceptance rate around $40\%$.

For the different kinds of regular sampling which make the samples more uncorrelated than applying multi-level sampling alone, we include moving beads one by one, moving all the beads at the same time, and shifting all the beads at the same time, according to section \ref{secregularsampling}.
By moving beads one by one, we randomly pick a direction, either propose moves from bead $N$ to bead $0$, or propose moves from bead $0$ to bead $N$.
For each bead, we move each particle in it and we choose the step parameter $\alpha$ to make the acceptance rate around $50\%$.
By moving all the beads at the same time, we move all the particles in each bead, the step parameter $\alpha$ needs to be smaller than when moving beads one by one, in order to make the acceptance rate of moving all the beads at the same time around $50\%$ again.
By shifting all the beads at the same time, we propose to move each of the beads by using Eq.(\ref{shiftmove}).
The step parameter in the common $\Lambda$ vector is again adjusted to make the acceptance rate of shifting all the beads around $50\%$.
The overall efficiency of the algorithm is not substantially affected by these acceptance rates as long as they are in the 20\% to 80\% range.

\chapter{Ground state calculations of light nuclei}
\label{RsDs}

\section{Ground State Energies}
\label{secGSE}

\subsection{Preparations}
\label{setup}

The exact PIMC expectation value of operator $\hat{O}$ is calculated by Eq.(\ref{PIMCopexpdetail}) and therefore Eq.(\ref{PIMCopexpdetailshorttime}), and we need to choose a proper total imaginary time $\tau$ and time step $\Delta \tau$ to project out the ground state $|\Phi_0\rangle$.
In this section, we discuss the choice of $\tau$ and $\Delta \tau$ and use $^4$He as an example.

From Eq.(\ref{eHPsiT}) we can see that, the residual amount of excited states depends on the ratio $a_i/a_0$ and the factor $e^{-(E_i-E_0)\tau_1}$.
We have to make sure that the coefficient of the first excited state $| \Phi_1 \rangle$ is negligible compared with the ground state $| \Phi_0 \rangle$.

The experimental ground state energy $E_0$ of $^4$He is about -28.3 MeV and the first excited state energy $E_1$ is the energy of a separated nucleon and triton, -8 MeV (\cite{Tiley92a}).
The energy difference $E_1-E_0$ is about -20 MeV.
If we choose $\tau_1=0.1 \textrm{ MeV}^{-1}$ then $e^{-(E_1-E_0)\tau_1}$ will be $e^{-2}\approx 0.135$.
If there is only $|\Phi_0\rangle$ and $|\Psi_1\rangle$ in $|\Psi_T\rangle$, and if $|a_1/a_0|$ is small enough,
then the percentage of the first excited state $|\Phi_1 \rangle$ in $|\Psi_T \rangle$ will be greatly suppressed by $|a_1/a_0 e^{-(E_1-E_0)\tau_1}|$,
therefore $e^{-H \tau_1} |\Psi_T \rangle$ is dominated by the ground state $|\Phi_0 \rangle$.
In fact, if there are other excited states in $|\Psi_T \rangle$, the ground state $|\Phi_0 \rangle$ will be projected out even sooner, because $|a_1/a_0|$ will be smaller for a good trial wave function and other excited states $|\Phi_i\rangle$ will die out faster than $|\Phi_1\rangle$ due to the factor $|a_i/a_0 e^{-(E_i-E_0)\tau_1}|$.
So, even a less than $0.1 \textrm{ MeV}^{-1}$ imaginary time $\tau_1$ may still be adequate.

The analysis above suggests that $\tau_1=0.1 \textrm{ MeV}^{-1}$ should be enough for our PIMC to project out the ground state.
Since we usually put operator $\hat{O}$ in the middle of two propagators as $\langle \Psi_T| e^{-H \tau_1}  \hat{O} e^{-H \tau_2} | \Psi_T \rangle $ as shown in Eq.(\ref{PIMCopexpdetail}), and we usually set $\tau_1=\tau_2$, so we can conclude that the total imaginary time $\tau$ for our PIMC calculations can be set as $\tau=2\tau_1=2\times 0.1=0.2 \textrm{ MeV}^{-1}$.
In our PIMC calculations, besides using $\tau=0.2 \textrm{ MeV}^{-1}$, and we also show results with $\tau=0.15 \textrm{ MeV}^{-1}$ and $\tau=0.3 \textrm{ MeV}^{-1}$ for comparisons.
The total imaginary time for chiral N$^2$LO interaction has a similar range and $\tau=0.2 \textrm{ MeV}^{-1}$ should be enough as well.

Given a total imaginary time $\tau$, for the time step $\Delta \tau$, we need to consider the computational cost and the accuracy.
In GFMC (\cite{Carlson87a}), $\Delta \tau$ is usually chosen to be $10^{-5} \sim 10^{-4} \textrm{ MeV}^{-1}$.
In our PIMC calculation, we find that a $10^{-4} \sim 10^{-3} \textrm{ MeV}^{-1}$ time step $\Delta \tau$ works reasonably well.
Depending on $\tau$ and $\Delta \tau$, the number of beads for PIMC ranges from 41 to 1501.
In the most time consuming case, for $\tau=0.3 \textrm{ MeV}^{-1}$ and $\Delta \tau=2\times 10^{-4} \textrm{ MeV}^{-1}$, we need 1500 short-time propagators and therefore 1501 beads.
Such a calculation (all the calculations are performed at the Agave Cluster at Research Computing at Arizona State University),
in order to reach a less than $0.02\textrm{ MeV}$ error in the ground state energy calculation, takes about 20 thousand core hours for the chiral N$^2$LO interaction and 13 thousand core hours for the AV6' interaction.
For 41 beads it takes about 240 core hours to reach less than $0.02\textrm{ MeV}$ error, which means it can be done on a modern laptop in few days or less.

\subsection{Calculations}
\label{secResultCalc}
The calculation of the ground state energy $E_0$ can be recast into the form of Eq.(\ref{OABP}) as discussed in section \ref{Recast}, with the operator $\hat{O}$ being the Hamiltonian $H=T+V$. And we calculate the ground state energy $E_0$ by
\begin{equation}
E_0 (\Delta \tau)
= \frac{
\frac{1}{2}  Re  \left\{  \langle \Psi_T| H [U(\Delta \tau)]^N  | \Psi_T \rangle
 +
 \langle \Psi_T|  [U(\Delta \tau)]^N H | \Psi_T \rangle
 \right\}
 }
{  Re \langle \Psi_T| [U(\Delta \tau)]^N  | \Psi_T \rangle} ,
\label{PIMCopexpdetailE0dtau}
\end{equation}
which in the limit of $\Delta \tau=0$ becomes the true expectation $E_0$ as shown in Eq.(\ref{PIMCopexpdetailOH}),
\begin{equation}
E_0
= \frac{
\frac{1}{2}  Re  \left(  \langle \Psi_T| H e^{-H \tau}  | \Psi_T \rangle
 +
 \langle \Psi_T|  e^{-H \tau} H | \Psi_T \rangle
 \right)
 }
{  Re \langle \Psi_T| e^{-H \tau} | \Psi_T \rangle} .
\label{PIMCopexpdetailE0}
\end{equation}
The the corresponding $g(\mathcal{R})_{lr}$ takes the form of Eq.(\ref{gRlrOH}) which can be written as,
\begin{equation}
  g(\mathcal{R})_{lr} = \frac{1}{2}
\left[
\sum\limits_{S} \langle \psi^l_T| H | R_0 S\rangle \langle S | \psi^r_0 (\mathcal{R}) \rangle
+ \sum\limits_{S} \langle \psi^l_N (\mathcal{R}) | S \rangle \langle R_N S| H | \psi^r_T \rangle
 \right]
g^F(\mathcal{R}),
\end{equation}
and the corresponding $g^V_{lr,M}(\mathcal{R})$ is,
\begin{align}
g^V_{lr,M}(\mathcal{R}) =
\begin{cases}
\sum\limits_{S} \langle \psi^l_T| H | R_0 S\rangle \langle S | \psi^r_0 (\mathcal{R}) \rangle,
 &  M=0. \\
\sum\limits_{S} \langle \psi^l_N (\mathcal{R}) | S \rangle \langle R_N S| H | \psi^r_T \rangle,
 & M=N .
\end{cases}
\label{gVRlrHdef}
\end{align}
All the other functions are the same as discussed in section \ref{Recast}.

The calculations of $\langle R S| H | \psi^r_T \rangle$ and $\langle \psi^l_T| H | R S\rangle$ are straightforward. First, they can be calculated as,
\begin{align}
\langle R S| H | \psi^r_T \rangle &= \langle R S| T | \psi^r_T \rangle + \langle R S| V | \psi^r_T \rangle,  \\
\langle \psi^l_T| H | R S\rangle &= \langle \psi^l_T| T | R S\rangle + \langle \psi^l_T| V | R S\rangle.
\label{RSHPsiTrcalc}
\end{align}
Then, with the help of Eqs.(\ref{PER}) and (\ref{RSPsiTr}), taking $^4$He as an example, the potential energy part becomes,
\begin{eqnarray}
\resizebox{.9\hsize}{!}{$
\!\!\!\!\!\!\!\!\! \langle R S| V | \psi^r_T \rangle \!=\!
\left\langle \!\! S \! \left| \left\{\! \sum\limits_{i<j} \left[V_{ij}^{NN}(r_{ij}) \!+\! V^{EM}_{ij}(r_{ij}) \right] \!\! \right\}
\! \! \sideset{}{^r} \prod\limits_{i<j} \! \!
\left[
\sum\limits_{p=1}^6   f_{ij}^p (r_{ij})  O_{ij}^p\!
\right]
\! \sum\limits_{N=1}^{96} \! \phi_N
\right| \! N \!\! \right\rangle$,} \\
\resizebox{.938\hsize}{!}{ $
\!\!\! \langle \psi^l_T| V | R S\rangle \!=\!
\! \sum \limits_{N=1}^{96}  \!\! \left\langle \!\! N \! \left| \phi^*_N \!
 \left\{\! \sum\limits_{i<j} \left[V_{ij}^{NN}(r_{ij}) \!+\! V^{EM}_{ij}(r_{ij}) \right]\!\! \right\}
\!\! \sideset{}{^l} \prod\limits_{i<j}\!
\left[
\sum\limits_{p=1}^6 {f_{ij}^p}^* (r_{ij})  O_{ij}^p\!
\right] \!
\right| \! S \! \right\rangle$.}
\end{eqnarray}

For the kinetic energy operator $T$ we have,
\begin{align}
\resizebox{.87222\hsize}{!}{ $
\!\!\!\langle R S| T | \psi^r_T \rangle \!=\!\!\sum \limits_{i=1}^4 \! \frac{-\hbar^2\nabla_i^2}{2m}
\langle R S | \psi^r_T \rangle
\!=\! -\frac{\hbar^2}{2m} \!\sum \limits_{i=1}^4 \!\! \left( \frac{d^2 }{dx_i^2} \!+\! \frac{d^2 }{dy_i^2} \!+\! \frac{d^2 }{dz_i^2} \right) \!\!
\langle R S | \psi^r_T \rangle$,}
\label{KE2ndrorder} \\
\resizebox{.87222\hsize}{!}{ $
\!\!\!\langle \psi^l_T  | T | R S \rangle \!=\!\!\sum \limits_{i=1}^4\! \frac{-\hbar^2\nabla_i^2}{2m}
\langle \psi^l_T | RS \rangle
\!=\! -\frac{\hbar^2}{2m} \!\sum \limits_{i=1}^4\!\! \left( \frac{d^2 }{dx_i^2} \!+\! \frac{d^2 }{dy_i^2} \!+\! \frac{d^2 }{dz_i^2} \right) \!\!
\langle \psi^l_T | RS \rangle .$}
\label{KE2ndlorder}
\end{align}
The \nth{2} order derivatives in Eqs.(\ref{KE2ndrorder}) and (\ref{KE2ndlorder}) are calculated numerically by using the \nth{2} order central finite difference formula (\cite{Fornberg88a})\footnote{For example, for $x_i$ the x coordinate of particle i, it is calculated as,
\begin{align}
\frac{d^2 \langle R S | \psi^r_T \rangle }{d x_i^2}  &=
\frac{  \langle R_{x_i^+} S | \psi^r_T \rangle
\! + \! \langle R_{x_i^-} S | \psi^r_T \rangle
\! - \! 2 \langle R S | \psi^r_T \rangle    }{\Delta^2}, \\
\frac{d^2 \langle \psi^l_T  | RS \rangle }{d x_i^2}  &=
\frac{  \langle \psi^l_T  | R_{x_i^+} S \rangle
\! + \! \langle \psi^l_T  |  R_{x_i^-} S  \rangle
\! - \! 2 \langle \psi^l_T  | RS \rangle    }{\Delta^2} .
\end{align}
where $\Delta$ is the spatial finite difference which is set as about $10^{-2}$ fm in our calculations. $R_{x_i^+}$ means keeping all the other coordinates in the configuration of the system $R$ the same and just change $x_i$ to $x_i+\Delta$, and $R_{x_i^-}$ means change $x_i$ to $x_i-\Delta$.}.

In our PIMC calculations for $^4$He, we include 2000 steps in one block.
A step means any one of the trial moves described in section \ref{secSamplingMethod}, it can be either accepted or rejected.
We adjusted the number of beads for multi-level sampling as well as the step parameters for regular sampling, and find out the samples are uncorrelated after 125 steps so that each block is uncorrelated with the others (we make sure the samples of the root mean square radius $r_m$ at the middle bead are uncorrelated). We typically calculate the ground-state energy $E_0$, $r_m^2$, potential $V^{NN}+V^{EM}$ along the beads, and one-particle number density $\rho(r)$. We begin to take samples after running several blocks from the initial condition in order to reach the equilibrium.

\begin{figure}[htbp!]
\centering
\includegraphics[scale=0.46]{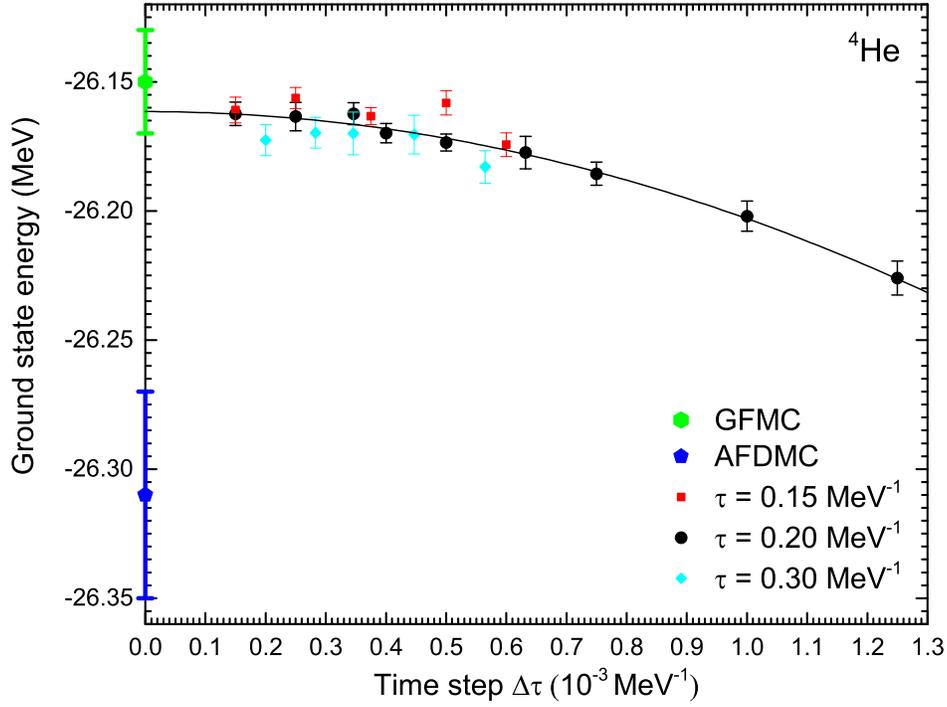}
\caption[Energy Vs. Time Step $\Delta \tau$ for Different Total Imaginary Time $\tau$ for $^4$He Based on the AV6' Potential (Quadratic Range)]{Energy vs. time step $\Delta \tau$ for different total imaginary time $\tau$ for $^4$He based on the AV6' potential (quadratic range). }
\label{energy}
\end{figure}

\begin{figure}[htbp!]
\centering
\includegraphics[scale=0.46]{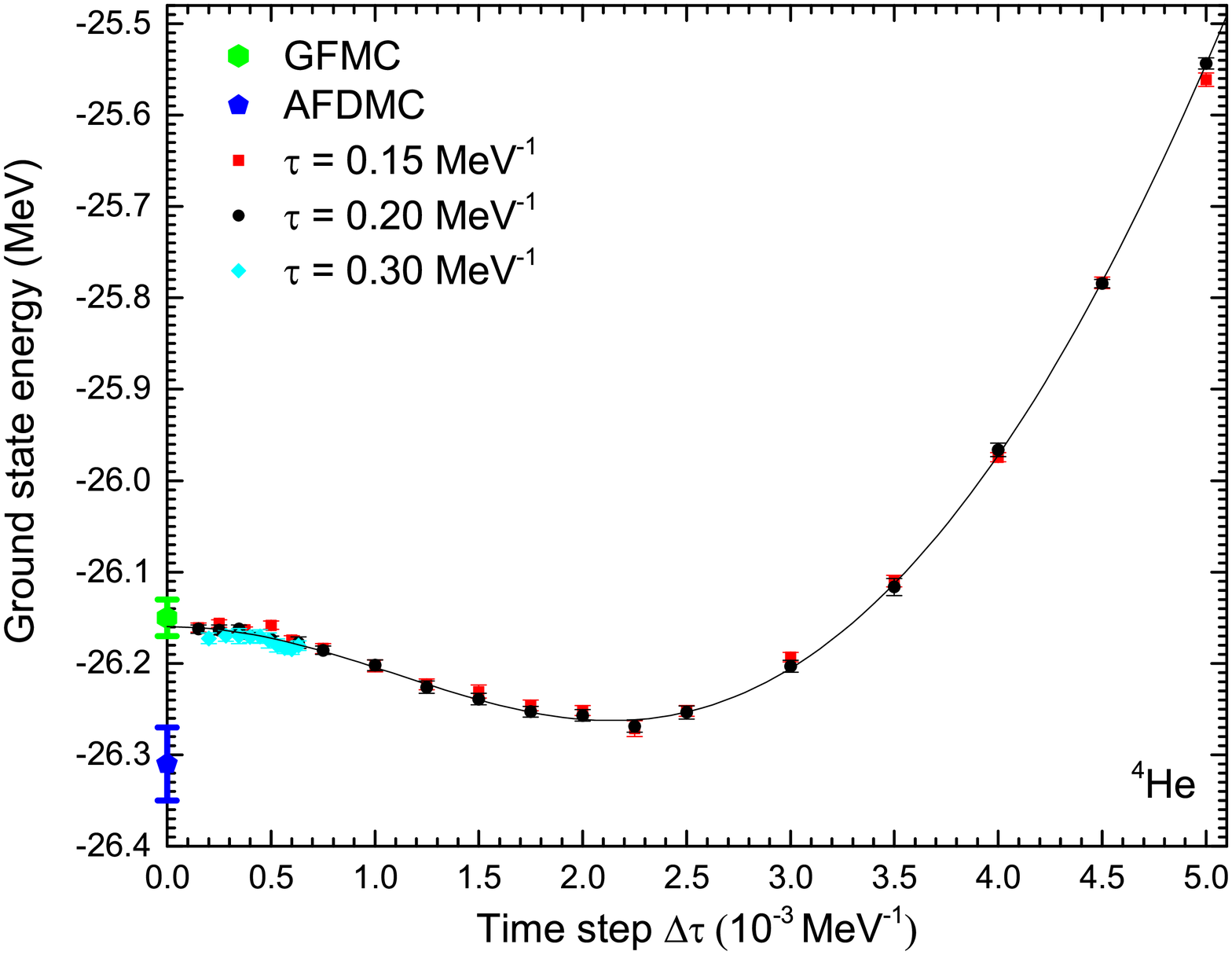}
\caption[Energy Vs. Time Step $\Delta \tau$ for Different Total Imaginary Time $\tau$ for $^4$He Based on the AV6' Potential (Wide Range)]{Energy vs. time step $\Delta \tau$ for different total imaginary time $\tau$ for $^4$He based on the AV6' potential (wide range). }
\label{energyall}
\end{figure}

\begin{figure}[htbp!]
\centering
\includegraphics[scale=0.46]{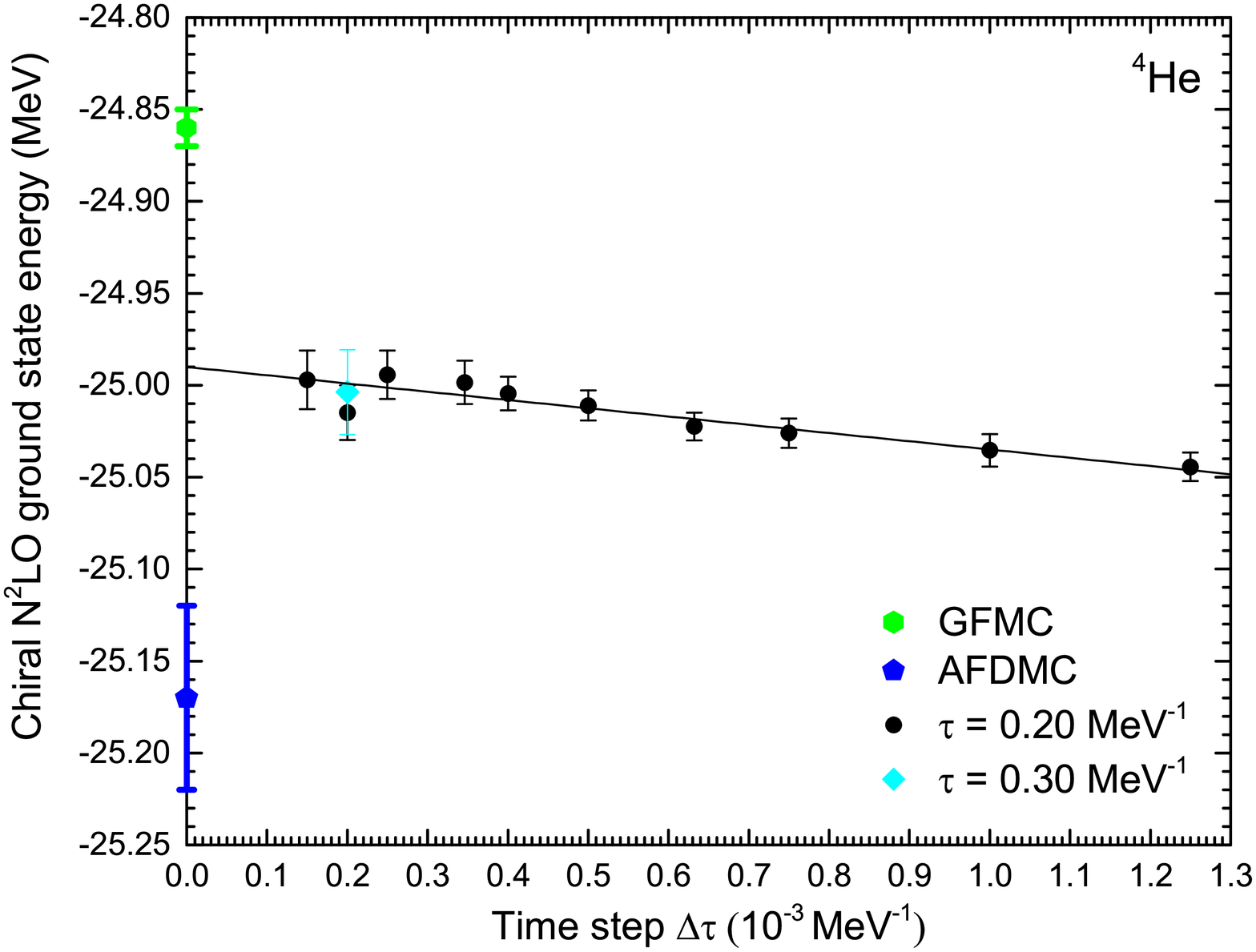}
\caption[Energy Vs. Time Step $\Delta \tau$ for Different Total Imaginary Time $\tau$ for $^4$He Based on the Local Chiral N$^2$LO Interaction (Linear Range)]{Energy vs. time step $\Delta \tau$ for different total imaginary time $\tau$ for $^4$He based on the local chiral N$^2$LO interaction (linear range). }
\label{energyN2LO}
\end{figure}

\begin{figure}[htbp!]
\centering
\includegraphics[scale=0.46]{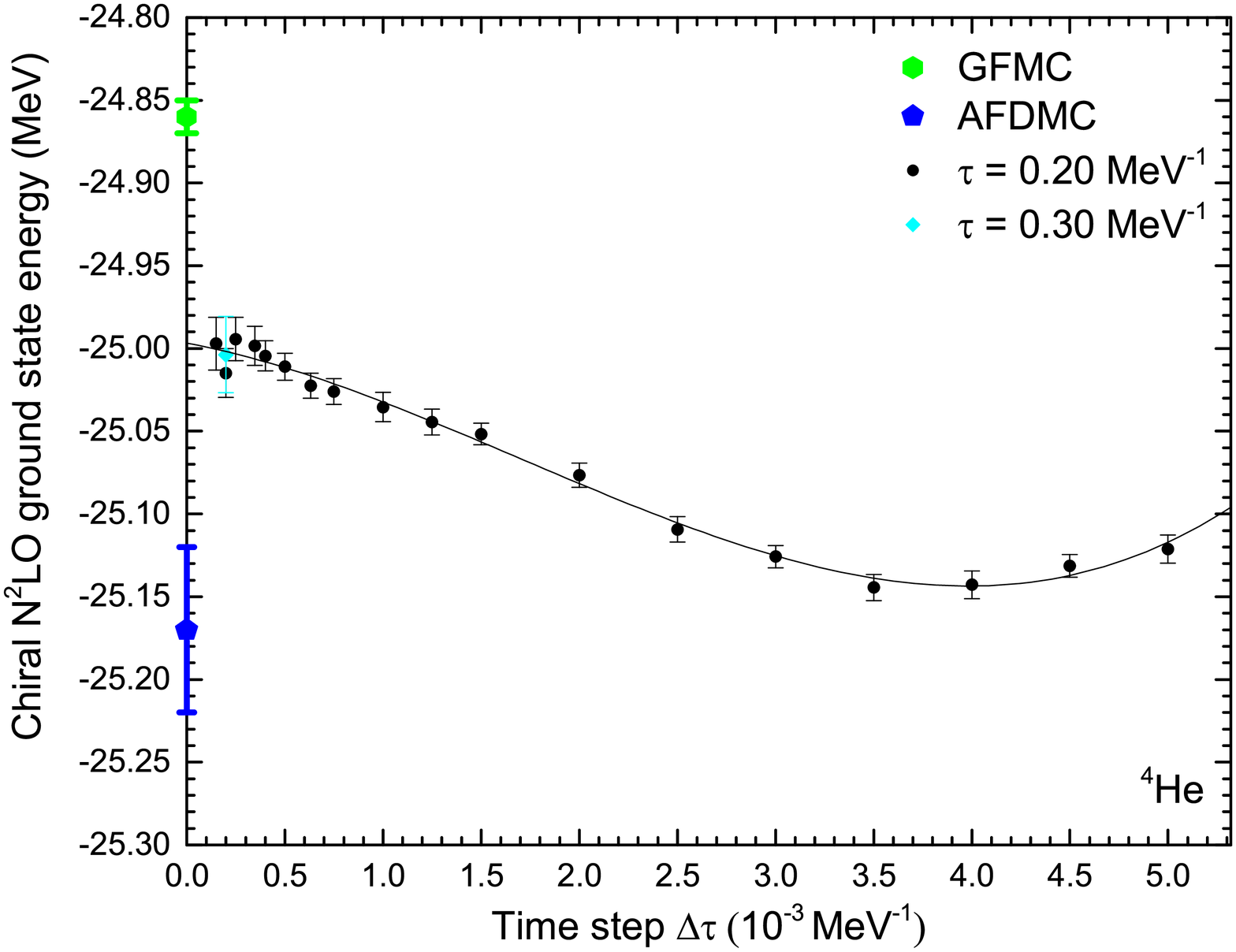}
\caption[Energy Vs. Time Step $\Delta \tau$ for Different Total Imaginary Time $\tau$ for $^4$He Based on the Local Chiral N$^2$LO Interaction (Wide Range)]{Energy vs. time step $\Delta \tau$ for different total imaginary time $\tau$ for $^4$He based on the local chiral N$^2$LO interaction (wide range). }
\label{energyallN2LO}
\end{figure}

In Fig. \ref{energy}, we show the calculation of the $^4$He ground state energy $E_0$ for different total imaginary time $\tau=0.15\textrm{ MeV}^{-1}$, $\tau=0.2\textrm{ MeV}^{-1}$ and $\tau=0.3\textrm{ MeV}^{-1}$,
with various time steps $\Delta \tau$.
As discussed in section \ref{setup},
at $\tau=0.2\textrm{ MeV}^{-1}$,
we can expect the true ground state energy $E_0$ for $^4$He to be located within our calculated ground state energy within its error bar.
We added $\tau=0.15\textrm{ MeV}^{-1}$ and $\tau=0.3\textrm{ MeV}^{-1}$ for comparison.
Particularly, we can see that for $\Delta \tau < 5 \times 10^{-4} \textrm{ MeV}^{-1}$, for $\tau=0.2\textrm{ MeV}^{-1}$ and $\tau=0.3\textrm{ MeV}^{-1}$, all of the calculated $E_0$ are almost the same within their error bars.
For $\tau=0.15\textrm{ MeV}^{-1}$, the smallest $\Delta \tau$ we use is $1.5 \times 10^{-4} \textrm{ MeV}^{-1}$ which means 1000 short-time propagators and 1001 beads, and its $E_0$ is $-26.155 \pm 0.008$ MeV.
For $\tau=0.2\textrm{ MeV}^{-1}$, the smallest $\Delta \tau$ we use is $1.5 \times 10^{-4} \textrm{ MeV}^{-1}$ which means 1333 short-time propagators and 1334 beads, and its $E_0$ is $-26.162 \pm 0.004$ MeV.
For $\tau=0.3\textrm{ MeV}^{-1}$, the smallest $\Delta \tau$ we use is $2 \times 10^{-4} \textrm{ MeV}^{-1}$ which means 1500 short-time propagators and 1501 beads, and its $E_0$ is $-26.173 \pm 0.006$ MeV.

Also, as pointed out in section \ref{propagator} and by Eq.(\ref{Oextrapolation}), for the AV6' interaction, our PIMC results should contain even orders of $\Delta \tau$ error comparing with the true ground state results for small time step $\Delta \tau$.
Usually we do an extrapolation at $\Delta \tau^2$ order,
\begin{equation}
\langle \hat{O} (\Delta \tau) \rangle
= \langle \hat{O} \rangle
+  C_{2} \Delta \tau^{2}.
\label{OextrapolationO2}
\end{equation}
In order to do extrapolation by using Eq.(\ref{OextrapolationO2}), we need to find the range of $\Delta \tau$ where the error of $\Delta \tau^2$ dominates, so that we can observe a clear $\Delta \tau^2$ dependence of $\langle \hat{O} (\Delta \tau) \rangle$. As a rough estimation, by looking at Eq.(\ref{Oextrapolation}) we would require $ |\frac{ C_2 \Delta \tau^2  }{C_4 \Delta \tau^4}| \gg 1 $.
There are combinations of commutator terms of kinetic energy $T$ and potential $V$ in coefficients $C_2$ and $C_4$. For example there can be $TTV$ in $C_2$ and $TTTTV$ in $C_4$ and we can roughly estimate $|\frac{ C_2 \Delta \tau^2  }{C_4 \Delta \tau^4}|$ as $|\frac{ Re \langle \Psi_T | TTV  | \psi_T \rangle  }{Re \langle \Psi_T | TTTTV  | \psi_T \rangle \Delta \tau^2 }|$.
For the AV6' interaction, the magnitude of $T$ and $V$ are $10^2$ MeV order, so $ |\frac{ C_2 \Delta \tau^2  }{C_4 \Delta \tau^4}| \approx |\frac{ Re \langle \Psi_T | TTV  | \psi_T \rangle  }{Re \langle \Psi_T | TTTTV  | \psi_T \rangle \Delta \tau^2 }|$ may be roughly estimated as $|\frac{ C_2 \Delta \tau^2  }{C_4 \Delta \tau^4}| \approx |\frac{ 10^6  }{10^{10} \Delta \tau^2 }|=|\frac{ 1  }{10^{4} \Delta \tau^2 }|$. In order to make sure $ |\frac{ C_2 \Delta \tau^2  }{C_4 \Delta \tau^4}| \gg 1 $, we may expect $\Delta \tau$ to be around $10^{-4} \sim 10^{-3} \textrm{ MeV}^{-1}$.

We find that in Fig. \ref{energy} we can observe a clear $\Delta \tau^2$ dependence when $\Delta \tau < 1.3 \times 10^{-3} \textrm{ MeV}^{-1}$, and by using Eq.(\ref{OextrapolationO2}), the true ground state energy $E_0$ is extrapolated to $-26.16(1)$ MeV. This is consistent with GFMC result (\cite{WiringaPieper02a}) which is $E_0=-26.15(2)$ MeV.
This is reasonable because in this work, we are summing over all the spin and isospin states like GFMC does. So our results should be comparable with GFMC's. The difference is that GFMC is a DMC method based on Eq.(\ref{DMCopexpdetail}), and we are doing a PIMC calculation based on Eq.(\ref{PIMCopexpdetail}).

In Fig. \ref{energyN2LO} we show the results of $^4$He based on the local chiral N$^2$LO two-body interaction. We choose the coordinate space cutoff (\cite{Gerzerlis14a}) as $R_0=1.2$ fm.
As discussed in section \ref{propagator} and Eq.(\ref{ON2LOextrapolation}), due to the existence of the spin-orbit operator, a $\Delta \tau$ error will exist in evaluating $\langle \hat{O} (\Delta \tau) \rangle$.
We do observe the $\Delta \tau$ error dominates when $\Delta \tau < 1.3 \times 10^{-3} \textrm{ MeV}^{-1}$. So we do a linear extrapolation in this range for the chiral interaction,
\begin{equation}
\langle \hat{O} (\Delta \tau) \rangle
= \langle \hat{O} \rangle
+  C_{1} \Delta \tau,
\label{OextrapolationO1}
\end{equation}
and the extrapolated ground-state energy is found to be $E_0=-24.99(1)$ MeV.

Fig. \ref{energyall} shows a wider range of $\Delta \tau$ than Fig. \ref{energy}, and we use Eq.(\ref{PIMCUSdetail}) up to $\Delta \tau^4$ in both the numerator and the denominator to do the fitting for all the data points.
In Fig. \ref{energyallN2LO} we fit the wide range of the data by using Eq.(\ref{ON2LOextrapolation}) up to $\Delta \tau^3$.

\begin{table}[htbp!]
\caption[PIMC Results for the Ground State Energy of Light Nuclei Based on the AV6' Interaction]{The AV6' interaction based light nuclei ground state energies.}
\label{TablePIMC2}%
\begin{tabular*}   {1.0\textwidth}{@{\extracolsep{\fill}}ccccc}
\toprule
$^A$Z   & $E^{\text{PIMC}}_0$ (MeV) & $\tau$ (MeV$^{-1}$) & $E^{\textrm{GFMC}}_0$ (MeV)  & $E^{\textrm{EXPT}}_0$ (MeV) \\
\midrule
$^2$H  & $-2.24(4)$  & 0.5 & &  -2.22 \\ \midrule
$^3$H  & $-7.953(5)$  & 0.4 & $-7.95(1)$ & -8.48 \\ \midrule
$^3$He  & $-7.336(6)$  & 0.5 & & -7.72  \\ \midrule
$^4$He  & $-26.162(4)$  & 0.2 & $-26.15(2)$ & -28.30  \\
\bottomrule
\end{tabular*}
\end{table}

\begin{table}[htbp!]
\caption[PIMC Results for the Ground State Energy of Light Nuclei Based on the Local Chiral N$^2$LO Interaction.]{The local chiral N$^2$LO interaction based light nuclei ground state energies.}
\label{TablePIMCN2LO}%
\begin{tabular*}   {1.0\textwidth}{@{\extracolsep{\fill}}cccccc}
\toprule
$^A$Z   & $E^{\text{PIMC}}_0$ (MeV) & $\tau$ (MeV$^{-1}$) & $E^{\textrm{GFMC}}_0$ (MeV) & $E^{\textrm{AFDMC}}_0$ (MeV) & $E^{\textrm{EXPT}}_0$ (MeV)   \\ \midrule
$^2$H  & $-2.203(2)$  & 0.5 & & & -2.22 \\ \midrule
$^3$H  & $-7.74(1)$  & 0.4 & $-7.74(1)$ & $-7.76(3)$ & -8.48  \\ \midrule
$^3$He  & $-7.11(1)$  & 0.5 & $-7.01(1)$ & $-7.12(3)$  & -7.72 \\ \midrule
$^4$He  & $-25.00(1)$  & 0.2 & $-24.86(1)$ & $-25.17(5)$ & -28.30 \\
\bottomrule
\end{tabular*}
\end{table}

The ground state energy for all the $A\leq 4$ nuclei are listed in Table \ref{TablePIMC2} for the AV6' interaction and in Table \ref{TablePIMCN2LO} for the chiral N$^2$LO interaction.
The GFMC and AFDMC results (\cite{Lonardoni2018PRC}) are listed for comparison.
Overall, all the PIMC, GFMC and AFDMC results are consistent with each other within $1\%$ error.
We also listed the experimental value of the binding energies (\cite{WiringaPieper02a,Lee2020FiP,wiki:NBE}) for comparison.
The difference between the experimental values $E^{\textrm{EXPT}}_0$ and our results is mainly due to the absence of the three body interactions in our calculations.

\subsection{Verifications}
\label{secVerify}

In order to further support our conclusion that $\tau=0.2\textrm{ MeV}^{-1}$ indeed projected out the ground state and therefore our PIMC results are reliable, we calculate the potential $V=V^{NN}+V^{EM}$ along all the beads on the path.
This is a typical quantity to calculate in PIMC calculations (\cite{Schmidt2000JCP}).
The calculation is done in the same way as introduced in section \ref{Recast}. We just need to use the potential operator $V$ as the operator $\hat{O}$, and we place the $V$ at each imaginary time position from bead 0 to the last bead $N$ on the path.
In other words, the $g^V_{lr,M}(\mathcal{R}) $ in Eq.(\ref{gVRlrdef2}) needs to be calculated from $M=0$ to $M=N={\tau}/{\Delta \tau}$.
In this way, we are actually calculating $\langle {V}(\tau_1) \rangle$ which is,
\begin{eqnarray}
\langle {V}(\tau_1) \rangle
= \frac{ Re  \langle \Psi_T| e^{-H \tau_1}  \left(V^{NN}+V^{EM}\right) e^{-H (\tau-\tau_1) } | \Psi_T \rangle  }
{ Re \langle \Psi_T| e^{-H \tau} | \Psi_T \rangle} ,
\label{potentialbeads}
\end{eqnarray}
where $\tau_1$ ranges from 0 MeV$^{-1}$ to $\tau$ as we put the $V$ operator from bead 0 to bead $N$.
$\langle {V}(\tau_1) \rangle$ is symmetric around $\tau_1=\tau/2$.
When $\tau/2$ is big enough to project out the ground state, $\langle {V}(\tau/2) \rangle$ is the ground state potential energy.

\begin{figure}[tbhp!]
\centering
\includegraphics[scale=0.46]{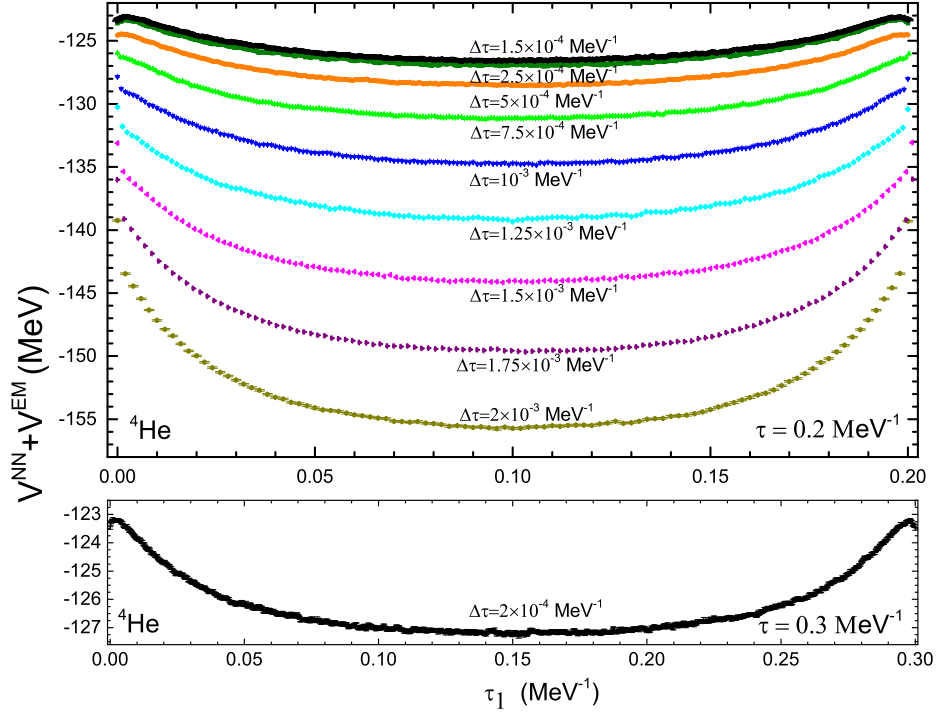}
\caption[The Potential Energy along the Beads for $\tau=0.2$ and $0.3 \textrm{ MeV}^{-1}$ for $^4$He Based on the AV6' Interaction]{The potential energy along the beads for $\tau=0.2$ and 0.3 $\textrm{MeV}^{-1}$ for $^4$He based on the AV6' interaction. }
\label{potential}
\end{figure}

\begin{figure}[tbhp!]
\centering
\includegraphics[scale=0.46]{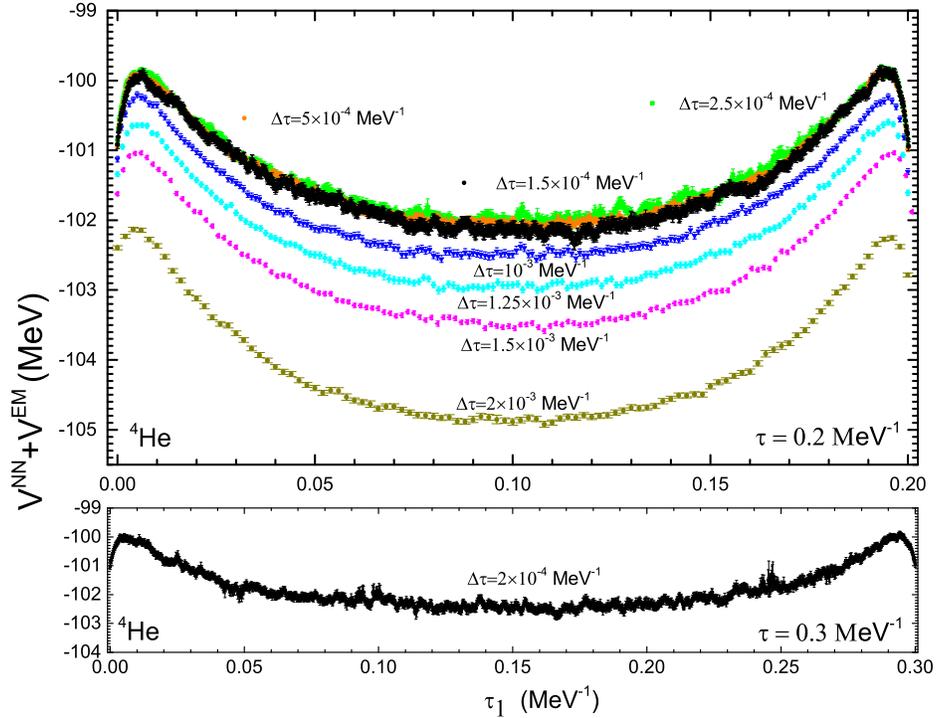}
\caption[The Potential Energy along the Beads for $\tau=0.2$ and 0.3 $\textrm{MeV}^{-1}$ for $^4$He Based on the Local Chiral N$^2$LO Interaction]{The potential energy along the beads for $\tau=0.2$ and 0.3 $\textrm{MeV}^{-1}$ for $^4$He based on the local chiral N$^2$LO interaction.}
\label{potentialN2LO}
\end{figure}

In Fig. \ref{potential}\footnote{Note that the potential operators and the Hamiltonian do not commute, so the eigenstates of the potential operators and the Hamiltonian are not the same. So even when the system reaches the ground state of its Hamiltonian, this ground state $| \Phi_0 \rangle$ is not an eigenstate of the potential operators. So $\langle {V}(\tau_1) \rangle$ has larger fluctuations than $\langle H(\tau_1) \rangle$. So in order to make Fig. \ref{potential} look `smooth', we need to take a longer computation time than computing the ground state energy.},
we show the results of $\langle {V}(\tau_1) \rangle$ as a functions of $\tau_1$ for $\tau=0.2$ MeV$^{-1}$ and $\tau=0.3$ MeV$^{-1}$.
On the upper panel of Fig. \ref{potential}, we show for different time step $\Delta \tau$ how $\langle {V}(\tau_1) \rangle$ change as $\tau_1$ from 0 MeV$^{-1}$ to the total imaginary time $\tau=0.2$ MeV$^{-1}$.
As expected, since PIMC is projecting out the Hamiltonian's ground state, and the Hamiltonian's ground state is not the same as the potential operator's eigenstate, the time step error introduced by $\Delta \tau$ will have noticeable impact on $\langle {V}(\tau_1) \rangle$.
That is why the curve of $\langle {V}(\tau_1) \rangle$ vs. $\tau_1$ are different for different $\Delta \tau$.
However, as $\Delta \tau$ decreases, time step error will become smaller and smaller.
And indeed, as we can see, the result for $\Delta \tau=1.5\times 10^{-4}$ MeV$^{-1}$ almost coincides with that of $\Delta \tau=2.5\times 10^{-4}$ MeV$^{-1}$.
$\langle {V}(\tau_1) \rangle$ is about $-127\sim-123$ MeV.
It looks like $\langle {V}(\tau_1) \rangle$ has converged when $\Delta \tau \leq 2.5\times 10^{-4}$ MeV$^{-1}$.
This is a clear signal that a stable eigenstate of the Hamiltonian for $\tau=0.2$ MeV$^{-1}$ has been reached, and smaller $\Delta \tau$ and therefore more beads will not improve the convergence.
Furthermore, in order to verify if this eigenstate is truly the ground state, we calculate for a larger total imaginary time of $\tau=0.3$ MeV$^{-1}$ with a similar time step $\Delta \tau=2 \times 10^{-4}$ MeV$^{-1}$, i.e. 1501 beads.
The result is in the lower panel of Fig. \ref{potential}.
Indeed, the curve of $\langle {V}(\tau_1) \rangle$ vs. $\tau_1$ is almost the same as the $\tau=0.2$ MeV$^{-1}$ and $\Delta \tau \leq 2.5\times 10^{-4}$ MeV$^{-1}$ cases.
This means further increasing the total time $\tau$ will not change the state, so the stable state reached for $\tau=0.2$ MeV$^{-1}$ must be the ground state.

Therefore, the ground state of the Hamiltonian is indeed reached when $\tau=0.2$ MeV$^{-1}$, and $\Delta \tau$ around $1.5\sim2.5\times 10^{-4}$ MeV$^{-1}$ is small enough.
Besides, from Fig. \ref{potential} we can see that in the area when $0.075\textrm{ MeV}^{-1}\leq\tau_1\leq0.125\textrm{ MeV}^{-1}$ for $\tau=0.2$ MeV$^{-1}$, and the area when $0.075\textrm{ MeV}^{-1}\leq\tau_1\leq0.225\textrm{ MeV}^{-1}$ for $\tau=0.3$ MeV$^{-1}$, the curves of $\langle {V}(\tau_1) \rangle$ vs. $\tau_1$ are almost flat.
So that even $\tau=2\times0.075=0.15$ MeV$^{-1}$ may be sufficient to project out the ground state.
Fig. \ref{potentialN2LO} is a similar graph for the local chiral interaction which also indicates $\tau=0.2$ MeV$^{-1}$ is sufficient\footnote{
In order to save computation time, we did not include spin-orbit interaction, but it does not alter the conclusion. Fig. \ref{potentialN2LO} is not as smooth as Fig. \ref{potential} simply because we used fewer core hours for it than we did for Fig. \ref{potential}.
}.

\begin{figure}[htbp!]
\centering
\includegraphics[width=0.9503\textwidth]{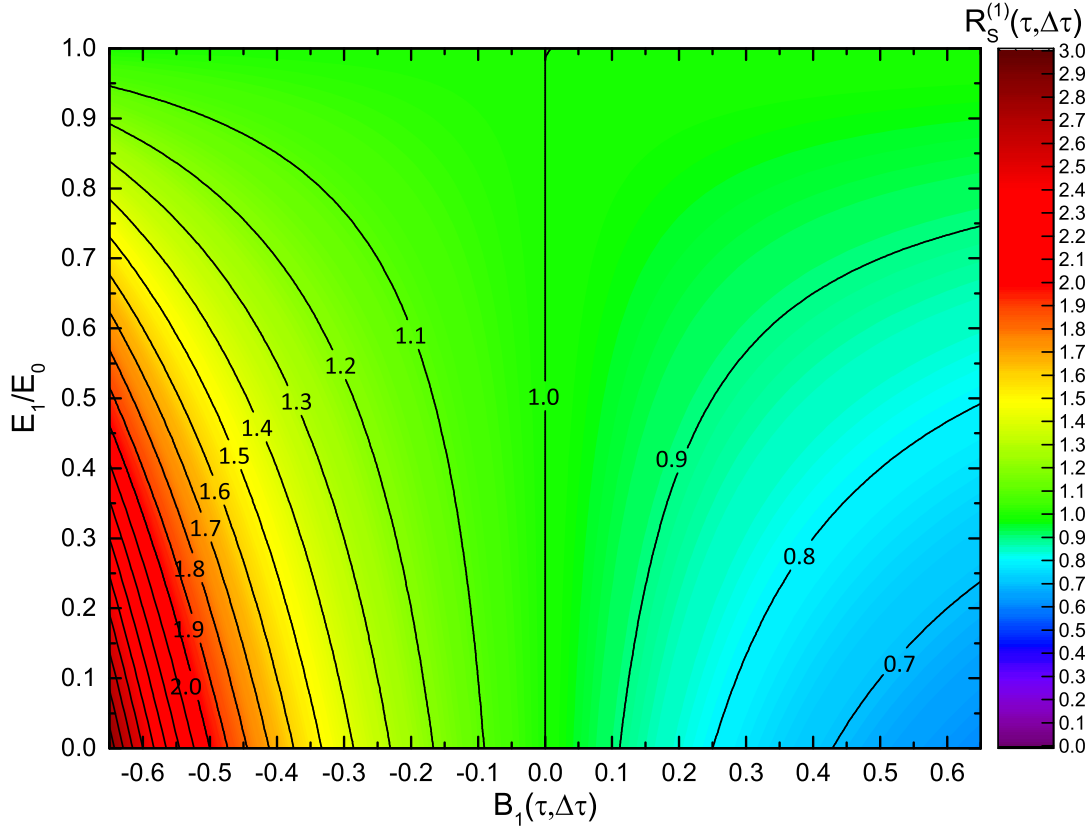} 
\caption[PIMC Leading Order Structure Function $R^{(1)}_{\text{S}}(\tau,\Delta \tau)$]{PIMC leading order structure function $R^{(1)}_{\text{S}}(\tau,\Delta \tau)$. }
\label{figPIMCstructR}
\end{figure}

Besides checking the potential along the beads,
the robustness of the ground state energy PIMC calculation can be understood as follows.

The trial state $| \Psi_T \rangle$, after operating with $\left[U(\Delta \tau)\right]^N$, can be expanded by the eigenstates of the Hamiltonian,
\begin{eqnarray}
\left[U(\Delta \tau)\right]^N |\Psi_T \rangle = \sum_{n=0}^{\infty} A_n(\tau,\Delta \tau) |\Phi_n\rangle
=A_0(\tau,\Delta \tau)\left[ | \Phi_0 \rangle  +  \sum_{i=1}^{+\infty} \frac{A_i(\tau,\Delta \tau)}{A_0(\tau,\Delta \tau)} | \Phi_i \rangle \right] . ~~~
\label{UdtNexp}
\end{eqnarray}
where $A_n(\tau,\Delta \tau)$ are the coefficients which depend on $\tau$ and $\Delta \tau$ for $|\Phi_n\rangle$.
We know that $\lim\limits_{\Delta \tau \rightarrow 0} \left[U(\Delta \tau)\right]^N= e^{-H \tau}$ and we also notice from Eq.(\ref{eHPsiT}) that,
\begin{equation}
 e^{-H \tau} |\Psi_T \rangle =
a_0 e^{-E_0 \tau}
\left[ | \Phi_0 \rangle +  \sum_{i=1}^{\infty}\frac{a_i}{a_0}e^{-(E_i-E_0) \tau} | \Phi_i \rangle  \right],
\label{eHPsiTau}
\end{equation}
So, comparing Eq.(\ref{UdtNexp}) and Eq.(\ref{eHPsiTau}) we find that for $\Delta \tau \rightarrow 0$,
\begin{flalign}
\lim\limits_{\Delta \tau \rightarrow 0} A_0(\tau,\Delta \tau) & = a_0 e^{-E_0\tau} , \\
\lim\limits_{\Delta \tau \rightarrow 0} \frac{A_n(\tau,\Delta \tau)}{A_0(\tau,\Delta \tau)} &= \frac{a_n}{a_0} e^{-(E_n-E_0)\tau}
\Rightarrow  \lim\limits_{\tau \rightarrow \infty} \left[ \lim\limits_{\Delta \tau \rightarrow 0} \frac{A_n(\tau,\Delta \tau)}{A_0(\tau,\Delta \tau)} \right] =0  \label{ratioAnA0} .
\end{flalign}

We calculate the ground state energy by using Eq.(\ref{PIMCopexpdetailE0dtau}), the two parts in it are essentially the same and we pick one of them to take a closer look. We find
\begin{equation}
\frac{\langle \Psi_T| H [U(\Delta \tau)]^N  | \Psi_T \rangle}{\langle \Psi_T| [U(\Delta \tau)]^N  | \Psi_T \rangle}
= E_0 \frac{1+ \sum_{i=1}^{+\infty}\frac{a_i^* A_i(\tau,\Delta \tau) E_i }{a_0^* A_0(\tau,\Delta \tau) E_0 }  }
{1+ \sum_{i=1}^{+\infty}\frac{a_i^* A_i(\tau,\Delta \tau) }{a_0^* A_0(\tau,\Delta \tau) }  }
\equiv E_0 R_{\text{S}}(\tau,\Delta \tau) ,
\label{PIMCstrucfull}
\end{equation}
since Eq.(\ref{PIMCstrucfull}) is real, we define a real function $R_{\text{S}}(\tau,\Delta \tau)$ as our PIMC error structure function,
\begin{equation}
R_{\text{S}}(\tau,\Delta \tau)
\equiv Re \frac{1+ \sum_{i=1}^{+\infty}\frac{a_i^* A_i(\tau,\Delta \tau) E_i }{a_0^* A_0(\tau,\Delta \tau) E_0 }  }
{1+ \sum_{i=1}^{+\infty}\frac{a_i^* A_i(\tau,\Delta \tau) }{a_0^* A_0(\tau,\Delta \tau) }  } ,
\label{PIMCstrucRs}
\end{equation}
and we can define its cutoff functions $R^{(n)}_{\text{S}}(\tau,\Delta \tau)$ such that
\begin{equation}
R^{(n)}_{\text{S}}(\tau,\Delta \tau)
\equiv Re \frac{1+ \sum_{i=1}^{n}\frac{a_i^* A_i(\tau,\Delta \tau) E_i }{a_0^* A_0(\tau,\Delta \tau) E_0 }  }
{1+ \sum_{i=1}^{n}\frac{a_i^* A_i(\tau,\Delta \tau) }{a_0^* A_0(\tau,\Delta \tau) }  } .
\label{PIMCstrucRsn}
\end{equation}
Considering Eq.(\ref{ratioAnA0}), $R_{\text{S}}(\tau,\Delta \tau)$ and $R^{(n)}_{\text{S}}(\tau,\Delta \tau)$ will converge to 1 when $\tau$ is big enough and $\Delta \tau$ is small enough, and the energy will converge to $E_0$.
We find that the structure Eq.(\ref{PIMCstrucfull}) may be much less sensitive to $\Delta \tau$ than one may expect.
We can assume the errors in Eq.(\ref{PIMCstrucfull}) are mostly coming from the leading order errors in its denominator and the numerator. If the leading order error is from the first excited state, Eq.(\ref{PIMCstrucfull}) can be approximated as
\begin{equation}
\frac{\langle \Psi_T| H [U(\Delta \tau)]^N  | \Psi_T \rangle}{\langle \Psi_T| [U(\Delta \tau)]^N  | \Psi_T \rangle}
\approx E_0 \times Re \frac{1+ \frac{a_1^* A_1(\tau,\Delta \tau) E_1 }{a_0^* A_0(\tau,\Delta \tau) E_0 }  }
{1+ \frac{a_1^* A_1(\tau,\Delta \tau) }{a_0^* A_0(\tau,\Delta \tau) }  }
=E_0 R^{(1)}_{\text{S}}(\tau,\Delta \tau) ,
\label{PIMCstruc1}
\end{equation}
and since Eq.(\ref{PIMCstrucfull}) is real, we can assume $\frac{a_1^* A_1(\tau,\Delta \tau) }{a_0^* A_0(\tau,\Delta \tau) }$ in Eq.(\ref{PIMCstruc1}) is mostly real and we can define a real function $B_1(\tau,\Delta \tau)$ as
\begin{equation}
B_1(\tau,\Delta \tau) = Re \frac{a_1^* A_1(\tau,\Delta \tau) }{a_0^* A_0(\tau,\Delta \tau) }
\label{PIMCstrucB1}
\end{equation}
such that
\begin{equation}
R_{\text{S}}(\tau,\Delta \tau) \approx
R^{(1)}_{\text{S}}(\tau,\Delta \tau) \approx
\frac{ 1+B_1(\tau,\Delta \tau) \frac{E_1}{E_0} }{1+B_1(\tau,\Delta \tau)}.
\label{PIMCstrucRsB1}
\end{equation}

Therefore, our result $E(\Delta \tau)$ in Eq.(\ref{PIMCopexpdetailE0dtau}) will be very close to the true expectation $E_0$,
as long as Eq.(\ref{PIMCstrucRsB1}) is close to 1.
In Fig. \ref{figPIMCstructR} we plot the contour graph and show how the value of the leading order structure function $R^{(1)}_{\text{S}}(\tau,\Delta \tau)$ changes as $B_1(\tau,\Delta \tau)$ and $E_1/E_0$ change.
We find as long as $B_1(\tau,\Delta \tau) \in [-0.1,0.1]$ (which is a reasonable range for a proper $\Delta \tau$),
$R^{(1)}_{\text{S}}(\tau,\Delta \tau) $ will be close to 1 (the green area),
it fluctuates from 0.9 to 1.1 so $E(\Delta \tau)$ will be around $E_0$ with at most $10\%$ error or so.
In fact, our results in section \ref{secResultCalc} shows that $E(\Delta \tau)$ in PIMC works even better.
From Figs. \ref{energyall} and \ref{energyallN2LO} we see that even with a wide range of $\Delta \tau$, the difference between $E(\Delta \tau)$ and $E_0$ is less than $3\%$ which is well below $10\%$.
The robust $E(\Delta \tau)$ value also indicates the ground state path is robust, so the expectation values of other operators should be reliable too.

\section{Radii}

The root mean square (rms) radius $r_m$ can be interpreted as the average distance between any one of the particles and the system's center of mass.
The label of the nucleon does not matter, so we can just take nucleon number 1 as reference.
The operator can be written as\footnote{In the code, we calculate the averaged operator
$ \hat{r}_m = \frac{1}{A} \sum_{i=1}^{A} \left| \hat{\bm{r}}_i - \frac{1}{A}\sum_{j=1}^A \hat{\bm{r}}_j \right|.
$},
\begin{eqnarray}
\hat{r}_m = \left| \hat{\bm{r}}_1 - \frac{1}{A}\sum_{j=1}^A \hat{\bm{r}}_j \right|,
\end{eqnarray}
and we calculate $\langle {r}_m^2 \rangle$ for its ground state expectation value as
\begin{flalign}
\langle {r}_m^2 \rangle &=
\frac{ Re \langle \Phi_0 | \hat{r}_m^2  | \Phi_0  \rangle }   { Re \langle \Phi_0 | \Phi_0  \rangle  } \nonumber \\
&= \frac{ Re  \langle \Psi_T| e^{-H \Delta \tau}...e^{-H \Delta \tau}
\Big| {\bm{r}}_1 - \frac{1}{A}\sum_{j=1}^A {\bm{r}}_j \Big|^2
e^{-H \Delta \tau} ...e^{-H \Delta \tau}| \Psi_T \rangle  }
{ Re \langle \Psi_T| e^{-H \Delta \tau}...e^{-H \Delta \tau} e^{-H \Delta \tau} ...e^{-H \Delta \tau} | \Psi_T \rangle} .
\label{rmdef}
\end{flalign}

Note that $\hat{r}_m$ does not commute with $H$, so PIMC is well suited to calculate it.
Eq.(\ref{rmdef}) can be recast into the form of Eq.(\ref{OABP}) in section \ref{Recast}.
We put the operator $\hat{r}_m^2$ at the middle bead M, the corresponding $g^V_{lr,M}(\mathcal{R})$ can be written as,
\begin{eqnarray}
g^V_{lr,M}(\mathcal{R})  = \langle \psi^l_{M\!-\!1}  (\mathcal{R}) | \bm{\mathcal{G}} e^{\frac{- \mathcal{V}_M  \Delta \tau }{2}}
\left| {\bm{r}}_1 - \frac{1}{A}\sum_{j=1}^A {\bm{r}}_j \right|^2
   e^{\frac{- \mathcal{V}^\dagger_M  \Delta \tau }{2}} \bm{\mathcal{G}} | \psi^r_{M\!+\!1} (\mathcal{R}) \rangle,
\label{gVRlrradiidef}
\end{eqnarray}
while other functions remain the same as in section \ref{Recast}.
Once we calculated $\langle {r}_m^2 \rangle$, the rms value is simply  $r_m=\sqrt{\langle {r}_m^2 \rangle}$.
Note that these are point nucleon radii.

Similarly, we can define the proton(neutron) radii operator $\hat{r}_{p(n)}$ as\footnote{
In the code, we calculate the averaged operator
$ \hat{r}_{p(n)} = \frac{1}{N_{p(n)}} \sum_{i=1}^{A} \left| \hat{\bm{r}}_i - \frac{1}{A}\sum_{j=1}^A \hat{\bm{r}}_j \right| P^i_{p(n)},$ where $N_{p(n)}$ is the number of protons (neutrons).},
\begin{equation}
\hat{r}_{p(n)} = \left| \hat{\bm{r}}_1 - \frac{1}{A}\sum_{j=1}^A \hat{\bm{r}}_j \right| P^1_{p(n)},
\end{equation}
where $P^1_{p}=\frac{1+\tau_{1z}}{2}$ and $P^1_{n}=\frac{1-\tau_{1z}}{2}$ are the proton and neutron projection operator respectively for nucleon whose label is 1.

\begin{table}[tbhp!]
\caption[PIMC Results for the RMS Radii of Light Nuclei Based on the AV6' Interaction]{ PIMC rms radii of light nuclei based on the AV6' interaction.}
\label{TablePIMC3}%
\begin{tabular*}   {1.0\textwidth}{@{\extracolsep{\fill}}cccc}
\toprule
$^A$Z   & $r_m$ (fm) & $r_p$ (fm) & $r_n$ (fm)  \\ \midrule
$^2$H  & $1.98(2)$  &  & \\ \midrule
$^3$H  & $1.7261(5)$  & 1.6240(4) & $1.7751(5)$ \\ \midrule
$^3$He  & $1.7426(6)$  & 1.7962(6) &  1.6289(5) \\ \midrule
$^4$He  & $1.4716(2)$  & 1.4736(2)  & 1.4693(2)  \\
\bottomrule
\end{tabular*}
\end{table}

\begin{table}[tbhp!]
\caption[PIMC Results for RMS Radii of Light Nuclei Based on the Local Chiral N$^2$LO Interaction]{ PIMC rms radii of light nuclei based on the local chiral N$^2$LO interaction. }
\label{TablePIMC4}%
\begin{tabular*}   {1.0\textwidth}{@{\extracolsep{\fill}}cccc}
\toprule
$^A$Z   & $r_m$ (fm) & $r_p$ (fm) & $r_n$ (fm)  \\ \midrule
$^2$H  & $1.991(1)$  &  & \\ \midrule
$^3$H  & $1.7497(7)$  & 1.6430(6) & $1.8007(7)$ \\ \midrule
$^3$He  & $1.7725(8)$  & 1.8292(9) &  1.6520(7) \\ \midrule
$^4$He  & $1.4860(4)$  & 1.4882(4)  &  1.4834(4)   \\
\bottomrule
\end{tabular*}
\end{table}

In Tables \ref{TablePIMC3} and \ref{TablePIMC4}, we list our PIMC ground state point nucleon rms radii of light nuclei based on the AV6' interaction and the N$^2$LO local chiral interaction.
For $^2$H, since there is no Coulomb interaction, $r_m=r_p=r_n$. For other nuclei however, due to the isospin dependent NN interaction (mainly due to the $\bm{\tau_i}\cdot \bm{\tau}_j$ part in the NN interaction) and the Coulomb interaction, $r_m$, $r_p$ and $r_n$ are not the same.
E.g., for $^3$H, the only proton is more likely in the center and the two neutrons are likely on the two sides, so the proton is more closer to the center of the mass of the system, so $r_p < r_n$.
$^3$He is the opposite.
The small difference between $r_p$ and $r_n$ for $^4$He is caused by the Coulomb interaction.

\section{Density Distributions}

The ground state of $^4$He is an angular momentum $J=0$ state and is spherically symmetric.
We can use the one-particle number density $\rho(r)$ to describe the probability density for one particle to be at distance $r$ from the nuclei's center of mass.
With this convention we normalize it as
\begin{equation}
\int 4 \pi r^2 \rho(r) dr = 1 .
\end{equation}
The distribution $\rho(r)$ can be extracted from high energy electron-nucleus scattering experiments, and is related with the electric form factor by Fourier transform (\cite{Xuzizong2009book,Lynn2019ann}).
The label of a particle does not matter, so we can just take particle 1 as the reference particle, and the operator $\hat{\rho}(r)$ can be written as\footnote{In the code, we actually calculate the averaged $\hat{\rho}(r)$ operator (\cite{Lynn2019ann}),
\begin{equation}
\hat{\rho}(r) = \frac{1}{A} \sum_{i=1}^{A} \delta \left( r - \left| \hat{\bm{r}}_i - \frac{1}{A}\sum_{j=1}^A \hat{\bm{r}}_j  \right|  \right).
\label{rhoravedef}
\end{equation}
We use Eq.(\ref{rhordef}) for illustration purpose.},
\begin{flalign}
\hat{\rho}(r) = \delta \left( r - \left| \hat{\bm{r}}_1 - \frac{1}{A}\sum_{j=1}^A \hat{\bm{r}}_j \right|   \right).
\label{rhordef}
\end{flalign}

Note that due to the kinetic energy operator $T$ in the Hamiltonian $H$, $\hat{\rho}(r)$ does not commute with $H$. Therefore it cannot be calculated very accurately by diffusion based QMC method. But it can be calculated directly by PIMC. The ground state expectation value for the one-particle number density $\rho(r)$ is,
\begin{flalign}
\rho(r)
&=\frac{ Re \langle \Phi_0 | \hat{\rho}(r)   | \Phi_0  \rangle }   { Re \langle \Phi_0 | \Phi_0  \rangle  } \nonumber \\
&= \frac{ Re  \langle \Psi_T| e^{-H \Delta \tau}...e^{-H \Delta \tau}
\delta \left(r - \big| \hat{\bm{r}}_1 - \frac{1}{A}\sum_{j=1}^A \hat{\bm{r}}_j   \big|   \right)
e^{-H \Delta \tau} ...e^{-H \Delta \tau}| \Psi_T \rangle  }
{ Re \langle \Psi_T| e^{-H \Delta \tau}...e^{-H \Delta \tau} e^{-H \Delta \tau} ...e^{-H \Delta \tau} | \Psi_T \rangle} .
\label{rhor1}
\end{flalign}
Eq.(\ref{rhor1}) can be recast into the form of Eq.(\ref{OABP}) in section \ref{Recast}.
We put the operator $\hat{\rho}(r)$ at the middle bead M, the corresponding $g^V_{lr,M}(\mathcal{R})$ can be written as,
\begin{eqnarray}
g^V_{lr,M}(\mathcal{R})  = \langle \psi^l_{M\!-\!1}  (\mathcal{R}) | \bm{\mathcal{G}} e^{\frac{- \mathcal{V}_M  \Delta \tau }{2}}
\delta \left( r - \big| \hat{\bm{r}}_1 - \frac{1}{A}\sum_{j=1}^A \hat{\bm{r}}_j   \big|   \right)
   e^{\frac{- \mathcal{V}^\dagger_M  \Delta \tau }{2}} \bm{\mathcal{G}}  | \psi^r_{M\!+\!1} (\mathcal{R}) \rangle,
\label{gVRlrrhodef}
\end{eqnarray}
while the other functions remain the same as in section \ref{Recast}.

The $\delta$ function in $\rho(r)$ cannot be directly sampled.
However it can be integrated.
We can expand in a histogram basis and the width of each bin is $h$,
and $r$ can be represented by $r_k$ when $r\in [kh,kh+h) $ and $k$ is an integer.
The basis of the histogram can be represented by $b_k$ which is 1 when $r\in [kh,kh+h)$ and 0 otherwise.
$\rho(r_k)$ can be represented by a number $a_k$ such that,
\begin{eqnarray}
a_k=\frac{\int \rho(r) b_k d\bm{r}}
{\frac{4}{3} \pi h^3 [(k+1)^3-k^3  ] }.
\label{rhobkintgrel}
\end{eqnarray}
Considering Eq.(\ref{rhor1}) and Eq.(\ref{OABP}) in section \ref{Recast}, $a_k$ can be written as,
\begin{eqnarray}
a_k =\frac{\sum_l\sum_r \int d \mathcal{R}
\left\{\frac{3\int A_{lr}  (\mathcal{R})  b_k d\bm{r}}{4 \pi h^3 [(k+1)^3-k^3  ] } \right\}
P_{lr} (\mathcal{R}) }{ \sum_l\sum_r \int d \mathcal{R} B_{lr} (\mathcal{R}) P_{lr} (\mathcal{R})  }
\label{akABP},
\end{eqnarray}
so we just need to sample $\frac{3\int A_{lr}  (\mathcal{R})  b_k d\bm{r}}{4 \pi h^3 [(k+1)^3-k^3  ] }$ to get $a_k$.
Note that $A_{lr} (\mathcal{R}) = \frac{Re[ g^V_{lr,M}(\mathcal{R})]}{|Re[ f^V_{lr}(\mathcal{R})]|  } $,
and Eq.(\ref{gVRlrrhodef}) of $ g^V_{lr,M}(\mathcal{R})$ contain the $\delta$ function, so the integral $\int A_{lr}  (\mathcal{R})  b_k d\bm{r}$ becomes,
\begin{flalign}
& \int A_{lr}  (\mathcal{R})  b_k d\bm{r}\nonumber \\
&= \frac{Re\left[  \displaystyle \int
\langle \psi^l_{M\!-\!1}  (\mathcal{R}) | \bm{\mathcal{G}}  e^{\frac{- \mathcal{V}_M  \Delta \tau }{2}}
\delta \left( r - \big| \hat{\bm{r}}_1 - \frac{\sum_{j=1}^A \hat{\bm{r}}_j}{A}   \big|   \right)
   e^{\frac{- \mathcal{V}^\dagger_M  \Delta \tau }{2}} \bm{\mathcal{G}}  | \psi^r_{M\!+\!1} (\mathcal{R}) \rangle
b_k d\bm{r} \right]  }
{|Re[ f(\mathcal{R})_{lr}]|  } \nonumber \\
&=
\begin{cases}
\frac{Re[ f(\mathcal{R})_{lr}]}{|Re[ f(\mathcal{R})_{lr}]|  }  , & \left|{\bm{r}}_1 - \frac{\sum_{j=1}^A {\bm{r}}_j}{A}\right| \in [kh,kh+h).  \\
0,  & \textrm{otherwise}.
\label{rhoAbdrintgral}
\end{cases} &&
\end{flalign}
Note that $\frac{Re[ f(\mathcal{R})_{lr}]}{|Re[ f(\mathcal{R})_{lr}]|  } $ is either 1 or -1, so $\frac{3\int A_{lr}  (\mathcal{R})  b_k d\bm{r}}{4 \pi h^3 [(k+1)^3-k^3  ] }$ just becomes either $\frac{3}{4 \pi h^3 [(k+1)^3-k^3  ] }$, $\frac{-3}{4 \pi h^3 [(k+1)^3-k^3  ] }$ or 0.

\begin{figure}[tbhp!]
\centering
\includegraphics[scale=0.46]{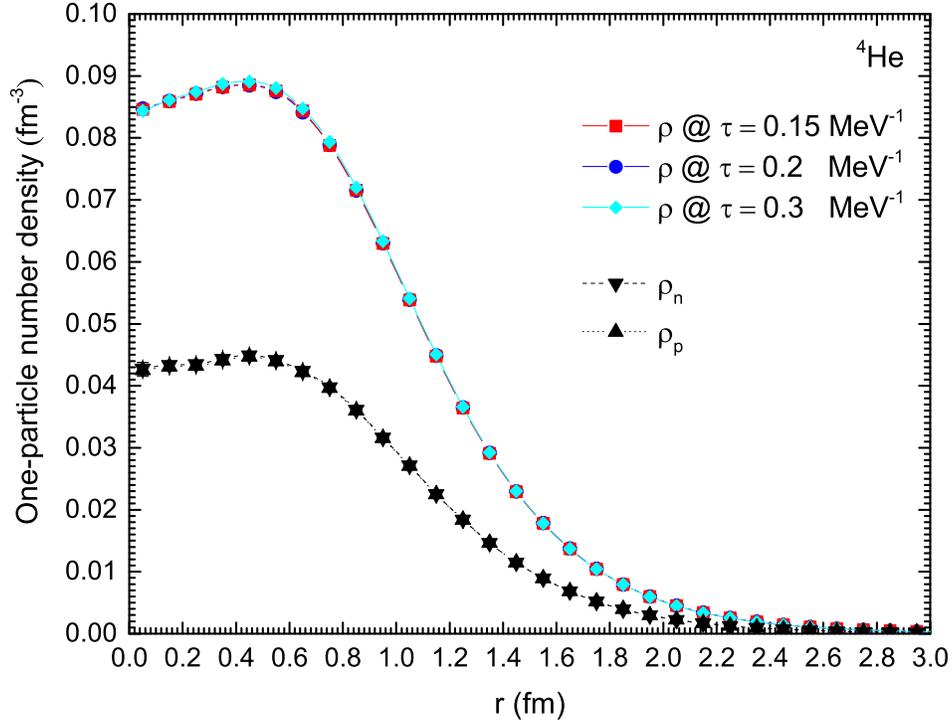}
\caption[The Ground State One-Particle Number Density Distributions of $^4$He Based on the AV6' Interaction]{The ground state one-particle number density distributions of $^4$He based on the AV6' interaction.}
\label{rho}
\end{figure}

\begin{figure}[tbhp!]
\centering
\includegraphics[scale=0.46]{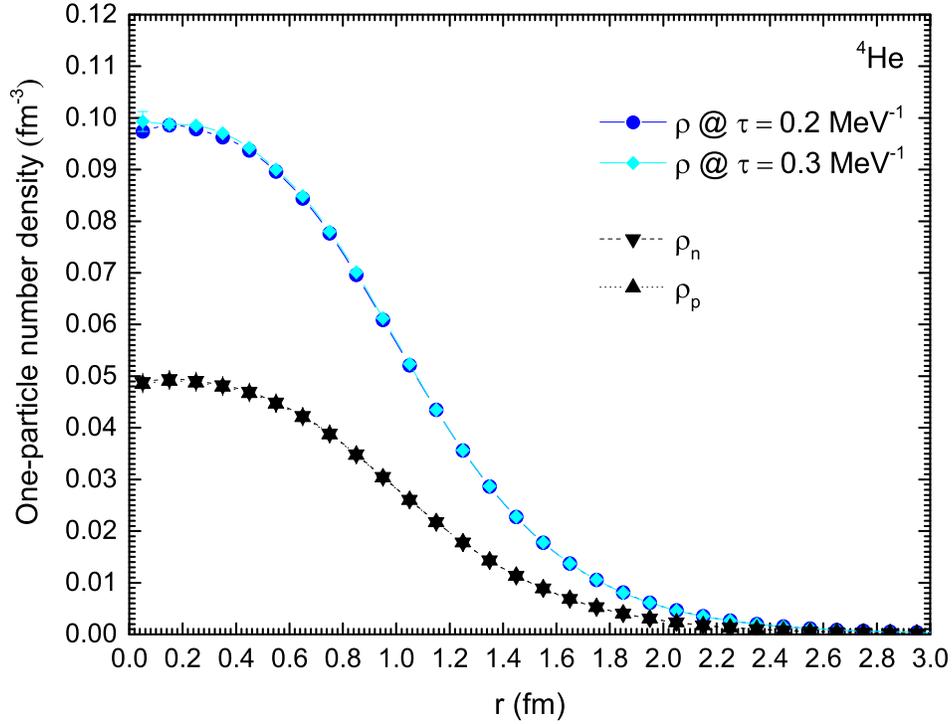}
\caption[The Ground State One-Particle Number Density Distributions of $^4$He Based on the Local Chiral N$^2$LO Interaction]{The ground state one-particle number density distributions of $^4$He based on the local chiral N$^2$LO interaction.}
\label{rhoN2LO}
\end{figure}

The sampling of $a_k$ histogram value is then straightforward.
We define a one-dimensional array $A(n)$ and initialize it to zero.
We sample according to $P_{lr} (\mathcal{R})$ as described in chapter \ref{compalgorithm},
and each time when we take a sample, we look at the configuration of the middle bead $M$.
At bead $M$, there always exists an integer $k$ such that $r_k=| {\bm{r}}_1 - \frac{1}{A}\sum_{j=1}^A {\bm{r}}_j|$.
At the same time, the $k$th element in $A(k)$ takes a sample of $\frac{3 {Re[ f(\mathcal{R})_{lr}]}/{|Re[ f(\mathcal{R})_{lr}]|  } }{4 \pi h^3 [(k+1)^3-k^3  ] }$, so $A(k)=A(k)+\frac{3 {Re[ f(\mathcal{R})_{lr}]}/{|Re[ f(\mathcal{R})_{lr}]|  } }{4 \pi h^3 [(k+1)^3-k^3  ] }$.
For any other element $j$ in array $A(n)$, they each take a sample of 0 and so $A(j)$ remain the same.
We continue sampling according to $P_{lr} (\mathcal{R})$ and continue.
Once we have enough $N$ samples, then each of the $a_k$ can be evaluated by $a_k=A(k)/N$ with corresponding error bar.
As shown in Fig. \ref{rho}, $\rho(r_k)$ are given by these values.
We can see from Fig. \ref{rho} for AV6' interaction and Fig. \ref{rhoN2LO} for the local chiral N$^2$LO interaction that for different total imaginary time, the particle number density distribution of $^4$He are almost identical.
These are further evidence that we have reached the ground state.

From Eq.(\ref{rhoravedef}) and Eq.(\ref{rhordef}), we can further define the single-proton number density $\hat{\rho}_p$ and the single-neutron number density $\hat{\rho}_n$ as
\begin{eqnarray}
\!\!\! \hat{\rho}_{p(n)}(r) \!=\! \delta \! \left( r - \left| \hat{\bm{r}}_1 - \frac{1}{A}\sum_{j=1}^A \hat{\bm{r}}_j  \right|   \right) \! P^1_{p(n)}
\!=\! \frac{1}{A} \sum_{i=1}^{A} \delta \! \left( r - \left| \hat{\bm{r}}_i - \frac{1}{A}\sum_{j=1}^A \hat{\bm{r}}_j  \right|   \right) \! P^i_{p(n)}.
\end{eqnarray}
where $P^i_{p}=\frac{1+\tau_{1z}}{2}$ and $P^i_{n}=\frac{1-\tau_{1z}}{2}$ are the proton and neutron projection operator for nucleon with label $i$. Since $P^i_{p}+P^i_{n}=1$, we have
\begin{eqnarray}
 \hat{\rho}(r) = \hat{\rho}_{p}(r) + \hat{\rho}_{n}(r) .
\end{eqnarray}
The sampling method of $\hat{\rho}_{p}(r)$ and $\hat{\rho}_{n}(r)$ is similar with that of $\hat{\rho}(r)$.
In Fig. \ref{rho} and Fig. \ref{rhoN2LO}, ${\rho}_{p}(r)$ and ${\rho}_{n}(r)$ are also presented.
As expected from the rms radii results, in $^4$He the two protons and the two neutrons are distributed evenly, and ${\rho}_{p}(r)$ and ${\rho}_{n}(r)$ are almost identical.

\begin{figure}[tbhp!]
\centering
\includegraphics[width=0.9503\textwidth]{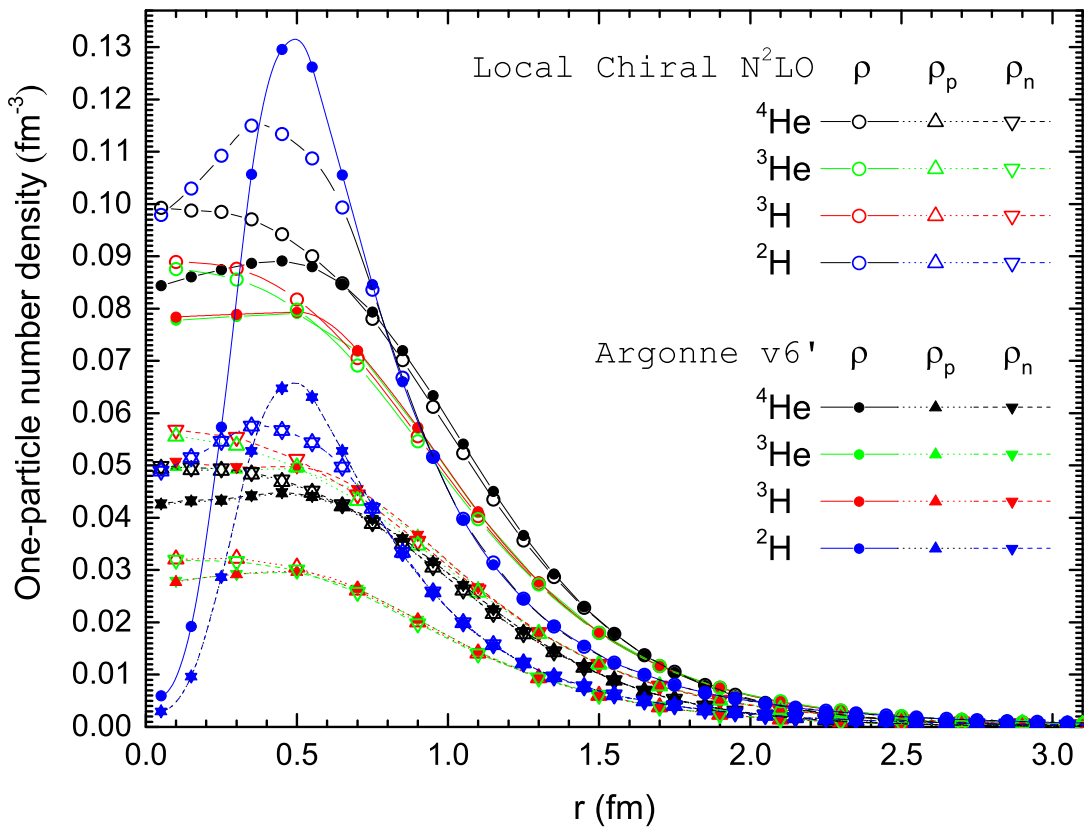}
\caption[The Ground State One-Particle Number Density Distributions of $A\leq 4$ Light Nuclei Based on the AV6' Interaction and the Local Chiral N$^2$LO Interaction.]{The ground state one-particle number density distributions of $A\leq 4$ light nuclei based on the AV6' interaction and the local chiral N$^2$LO interaction. The lines are a guide to the eye.}
\label{rho234AVN2LO}
\end{figure}

In Fig. \ref{rho234AVN2LO} all the density distributions for $A \leq 4$ light nuclei are presented.
For $^2$H the ${\rho}_{p}(r)$ and ${\rho}_{n}(r)$ are identical,
and as $r$ decreases, the density distributions based on the AV6' interaction decrease rapidly which are different from those based on the local chiral N$^2$LO interaction.
This is because, as shown in Fig. \ref{AV6N2LOradial}, for the two-nucleon system the central part becomes the dominant part as $r$ decreases, and the AV6' interaction has a much more repulsive central radial function than the chiral one, this makes the proton and the neutron highly unlikely to be very close to each other and the center of mass, which means density distributions will rapidly decrease as $r$ decreases.
For $^3$H there is one proton and two neutrons, and at a given $r$, it is reasonable that the probability of finding a neutron ${\rho}_{n}(r)$  is always higher than that of finding a proton ${\rho}_{p}(r)$. $^3$H and its mirror nuclei $^3$He shows nearly the same ${\rho}(r)$ distribution, however their ${\rho}_{p}(r)$ and ${\rho}_{n}(r)$ distributions are as if they are exchanged, i.e., $^3$H's ${\rho}_{p}(r)$ is like the $^3$He's ${\rho}_{n}(r)$ and $^3$H's ${\rho}_{n}(r)$ is like $^3$He's ${\rho}_{p}(r)$.

\section{Euclidean Response Functions}
\label{SECresponse}

Response functions are important quantities (\cite{Carlson92a, Benhar1994book, Carlson94a, Carlson02a,sick2003meson,Lovato13a,Lovato15a}) and the related operators do not commute with $H$. Using PIMC to calculate the response functions thus has a natural advantage over diffusion methods.

In an electron-nucleus scattering experiment, the response of a weakly coupled external probe can be written as the function $S(k,\omega)$ which can be expanded in the energy eigenstates (\cite{Carlson94a}), as
\begin{eqnarray}
S(k,\omega)=\frac{\sum_n \langle \Phi_0 | \rho^\dagger(\bm{k}) | \Phi_n \rangle \langle \Phi_n | \rho(\bm{k})| \Phi_0 \rangle  \delta(\omega+E_0-E_n)  }{\langle \Phi_0 | \Phi_0 \rangle} ,
\label{Skwdef}
\end{eqnarray}
where $\bm{k}$ is the momentum transfer between the final and initial momentum of the nucleus, and $\omega$ is the energy transfer which is the difference between the final and initial energy of the nucleus.
$(k,\omega)$ is the four-momentum carried by the virtual photon (\cite{Bacca2014JPG}) which is exchanged between the electron and the nucleus. $E_n$ is eigenenergy of the excited states $|\Phi_n \rangle$, and the $\rho(\bm{k})$ is the coupling operator.

The $S(k,\omega)$ in Eq.(\ref{Skwdef}) is called the response function.
It is useful because it is related with the scattering cross section therefore directly connects theory and experiment.
For different scattering processes the couplings of the probe to the nucleus gives different $\rho(\bm{k})$ operators.
However, it is not easy to directly calculate the response function itself.
Instead, we calculate the Euclidean response function $E(k,\tau)$ of $S(k,\omega)$ by using QMC, and then find ways (\cite{GShen2012PRC,Lovato15a}) to invert $E(k,\tau)$ to $S(k,\omega)$.

The Euclidean response function $E(k,\tau)$ is related to $S(k,\omega)$ by the following Laplace transform,
\begin{align}
E(k,\tau) &=
\int_{0}^{\infty} e^{-\tau(\omega-\omega_{qe})} S(k,\omega) d\omega \nonumber \\
&= \frac{e^{\omega_{qe}\tau} \sum_n \langle \Phi_0 | \rho^\dagger(\bm{k}) e^{-H \tau}  | \Phi_n \rangle \langle \Phi_n | \rho(\bm{k})| \Phi_0 \rangle   }
{ e^{-E_0 \tau} | \langle \Phi_0 | \Phi_0 \rangle} \nonumber \\
&=\frac{ e^{\omega_{qe}\tau} \langle \Phi_0| \rho^\dagger(\bm{k}) e^{-H\tau} \rho(\bm{k}) | \Phi_0 \rangle}{\langle \Phi_0| e^{-H\tau} | \Phi_0 \rangle} \nonumber \\
&=\frac{ e^{\omega_{qe}\tau} \langle \Psi_T | e^{-H \tau_1}
\left[\rho^\dagger(\bm{k}) e^{-H\tau} \rho(\bm{k})\right]
e^{-H \tau_1} | \Psi_T \rangle}
{\langle \Psi_T | e^{-H \tau_1} e^{-H\tau} e^{-H \tau_1} | \Psi_T \rangle},
\label{Ektauresponse}
\end{align}
where $\omega_{qe}=k^2/2m$ is the quasielastic energy transfer (\cite{lovato2020ab})  and $\tau_1$ is chosen to be large enough to project out the ground state $\Phi_0$.

In this work, we calculate the $^4$He Euclidean response functions that correspond to several different single-nucleon couplings $\rho(\bm{k})$, which are similar to those in Carlson and Schiavilla's 1994 paper (\cite{Carlson94a}).
Those include the nucleon coupling $\rho_N (\bm{k})$ , proton coupling $\rho_p (\bm{k})$, isovector coupling $\rho_\tau (\bm{k})$, spin-longitudinal coupling $\rho_{\sigma \tau L} (\bm{k})$ and spin-transverse coupling $\rho_{\sigma \tau T} (\bm{k})$. They are defined as,
\begin{align}
\rho_N (\bm{k}) & = \sum_{i=1}^{A} e^{i \bm{k} \cdot \bm{r}_i } , \\
\rho_p (\bm{k}) & = \sum_{i=1}^{A} e^{i \bm{k} \cdot \bm{r}_i } \frac{1+\tau_{iz}}{2} , \\
\rho_\tau (\bm{k}) & = \sum_{i=1}^{A} e^{i \bm{k} \cdot \bm{r}_i } \tau_{iz} , \\
\rho_{\sigma \tau L} (\bm{k}) & = \sum_{i=1}^{A} e^{i \bm{k} \cdot \bm{r}_i } (\bm{\sigma}_i \cdot \hat{\bm{k}}) \tau_{iz} , \\
\rho_{\sigma \tau T} (\bm{k}) & = \sum_{i=1}^{A} e^{i \bm{k} \cdot \bm{r}_i } (\bm{\sigma}_i \times \hat{\bm{k}}) \tau_{iz} .
\end{align}

Note that $\tau_{iz}$ is the difference between the proton and neutron projection operators, so it is related with the difference between protons and neutrons. $\bm{\sigma}_i \cdot \hat{\bm{k}}$ and $\bm{\sigma}_i \times \hat{\bm{k}}$ can be related with different mesons including pion, EM scattering, magnetic fields, etc (\cite{CarlsonRMP1998}).

In diffusion QMC such as GFMC and AFDMC,
it is often $\langle \Psi_T |\rho^\dagger(\bm{k}) e^{-H\tau} \rho(\bm{k}) |\Phi_0 \rangle$  that is calculated.
Since the operator $\rho^\dagger(\bm{k}) e^{-H\tau} \rho(\bm{k})$ does not commute with the Hamiltonian, $\langle \Psi_T |\rho^\dagger(\bm{k}) e^{-H\tau} \rho(\bm{k}) |\Phi_0 \rangle$ is mostly a mixed estimator (although forward walking extrapolation can improve this) instead of the true ground state estimator $\langle \Phi_0 |\rho^\dagger(\bm{k}) e^{-H\tau} \rho(\bm{k}) |\Phi_0 \rangle$.
This disadvantage of diffusion QMC is exactly the advantage of PIMC method, because PIMC can project out the ground state $\Phi_0$ from both $\langle \Psi_T |$ and $| \Psi_T \rangle$ by using the projection operator $e^{-H \tau_1}$ with the short-time approximation.
PIMC really calculates the ground state estimator $ \langle \Phi_0| \rho^\dagger(\bm{k}) e^{-H\tau} \rho(\bm{k}) | \Phi_0 \rangle$, whether $\rho^\dagger(\bm{k}) e^{-H\tau} \rho(\bm{k})$ commutes with the Hamiltonian or not.

\begin{figure}[htbp!]
\centering
\includegraphics[scale=0.45]{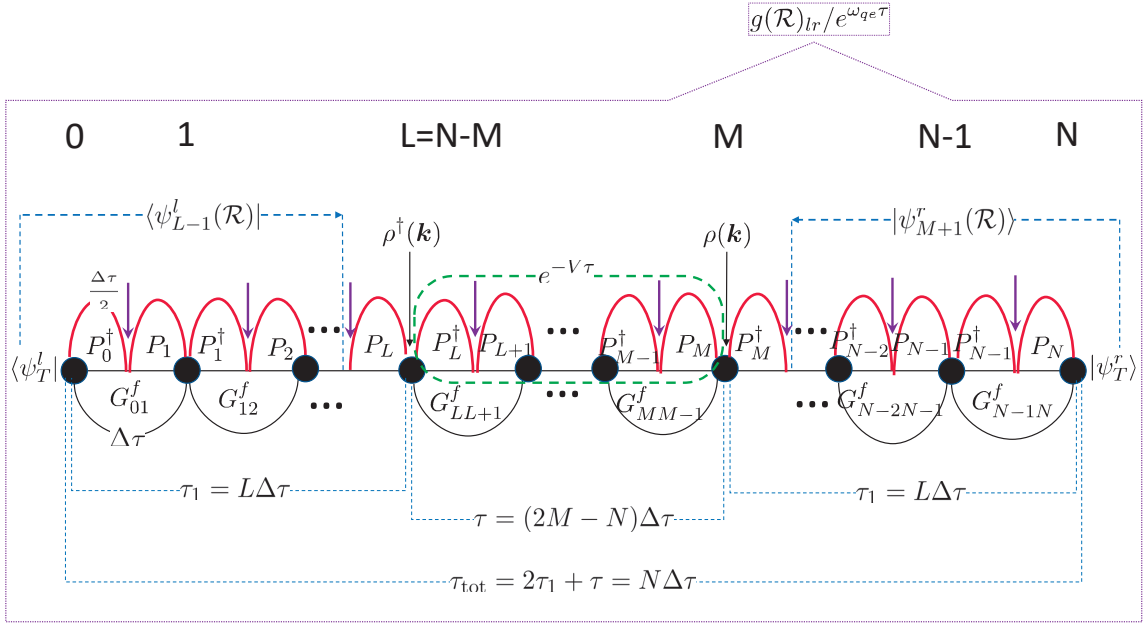}
\caption[Illustration of $g^V_{lr,M}(\mathcal{R})$ for Euclidean Response Functions]{Illustration of $g^V_{lr,M}(\mathcal{R})$ in Eq.(\ref{gVRlrresponse}) for Euclidean response functions.}
\label{PIMCBeadsResponse}
\end{figure}

The calculations of $E(k,\tau)$ in PIMC is straightforward. Eq.(\ref{Ektauresponse}) can be recast into the form of Eq.(\ref{OABP}) in section \ref{Recast}. As illustrated in Fig. \ref{PIMCBeadsResponse}, we can denote the total imaginary time as $\tau_{\textrm{tot}}=N\Delta \tau$, and $\tau_1=L \Delta \tau$, $L=N-M$, $\tau=(M-L)\Delta \tau =(2M-N) \Delta \tau$.
We just need to place the operator $\rho^\dagger(\bm{k})$ at bead $L$, and place $\rho (\bm{k})$ at bead $M$, and $g^V_{lr,M}(\mathcal{R})$ can be written as,
\begin{equation}
 g^V_{lr,M}(\mathcal{R}) = e^{\omega_{qe}\tau}   \langle \psi^l_{L\!-\!1}  (\mathcal{R}) |
 e^{\frac{- \mathcal{V}_L  \Delta \tau }{2}} \rho^\dagger(\bm{k}) e^{-V\tau} \rho(\bm{k}) e^{\frac{- \mathcal{V}^\dagger_M  \Delta \tau }{2}}
 | \psi^r_{M\!+\!1} (\mathcal{R}) \rangle ,
\label{gVRlrresponse}
\end{equation}
where $ e^{-V\tau}$ is defined as,
\begin{eqnarray}
e^{-V\tau} \! \equiv
 e^{\frac{- \mathcal{V}^\dagger_L  \Delta \tau }{2}} \bm{\mathcal{G}}
 \left(e^{\frac{- \mathcal{V}_{L+1}  \Delta \tau }{2}}  e^{\frac{- \mathcal{V}^\dagger_{L+1}  \Delta \tau }{2}}\right) \bm{\mathcal{G}} ~...~
 \bm{\mathcal{G}} \left(e^{\frac{- \mathcal{V}_{M-1}  \Delta \tau }{2}}  e^{\frac{- \mathcal{V}^\dagger_{M-1}  \Delta \tau }{2}} \right)\bm{\mathcal{G}}
 e^{\frac{- \mathcal{V}_M  \Delta \tau }{2}}  .
\label{eVtaudef}
\end{eqnarray}
All the other functions and calculations are the same as those in section \ref{Recast}.

The ground state of $^4$He has total isospin and its z component both 0,
and it is a spherically symmetric object.
Therefore the response functions do not depend on the direction of $\bm{k}$.
So when calculating the $E(k,\tau)$, the momentum $\bm{k}$ can be averaged over all the directions.
The angle averaged response functions are not calculated on the fly because they are relatively more time-consuming comparing with other quantities, so they are calculated after we have enough paths stored on the storage systems.
For a given amount of computation time, the angle averaged $E(k,\tau)$ have less variance than the non angle averaged $E(k,\tau)$.
The angle averaged $E(k,\tau)$ is calculated as,
\begin{eqnarray}
E(k,\tau) = \frac{ e^{\omega_{qe}\tau} \langle \Psi_T | e^{-H \tau_1}
\left[\frac{1}{4\pi}\int \rho^\dagger(\bm{k}) e^{-H\tau} \rho(\bm{k}) d\Omega\right]
e^{-H \tau_1} | \Psi_T \rangle}
{\langle \Psi_T | e^{-H \tau_1} e^{-H\tau} e^{-H \tau_1} | \Psi_T \rangle},
\label{Ektauresponseave}
\end{eqnarray}
so that the correspondingly $g^V_{lr,M}(\mathcal{R})$ is calculated as\footnote{The detailed angle averaged calculations of $\int \rho^\dagger(\bm{k}) \hat{O} \rho(\bm{k}) d \Omega/4\pi$ for each response function can be found in Appendix \ref{secAngleAveResponse}.},
\begin{eqnarray}
\!\! g^V_{lr,M}(\mathcal{R}) = e^{\omega_{qe}\tau}   \langle \psi^l_{L\!-\!1}  (\mathcal{R}) | \bm{\mathcal{G}}  e^{\frac{- \mathcal{V}_L  \Delta \tau }{2}}
 \left[ \frac{  \int \rho^\dagger(\bm{k}) e^{-V\tau} \rho(\bm{k})  d \Omega}{4\pi} \right]
 e^{\frac{- \mathcal{V}^\dagger_M  \Delta \tau }{2}} \bm{\mathcal{G}}
 | \psi^r_{M\!+\!1} (\mathcal{R}) \rangle,
\label{gVRlrresponseave}
\end{eqnarray}
where $e^{-V\tau}$ is defined in Eq.(\ref{eVtaudef}).
For any operator $\hat{O}$ which is independent of $\bm{k}$, e.g., $\hat{O}$ can be $e^{-H\tau}$ or $e^{-V\tau}$ as defined in Eq.(\ref{eVtaudef}).

\begin{figure}[htbp!]
\centering
\includegraphics[scale=0.46]{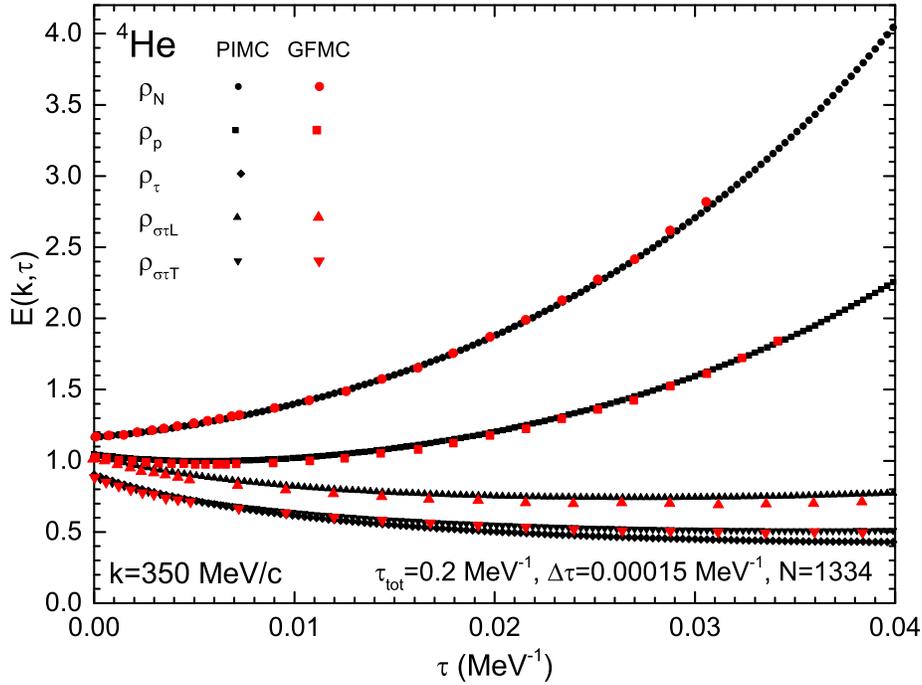}
\caption[Angle Averaged Euclidean Response Functions of $^4$He at $k=350\textrm{ MeV/c}$ for the AV6' Interaction]{Angle averaged Euclidean response functions of $^4$He at $k=350 \textrm{ MeV/c}$ for the AV6' interaction.}
\label{response}
\end{figure}

\begin{figure}[htbp!]
\centering
\includegraphics[scale=0.46]{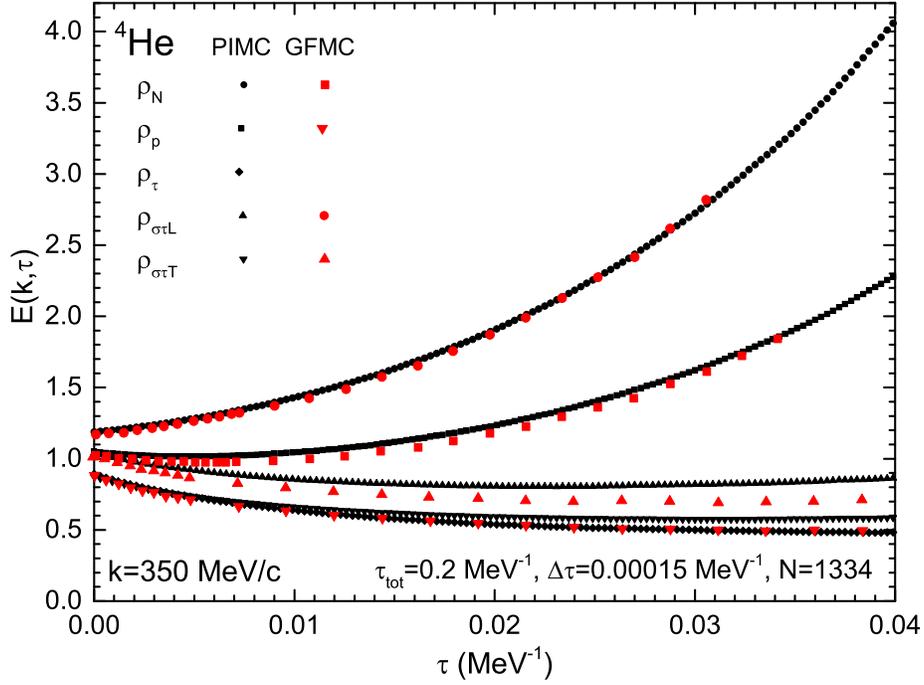}
\caption[Angle Averaged Euclidean Response Functions of $^4$He at $k=350\textrm{ MeV/c}$ for the Local Chiral N$^2$LO Interaction]{Angle averaged Euclidean response functions of $^4$He at $k=350 \textrm{ MeV/c}$ for the local chiral N$^2$LO interaction.}
\label{responseN2LO}
\end{figure}

Fig. \ref{response} and Fig \ref{responseN2LO} shows all the angle averaged Euclidean response functions $E_\alpha$ for each $\rho_\alpha$ defined in this work.
Fig. \ref{response} is for the AV6' interaction and Fig \ref{responseN2LO} is for the local chiral N$^2$LO interaction.
Each of the $E_\alpha$ has been normalized such that $\lim_{k\rightarrow \infty} E_\alpha(k,\tau=0)=1$.
Our calculations are done in the center of mass frame,
but we have converted our results to the lab frame\footnote{See Appendix \ref{secResconvert} for the conversion process.}.
We calculate our response functions in the center of mass frame first because it can give lower variance compare with lab frame. We then multiply our results by $e^{- 16.3\tau}$ (the 16.3 MeV is the center of mass energy for this $k=350\textrm{ MeV/c}$) in order to compare with the results in Carlson and  Schiavilla's paper (\cite{Carlson94a}) which are digitized as the red dots in Figs. \ref{response} and \ref{responseN2LO}.
The small difference between ours and theirs comes from several sources.
First, we use PIMC, they use GFMC.
Second, the interactions are somewhat different, they used the Argonne $v8$ interaction combined with the Urbana model-\uppercase\expandafter{\romannumeral8}.
Third, we use $\tau_{iz}$ in stead of $\tau_+(i)$ for the isovector, spin-longitudinal and spin-transverse couplings, this is suggested in their paper when dealing with an isoscalar target like the ground state $^4$He, and with interactions that conserve the number of protons and neutrons. The small difference can also indicate our results are more accurate because in PIMC we do not need forward walking as DMC does.

\section{Parallel Scalability of the Code}

\begin{figure}[hptb!]
\centering
\includegraphics[scale=0.46]{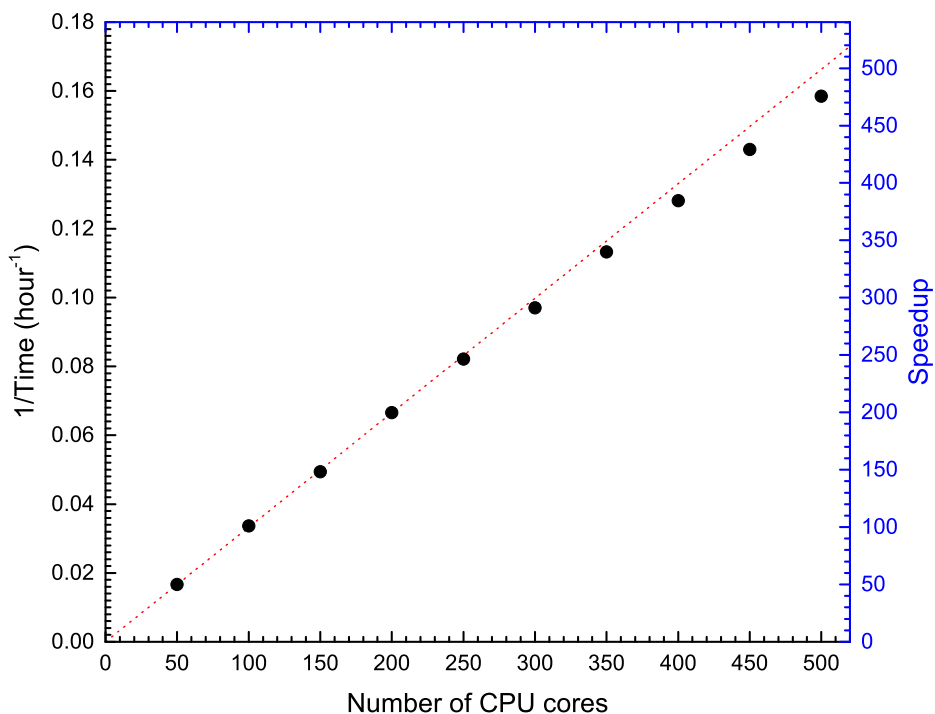}
\caption[The Scaling Efficiency of Our PIMC Code]{The scaling efficiency of our PIMC code. It is based on sampling $4 \times 10^5$ samples of $^4$He for $\tau=0.3 \textrm{ MeV}^{-1} $ with $\Delta \tau=2\times 10^{-4} \textrm{ MeV}^{-1}$.}
\label{figscaling}
\end{figure}

Our code is parallel and uses the Message Passing Interface (MPI).
Each CPU core is assigned the same amount of blocks of steps.
At every about 125 steps, all the cores collect the values of the samples and send them to the rank 0 core.
They also send the information about the current state $s$ to rank 0 for it to write them to a file which includes all the information of all these independent paths\footnote{
Depending on the number of CPU cores, the number of beads in the PIMC calculation and the total computation time involved, a complete path file can take from tens of GB to several TB storage space.
In the future we may use MPI/IO which can further improve the parallel scaling.
}.
The information of a state (path) as shown in Eq.(\ref{state_s}), includes the spatial configuration of all the beads $\mathcal{R}$ and $l$ and $r$ order of the trial wave functions. The calculations are stopped when we have enough samples to make the error bar of each quantity smaller than the desired values.

Since the code is parallel, we checked the scaling of our PIMC code\footnote{
All the calculations in this work are performed at the Agave cluster at Research Computing at Arizona State University during 2019 and 2020. The Cluster is mostly configured with Intel Xeon E5-2680v4 CPUs (CPU Architecture Broadwell) running at 2.40GHz at each node. Each node has 28 cores. Each core is configured with about 4.5GB of RAM. There are 297 nodes so there are 8316 cores ($\approx 320$ FP64 TFLOPS) and 4.5PB storage space. Sometimes we also use some Intel Xeon Gold 6230 and Intel Xeon Gold 6252 (CPU Architectures are both CascadeLake). Occasionally, we use several AMD EPYC 7551 (CPU Architecture is Naples) too. Our code currently does not run on coprocessors such as Intel Xeon Phi 7210 (CPU Architecture Knights Landing), because although each Xeon Phi CPU has hundreds of cores, each core is slow and does not have enough RAM. We checked the scaling only up to 500 cores simply because it is difficult to request much more than that number of cores to actually run the code on the Agave cluster.
}
which is shown in Fig. \ref{figscaling}. We plot the inverse of the total time it takes vs. the number of CPU cores involved.
We expect that based on doing the same amount of work, the inverse of the total time it takes should be proportional to the number of the CPU cores.
The labels on the right y-axis is the corresponding speedup, i.e., ideally the speedup for $n$ CPU cores should be $n$.
The dotted line is the prediction based on the time 50 CPU cores takes.
We can conclude from Fig. \ref{figscaling} that our scaling is in the linear range so our MPI code is efficient.

\chapter{Conclusion}
\label{secSumOutlook}

\section{Summary}
\label{summary}

This work is mainly to answer one of the questions in section \ref{secBackground}: how to solve the many-body problem?
Our answer is, using the path integral quantum Monte Carlo method.
In nuclear physics, although continuous space light nuclei calculations for $A\leq 4$ have been studied before by GFMC (\cite{Carlson88a,WiringaPieper02a}), this is the first work using PIMC. The code is now publicly available and open source (\cite{CRgitrepov6pimc1}).

By using multiple sampling algorithms, multi-level sampling most of the time,  and occasionally the regular sampling described in section \ref{secSamplingMethod},
as well as the optimized calculation and updating strategies described in section \ref{secCalcPath} and section \ref{secUpdateStrat}, we successfully performed our ground state nuclear PIMC calculation of light nuclei, based on the local chiral N$^2$LO interaction and the AV6' interaction.
From our analysis in section \ref{setup} and as well as calculation of the potential along the beads shown in Fig. \ref{potential}, we can conclude that our choice of the total imaginary time $\tau=0.2$ MeV$^{-1}$ indeed projected out the ground state, so our results are reliable.
As discussed in section \ref{setup}, the time step $\Delta \tau$ we use is $10^{-4} \sim 10^{-3} \textrm{ MeV}^{-1}$ which are about 5 times bigger than GFMC calculations. This indicates that PIMC calculations are less sensitive to $\Delta \tau$ than GFMC is, due to the time step error structure of the PIMC expectation value such as Eq.(\ref{PIMCUSdetail}) or Eq.(\ref{PIMCUN2LOdetail}).

Our PIMC ground state energies of light nuclei are consistent with the results from GFMC and AFDMC as shown in Tables \ref{TablePIMC2} and \ref{TablePIMCN2LO}.
For operators which do not commute with Hamiltonian and therefore usually cannot be obtained from DMC based methods such as GFMC and AFDMC without tradeoffs, we calculated the particle number density distribution $\rho(r)$, rms radii, and the angle averaged response functions.
Our results are comparable with the GFMC results.

\section{Outlook}

We hope our $^4$He calculation may be served as a benchmark test for future PIMC calculations of larger nuclei.
The reason to do larger nuclei is because we can first test how good the nucleon-nucleon interaction is, then we can try to predict the properties of neutron rich nuclei which are most likely generated in supernovae,
or in the laboratories\footnote{Such as the Facility for Rare Isotope Beams (FRIB) at Michigan State University.}.
However, larger nuclei PIMC calculations will be more complicated and computational expensive than the light nuclei calculations presented in this work.
We estimate that without significant improvement,
our PIMC may take similar computational cost as GFMC does.
We can try to do several things to make the calculations more accessible.

We can try to increase the time step and reduce the beads needed in PIMC and therefore save computation time.
In fact, in our PIMC calculations, the time step $\Delta \tau$ are about an order of magnitude larger than those used in DMC based methods.
If a pair-product propagator (\cite{Ceperley95a,SchmidtLee95a,Pudliner1997PRC,CarlsonRMP15a}) is used, it may enable us to use even bigger time steps, this can reduce the number of beads in PIMC calculation and save computation time.

We can try to reduce the size of the spin-isospin basis.
We may separate the interaction into two parts, one part conserves total isospin such as the AV6' NN operators,
and other part includes isospin breaking terms, such as the Coulomb interaction.
We can first use a good isospin basis for the Hamiltonian with these nuclear interactions only,
because the number of isospin basis states can be somewhat smaller than the charge basis we use now.
We can save the spatial and other relevant configurations,
and then, we can use them to calculate the isospin breaking parts perturbatively (\cite{Lonardoni2018PRC}).

We may also try to combine AFDMC ideas to efficiently sample the spin isospin states instead of summing over them.
If this can be made possible it will significantly reduce the computation time for PIMC calculations of larger nuclei.

A nuclear PIMC finite temperature calculation of $^4$He or neutron matter including three-body interactions is also possible.
That can be a pioneering step towards QMC calculations in continuous space for a finite temperature nuclear equation of state (\cite{LWChen2008a,Chen2010PRC}).

However, large nuclei and finite temperature nuclear matter PIMC calculations will inevitably lead to large variance due to sign problem unless approximate methods like adding fixed phase or fixed node constraints are used.
The sign problem has no known general fix yet.
Improving the trial wave function and the short-time propagator, and constraining the path can somewhat make the sign problem less severe or eliminate it giving an approximate result.

Last but not least, even though we will face sign problem, and, even with the most powerful supercomputers today, there are still problems in QMC that seem to make calculating large nuclei or the equation of state at finite temperature not accessible at the current stage, those messes simply mean there are big opportunities.
I simply cannot imagine human beings cannot go far beyond the current situation (\cite{Laughlin2000}).
And I would like to use Steven Weinberg's advice (\cite{Weinberg2003}) to end my dissertation, which is also what I learned from my advisor professor Kevin E. Schmidt these years:

`` ... go for the messes --- that's where the action is. ''

\newpage
{\singlespace
\phantomsection
\addcontentsline{toc}{part}{REFERENCES}
\bibliographystyle{asudis}

\begin{thebibliography}{0}
\newcommand{\enquote}[1]{``#1''}
\expandafter\ifx\csname natexlab\endcsname\relax\def\natexlab#1{#1}\fi
\expandafter\ifx\csname url\endcsname\relax
  \def\url#1{\texttt{#1}}\fi
\expandafter\ifx\csname urlprefix\endcsname\relax\def\urlprefix{URL }\fi

\end{thebibliography}


\begin{thebibliography}{111}
\newcommand{\enquote}[1]{``#1''}
\expandafter\ifx\csname natexlab\endcsname\relax\def\natexlab#1{#1}\fi
\expandafter\ifx\csname url\endcsname\relax
  \def\url#1{\texttt{#1}}\fi
\expandafter\ifx\csname urlprefix\endcsname\relax\def\urlprefix{URL }\fi

\bibitem[Abramowitz and Stegun(1964)]{Handbook_abramowitz}
Abramowitz, M. and I.~A. Stegun, \emph{Handbook of Mathematical Functions with
  Formulas, Graphs, and Mathematical Tables} (Dover, New York, 1964), ninth
  dover printing, tenth gpo printing edn.

\bibitem[Alhassid \emph{et~al.}(1994)]{Alhassid1994PRL}
Alhassid, Y., D.~J. Dean, S.~E. Koonin, G.~Lang and W.~E. Ormand,
  \enquote{Practical solution to the {Monte} {Carlo} sign problem: Realistic
  calculations of $^{54}\mathrm{Fe}$}, Physical Review Letters \textbf{72},
  613--616, \urlprefix\url{https://link.aps.org/doi/10.1103/PhysRevLett.72.613}
  (1994).

\bibitem[Anderson(1995)]{JBAnderson2008}
Anderson, J.~B., \enquote{Fixed-node quantum {Monte} {Carlo}}, International
  Reviews in Physical Chemistry \textbf{14}, 1, 85--112,
  \urlprefix\url{https://doi.org/10.1080/01442359509353305} (1995).

\bibitem[Apaja(2009)]{VesaPDF}
Apaja, V., \enquote{{M}onte {C}arlo methods version 2.23},
  \urlprefix\url{http://users.jyu.fi/~veapaja/Monte_Carlo/MC.lecture.vesa.pdf}
  (2009).

\bibitem[Bacca and Pastore(2014)]{Bacca2014JPG}
Bacca, S. and S.~Pastore, \enquote{Electromagnetic reactions on light nuclei},
  Journal of Physics G: Nuclear and Particle Physics \textbf{41}, 12, 123002,
  \urlprefix\url{https://doi.org/10.1088%2F0954-3899%2F41%2F12%2F123002}
  (2014).

\bibitem[Bajdich \emph{et~al.}(2006)]{Bajdich2006PRL}
Bajdich, M., L.~Mitas, G.~Drobn\'y, L.~K. Wagner and K.~E. Schmidt,
  \enquote{{Pfaffian} pairing wave functions in electronic-structure quantum
  {Monte} {Carlo} simulations}, Physical Review Letters \textbf{96}, 130201,
  \urlprefix\url{https://link.aps.org/doi/10.1103/PhysRevLett.96.130201}
  (2006).

\bibitem[Baroni and Moroni(1999)]{Baroni99a}
Baroni, S. and S.~Moroni, \enquote{Reptation quantum {Monte} {Carlo}: A method
  for unbiased ground-state averages and imaginary-time correlations}, Physical
  Review Letters \textbf{82}, 24, 4745--4748,
  \urlprefix\url{http://link.aps.org/doi/10.1103/PhysRevLett.82.4745} (1999).

\bibitem[Barrett \emph{et~al.}(2013)]{BARRETT2013131}
Barrett, B.~R., Navr{\'{a}}til and J.~P. Vary, \enquote{Ab initio no core shell
  model}, Progress in Particle and Nuclear Physics \textbf{69}, 131 -- 181,
  \urlprefix\url{http://www.sciencedirect.com/science/article/pii/S0146641012001184}
  (2013).

\bibitem[Beane \emph{et~al.}(2008)]{BEANE2008}
Beane, S.~R., K.~Orginos and M.~J. Savage, \enquote{Hadronic interactions from
  lattice {QCD}}, International Journal of Modern Physics E \textbf{17}, 07,
  1157--1218, \urlprefix\url{https://doi.org/10.1142/s0218301308010404} (2008).

\bibitem[Benhar \emph{et~al.}(1994)]{Benhar1994book}
Benhar, O., A.~Fabrocini and R.~Schiavilla, \emph{Electron-nucleus Scattering}
  (WORLD SCIENTIFIC, 1994),
  \urlprefix\url{https://www.worldscientific.com/doi/abs/10.1142/2278}.

\bibitem[Bouchaud \emph{et~al.}(1988)]{Bouchaud1988a}
Bouchaud, J., A.~Georges and C.~Lhuillier, \enquote{Pair wave functions for
  strongly correlated fermions and their determinantal representation}, Journal
  de Physique \textbf{49}, 4, 553--559,
  \urlprefix\url{https://doi.org/10.1051/jphys:01988004904055300} (1988).

\bibitem[Carleo \emph{et~al.}(2010)]{Baroni10a}
Carleo, G., F.~Becca, S.~Moroni and S.~Baroni, \enquote{Reptation quantum
  {Monte} {Carlo} algorithm for lattice {Hamiltonians} with a directed-update
  scheme}, Physical Review E \textbf{82}, 046710,
  \urlprefix\url{https://link.aps.org/doi/10.1103/PhysRevE.82.046710} (2010).

\bibitem[Carlson(1987)]{Carlson87a}
Carlson, J., \enquote{{Green}'s function {Monte} {Carlo} study of light
  nuclei}, Physical Review C \textbf{36}, 2026--2033,
  \urlprefix\url{https://link.aps.org/doi/10.1103/PhysRevC.36.2026} (1987).

\bibitem[Carlson(1988)]{Carlson88a}
Carlson, J., \enquote{Alpha particle structure}, Physical Review C \textbf{38},
  4, 1879--1885, \urlprefix\url{https://doi.org/10.1103/PhysRevC.38.1879}
  (1988).

\bibitem[Carlson \emph{et~al.}(2015)]{CarlsonRMP15a}
Carlson, J., S.~Gandolfi, F.~Pederiva, S.~C. Pieper, R.~Schiavilla, K.~E.
  Schmidt and R.~B. Wiringa, \enquote{Quantum {Monte} {Carlo} methods for
  nuclear physics}, Reviews of Modern Physics \textbf{87}, 1067--1118,
  \urlprefix\url{https://link.aps.org/doi/10.1103/RevModPhys.87.1067} (2015).

\bibitem[Carlson \emph{et~al.}(2002)]{Carlson02a}
Carlson, J., J.~Jourdan, R.~Schiavilla and I.~Sick, \enquote{Longitudinal and
  transverse quasielastic response functions of light nuclei}, Physical Review
  C \textbf{65}, 024002,
  \urlprefix\url{https://link.aps.org/doi/10.1103/PhysRevC.65.024002} (2002).

\bibitem[Carlson \emph{et~al.}(1983)]{CARLSON198359}
Carlson, J., V.~Pandharipande and R.~Wiringa, \enquote{Three-nucleon
  interaction in 3-, 4- and $\infty$-body systems}, Nuclear Physics A
  \textbf{401}, 1, 59 -- 85,
  \urlprefix\url{http://www.sciencedirect.com/science/article/pii/0375947483903366}
  (1983).

\bibitem[Carlson and Schiavilla(1992)]{Carlson92a}
Carlson, J. and R.~Schiavilla, \enquote{{Euclidean} proton response in light
  nuclei}, Physical Review Letters \textbf{68}, 3682--3685,
  \urlprefix\url{https://link.aps.org/doi/10.1103/PhysRevLett.68.3682} (1992).

\bibitem[Carlson and Schiavilla(1994)]{Carlson94a}
Carlson, J. and R.~Schiavilla, \enquote{Inclusive electron scattering and pion
  degrees of freedom in light nuclei}, Physical Review C \textbf{49},
  R2880--R2884,
  \urlprefix\url{https://link.aps.org/doi/10.1103/PhysRevC.49.R2880} (1994).

\bibitem[Carlson and Schiavilla(1998)]{CarlsonRMP1998}
Carlson, J. and R.~Schiavilla, \enquote{Structure and dynamics of few-nucleon
  systems}, Reviews of Modern Physics \textbf{70}, 743--841,
  \urlprefix\url{https://link.aps.org/doi/10.1103/RevModPhys.70.743} (1998).

\bibitem[Carlson and Wiringa(1991)]{Koonin91}
Carlson, J. and R.~B. Wiringa, \enquote{Variational {Monte}-{Carlo} techniques
  in nuclear physics}, in \enquote{Computational Nuclear Physics 1: Nuclear
  Structure}, edited by K.~Langanke, J.~A. Maruhn and S.~E. Koonin, chap.~9,
  pp. 171--187 (Springer-Verlag, 1991),
  \urlprefix\url{https://www.springer.com/gp/book/9783642763588}.

\bibitem[Casulleras and Boronat(1995)]{Casulleras1995PRB}
Casulleras, J. and J.~Boronat, \enquote{Unbiased estimators in quantum {Monte}
  {Carlo} methods: Application to liquid $^{4}\mathrm{He}$}, Physical Review B
  \textbf{52}, 3654--3661,
  \urlprefix\url{https://link.aps.org/doi/10.1103/PhysRevB.52.3654} (1995).

\bibitem[Ceperley(1995)]{Ceperley95a}
Ceperley, D.~M., \enquote{Path integrals in the theory of condensed helium},
  Reviews of Modern Physics \textbf{67}, 2, 279--355,
  \urlprefix\url{https://doi.org/10.1103/RevModPhys.67.279} (1995).

\bibitem[Chen \emph{et~al.}(2010)]{Chen2010PRC}
Chen, L.-W., C.~M. Ko, B.-A. Li and J.~Xu, \enquote{Density slope of the
  nuclear symmetry energy from the neutron skin thickness of heavy nuclei},
  Physical Review C \textbf{82}, 024321,
  \urlprefix\url{https://link.aps.org/doi/10.1103/PhysRevC.82.024321} (2010).

\bibitem[Chen \emph{et~al.}(2012)]{ChenRong2012PRC}
Chen, R., B.-J. Cai, L.-W. Chen, B.-A. Li, X.-H. Li and C.~Xu,
  \enquote{Single-nucleon potential decomposition of the nuclear symmetry
  energy}, Physical Review C \textbf{85}, 024305,
  \urlprefix\url{https://link.aps.org/doi/10.1103/PhysRevC.85.024305} (2012).

\bibitem[Chen and Schmidt(2020)]{CRgitrepov6pimc1}
Chen, R. and K.~E. Schmidt, \enquote{Project v6pimc1},
  \url{https://gitlab.com/CRquantum/v6pimcmpi}, {GitLab repository} (2020).

\bibitem[Chin(1990)]{Chin90}
Chin, S.~A., \enquote{Quadratic diffusion {Monte} {Carlo} algorithms for
  solving atomic many-body problems}, Physical Review A \textbf{42},
  6991--7005, \urlprefix\url{https://link.aps.org/doi/10.1103/PhysRevA.42.6991}
  (1990).

\bibitem[Chodosh(2017)]{ChodoshEVO1946}
Chodosh, S., \enquote{The incredible evolution of supercomputers' powers, from
  1946 to today},
  \urlprefix\url{https://www.popsci.com/supercomputers-then-and-now/} (2017).

\bibitem[Coon and Gl\"ockle(1981)]{Coon1981PRC}
Coon, S.~A. and W.~Gl\"ockle, \enquote{Two-pion-exchange three-nucleon
  potential: Partial wave analysis in momentum space}, Physical Review C
  \textbf{23}, 1790--1802,
  \urlprefix\url{https://link.aps.org/doi/10.1103/PhysRevC.23.1790} (1981).

\bibitem[Davies(2002)]{davies2002lattice}
Davies, C., \enquote{{Lattice QCD}}, arXiv e-prints
  \urlprefix\url{https://arxiv.org/abs/hep-ph/0205181} (2002).

\bibitem[Decharg\'e and Gogny(1980)]{Gogny1980PRC}
Decharg\'e, J. and D.~Gogny, \enquote{{Hartree}-{Fock}-{Bogolyubov}
  calculations with the {$D1$} effective interaction on spherical nuclei},
  Physical Review C \textbf{21}, 1568--1593,
  \urlprefix\url{https://link.aps.org/doi/10.1103/PhysRevC.21.1568} (1980).

\bibitem[Epelbaum \emph{et~al.}(2009)]{Epelbaum2009a}
Epelbaum, E., H.-W. Hammer and U.-G. Mei\ss{}ner, \enquote{Modern theory of
  nuclear forces}, Reviews of Modern Physics \textbf{81}, 1773--1825,
  \urlprefix\url{https://link.aps.org/doi/10.1103/RevModPhys.81.1773} (2009).

\bibitem[Epelbaum \emph{et~al.}(2012)]{Lee2012PRL}
Epelbaum, E., H.~Krebs, T.~A. L\"ahde, D.~Lee and U.-G. Mei\ss{}ner,
  \enquote{Structure and rotations of the {Hoyle} state}, Physical Review
  Letters \textbf{109}, 252501,
  \urlprefix\url{https://link.aps.org/doi/10.1103/PhysRevLett.109.252501}
  (2012).

\bibitem[Epelbaum \emph{et~al.}(2002)]{Epelbaum2002PRC}
Epelbaum, E., A.~Nogga, W.~Gl\"ockle, H.~Kamada, U.-G. Mei\ss{}ner and
  H.~Wita\l{}a, \enquote{Three-nucleon forces from chiral effective field
  theory}, Physical Review C \textbf{66}, 064001,
  \urlprefix\url{https://link.aps.org/doi/10.1103/PhysRevC.66.064001} (2002).

\bibitem[Feynman(1948)]{Feynman1948RMP}
Feynman, R.~P., \enquote{Space-time approach to non-relativistic quantum
  mechanics}, Reviews of Modern Physics \textbf{20}, 367--387,
  \urlprefix\url{https://link.aps.org/doi/10.1103/RevModPhys.20.367} (1948).

\bibitem[Fornberg(1988)]{Fornberg88a}
Fornberg, B., \enquote{Generation of finite difference formulas on arbitrarily
  spaced grids}, Mathematics of Computation \textbf{51}, 699--706,
  \urlprefix\url{https://www.ams.org/journals/mcom/1988-51-184/S0025-5718-1988-0935077-0/}
  (1988).

\bibitem[Frame \emph{et~al.}(2018)]{DLee2018PRL}
Frame, D., R.~He, I.~Ipsen, D.~Lee, D.~Lee and E.~Rrapaj, \enquote{Eigenvector
  continuation with subspace learning}, Physical Review Letters \textbf{121},
  032501,
  \urlprefix\url{https://link.aps.org/doi/10.1103/PhysRevLett.121.032501}
  (2018).

\bibitem[Gandolfi(2007)]{gandolfi2007auxiliary}
Gandolfi, S., \enquote{The auxiliary field diffusion {Monte} {Carlo} method for
  nuclear physics and nuclear astrophysics}, arXiv e-prints
  \urlprefix\url{https://arxiv.org/abs/0712.1364} (2007).

\bibitem[Gandolfi \emph{et~al.}(2009)]{Gandolfi2009PRCa}
Gandolfi, S., A.~Y. Illarionov, K.~E. Schmidt, F.~Pederiva and S.~Fantoni,
  \enquote{Quantum {Monte} {Carlo} calculation of the equation of state of
  neutron matter}, Physical Review C \textbf{79}, 054005,
  \urlprefix\url{https://link.aps.org/doi/10.1103/PhysRevC.79.054005} (2009).

\bibitem[Gezerlis \emph{et~al.}(2014)]{Gerzerlis14a}
Gezerlis, A., I.~Tews, E.~Epelbaum, M.~Freunek, S.~Gandolfi, K.~Hebeler,
  A.~Nogga and A.~Schwenk, \enquote{Local chiral effective field theory
  interactions and quantum {Monte} {Carlo} applications}, Physical Review C
  \textbf{90}, 054323,
  \urlprefix\url{https://link.aps.org/doi/10.1103/PhysRevC.90.054323} (2014).

\bibitem[Gezerlis \emph{et~al.}(2013)]{Gerzerlis13a}
Gezerlis, A., I.~Tews, E.~Epelbaum, S.~Gandolfi, K.~Hebeler, A.~Nogga and
  A.~Schwenk, \enquote{Quantum {Monte} {Carlo} calculations with chiral
  effective field theory interactions}, Physical Review Letters \textbf{111},
  032501,
  \urlprefix\url{https://link.aps.org/doi/10.1103/PhysRevLett.111.032501}
  (2013).

\bibitem[Gross and Wilczek(1973)]{Wilczek73a}
Gross, D.~J. and F.~Wilczek, \enquote{Ultraviolet behavior of non-abelian gauge
  theories}, Physical Review Letters \textbf{30}, 1343--1346,
  \urlprefix\url{https://link.aps.org/doi/10.1103/PhysRevLett.30.1343} (1973).

\bibitem[Hagen \emph{et~al.}(2014)]{Hagen_2014}
Hagen, G., T.~Papenbrock, M.~Hjorth-Jensen and D.~J. Dean,
  \enquote{Coupled-cluster computations of atomic nuclei}, Reports on Progress
  in Physics \textbf{77}, 9, 096302,
  \urlprefix\url{https://doi.org/10.1088%2F0034-4885%2F77%2F9%2F096302} (2014).

\bibitem[Hammond \emph{et~al.}(1994)]{HammondMCbook}
Hammond, B.~L., W.~A. Lester and P.~J. Reynolds, \emph{{Monte} {Carlo} Methods
  in Ab Initio Quantum Chemistry} (WORLD SCIENTIFIC, 1994),
  \urlprefix\url{https://www.worldscientific.com/doi/abs/10.1142/1170}.

\bibitem[Jackson(1999)]{jackson3rd}
Jackson, J.~D., \emph{Classical electrodynamics} (Wiley, New York, {NY}, 1999),
  3rd ed. edn., \urlprefix\url{http://cdsweb.cern.ch/record/490457}.

\bibitem[Kalos \emph{et~al.}(1974)]{Kalos1974PRA}
Kalos, M.~H., D.~Levesque and L.~Verlet, \enquote{Helium at zero temperature
  with hard-sphere and other forces}, Physical Review A \textbf{9}, 2178--2195,
  \urlprefix\url{https://link.aps.org/doi/10.1103/PhysRevA.9.2178} (1974).

\bibitem[Kalos and Whitlock(1986)]{Kalos1986a}
Kalos, M.~H. and P.~A. Whitlock, \emph{{Monte} {Carlo} Methods. Vol. 1: Basics}
  (Wiley-Interscience, New York, NY, USA, 1986),
  \urlprefix\url{https://onlinelibrary.wiley.com/doi/abs/10.1002/bbpc.198800128}.

\bibitem[Kalos and Whitlock(2008)]{Kalosbook2008}
Kalos, M.~H. and P.~A. Whitlock, \emph{{Monte} {Carlo} Methods} (Wiley, 2008),
  \urlprefix\url{https://doi.org/10.1002/9783527626212}.

\bibitem[Lattes \emph{et~al.}(1947)]{Meson1947a}
Lattes, C. M.~G., H.~Muirhead, G.~P.~S. Occhialini and C.~F. Powell,
  \enquote{Processes involving charged mesons}, Nature \textbf{159}, 4047,
  694--697, \urlprefix\url{https://doi.org/10.1038/159694a0} (1947).

\bibitem[Laughlin and Pines(2000)]{Laughlin2000}
Laughlin, R.~B. and D.~Pines, \enquote{The theory of everything}, Proceedings
  of The National Academy of Sciences \textbf{97}, 1, 28--31,
  \urlprefix\url{https://www.pnas.org/content/97/1/28} (2000).

\bibitem[Lee(2009)]{LEE2009117}
Lee, D., \enquote{Lattice simulations for few- and many-body systems}, Progress
  in Particle and Nuclear Physics \textbf{63}, 1, 117 -- 154,
  \urlprefix\url{http://www.sciencedirect.com/science/article/pii/S014664100800094X}
  (2009).

\bibitem[Lee(2020)]{Lee2020FiP}
Lee, D., \enquote{Recent progress in nuclear lattice simulations}, Frontiers in
  Physics \textbf{8}, 174,
  \urlprefix\url{https://www.frontiersin.org/article/10.3389/fphy.2020.00174}
  (2020).

\bibitem[Li \emph{et~al.}(2008)]{LWChen2008a}
Li, B.-A., L.-W. Chen and C.~M. Ko, \enquote{Recent progress and new challenges
  in isospin physics with heavy-ion reactions}, Physics Reports \textbf{464},
  4, 113 -- 281,
  \urlprefix\url{http://www.sciencedirect.com/science/article/pii/S0370157308001269}
  (2008).

\bibitem[Li and Yao(2019)]{LiZiXiang2019a}
Li, Z.-X. and H.~Yao, \enquote{Sign-problem-free fermionic quantum {Monte}
  {Carlo}: Developments and applications}, Annual Review of Condensed Matter
  Physics \textbf{10}, 1, 337--356,
  \urlprefix\url{https://doi.org/10.1146/annurev-conmatphys-033117-054307}
  (2019).

\bibitem[Lomnitz-Adler \emph{et~al.}(1981)]{Lomnitz1981NPA}
Lomnitz-Adler, J., V.~Pandharipande and R.~Smith, \enquote{{Monte} {Carlo}
  calculations of triton and {$^4$He} nuclei with the {Reid} potential},
  Nuclear Physics A \textbf{361}, 2, 399 -- 411,
  \urlprefix\url{http://www.sciencedirect.com/science/article/pii/0375947481906424}
  (1981).

\bibitem[Lonardoni \emph{et~al.}(2018{\natexlab{a}})]{Lonardoni2018PRL}
Lonardoni, D., J.~Carlson, S.~Gandolfi, J.~E. Lynn, K.~E. Schmidt, A.~Schwenk
  and X.~B. Wang, \enquote{Properties of nuclei up to {$A=16$} using local
  chiral interactions}, Physical Review Letters \textbf{120}, 122502,
  \urlprefix\url{https://link.aps.org/doi/10.1103/PhysRevLett.120.122502}
  (2018{\natexlab{a}}).

\bibitem[Lonardoni \emph{et~al.}(2018{\natexlab{b}})]{Lonardoni2018PRC}
Lonardoni, D., S.~Gandolfi, J.~E. Lynn, C.~Petrie, J.~Carlson, K.~E. Schmidt
  and A.~Schwenk, \enquote{Auxiliary field diffusion {Monte} {Carlo}
  calculations of light and medium-mass nuclei with local chiral interactions},
  Physical Review C \textbf{97}, 044318,
  \urlprefix\url{https://link.aps.org/doi/10.1103/PhysRevC.97.044318}
  (2018{\natexlab{b}}).

\bibitem[{Lovato} \emph{et~al.}(2020)]{lovato2020ab}
{Lovato}, A., J.~{Carlson}, S.~{Gandolfi}, N.~{Rocco} and R.~{Schiavilla},
  \enquote{{Ab initio study of $\boldsymbol{(\nu_\ell,\ell^-)}$ and
  $\boldsymbol{(\overline{\nu}_\ell,\ell^+)}$ inclusive scattering in $^{12}$C:
  confronting the MiniBooNE and T2K CCQE data}}, arXiv e-prints
  \urlprefix\url{https://arxiv.org/abs/2003.07710} (2020).

\bibitem[Lovato \emph{et~al.}(2013)]{Lovato13a}
Lovato, A., S.~Gandolfi, R.~Butler, J.~Carlson, E.~Lusk, S.~C. Pieper and
  R.~Schiavilla, \enquote{Charge form factor and sum rules of electromagnetic
  response functions in $^{12}\mathbf{C}$}, Physical Review Letters
  \textbf{111}, 092501,
  \urlprefix\url{https://link.aps.org/doi/10.1103/PhysRevLett.111.092501}
  (2013).

\bibitem[Lovato \emph{et~al.}(2015)]{Lovato15a}
Lovato, A., S.~Gandolfi, J.~Carlson, S.~C. Pieper and R.~Schiavilla,
  \enquote{Electromagnetic and neutral-weak response functions of
  $^{4}\mathrm{He}$ and $^{12}\mathrm{C}$}, Physical Review C \textbf{91},
  062501, \urlprefix\url{https://link.aps.org/doi/10.1103/PhysRevC.91.062501}
  (2015).

\bibitem[Lynn \emph{et~al.}(2019)]{Lynn2019ann}
Lynn, J., I.~Tews, S.~Gandolfi and A.~Lovato, \enquote{Quantum {Monte} {Carlo}
  methods in nuclear physics: Recent advances}, Annual Review of Nuclear and
  Particle Science \textbf{69}, 1, 279--305,
  \urlprefix\url{https://doi.org/10.1146/annurev-nucl-101918-023600} (2019).

\bibitem[Lynn \emph{et~al.}(2014)]{Lynn14a}
Lynn, J.~E., J.~Carlson, E.~Epelbaum, S.~Gandolfi, A.~Gezerlis and A.~Schwenk,
  \enquote{Quantum {Monte} {Carlo} calculations of light nuclei using chiral
  potentials}, Physical Review Letters \textbf{113}, 192501,
  \urlprefix\url{https://link.aps.org/doi/10.1103/PhysRevLett.113.192501}
  (2014).

\bibitem[Lynn \emph{et~al.}(2017)]{Lynn2017PRC}
Lynn, J.~E., I.~Tews, J.~Carlson, S.~Gandolfi, A.~Gezerlis, K.~E. Schmidt and
  A.~Schwenk, \enquote{Quantum {Monte} {Carlo} calculations of light nuclei
  with local chiral two- and three-nucleon interactions}, Physical Review C
  \textbf{96}, 054007,
  \urlprefix\url{https://link.aps.org/doi/10.1103/PhysRevC.96.054007} (2017).

\bibitem[Machleidt and Entem(2011)]{MACHLEIDT20111}
Machleidt, R. and D.~Entem, \enquote{Chiral effective field theory and nuclear
  forces}, Physics Reports \textbf{503}, 1, 1 -- 75,
  \urlprefix\url{http://www.sciencedirect.com/science/article/pii/S0370157311000457}
  (2011).

\bibitem[Machleidt \emph{et~al.}(1996)]{Machleidt1996PRC}
Machleidt, R., F.~Sammarruca and Y.~Song, \enquote{Nonlocal nature of the
  nuclear force and its impact on nuclear structure}, Physical Review C
  \textbf{53}, R1483--R1487,
  \urlprefix\url{https://link.aps.org/doi/10.1103/PhysRevC.53.R1483} (1996).

\bibitem[Machleidt and Slaus(2001)]{Machleidt_2001}
Machleidt, R. and I.~Slaus, \enquote{The nucleon-nucleon interaction}, Journal
  of Physics G: Nuclear and Particle Physics \textbf{27}, 5, R69--R108,
  \urlprefix\url{https://doi.org/10.1088%2F0954-3899%2F27%2F5%2F201} (2001).

\bibitem[Malfliet and Tjon(1969)]{MALFLIET1969161}
Malfliet, R. and J.~Tjon, \enquote{Solution of the {Faddeev} equations for the
  triton problem using local two-particle interactions}, Nuclear Physics A
  \textbf{127}, 1, 161 -- 168,
  \urlprefix\url{http://www.sciencedirect.com/science/article/pii/0375947469907751}
  (1969).

\bibitem[Mayer(1964)]{mayer1964shell}
Mayer, M.~G., \enquote{The shell model}, Science \textbf{145}, 3636, 999--1006,
  \urlprefix\url{https://science.sciencemag.org/content/145/3636/999} (1964).

\bibitem[Messiah(2014)]{messiah2014quantum}
Messiah, A., \emph{Quantum Mechanics} (Dover Publications, City, 2014).

\bibitem[Metropolis \emph{et~al.}(1953)]{Metropolis53a}
Metropolis, N., A.~W. Rosenbluth, M.~N. Rosenbluth, A.~H. Teller and E.~Teller,
  \enquote{Equation of state calculations by fast computing machines}, Journal
  of Chemical Physics \textbf{21}, 6, 1087--1092,
  \urlprefix\url{https://aip.scitation.org/doi/abs/10.1063/1.1699114} (1953).

\bibitem[Moskowitz \emph{et~al.}(1982)]{Moskowitz1982a}
Moskowitz, J.~W., K.~E. Schmidt, M.~A. Lee and M.~H. Kalos, \enquote{A new look
  at correlation energy in atomic and molecular systems. {\Romannum{2}}. the
  application of the {Green}'s function {Monte} {Carlo} method to {LiH}},
  Journal of Chemical Physics \textbf{77}, 1, 349--355,
  \urlprefix\url{https://doi.org/10.1063/1.443612} (1982).

\bibitem[Motta and Zhang(2018)]{Motta2018}
Motta, M. and S.~Zhang, \enquote{Ab initio computations of molecular systems by
  the auxiliary-field quantum {Monte} {Carlo} method}, WIREs Computational
  Molecular Science \textbf{8}, 5, e1364,
  \urlprefix\url{https://onlinelibrary.wiley.com/doi/abs/10.1002/wcms.1364}
  (2018).

\bibitem[Navr{\'{a}}til \emph{et~al.}(2009)]{Navr_til_2009}
Navr{\'{a}}til, P., S.~Quaglioni, I.~Stetcu and B.~R. Barrett, \enquote{Recent
  developments in no-core shell-model calculations}, Journal of Physics G:
  Nuclear and Particle Physics \textbf{36}, 8, 083101,
  \urlprefix\url{https://doi.org/10.1088%2F0954-3899%2F36%2F8%2F083101} (2009).

\bibitem[Negele and Orland(1998)]{Negele1998Quantum}
Negele, J.~W. and H.~Orland, \emph{Quantum Many-particle Systems} (Westview
  Press, 1998), \urlprefix\url{http://www.worldcat.org/isbn/0738200522}.

\bibitem[Nogga \emph{et~al.}(2000)]{Nogga2000PRL}
Nogga, A., H.~Kamada and W.~Gl\"ockle, \enquote{Modern nuclear force
  predictions for the $\mathit{\ensuremath{\alpha}}$ particle}, Physical Review
  Letters \textbf{85}, 944--947,
  \urlprefix\url{https://link.aps.org/doi/10.1103/PhysRevLett.85.944} (2000).

\bibitem[Politzer(1973)]{Polizer73a}
Politzer, H.~D., \enquote{Reliable perturbative results for strong
  interactions?}, Physical Review Letters \textbf{30}, 1346--1349,
  \urlprefix\url{https://link.aps.org/doi/10.1103/PhysRevLett.30.1346} (1973).

\bibitem[Pudliner \emph{et~al.}(1997)]{Pudliner1997PRC}
Pudliner, B.~S., V.~R. Pandharipande, J.~Carlson, S.~C. Pieper and R.~B.
  Wiringa, \enquote{Quantum {Monte} {Carlo} calculations of nuclei with {$A <
  \sim 7$}}, Physical Review C \textbf{56}, 1720--1750,
  \urlprefix\url{https://link.aps.org/doi/10.1103/PhysRevC.56.1720} (1997).

\bibitem[Reinhard(1989)]{Reinhard_1989}
Reinhard, P.~G., \enquote{The relativistic mean-field description of nuclei and
  nuclear dynamics}, Reports on Progress in Physics \textbf{52}, 4, 439--514,
  \urlprefix\url{https://doi.org/10.1088%2F0034-4885%2F52%2F4%2F002} (1989).

\bibitem[Reinsch(2000)]{Reinsch2000JMP}
Reinsch, M.~W., \enquote{A simple expression for the terms in the
  {B}aker-{C}ampbell-{H}ausdorff series}, Journal of Mathematical Physics
  \textbf{41}, 4, 2434--2442, \urlprefix\url{https://doi.org/10.1063/1.533250}
  (2000).

\bibitem[Ring(1996)]{RING1996193}
Ring, P., \enquote{Relativistic mean field theory in finite nuclei}, Progress
  in Particle and Nuclear Physics \textbf{37}, 193 -- 263,
  \urlprefix\url{http://www.sciencedirect.com/science/article/pii/0146641096000543}
  (1996).

\bibitem[Runge(1992)]{Runge92PRB}
Runge, K.~J., \enquote{Quantum {Monte} {Carlo} calculation of the long-range
  order in the {Heisenberg} antiferromagnet}, Physical Review B \textbf{45},
  7229--7236, \urlprefix\url{https://link.aps.org/doi/10.1103/PhysRevB.45.7229}
  (1992).

\bibitem[Sakurai(1993)]{sakurai1993}
Sakurai, J.~J., \emph{Modern Quantum Mechanics (Revised Edition)} (Addison
  Wesley, 1993), \urlprefix\url{https://www.xarg.org/ref/a/0201539292/}.

\bibitem[Samaras and Hamer(1999)]{Samaras1999ForwardwalkingGF}
Samaras, M. and C.~J. Hamer, \enquote{Forward-walking {Green}'s function
  {Monte} {Carlo} method for correlation functions}, Australian Journal of
  Physics \textbf{52}, 637--657,
  \urlprefix\url{https://doi.org/10.1071/PH98092} (1999).

\bibitem[Sarsa \emph{et~al.}(2000)]{Schmidt2000JCP}
Sarsa, A., K.~E. Schmidt and W.~R. Magro, \enquote{A path integral ground state
  method}, Journal of Chemical Physics \textbf{113}, 4, 1366--1371,
  \urlprefix\url{https://doi.org/10.1063/1.481926} (2000).

\bibitem[Savage(2015)]{savage2015nuclear}
Savage, M.~J., \enquote{Nuclear physics from lattice quantum chromodynamics},
  arXiv e-prints \urlprefix\url{https://arxiv.org/abs/1510.01787} (2015).

\bibitem[Schmidt and Fantoni(1999)]{Schmidt99a}
Schmidt, K. and S.~Fantoni, \enquote{A quantum {Monte} {Carlo} method for
  nucleon systems}, Physics Letters B \textbf{446}, 2, 99 -- 103,
  \urlprefix\url{http://www.sciencedirect.com/science/article/pii/S0370269398015226}
  (1999).

\bibitem[Schmidt and Lee(1995)]{SchmidtLee95a}
Schmidt, K.~E. and M.~A. Lee, \enquote{High-accuracy {Trotter}-formula method
  for path integrals}, Physical Review E \textbf{51}, 5495--5498,
  \urlprefix\url{https://link.aps.org/doi/10.1103/PhysRevE.51.5495} (1995).

\bibitem[Shen \emph{et~al.}(2012)]{GShen2012PRC}
Shen, G., L.~E. Marcucci, J.~Carlson, S.~Gandolfi and R.~Schiavilla,
  \enquote{Inclusive neutrino scattering off the deuteron from threshold to
  {GeV} energies}, Physical Review C \textbf{86}, 035503,
  \urlprefix\url{https://link.aps.org/doi/10.1103/PhysRevC.86.035503} (2012).

\bibitem[Sick(2003)]{sick2003meson}
Sick, I., \enquote{Meson exchange currents in quasi-elastic electron-nucleus
  scattering}, in \enquote{Nuclear Dynamics: From Quarks to Nuclei}, edited by
  M.~T. Pe{\~{n}}a, A.~Stadler, A.~M. Eir{\'o} and J.~Adam, pp. 1--12 (Springer
  Vienna, Vienna, 2003),
  \urlprefix\url{https://link.springer.com/chapter/10.1007/978-3-7091-6014-5_1}.

\bibitem[Siemens and Jensen(2018)]{SiemensJensenbook}
Siemens, P.~J. and A.~S. Jensen, \emph{Elements of Nuclei} ({CRC} Press, 2018),
  \urlprefix\url{https://doi.org/10.1201/9780429493904}.

\bibitem[Skyrme(1958)]{SKYRME1958615}
Skyrme, T., \enquote{The effective nuclear potential}, Nuclear Physics
  \textbf{9}, 4, 615 -- 634,
  \urlprefix\url{http://www.sciencedirect.com/science/article/pii/0029558258903456}
  (1958).

\bibitem[Stoks \emph{et~al.}(1994)]{Stoks1994PRC}
Stoks, V. G.~J., R.~A.~M. Klomp, C.~P.~F. Terheggen and J.~J. de~Swart,
  \enquote{Construction of high-quality {NN} potential models}, Physical Review
  C \textbf{49}, 2950--2962,
  \urlprefix\url{https://link.aps.org/doi/10.1103/PhysRevC.49.2950} (1994).

\bibitem[Tilley \emph{et~al.}(1992)]{Tiley92a}
Tilley, D., H.~Weller and G.~Hale, \enquote{Energy levels of light nuclei {A =
  4}}, Nuclear Physics A \textbf{541}, 1, 1 -- 104,
  \urlprefix\url{http://www.sciencedirect.com/science/article/pii/037594749290635W}
  (1992).

\bibitem[Umrigar \emph{et~al.}(2007)]{Umrigar2007PRL}
Umrigar, C.~J., J.~Toulouse, C.~Filippi, S.~Sorella and R.~G. Hennig,
  \enquote{Alleviation of the fermion-sign problem by optimization of many-body
  wave functions}, Physical Review Letters \textbf{98}, 110201,
  \urlprefix\url{https://link.aps.org/doi/10.1103/PhysRevLett.98.110201}
  (2007).

\bibitem[van Kolck(1994)]{Klock1994a}
van Kolck, U., \enquote{Few-nucleon forces from chiral {Lagrangians}}, Physical
  Review C \textbf{49}, 2932--2941,
  \urlprefix\url{https://link.aps.org/doi/10.1103/PhysRevC.49.2932} (1994).

\bibitem[Veerasamy and Polyzou(2011)]{Veerasamy2011PRC}
Veerasamy, S. and W.~N. Polyzou, \enquote{Momentum-space {Argonne} v18
  interaction}, Physical Review C \textbf{84}, 034003,
  \urlprefix\url{https://link.aps.org/doi/10.1103/PhysRevC.84.034003} (2011).

\bibitem[Weinberg(1979)]{WEINBERG1979327}
Weinberg, S., \enquote{Phenomenological {Lagrangians}}, Physica A: Statistical
  Mechanics and its Applications \textbf{96}, 1, 327 -- 340,
  \urlprefix\url{http://www.sciencedirect.com/science/article/pii/0378437179902231}
  (1979).

\bibitem[Weinberg(1990)]{WEINBERG1990288}
Weinberg, S., \enquote{Nuclear forces from chiral lagrangians}, Physics Letters
  B \textbf{251}, 2, 288 -- 292,
  \urlprefix\url{http://www.sciencedirect.com/science/article/pii/0370269390909383}
  (1990).

\bibitem[Weinberg(1991)]{WEINBERG19913}
Weinberg, S., \enquote{Effective chiral lagrangians for nucleon-pion
  interactions and nuclear forces}, Nuclear Physics B \textbf{363}, 1, 3 -- 18,
  \urlprefix\url{http://www.sciencedirect.com/science/article/pii/055032139190231L}
  (1991).

\bibitem[Weinberg(2003)]{Weinberg2003}
Weinberg, S., \enquote{Four golden lessons}, Nature \textbf{426}, 6965,
  389--389, \urlprefix\url{https://doi.org/10.1038/426389a} (2003).

\bibitem[{Wikipedia contributors}(2020{\natexlab{a}})]{wikiCompcost}
{Wikipedia contributors}, \enquote{Computational complexity of mathematical
  operations --- {Wikipedia}{,} the free encyclopedia},
  \url{https://en.wikipedia.org/w/index.php?title=Computational_complexity_of_mathematical_operations&oldid=944589767},
  [Online; accessed 15-May-2020] (2020{\natexlab{a}}).

\bibitem[{Wikipedia contributors}(2020{\natexlab{b}})]{wiki:Fugaku2020}
{Wikipedia contributors}, \enquote{Fugaku (supercomputer) --- {Wikipedia}{,}
  the free encyclopedia},
  \url{https://en.wikipedia.org/w/index.php?title=Fugaku_(supercomputer)&oldid=964639636},
  [Online; accessed 26-June-2020] (2020{\natexlab{b}}).

\bibitem[{Wikipedia contributors}(2020{\natexlab{c}})]{wiki:NBE}
{Wikipedia contributors}, \enquote{Nuclear binding energy --- {Wikipedia}{,}
  the free encyclopedia},
  \url{https://en.wikipedia.org/w/index.php?title=Nuclear_binding_energy&oldid=957017239},
  [Online; accessed 10-July-2020] (2020{\natexlab{c}}).

\bibitem[Wiringa and Pieper(2002)]{WiringaPieper02a}
Wiringa, R.~B. and S.~C. Pieper, \enquote{Evolution of nuclear spectra with
  nuclear forces}, Physical Review Letters \textbf{89}, 182501,
  \urlprefix\url{https://link.aps.org/doi/10.1103/PhysRevLett.89.182501}
  (2002).

\bibitem[Wiringa \emph{et~al.}(1995)]{Wiringa95PRC}
Wiringa, R.~B., V.~G.~J. Stoks and R.~Schiavilla, \enquote{Accurate
  nucleon-nucleon potential with charge-independence breaking}, Physical Review
  C \textbf{51}, 38--51,
  \urlprefix\url{https://link.aps.org/doi/10.1103/PhysRevC.51.38} (1995).

\bibitem[Xu(2009)]{Xuzizong2009book}
Xu, Z.-Z., \emph{\begin{CJK}{UTF8}{gkai}核与粒子物理导论\end{CJK}}
  (\begin{CJK}{UTF8}{gbsn}中国科学技术大学出版社\end{CJK}, 2009).

\bibitem[Yang(1980)]{Yang1980a}
Yang, C.-N., \enquote{{Einstein}'s impact on theoretical physics}, Physics
  Today \textbf{33}, 6, 42--49,
  \urlprefix\url{https://doi.org/10.1063/1.2914117} (1980).

\bibitem[Yukawa(1935)]{Yukawa1935}
Yukawa, H., \enquote{On the interaction of elementary particles. {I}},
  Proceedings of the Physico-Mathematical Society of Japan. 3rd Series
  \textbf{17}, 48--57,
  \urlprefix\url{https://doi.org/10.11429/ppmsj1919.17.0_48} (1935).

\bibitem[Zabolitzky and Kalos(1981)]{ZABOLITZKY1981114}
Zabolitzky, J. and M.~Kalos, \enquote{Solution of the four-nucleon
  {S}chr\"{o}dinger equation}, Nuclear Physics A \textbf{356}, 1, 114 -- 128,
  \urlprefix\url{http://www.sciencedirect.com/science/article/pii/0375947481901214}
  (1981).

\bibitem[Zeng(2013)]{zengjinyanQMbook}
Zeng, J.-Y.,
  \emph{\begin{CJK}{UTF8}{gkai}量子力学：卷{$\mathrm{\Romannum{1}}$}\end{CJK}}
  (\begin{CJK}{UTF8}{gbsn}科学出版社\end{CJK}, 2013).

\bibitem[Zhang and Krakauer(2003)]{ZhangShiwei2003PRL}
Zhang, S. and H.~Krakauer, \enquote{Quantum {Monte} {Carlo} method using
  phase-free random walks with {Slater} determinants}, Physical Review Letters
  \textbf{90}, 136401,
  \urlprefix\url{https://link.aps.org/doi/10.1103/PhysRevLett.90.136401}
  (2003).

\end{thebibliography}


\newpage
\phantomsection
\bookmark[dest=appendixhyper,startatroot]{APPENDIX}

\renewcommand{\chaptername}{APPENDIX}
\addtocontents{toc}{\APPENDIXcontinue}

\appendix \hypertarget{appendixhyper}{}


\chapter{Variational Monte Carlo Brief}
\label{secVMC}
\clearpage
The time-independent Schr\"{o}dinger equation can be written as
$H \psi_0 = E_0 \psi_0$ if the system is at its lowest energy $E_0$ with wave function $\psi_0$.
We are interested in the ground state energy $E_0$ because a system will naturally go to its lowest energy state $\psi_0$. The question is how to find $\psi_0$ and $E_0$.

Variational Monte Carlo(VMC), based on a trial wave function $\psi_T$ (usually not normalized) with parameters, provides a way to find the upper limit of $E_0$. Its foundation is
\begin{eqnarray}
E_{var}  \equiv  \frac{\langle \psi_T | H |\psi_T\rangle }{\langle \psi_T | \psi_T \rangle}
\equiv \frac{\displaystyle{\int \frac{H \psi_T(R)}{\psi_T(R)} |\psi_T(R)|^2 dR}}{\displaystyle{\int |\psi_T(R)|^2 dR}}
\equiv \int E_L (R) p(R) dR
\geq E_0,
 \label{Evar}
\end{eqnarray}
where the local energy is defined as
\begin{eqnarray}
E_L(R) = \frac{H \psi_T(R)}{\psi_T(R)},
\end{eqnarray}
the probability density is given by
\begin{eqnarray}
p(R)= \frac{ |\psi_T(R)|^2  }{\displaystyle{\int |\psi_T(R)|^2 dR}}. \label{PrVMC}
\end{eqnarray}

The closer $\psi_T$ is to $\psi_0$, the closer $E_{var}$ is to $E_0$ from above.
For bosons $\psi_T$ has to be symmetric while for fermions $\psi_T$ has to be anti-symmetric. For example, a possible trial wave function for the system of bosons might be written in the form of
\begin{eqnarray}
\psi_T(R) = \prod_{i<j} f_2(r_{ij}) \prod_{i<j<k} f_3(i,j,k) \prod_{i<j<k<l} f_4(i,j,k,l) ... \prod_{k} g(\bm{r}_k-\bm{r}_c),
\end{eqnarray}
where $f_n$ denotes the correlation function between $n$ particles, and $ g(\bm{r}_k-\bm{r}_c)$ denotes the correlation between the position of one particle ($\bm{r}_k$) and the system's center of mass($\bm{r}_c$).
For fermions, the following form (\cite{Kalosbook2008}) may be used as a simple example,
\begin{eqnarray}
\psi_T(R) = \exp \bigg[ - \sum_{i<j} f(r_{ij}) \bigg] \times
\left|\begin{matrix}
                    e^{i \bm{k}_1 \cdot \bm{r}_1 } &  e^{i \bm{k}_2 \cdot \bm{r}_1 } & ...\\
                    e^{i \bm{k}_1 \cdot \bm{r}_2 } &  e^{i \bm{k}_2 \cdot \bm{r}_2 } & ...\\
                    e^{i \bm{k}_1 \cdot \bm{r}_3 } &  e^{i \bm{k}_2 \cdot \bm{r}_3 } & ...\\
                    \vdots & \vdots & \ddots
                  \end{matrix}\right|,
\end{eqnarray}
where $\bm{k}_i$ is momentum of the particle $i$ in three dimensional space.

The VMC trial wave function $\psi_T$ usually contains some parameters which can be adjusted to minimize $E_{var}$, so that $\psi_T$ can be optimized as close to the ground state as its structure allows.
Then DMC and PIMC can be performed based on this optimized trial wave function.

\chapter{Diffusion Monte Carlo Brief}
\label{secDMC}
\clearpage
Diffusion quantum Monte Carlo (DMC) method (\cite{Negele1998Quantum,Kalosbook2008,HammondMCbook,VesaPDF}) is based on the time-dependent Green's function.
Beginning with a trial wave function which is non-orthogonal to the true ground state,
by treating the imaginary time as a real number, it provides a way to project out the true ground state wave function through diffusion and branching process. Unlike VMC which uses trail wave function with adjustable parameters, DMC is able to numerically find the true ground state wave function and the corresponding ground state energy with time step error.

\section{ Time-Dependent Green's Function}
The solution of the time-dependent Schr\"{o}dinger equation
$i \frac{\partial \Psi(\bm{x},t)}{\partial t}=H \Psi(\bm{x},t)$ in natural units can be written as the superposition of the eigenstates $\Phi_k(\bm{x})$ with time evolution,
\begin{eqnarray}
\Psi(\bm{x},t) = \sum_{i=0}^{\infty} a_k \Phi_k(\bm{x}) e^{-iE_k t}.
 \end{eqnarray}

In Monte Carlo method, we care more about the energy shifted Schr\"{o}dinger equation,
\begin{eqnarray}
i \frac{\partial \phi(\bm{x},t)}{\partial t}= (H-E_T) \phi(\bm{x},t), \label{dmcschtimesht}
 \end{eqnarray}
and the solution can be written as
\begin{eqnarray}
\phi(\bm{x},t) = \sum_{k=0}^{\infty} c_k \Phi_k(\bm{x}) e^{-i (E_k - E_T) t},
 \end{eqnarray}
where $c_k=\langle \Phi_k | \phi(t=0) \rangle$ are the coefficients which only depend on the initial condition.

In DMC, we want to find connections between Eq.(\ref{dmcschtimesht}) and the diffusion equation. In order to do this, we use imaginary time $\tau = i t$ and rewrite Eq.(\ref{dmcschtimesht}) as
\begin{eqnarray}
-\frac{\partial \phi(\bm{x},\tau)}{\partial \tau} = (H - E_T) \phi(\bm{x},\tau), \label{dmcimgt}
 \end{eqnarray}
with the solution
\begin{eqnarray}
\phi(\bm{x},\tau) = \sum_{k=0}^{\infty} c_k \Phi_k(\bm{x}) e^{- (E_k - E_T) \tau}. \label{dmcimgtsol}
 \end{eqnarray}

If we merely look at Eq.(\ref{dmcimgt}) and Eq.(\ref{dmcimgtsol}) and take $\tau$ as a real number,
as $\tau \rightarrow \infty$, there are three situations. First,  if $E_T>E_0$, then $\lim_{\tau \rightarrow \infty} \phi(\bm{x},\tau) = \infty $. Second,  if $E_T<E_0$, then $\lim_{\tau \rightarrow \infty} \phi(\bm{x},\tau) = 0$. Third, if $E_T=E_0$, then $ \lim_{\tau \rightarrow \infty} \phi(\bm{x},\tau) = c_0 \Phi_0 (\bm{x})$, which means if $c_0 \neq 0$, then only the ground state $\Phi_0$ contributes. However if $c_0$ happen to be zero, then with proper $E_T$ we should be left with the first excited state $\Phi_1$.

Eq.(\ref{dmcimgt}) tells us that
\begin{equation}
  | \phi(\tau_2) \rangle = e^{-(H-E_T)(\tau_2-\tau_1)} | \phi(\tau_1) \rangle,
\end{equation}
acting $\langle \bm{y} |$ on both side leads to
\begin{eqnarray}
\phi(\bm{y},\tau_2) = \int \langle \bm{y} | e^{-(H-E_T)(\tau_2-\tau_1)} | \bm{x} \rangle  \phi(\bm{x},\tau_1)  d\bm{x}, \label{dmcphiyt2}
\end{eqnarray}
where
\begin{equation}
  \langle \bm{y} | e^{-(H-E_T)(\tau_2-\tau_1)} | \bm{x} \rangle \equiv G(\bm{y},\tau_2; \bm{x}, \tau_1)
\end{equation}
is the Green's function.
It only depends on the time difference $\tau_2-\tau_1$, so, for the small step $\delta \tau$, Eq.(\ref{dmcphiyt2}) can be written as
\begin{eqnarray}
\phi(\bm{y},\tau + \delta \tau) = \int G(\bm{y},\bm{x}; \delta \tau) \phi(\bm{x},\tau) d\bm{x}. \label{dmcphiydt}
\end{eqnarray}
The Green's function can be expanded as
\begin{eqnarray}
 G(\bm{y},\bm{x}; \delta \tau) = \sum_{i=0}^{\infty} \Phi_i(\bm{y})\Phi_i(\bm{x}) e^{-(E_i-E_T)\delta \tau}. \label{dmcGexp}
\end{eqnarray}
Assuming $\phi(\bm{y},\tau=0)=\sum_{k}^{\infty} c_k \Phi_k$ and substitute Eq.(\ref{dmcGexp}) into Eq.(\ref{dmcphiydt}) we have
\begin{eqnarray}
\phi(\bm{y},\delta \tau) = \sum_{k=0}^{\infty} c_k \Phi_k(\bm{y}) e^{-(E_k-E_T)\delta \tau}   . \label{dmcphiydt2}
\end{eqnarray}
After $n$ iterations we have
\begin{eqnarray}
\phi(\bm{y}, n \delta \tau) = \sum_{k=0}^{\infty} c_k \Phi_k(\bm{y}) e^{-(E_k-E_T) n \delta \tau}  . \label{dmcphiydtn}
\end{eqnarray}
As $n \rightarrow \infty$, and as long as $c_0 \neq 0$, the final state will be dominated by the ground state $\Psi_0$. Furthermore, if we set $E_T=E_0$, the exponential term will be just a constant.

It needs to pointed out that since we treat $\tau$ as a real number, for finite $\tau$, $\phi(\bm{x}, \tau)$ does not have the physical meaning as the time-dependent wave function anymore.

\section{Diffusion Monte Carlo} \label{subsecDMC}

The exact Green's function is usually hard to obtain. But we can use short-time approximation to obtain an approximated analytic Green's function. The Green's function tells us how to make the configurations (walkers) evolve into the desired distribution.

\subsubsection{ Theoretical foundation }
Let us begin with a system of $N$ particles with mass $m$. The energy shifted imaginary time Schr\"{o}dinger equation is
\begin{eqnarray}
\frac{\partial \phi(\bm{x},\tau)}{\partial \tau} = D\nabla^2 \phi(\bm{x},\tau) + (E_T - V(\bm{x})) \phi(\bm{x},\tau), \label{dmcimgtexp}
 \end{eqnarray}
where $D=\hbar^2/{2 m}$ is called the diffusion constant, which is $1/2$ in atomic unit. One can see from Eq.(\ref{dmcimgtexp}) that, if $(E_T - V(\bm{x})) \phi(\bm{x},\tau) = 0$, we have a diffusion equation
\begin{eqnarray}
\frac{\partial \phi(\bm{x},\tau)}{\partial \tau} = D\nabla^2 \phi(\bm{x},\tau). \label{dmcimgtdiff}
 \end{eqnarray}
If $D\nabla^2 \phi(\bm{x},\tau) = 0$, we have a rate equation which can be simulated by the branching process,
\begin{eqnarray}
\frac{\partial \phi(\bm{x},\tau)}{\partial \tau} = (E_T - V(\bm{x})) \phi(\bm{x},\tau). \label{dmcimgtbran}
 \end{eqnarray}

By using short-time approximation, we can decompose the operator of the Green's function $G$ into a diffusion part operator $G_{\textrm{diff}}$ and a branching part operator $G_{\textrm{B}}$. We know that, as an operator,
\begin{eqnarray}
G = e^{-H\tau} = e^{-(J+V-E_T)\tau} \neq e^{-J\tau} e^{-(V-E_T)\tau} = G_{\textrm{diff}} G_{\textrm{B}},
 \end{eqnarray}
where $G_{\textrm{diff}} \equiv  e^{-J\tau}, G_{\textrm{B}} \equiv  e^{-(V-E_T)\tau}, J \equiv - D \nabla ^2$.
However, if $\tau$ is very small, doing Taylor expansion around $\tau$ for $G$, we obtain $G - G_{\textrm{diff}} G_{\textrm{B}} = \frac{1}{2} [V,J] \tau^2 + \mathcal{O}(\tau^3)$. Therefore we can write down the approximation
\begin{eqnarray}
 \lim_{\tau \rightarrow 0}G = G_{\textrm{diff}} G_{\textrm{B}} .
\end{eqnarray}
According to Trotter formula (\cite{Kalosbook2008}), we can further write it in a symmetrical way such that
\begin{eqnarray}
G(\bm{y},\bm{x},\tau) &=& \langle \bm{y} |e^{-(J+V-E_T)\tau} | \bm{x} \rangle \nonumber \\
& \approx & \langle \bm{y} | e^{-\frac{V-E_T}{2}} e^{-J \tau} e^{-\frac{V-E_T}{2}} | \bm{x} \rangle \nonumber \\
& \equiv & G_{\textrm{diff}}(\bm{y},\bm{x},\tau) G_{\textrm{B}}(\bm{y},\bm{x},\tau). \label{eq204}
 \end{eqnarray}

The diffusion part $G_{\textrm{diff}}(\bm{y},\bm{x},\tau) \equiv \langle \bm{y} |e^{-J \tau} | \bm{x} \rangle$ has the form of the free particle propagator,
\begin{eqnarray}
 G_{\textrm{diff}}(\bm{y},\bm{x},\tau) = (4\pi D \tau)^{-\frac{3N}{2}} \prod_{i=1}^N e^{- \frac{(\bm{y}_i - \bm{x}_i)^2}{4D\tau} }
 \equiv  \prod_{i=1}^N G^{(i)}_{\textrm{diff}}(\bm{y}_i,\bm{x}_i,\tau) .
 \end{eqnarray}
Note that
$\int {G}_{\textrm{diff}}(\bm{y},\bm{x},\tau) d\bm{x} = 1$ and
$\int {G}_{\textrm{diff}}(\bm{y},\bm{x},\tau) d\bm{y} = 1$
are Gaussian integrals.
So, actually ${G}_{\textrm{diff}}(\bm{y},\bm{x},\tau)$ is a probability density which tells us that for a given $\bm{x}_i$ or $\bm{y}_i$,
\begin{eqnarray}
(\bm{y}_i)_{\alpha} - (\bm{x}_i)_{\alpha} = \chi, \label{DMCChimove}
\end{eqnarray}
where $\chi$ is a Gaussian random number with zero mean value and variance $2D\tau$, and the subscript $\alpha$ means the x, y or z component of the $i^{\text{th}}$ particle.
Eq.(\ref{DMCChimove}) already tells us how to move the $i^{\text{th}}$ particle from $\bm{x}_i$($\bm{y}_i$) to $\bm{y}_i$($\bm{x}_i$) for fixed $\bm{x}_i$($\bm{y}_i$), as long as we have a gaussian random number generator. We can denote ${G}_{\textrm{diff}}(\bm{y},\bm{x},\tau)$ as the transition probability from $\bm{x}$ to $\bm{y}$ as $P(\bm{y},\bm{x})$. $ G_{\textrm{diff}}(\bm{y},\bm{x},\tau)$ is called diffusion part because it satisfies the diffusion equation
\begin{eqnarray}
\frac{\partial G_{\textrm{diff}}(\bm{y},\bm{x},\tau)}{\partial \tau} \equiv D \nabla^2 G_{\textrm{diff}}(\bm{y},\bm{x},\tau).
\end{eqnarray}
The solution of the branching part $G_{\textrm{B}}(\bm{y},\bm{x},\tau)$ is
\begin{eqnarray}
 G_{\textrm{B}}(\bm{y},\bm{x},\tau) = e^{ - \left[  \frac{ V(\bm{x}) + V(\bm{y})  }{2} - E_T \right] \tau  }.
\end{eqnarray}
It is called branching part because
\begin{eqnarray}
 \frac{\partial G_{\textrm{B}}(\bm{y},\bm{x},\tau)}{\partial \tau} = (E_T-V) G_{\textrm{B}}(\bm{y},\bm{x},\tau),
\end{eqnarray}
which is a rate equation. The branching part ${G}_{\textrm{B}}(\bm{y},\bm{x},\tau)$ serves as a weight which can be denoted as $W(\bm{y},\bm{x})$. If $W(\bm{y},\bm{x})<1$, we delete the new configuration $\bm{y}$ with the probability $1-W(\bm{y},\bm{x})$. If $W(\bm{y},\bm{x})>1$, we replicate the new configuration $\bm{y}$ with $\textrm{INT}[W(\bm{y},\bm{x})]$ copies, and with the probability $W(\bm{y},\bm{x})-\textrm{INT}[W(\bm{y},\bm{x})]$ to replicate it once more.

Now, the short-time evolution of $\phi$ can be written as
\begin{eqnarray}
\phi(\bm{y},\tau) =
\int
 {G}_{\textrm{diff}}(\bm{y},\bm{x},\tau)  {G}_{\textrm{B}}(\bm{y},\bm{x},\tau)
\phi(\bm{x}) d\bm{x}
\equiv \int d\bm{x} P(\bm{y},\bm{x}) W(\bm{y},\bm{x}) \phi(\bm{x}),
\label{fGdiffGBwoimp}
 \end{eqnarray}
this expression provides the foundation for the simulation.

\subsubsection{ Interpretation}

A walker means a configuration, $\bm{x}$, for the N-particle system. If M walkers are distributed according $\phi$, the number of them around the configuration $\bm{x}+d\bm{x}$ is $M \phi(\bm{x})$.
In other words the normalized probability (density) for a walker to be around $\bm{x}+d\bm{x}$ is $\phi(\bm{x})$. Notice that now $\phi$ means a distribution, so it must be non-negative.

Assuming there are $M^{(1)}$ walkers in the first generation, denoted as $\bm{x}^{(1)}$, $\bm{x}^{(2)}$, ..., $\bm{x}^{(  M^{(1)}  )}$, and are distributed according to $\phi$. For a short-time $\tau$, we move each configuration according to $P(\bm{y},\bm{x})$ and branch them according to the weight $W(\bm{y},\bm{x})$. The number of walkers around a particular configuration $\bm{y}$ is,
\begin{equation}
\sum_{i=1}^{M^{(1)}} P(\bm{y},\bm{x}^{(i)})W(\bm{y},\bm{x}^{(i)}) M^{(1)} \phi(\bm{x}^{(i)}) . \nonumber
\end{equation}
If $M^{(1)}$ is large enough, this sum becomes the following estimator of the integral,
\begin{equation}
M^{(1)}  \int d\bm{x} P(\bm{y},\bm{x}) W(\bm{y},\bm{x}) \phi(\bm{x}) . \nonumber
\end{equation}
On the other hand, summing over all the weight, we obtain the number of the second generation of walkers, $M^{(2)}$. We denote the probability that they are around $\bm{y}$ as $\phi'(\bm{y},\tau)$, so the number of them is $M^{(2)} \phi'(\bm{y},\tau)$, and we have
\begin{equation}
M^{(2)} \phi'(\bm{y},\tau) = M^{(1)}  \int d\bm{x} P(\bm{y},\bm{x}) W(\bm{y},\bm{x}) \phi(\bm{x}) . \nonumber
\end{equation}
In other words,
\begin{equation}
\phi'(\bm{y},\tau) = \frac{ M^{(1)} }{M^{(2)} }  \int d\bm{x} P(\bm{y},\bm{x}) W(\bm{y},\bm{x}) \phi(\bm{x})
= \frac{ M^{(1)} }{M^{(2)} } \phi(\bm{y},\tau) . \nonumber
\end{equation}
So, the ratio of number of walkers between any two configurations in the next generation $\bm{y}$ and $\bm{y}'$ is $\phi'(\bm{y},\tau)/\phi'(\bm{y}',\tau)=\phi(\bm{y},\tau)/\phi(\bm{y}',\tau)$. This means that the walkers will indeed be distributed according to $ \phi(\bm{y},\tau)$.
This is remarkable because now an integral with high dimensions can be equivalently done through a stochastic process! In the spirit of Eq.(\ref{dmcphiydtn}), by repeating the short-time diffusion process for sufficient number of generations, finally the configurations will be distributed according to ground state wave function $\Psi_0$ (not  $\Psi^2_0$).

\section{ Diffusion Monte Carlo with Importance Sampling} \label{secDMCimp}

DMC with constant short-time step without importance sampling is seldom used.
Because the walkers may explore regions in which the short-time approximation is poor.
When this happens, the branch term $e^{ - \left[  \frac{ V(\bm{x}) + V(\bm{y})  }{2} - E_T \right] \tau  }$ can become abnormally large, and more and more walkers (configurations) will sink into such regions, the walkers will not evolve into the desired distribution, and the variance of estimators can be too big. Therefore we have to use importance sampling to guide the movement of walkers and thus reduce the variance.

\subsubsection{ Theoretical foundation }
We use an analytical function $\Psi_T(\bm{x})$ as the importance sampling function for the $N$-particle system. We define a new distribution function $f(\bm{x},\tau)$,
\begin{eqnarray}
f(\bm{x},\tau) \equiv \Psi_T(\bm{x}) \phi(\bm{x},\tau).
 \end{eqnarray}
The key is, we want to find the short-time Green's function for the distribution $f(\bm{x},\tau)$, such that after many time steps, the distribution will evolve into $\Psi_T \Psi_0$(not $\Psi_0$), which will be convenient for further calculations.

By multiplying $\Psi_T(\bm{x})$ to both sides of Eq.(\ref{dmcimgtexp}), we have Fokker-Planck equation,
\begin{align}
\frac{\partial f}{\partial \tau} &= D \Psi_T \nabla^2 \phi + [E_T-V(\bm{x})]f \nonumber \\
&= \underbrace{\big[ D\nabla^2 f - D \sum_{i=1}^N \nabla_{i} \cdot (f {\bm{F}_i}_Q) \big]}_{\textrm{drifting}}
+ \underbrace{[ E_T - E_L(\bm{x}) ]f}_{\textrm{branching}},  \label{DMCimp}
 \end{align}
where $\nabla^2 \equiv  \sum_{i=1}^{N} \nabla_i^2$, and the local energy is defined as $E_L(\bm{x})  \equiv  \frac{H \Psi_T}{\Psi_T} $. In particular, the so called quantum force for particle $i$ is defined as
\begin{eqnarray}
{\bm{F}_i}_Q  \equiv  \frac{2 \nabla_i \Psi_T}{\Psi_T}.
\end{eqnarray}
We notice that by using the property $\nabla \cdot (f \bm{F})=(\bm{F} \cdot \nabla + \nabla \cdot \bm{F})f$, the drifting part in the right-hand side (RHS) of Eq.(\ref{DMCimp}) can be rewritten as
\begin{eqnarray}
D\nabla^2 f - \sum_{i=1}^N D  ({\bm{F}_i}_Q \cdot \nabla_i + \nabla_i \cdot
{\bm{F}_i}_Q) f \equiv  - \tilde{J} f ,
\end{eqnarray}
where
\begin{eqnarray}
\tilde{J} \equiv - D\nabla^2 + \sum_{i=1}^N D  ({\bm{F}_i}_Q \cdot \nabla_i + \nabla_i \cdot
{\bm{F}_i}_Q)
\end{eqnarray}
is the modified kinetic energy operator. So, Eq.(\ref{DMCimp}) can be further written as
\begin{eqnarray}
\frac{\partial f}{\partial \tau} = \Big\{ -\tilde{J}
+ [ E_T - E_L(\bm{x}) ] \Big\} f, \label{eqdmcimp}
 \end{eqnarray}
which means the short-time evolution operator is just $e^{  \{ -\tilde{J}
+ [ E_T - E_L(\bm{x}) ] \}  \tau  }$. Consider that $\tau$ is very small, we further have
$e^{  \{ -\tilde{J}
+ [ E_T - E_L(\bm{x}) ] \}  \tau  }
\approx
e^{  -\tilde{J}\tau }
e^{ [ E_T - E_L(\bm{x}) ]   \tau  } $. Therefore
\begin{eqnarray}
f(\bm{x},t+\tau) = e^{  -\tilde{J}\tau } e^{ [ E_T - E_L(\bm{x}) ] \tau  }f(\bm{x},t)
=e^{  -\tilde{J}\tau } f(\bm{x},t) e^{ [ E_T - E_L(\bm{x}) ] \tau  }. \label{fGdiffGB0}
 \end{eqnarray}
By sandwiching Eq.(\ref{fGdiffGB0}) between $\langle \bm{y} |$ and $| \bm{x} \rangle$, we have
\begin{flalign}
\langle \bm{y} | f(\bm{x},t+\tau) | \bm{x} \rangle
&= \int d\bm{x}'
\langle \bm{y} | e^{  -\tilde{J}\tau } | \bm{x}' \rangle     \langle \bm{x}' | f(\bm{x},t)| \bm{x} \rangle  e^{ [ E_T - E_L(\bm{x}) ] \tau  } \nonumber \\
\Rightarrow
 f(\bm{x},t+\tau) \delta( \bm{y} - \bm{x} )
&= \int d\bm{x}'
\langle \bm{y} | e^{  -\tilde{J}\tau } | \bm{x}' \rangle  f(\bm{x},t) \delta( \bm{x} - \bm{x}') e^{ [ E_T - E_L(\bm{x}) ] \tau  }.
\label{fGdiffGB01}
 \end{flalign}
By doing integral over $\bm{x}$ on both sides of Eq.(\ref{fGdiffGB01}) we have
\begin{eqnarray}
&& \int
 f(\bm{x},t+\tau) \delta( \bm{y} - \bm{x} ) d \bm{x}
=\int d\bm{x}' \int d\bm{x}
\langle \bm{y} | e^{  -\tilde{J}\tau } | \bm{x}' \rangle  f(\bm{x},t) \delta( \bm{x} - \bm{x}') e^{ [ E_T - E_L(\bm{x}) ] \tau  } \nonumber \\
&& \Rightarrow
 f(\bm{y},t+\tau)
= \int d\bm{x}
\underbrace{
\langle \bm{y} | e^{  -\tilde{J}\tau } | \bm{x} \rangle }_{ \tilde{G}_{\textrm{diff}}(\bm{y},\bm{x},\tau)  }  f(\bm{x},t)
\underbrace{ e^{ [ E_T - E_L(\bm{x}) ] \tau  } }_{ \tilde{G}_{\textrm{B}}(\bm{y},\bm{x},\tau)}.
\label{fGdiffGB02}
 \end{eqnarray}

$\tilde{G}_{\textrm{diff}}(\bm{y},\bm{x},\tau)$ satisfies the diffusion equation $\frac{\partial \tilde{G}_{\textrm{diff}}(\bm{y},\bm{x},\tau)}{\partial \tau} = -\tilde{J} \tilde{G}_{\textrm{diff}}(\bm{y},\bm{x},\tau)$, which means that if we set the branch part as zero, the distribution $f(\bm{x},\tau)$ will evolve into $\Psi^2_T(\bm{x})$, and $\langle E_L \rangle$ will just be variational energy. Assuming the quantum force ${\bm{F}_i}_{Q}(\bm{x})$ is constant between $\bm{x}$ and $\bm{y}$, the solution is
\begin{eqnarray}
\tilde{G}_{\textrm{diff}}(\bm{y},\bm{x},\tau) = (4\pi D \tau)^{-\frac{3N}{2}} \prod_{i=1}^N e^{- \frac{| \bm{y}_i - \bm{x}_i -D\tau {\bm{F}_i}_{Q}(\bm{x}) | ^2}{4D\tau} }
 \equiv  \prod_{i=1}^N \tilde{G}^{(i)}_{\textrm{diff}}(\bm{y}_i,\bm{x}_i,\tau) , \label{Gdifftilde}
\end{eqnarray}
note that $\int \tilde{G}_{\textrm{diff}}(\bm{y},\bm{x},\tau) d\bm{x} = 1$ which is a Gaussian integral. So, $\tilde{G}_{\textrm{diff}}(\bm{y},\bm{x},\tau)$ can be treat as a probability density.

The branch part $\tilde{G}_{\textrm{B}}(\bm{x},\tau) =e^{ - [ E_L(\bm{x}) - E_T ] \tau  } $ is usually written symmetrically (\cite{HammondMCbook,Kalosbook2008}) as
\begin{eqnarray}
\tilde{G}_{\textrm{B}}(\bm{y},\bm{x},\tau) = e^{ - [  \frac{ E_L(\bm{x}) + E_L(\bm{y})  }{2} - E_T ] \tau  }. \label{GBsymmetric}
\end{eqnarray}

Now the short-time approximated Green's function for $f$ can be written as
\begin{equation}
\tilde{G}(\bm{y},\bm{x},\tau) = \tilde{G}_{\textrm{diff}}(\bm{y},\bm{x},\tau)  \tilde{G}_{\textrm{B}}(\bm{y},\bm{x},\tau) ,
\end{equation}
such that,
\begin{eqnarray}
f(\bm{y},t+\tau) = \int \tilde{G}(\bm{y},\bm{x};\tau) f(\bm{x},t) d\bm{x} =
\int
 \tilde{G}_{\textrm{diff}}(\bm{y},\bm{x},\tau)  \tilde{G}_{\textrm{B}}(\bm{y},\bm{x},\tau)
f(\bm{x},t) d\bm{x}
\label{fGdiffGB}.
 \end{eqnarray}

Similar with the non-importance sampling DMC, Eq.(\ref{DMCimp}) is the combination of drifting and branching terms. The entire simulation can be achieved by drifting and branching process described in Sec.\ref{secDMC}, just use the modified Green's function instead. Finally $f$ will converge to $\Psi_T \Psi_0$. This is because that we are actually still solving Eq.(\ref{dmcimgtexp}), but just by using Eq.(\ref{DMCimp}). The $\phi$ in Eq.(\ref{dmcimgtexp}) converges to $\Psi_0$, which exactly means that solution of Eq.(\ref{DMCimp}) will converge to $\Psi_T \Psi_0$. Moreover, if we neglect the branch part, the diffusion process will evolve the distribution into $\Psi_T^2$, and DMC then just becomes VMC.

\subsubsection{Time step error estimating}

If all the configurations are taken from the final distribution
\begin{eqnarray}
f(\bm{x})=\frac{\Psi_T \Psi_0}{ { \int \Psi_T \Psi_0 d \bm{x}}},
 \end{eqnarray}
the expectation value of the local energy $E_L$ then becomes
\begin{eqnarray}
\langle E_L \rangle =  \int  \frac{H \Psi_T}{\Psi_T} f(\bm{x}) d\bm{x}
=\frac{ { \int   \Psi_0 H\Psi_T d\bm{x} } }{{ \int \Psi_T \Psi_0 d \bm{x}}}
=\frac{\langle \Psi_0 | H | \Psi_T \rangle}{\langle \Psi_0 | \Psi_T \rangle}
= E_0. \label{ELaveE0}
 \end{eqnarray}
Clearly, the local energy serves as the estimator for the ground state energy $E_0$,
and $\langle E_L \rangle$ is called the variational estimator.

However, the Green's function we use is short-time approximated, so time-step error is always there. We need to do extrapolation. Just say the local energy $\langle E_L \rangle$, what equation do we use to extrapolate the true $E_0$?
A closer look at Eq.(\ref{eq204}) reveals that the short-time approximated Green's function (now call it $G'$, neglect the constant $E_T$) to the \nth{3} order of $\tau$ can be written as (by using CBH formula)
\begin{eqnarray}
G'=  e^{-\frac{1}{2}\tau V} e^{-\frac{1}{2}\tau J} e^{-\frac{1}{2}\tau J} e^{-\frac{1}{2}\tau V}
= e^{-\tau H + \frac{\tau^3}{48}[H-2V, [H,V]] }
 \equiv  e^{-\tau H'},
\end{eqnarray}
compare this $G'$ with the exact Green's function $G=e^{-\tau H}$, we find that
\begin{eqnarray}
H' = H + \frac{\tau^2}{48} \big[  H-2V, [H,V]  \big].
\end{eqnarray}
This result seems indicating that $E_0^{DMC}(\tau) = E_0 + c_2 \tau^2$. However, since we are actually using importance sampling, the linear term still exists (\cite{VesaPDF,Chin90}). So, full second order extrapolation
\begin{eqnarray}
E_0^{DMC}(\tau) = E_0 + c_1 \tau + c_2 \tau^2 \label{2ndextrap}
 \end{eqnarray}
should be used. As pointed out in (\cite{VesaPDF}), any extrapolation with higher than second order must be based on different decompositions from Eq.(\ref{eq204}).

\subsubsection{Approximated pure estimator $\tilde{A}_p$} \label{secDMCests}

Let us denote an operator as $A(\bm{x})$ in the coordinate space which does not commute with the Hamiltonian. Just consider the ground state, now we have three Monte Carlo estimators for the expectation value of $A(\bm{x})$. Namely the pure estimator $A_p$, the variational estimator $A_v$ and mixed estimator $A_m$ (\cite{HammondMCbook}),
\begin{flalign}
 A_p &= \frac{\int \Psi_0 A(\bm{x})  \Psi_0 d \bm{x}  }{ \int \Psi_0^2 d\bm{x} }, \\
 A_v &= \frac{\int \Psi_T A(\bm{x})  \Psi_T d \bm{x}  }{ \int \Psi_T^2 d\bm{x} }, \\
 A_m &= \frac{\int \Psi_0 A(\bm{x})  \Psi_T d \bm{x}  }{ \int \Psi_0 \Psi_T d\bm{x} }.
\end{flalign}
We define a function $\Delta(\bm{x})\equiv \Psi_0(\bm{x}) - \Psi_T(\bm{x})$, and rewrite $A_m$ and $A_v$ by functional Taylor expansion series in $\Delta$. Now we have
\begin{flalign}
 A_m &= A_p + \int d\bm{x} \Psi_0 [A_p - A(\bm{x})] \Delta + \mathcal{O}(\Delta^2), \\
 A_v &= A_p + 2\int d\bm{x} \Psi_0 [A_p - A(\bm{x})] \Delta + \mathcal{O}(\Delta^2),
\end{flalign}
combining them we have
\begin{equation}
A_p=2 A_m - A_v + \mathcal{O}(\Delta^2) \approx 2 A_m - A_v \equiv \tilde{A}_p. \label{Apest}
\end{equation}
Eq.(\ref{Apest}) the approximated estimator $\tilde{A}_p$ provides a better way to evaluate the pure estimator $A_p$ than just the mixed or the variational estimators, as long as $\Delta$ is small.

\section{ Diffusion Monte Carlo Algorithm}
The algorithm for $N$-particle system as illustrated by Fig. \ref{Algorithm1} mainly consists the following steps (\cite{HammondMCbook,VesaPDF}).
\begin{figure} [!htbp]
\centering
\includegraphics[scale=0.6]{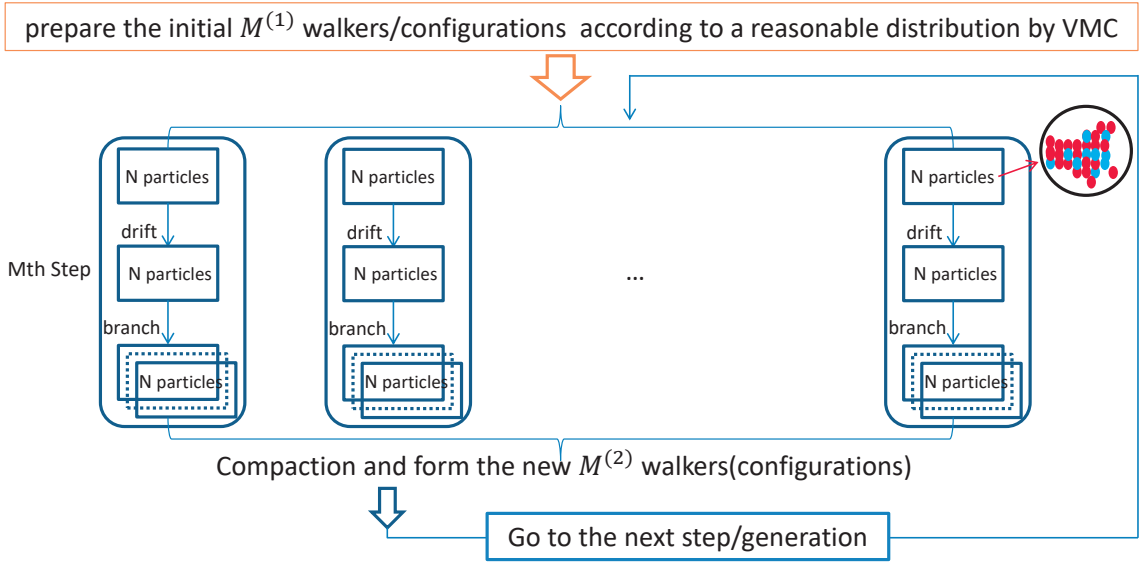}
\caption[Illustration of the Parallel DMC Algorithm]{The illustration of the parallel DMC algorithm. Each walker means a computer core which can store the configuration of the $N$-particle system.}
\label{Algorithm1}
\end{figure}
\begin{description}
  \item[Initialization] For the $N$-particle system, first we need to choose the short-time step $\tau$ and M configurations (walkers) as the first generation, let us denote the M as $M^{(1)}$. We then use M(RT)$^2$ algorithm to make this $M^{(1)}$ configurations distributed according to the importance sampling wave function $\Psi$. In other words, we set the initial wave function $\phi$ as the importance sampling function $\Psi$. This step is carried out by the VMC procedure in which we also calculate the $E_{var}  \equiv  \frac{\langle \psi_T | H |\psi_T\rangle }{\langle \psi_T | \psi_T \rangle}=\langle E_L \rangle$, and take this $E_{var} $ as the trial energy $E_T$ for the first generation.
  \item[Drifting] Then, let us begin with the first walker in the initial $M^{(1)}$ configurations. In this walker, we drift all the $N$ particles in this walker one by one. The diffusion part of the Green's function tells us that, for the $i^{\text{th}}$ particle, we can drift it from the old position $\bm{x}_i$ to its possible new position $\bm{y}_i$ according to
      \begin{eqnarray}
      (\bm{y}_i)_{\alpha}=(\bm{x}_i)_{\alpha} + D\tau ({\bm{F}_i}_Q)_{\alpha} + \chi,
      \end{eqnarray}
      where $\chi$ is a Gaussian random number with zero mean value and $2D\tau$ as variance, the subscript $\alpha$ means the x, y or z component of the $i^{\text{th}}$ particle. The acceptance probability $A(\bm{y},\bm{x},\tau) = \textrm{min}[1, q(\bm{y},\bm{x},\tau)]$ in the M(RT)$^2$ algorithm should have $q(\bm{y},\bm{x},\tau)= [ \tilde{G}_{\textrm{diff}}(\bm{x},\bm{y},\tau) |\Psi(\bm{y})|^2  ]/[\tilde{G}_{\textrm{diff}}(\bm{y},\bm{x},\tau) |\Psi(\bm{x})|^2  ]$. Now we use $A(\bm{y},\bm{x},\tau)$ to judge if this move is accepted.
      In principle we should move all the particles (not one by one) in a walker at the same time, then judge if the new walker is accept. If the time step $\tau$ is small, the acceptance ratio of a good DMC run should be close to $100\%$. In principle we should make sure the drift force is a constant during the time step (however this condition is actually violated, and that is why there is a linear term of time-step error in the extrapolation function). Anything violating the short-time approximation should be well-treated.
  \item[Branching] Until here, the old walker $\bm{x}$ with $M^{(1)}$ particles has been `drifted' to the new walker $\bm{y}$. Now we need to do branching. Namely, we need judge how many of the new walkers do we need, zero (means we destroy the new walker), one, two, three or more? We judge this by first calculating
\begin{eqnarray}
\tilde{G}_{\textrm{B}}(\bm{y},\bm{x},\tau) = e^{ - [  \frac{ E_L(\bm{x}) + E_L(\bm{y})  }{2} - E_T ] \tau_{\textrm{eff}}  },
 \end{eqnarray}
and then
 \begin{eqnarray}
 \textrm{MB} = \textrm{MIN}\big[\textrm{INT}[ \tilde{G}_{\textrm{B}} + \textrm{RAN}], 2\big],
 \end{eqnarray}
here $\textrm{RAN}$ means a uniform random number and INT means the integer part. Here the 2 is put by hand, we can use other values. This upper limit is to prevent more and more walkers sinking into abnormal regions where MB is very big which may cause incorrect distributions.
If MB is 0, we destroy this walker.
If MB is 1, we keep the walker.
If $\textrm{MB} \geq 2$, we keep MB copies of the new walker. Usually MB is around 1.

  \item[Ergodicity] Until here, in the first generation of $M^{(1)}$ walkers, we have drifted and branched the first walker. Now we need to drift and branch all the rest of the walkers in the first generation. So we do it in this stage. Once we finished this, we go to compaction.
  \item[Compaction] Now, all the $M^{(1)}$ walkers in the first generation has been drifted and branched. Some were killed, some remain, some were replicated. In this procedure we just remove those dead walkers and form the new $M^{(2)}$ number of walkers for the second generation.
  \item[Population control] Once we obtain the $M^{(2)}$ walkers, we need to keep the number of walkers in each generations almost the same as the initial $M^{(1)}$. The feedback mechanism I use is direct shrinking.

  \item[Loop] Now we have the new generation of $M^{(2)}$ walkers, we then go to the loop, Drifting$\rightarrow$Branching$\rightarrow$Ergodicity$\rightarrow$Compaction$\rightarrow$Population control$\rightarrow$Drifting, and the number of the loops is the number of the total generations. The number of generations is large enough such that all the excited states are removed, and the distribution of walkers are stabilized and is $\Psi \phi_0$.
  \item[Calculations] After enough generations of diffusion for the given time step $\tau$, we sample $\langle E_L \rangle$ from the stabilized generations. We take the average of those $\langle E_L \rangle$ and find its error. We then do extrapolation to obtain the ground state energy with error bar.
\end{description}

\chapter{Auxiliary Field Diffusion Monte Carlo Brief}
\label{secAFDMC}
\clearpage
\section{Formalism}
DMC with or without importance sampling mentioned previously, can be taken as a special case of AFDMC (\cite{Schmidt99a,gandolfi2007auxiliary,CarlsonRMP15a}). In general, in DMC simulations, the evolution of a state can be cast into such a form,
\begin{eqnarray}
|\Psi(t+\delta t) \rangle = \int dX P(X) T(X) |\Psi(t) \rangle, \label{eq53}
\end{eqnarray}
where X are a series of variables called auxiliary fields. Note that X are not necessarily coordinates. $P(X)$ is the normalized probability density. $T(X)$ is an operator acting on $| \Psi(t) \rangle$, it basically tells us how to move the walkers which form the state $| \Psi(t) \rangle$. As $\delta t \rightarrow 0$, we expect that the propagator is given by
\begin{eqnarray}
e^{-(H-E_T) \delta t} = \int dX P(X) T(X).
\end{eqnarray}

We can think of $| \Psi(t) \rangle$ as a linear combination of $N_W$ walker states $| R_i S_i \rangle$ with corresponding weight $w_i$ such that
\begin{eqnarray}
| \Psi(t) \rangle = \sum_{i=1}^{N_W} w_i | R_i S_i \rangle,
\end{eqnarray}
where $R$ and $S$ denote the spatial and spinor (spin-isospin) part in a walker. Here the amplitude of different walkers are related by Kronecker delta $\langle R_j S_j | R_i S_i \rangle = \delta_{ij}$.

The key is that, we sample $X$ from $P(X)$, then use $T(X)$ to move each walker $| R_i S_i \rangle$ into the new walker $| R'_i S'_i \rangle$ with its weight $W(X,R_i,S_i)$,
\begin{eqnarray}
T(X) | R_i S_i \rangle = W(X, R_i, S_i) | R'_i S'_i \rangle.
\end{eqnarray}
With importance sampling, we denote $|\Psi_T \Psi(t)\rangle$ as a state such that we define
\begin{eqnarray}
\langle R_i S_i | \Psi_T \Psi(t) \rangle \equiv \langle \Psi_T | R_i S_i \rangle \langle R_i S_i | \Psi(t) \rangle,
\end{eqnarray}
and that
\begin{eqnarray}
| \Psi_T \Psi(t) \rangle = \sum_{i=1}^{N_W} w_i | R_i S_i \rangle.
\end{eqnarray}
In other words, it is
\begin{eqnarray}
| \Psi(t) \rangle = \sum_{i=1}^{N_W} \frac{w_i | R_i S_i \rangle}{ \langle \Psi_T | R_i S_i \rangle   }. \label{eq62}
\end{eqnarray}
Consider that actually $w_i=1$, $O_m$ the mixed average of operator $O$ can be written as
\begin{eqnarray}
O_m = \frac{ \langle \Psi_T | O | \Psi(t) \rangle }{ \langle \Psi_T | \Psi(t) \rangle }
= \frac{1}{N_W} \sum_{i=1}^{N_W} \frac{\langle \Psi_T | O |R_i S_i \rangle  }{\langle \Psi_T | R_i S_i\rangle}
= \frac{1}{N_W} \sum_{i=1}^{N_W} \frac{ \hat{O}^\dag \langle \Psi_T |R_i S_i \rangle  }{\langle \Psi_T | R_i S_i\rangle}.
\end{eqnarray}
Putting Eq.(\ref{eq62}) into Eq.(\ref{eq53}) we have
\begin{eqnarray}
|\Psi_T \Psi(t+\delta t) \rangle = \sum_{i=1}^{N_W} w_i \int dX P(X) \frac{\langle \Psi_T |T(X)|R_i S_i\rangle}{\langle \Psi_T | R_i S_i \rangle}\frac{T(X)| R_i S_i \rangle}{W(X,R_i,S_i)}. \label{eq65}
\end{eqnarray}
Note that
\begin{eqnarray}
|R'_i S'_i \rangle = \frac{T(X)| R_i S_i \rangle}{   W(X,R_i,S_i)}.
\end{eqnarray}
Now we need a normalized distribution
\begin{eqnarray}
\tilde{P}(X)=\frac{P(X)}{N} \frac{\langle \Psi_T |T(X)|R_i S_i\rangle}{\langle \Psi_T | R_i S_i \rangle},
\end{eqnarray}
in which the normalization factor
\begin{eqnarray}
N=\int dX P(X) \frac{\langle \Psi_T|T(X) |R_i S_i \rangle }{\langle \Psi_T | R_i S_i \rangle},
\end{eqnarray}
so that we can collect all the new walkers $|R'_i S'_i \rangle$ with less fluctuation. Neglecting $\mathcal{O}(\delta t^2)$, we have
\begin{eqnarray}
N=e^{-[E_L(R_i S_i) - E_T]\delta t  },
\end{eqnarray}
where the local energy is given by
\begin{eqnarray}
E_L(R_i, S_i) = \frac{\langle \Psi_T | H |R_i S_i\rangle}{ \langle \Psi_T |R_i S_i \rangle  }
=\frac{\hat{H} \langle \Psi_T | R_i S_i\rangle}{ \langle \Psi_T |R_i S_i \rangle  }.
\end{eqnarray}
Now Eq.(\ref{eq65}) can be written as
\begin{eqnarray}
|\Psi_T \Psi(t+\delta t) \rangle = \sum_{i=1}^{N_W} w_i \int dX \tilde{P}(X) e^{-[E_L(R_i S_i) - E_T]\delta t  }
\frac{T(X)| R_i S_i \rangle}{W(X,R_i,S_i)}.
\end{eqnarray}

\section{Hubbard-Stratonovich Transformation}
The key of AFDMC is to reduce the quadratic dependence of operators to linear, with the Hubbard-Stratonovich transformation,
\begin{eqnarray}
e^{\frac{O^2}{2}} = \frac{1}{\sqrt{2\pi}} \int dx e^{-\frac{x^2}{2}} e^{xO}, \label{HST}
\end{eqnarray}
such that the interaction between nucleons can be conveniently cast into the form of DMC.

If possible, we can find ways to write a Hamiltonian into $N$ operators with quadratic dependence, that is
\begin{eqnarray}
H=\frac{1}{2} \sum_{j=1}^{N} \lambda_j O^2_j,
\end{eqnarray}
and then the DMC propagator(neglecting $\mathcal{O}(\delta t^2)$) can be written as
\begin{eqnarray}
e^{-(H-E_T)\delta t} = \int \underbrace{dx_1...dx_N}_{dX}
\underbrace{\frac{1}{ (2\pi)^{N/2} } e^{-\frac{1}{2}\sum_{j=1}^{N}x_j^2  } }_{P(X)} \underbrace{e^{-i\sum_{j=1}^N x_j \sqrt{\lambda_j \delta t}O_j}e^{E_T \delta t}}_{T(X)},
\end{eqnarray}
neglecting $\mathcal{O}(\delta t^{3/2})$, the normalized distribution $\tilde{P}(X)$ can be written as
\begin{eqnarray}
\tilde{P}(X) = \frac{1}{ (2\pi)^{N/2} } \exp \left\{-\frac{1}{2} \sum_{j=1}^N \bigg[ x_j + i\sqrt{\lambda \delta t} \underbrace{\frac{\langle \Psi_T | O_j |R_i S_i \rangle}{\langle \Psi_T | R_i S_i \rangle}}_{\langle O_j \rangle}  \bigg]^2   \right\},
\end{eqnarray}
which means that $\chi_j \equiv x_j + i\sqrt{\lambda \delta t} \langle O_j \rangle $ is a gaussian random number with unit variance. In other words, we choose each auxiliary field $x_j$ by
\begin{eqnarray}
x_j = \chi_j - i\sqrt{\lambda_j \delta t} \langle O_j \rangle.
\end{eqnarray}
We move each walker by such an operation
\begin{eqnarray}
|R'_i S'_i \rangle = e^{-i\sum_{j=1}^N x_j \sqrt{\lambda_j \delta t}O_j} |R_i S_i \rangle,
\end{eqnarray}
and the weight of the new walker $|R'_i S'_i \rangle$ is given by $e^{-[E_L(R_i S_i) - E_T]\delta t  }$.

Hubbard-Stratonovich transformation can be mathematically understood as a Fourier transform.
Consider that if we define
\begin{align}
O & \equiv -i\tilde{O}, \\
F(\tilde{O}) & \equiv e^{-\tilde{O}^2/2},
\end{align}
the Fourier transform pair is
\begin{flalign}
F(\tilde{O}) &= \int \frac{dx}{2\pi} e^{-ix\tilde{O}}\tilde{F}(x), \label{FTo1} \\
\tilde{F}(x) &= \int d \tilde{O} e^{ix \tilde{O}} F(\tilde{O}) = \int d \tilde{O} e^{ix \tilde{O}} e^{-\tilde{O}^2/2} = \sqrt{2\pi} e^{-x^2/2}. \label{FTx2}
\end{flalign}
Substituting Eq.(\ref{FTx2}) back to Eq.(\ref{FTo1}) we have
\begin{eqnarray}
e^{-\tilde{O}^2/2} = \int \frac{dx}{\sqrt{2\pi}}  e^{-x^2/2}  e^{-ix\tilde{O}},\label{FTx2}
\end{eqnarray}
which is essentially the same as Eq.(\ref{HST}).

\section{AFDMC and the Standard DMC with Importance Sampling}
Let us consider the Hamiltonian
\begin{eqnarray}
H = \sum_{j=1}^A \sum_{\alpha=1}^3 \frac{p_{j\alpha}^2}{2m} + V(R),
\end{eqnarray}
the approximated propagator $e^{-(H-E_T)\delta t}$ after Hubbard-Stratonovich transformation is
\begin{eqnarray}
\!\!\!\!\! e^{-\sum_j\sum_\alpha \frac{p_{j\alpha}^2}{2m} \delta t}e^{-[V(R)-E_T]\delta t}
\! = \! \!\! \int \underbrace{  \prod_{j \alpha}  dx_{j\alpha} \frac{1}{\sqrt{2\pi}} e^{-\frac{x^2_{j\alpha}}{2}} }_{dX P(X)}
\underbrace{
e^{-\frac{i}{\hbar} p_{j\alpha} x_{j \alpha} \sqrt{\frac{\hbar^2 \delta t}{m}} }
\overbrace{
e^{-[V(R)-E_T]\delta t}}^{weight}  }_{T(X)}.
\end{eqnarray}
For the importance sampling case,
\begin{flalign}
i\sqrt{\lambda_j \delta t} \langle O_j \rangle
&= i \sqrt{\frac{\delta t}{m}} \frac{\langle \Psi_T | p_{j \alpha} | RS\rangle}{\langle \Psi_T | RS \rangle} \nonumber \\
&= i \sqrt{\frac{\delta t}{m}} \frac{p^\dag_{j \alpha} \langle \Psi_T |  RS\rangle}{\langle \Psi_T | RS \rangle} \nonumber \\
&= - \sqrt{\frac{\hbar^2 \delta t}{m}} \frac{\partial_{j \alpha} \langle \Psi_T |  RS\rangle}{\langle \Psi_T | RS \rangle},
\end{flalign}
and therefore the sampled auxiliary field $x_{j\alpha}$ are
\begin{eqnarray}
x_{j\alpha} = \chi_{j\alpha} + \sqrt{\frac{\hbar^2 \delta t}{m}} \frac{\partial_{j \alpha} \langle \Psi_T |  RS\rangle}{\langle \Psi_T | RS \rangle},
\end{eqnarray}
the new walker $|R'S'\rangle$ with weight $e^{-[E_L(R_i S_i) - E_T]\delta t }$ is moved by the translation operator,
\begin{eqnarray}
|R'S'\rangle = e^{-\frac{i}{\hbar}\sum_{j\alpha} x_{j\alpha} \sqrt{\frac{\hbar^2 \delta t}{m}} p_{j\alpha} } |RS \rangle,
\end{eqnarray}
which means that $S'=S$ and
\begin{align}
R'_{j\alpha} &= R_{j\alpha} + x_{j\alpha} \sqrt{\frac{\hbar^2 \delta t}{m}} \nonumber \\
& = R_{j\alpha} + \chi_{j\alpha} \sqrt{\frac{\hbar^2 \delta t}{m}}
+ \frac{\hbar^2 \delta t}{m} \frac{\partial_{j \alpha} \langle \Psi_T |  RS\rangle}{\langle \Psi_T | RS \rangle}.
\end{align}
This is exactly the standard DMC with importance sampling introduced in Sec.\ref{secDMCimp}.

\section{Quadratic Form of the AV6' Interaction}

The Hamiltonian with the AV6' interaction for A-particle system can be written as
\begin{flalign}
H &= \sum_{i\alpha} \frac{p^2_{i\alpha}}{2m} + \sum_{i<j} v^c (r_{ij})
+ \frac{1}{2}\sum_{i\alpha, j\beta} C^\sigma_{i\alpha, j\beta} \sigma_{i\alpha}\sigma_{j\beta} \nonumber \\
&+ \frac{1}{2}\sum_{i\alpha, j\beta, \gamma} C^{\sigma\tau}_{i\alpha, j\beta} (\sigma_{i\alpha}\tau_{i\gamma})(\sigma_{j\beta}\tau_{j\gamma})
+ \frac{1}{2}\sum_{i\gamma,j\gamma} C_{i,j}^\tau \tau_{i\gamma}\tau_{j\gamma}
\end{flalign}
where the matrices are defined as
\begin{flalign}
C^\sigma_{i\alpha, j\beta} &= v^\sigma (r_{ij}) \delta_{\alpha \beta} + v^t(r_{ij})( 3 \hat{\alpha}\cdot \hat{r}_{ij} \hat{\beta}\cdot \hat{r}_{ij} - \delta_{\alpha \beta}   ) , \nonumber \\
C^{\sigma\tau}_{i\alpha, j\beta} &= v^{\sigma\tau} (r_{ij}) \delta_{\alpha \beta} + v^{t\tau}(r_{ij})( 3 \hat{\alpha}\cdot \hat{r}_{ij} \hat{\beta}\cdot \hat{r}_{ij} - \delta_{\alpha \beta}   ) , \nonumber \\
C_{i,j}^\tau &= v^\tau(r_{ij}) .
\end{flalign}
Here $\alpha$, $\beta$, $\gamma$ refer to $x$, $y$ and $z$ components and $\hat{\alpha}$, $\hat{\beta}$ refer to the corresponding unit vector, and their $n^{\text{th}}$ eigenvectors and eigenvalues can be written as
\begin{flalign}
\sum_{j\beta} C^\sigma_{i\alpha, j\beta} \psi^{\sigma(n)}_{j\beta}&= \lambda_n^\sigma \psi^{\sigma(n)}_{i\alpha} , \nonumber \\
\sum_{j\beta} C^{\sigma\tau}_{i\alpha, j\beta} \psi^{\sigma(n)}_{j\beta} &= \lambda_n^{\sigma\tau} \psi^{\sigma\tau(n)}_{i\alpha} , \nonumber \\
\sum_j C_{i,j}^\tau \psi^{\tau(n)}_j &= \lambda^\tau_n \psi^{\tau(n)}_i.
\end{flalign}
We define the following operators,
\begin{flalign}
O^\sigma_n &= \sum_{i\alpha} \psi^{\sigma(n)}_{i\alpha} \sigma_{i\alpha} , \nonumber \\
O^{\sigma\tau}_{n\beta} &= \sum_{i\alpha} \psi^{\sigma\tau(n)}_{i\alpha} \sigma_{i\alpha} \tau_{i\beta} , \nonumber \\
O^\tau_{n\alpha} &= \sum_{i} \psi^{\tau(n)}_{i} \tau_{i\alpha},
\end{flalign}
such that the Hamiltonian can now be written in a quadratic form
\begin{flalign}
H &= \sum_{i=1}^A \sum_{\alpha=1}^3 \frac{p^2_{i\alpha}}{2m}
+ \sum_{i<j} v^c (r_{ij})
+ \frac{1}{2}\sum_{n=1}^{3A} \lambda_n^\sigma (O_n^\sigma)^2 \nonumber \\
&+ \frac{1}{2}\sum_{n=1}^{A}\sum_{\alpha=1}^3 \lambda_n^\tau (O_{n\alpha}^\tau)^2
+ \frac{1}{2}\sum_{n=1}^{3A}\sum_{\alpha=1}^3 \lambda_n^{\sigma\tau} (O_{n\alpha}^{\sigma\tau})^2
\end{flalign}
which can be conveniently used in AFDMC.

\section{Trial Wave Function}

The trial wave function used in AFDMC for the $N$-particle system has the following form,
\begin{eqnarray}
\psi_T(R,S) = \Phi_S(R) \Phi_A(R,S),
\end{eqnarray}
where $R\equiv(\bm{r}_1,...,\bm{r}_N)$ are the spatial coordinates and $S\equiv(s_1,...,s_N)$ are the spin-isospin part of the system. For example $s_i$ can be written as
\begin{eqnarray}
s_i = \left(\begin{matrix}
                   a_i   \\
                   b_i  \\
                   c_i  \\
                   d_i
                  \end{matrix}\right)
= a_i |p \uparrow \rangle + b_i |p \downarrow \rangle + c_i|n \uparrow \rangle + d_i|n \downarrow \rangle,
\end{eqnarray}
where $a_i$, $b_i$, $c_i$ and $d_i$ are complex coefficients for each basis.

The $\Phi_S(R)$ is composed of the Jastrow two-particle correlation functions and is given by
\begin{eqnarray}
\Psi_S(R) = \prod_{i<j} f_J(r_{ij}),
\end{eqnarray}
where $f_J$ is taken from the scalar component of the correlation operator $\hat{F}_{ij}$ in the Fermi-Hyper-Netted-Chain in the Single Operator Chain approximation (FHNC/SOC).
The antisymmetric part of the trial wave function depends on the particular system to be studied.
The general form is given by the non-interacting fermion wave functions in the Slater determinant,
\begin{eqnarray}
\Phi_A(R,S)=\mathcal{A} \prod_{i=1}^N \phi_\alpha(\bm{r}_i,s_i)=\det[\phi_\alpha(\bm{r}_i,s_i)],
\end{eqnarray}
where $\alpha$ is a set of quantum numbers of single-particle wave function. It can be label as $\alpha=\{n,j,m_j\}$ such that
\begin{eqnarray} \phi_\alpha(\bm{r}_i,s_i) = R_{n,j}(r_i)[Y_{l,m_l}(\Omega)\chi_{s,m_s}(s_i)]_{j,m_j},
\end{eqnarray}
where $R_{n,j}$ is a radial function, $Y_{l,m_l}$ are spherical harmonics and $\chi_{s,m_s}$ are spinor part.

For uniform infinite nuclear matter, we can simulate the system by putting the $N$ particles in the periodic box of volume $L^3$, and the single-particle function is given by the plane wave,
\begin{eqnarray}
\phi_\alpha(\bm{r}_i,s_i) = e^{i \bm{k}_\alpha \cdot \bm{r}_i} \chi_{s,m_s}(s_i),
\end{eqnarray}
with integers $n_{\alpha x}, n_{\alpha y}, n_{\alpha z}$, and the quantized momentum vector is given by
\begin{eqnarray}
\bm{k}_{\alpha}=\frac{2\pi}{L}(n_{\alpha x}, n_{\alpha y}, n_{\alpha z}).
\end{eqnarray}

\chapter{AV6' Linearizing Trick}
\label{secAV6'linear}
\clearpage

Here I follow the note written by my advisor, and present how the six AV6' operators in the propagator are linearized.
First of all, notice that for non-negative integer $n$, $\left[\sum_{p=1}^6 v_p(r_{ij}) O_{ij}^p\right]^n$ can still be express by the linear combination of the same 6 operators (just with different coefficients in front of the each operator). The 6 operators form a group. This is the foundation.

We then follow my advisor's way of labeling the order of the operators.
For a given $ij$ pair, label them 1 and 2, we wish to calculate the AV6' propagator linear in the 6 of its operators such that,
\begin{eqnarray}
&& e^{- \left[
v_c + v_\sigma \vec \sigma_1 \cdot \vec \sigma_2
+ v_t S_{12}
+ v_\tau \vec \tau_1 \cdot \vec \tau_2
+ v_{\sigma \tau} \vec \tau_1 \cdot \vec \tau_2 \vec \sigma_1 \cdot \vec\sigma_2
+v_{t \tau} S_{12} \vec \tau_1 \cdot \vec \tau_2
\right] \Delta \tau}
\nonumber\\
&=&
p_c + p_\sigma \vec \sigma_1 \cdot \vec \sigma_2
+ p_t t_{12}
+ p_\tau \vec \tau_1 \cdot \vec \tau_2
+ p_{\sigma \tau} \vec \tau_1 \cdot \vec \tau_2 \vec \sigma_1 \cdot \vec\sigma_2
+p_{t \tau} t_{12} \vec \tau_1 \cdot \vec\tau_2
\label{av6linearizing}
\end{eqnarray}
where $S_{12}$ is the tensor operator in Eq.(\ref{AV6'O}).
We need to solve for the coefficients $p_c, p_\sigma, p_t, p_\tau, p_{\sigma \tau}, p_{t \tau}$.
We choose $z$ axis to be along $\hat r$ in $S_{12}$, and for the 6 operators we can find 6 states in terms of the total isospin quantum number $T$ and spin quantum number $S$ and its z component $S_z$. We list the states and the corresponding eigenvalues for certain operators in Table \ref{Table6statesAV6'}.
\begin{table}[tbhp!]
\caption[The 6 Total Isospin and Spin States of the AV6' Interaction]{The 6 total isospin and spin states of the AV6' interaction.}
\label{Table6statesAV6'}%
\begin{tabular*}{1.0\textwidth}{@{\extracolsep{\fill}}cccccc}
\toprule
$\textrm{Label}$ & $\textrm{State}$ & $\bm{\tau}_1 \cdot \bm{\tau}_2$  &  $\bm{\sigma}_1 \cdot \bm{\sigma}_2$ &  $S_{12}$ &
$\textrm{Degeneracy}$ \\ \midrule
1 & $|S=1, |S_z|=1 \rangle \otimes | T=1 \rangle$ & 1 & 1 & 2 & 6 \\ [\smallskipamount]
2 & $| S=1, S_z=0\rangle \otimes | T=1 \rangle$  & 1 & 1 & -4 & 3\\ [\smallskipamount]
3 & $|S=1, |S_z|=1 \rangle \otimes | T=0 \rangle$ & -3 & 1 & 2 & 2 \\ [\smallskipamount]
4 & $| S=1, S_z=0\rangle \otimes | T=0 \rangle$  & -3 & 1 & -4 & 1 \\ [\smallskipamount]
5 & $|S=0 \rangle \otimes | T=0 \rangle$ & 1 & -3 & 0 & 3\\ [\smallskipamount]
6 & $|S=0 \rangle \otimes | T=0 \rangle$ & -3 & -3 & 0 & 1 \\
\bottomrule
\end{tabular*}
\end{table}

Next, we sandwich $e^{-[
v_c + v_\sigma \vec \sigma_1 \cdot \vec \sigma_2
+ v_t S_{12}
+ v_\tau \vec \tau_1 \cdot \vec \tau_2
+ v_{\sigma \tau} \vec \tau_1 \cdot \vec \tau_2 \vec \sigma_1 \cdot \vec\sigma_2
+v_{t \tau} S_{12} \vec \tau_1 \cdot \vec \tau_2
] \Delta \tau}$ between each of the 6 labeled states listed in Table \ref{Table6statesAV6'}, and we get the 6 equations correspondingly,
\begin{eqnarray}
&& \langle 1 |  e^{-[
v_c + v_\sigma \vec \sigma_1 \cdot \vec \sigma_2
+ v_t S_{12}
+ v_\tau \vec \tau_1 \cdot \vec \tau_2
+ v_{\sigma \tau} \vec \tau_1 \cdot \vec \tau_2 \vec \sigma_1 \cdot \vec\sigma_2
+v_{t \tau} S_{12} \vec \tau_1 \cdot \vec \tau_2
] \Delta \tau}  | 1 \rangle \nonumber \\
&\Rightarrow {\large \textcircled{\small 1}} &
e_1 = \exp \left[-(v_c+v_\sigma+2 v_t + v_\tau +v_{\sigma\tau} +2v_{t\tau})\Delta \tau \right]
\nonumber\\
&& \langle 2 |  e^{-[
v_c + v_\sigma \vec \sigma_1 \cdot \vec \sigma_2
+ v_t S_{12}
+ v_\tau \vec \tau_1 \cdot \vec \tau_2
+ v_{\sigma \tau} \vec \tau_1 \cdot \vec \tau_2 \vec \sigma_1 \cdot \vec\sigma_2
+v_{t \tau} S_{12} \vec \tau_1 \cdot \vec \tau_2
] \Delta \tau}  | 2 \rangle \nonumber \\
&\Rightarrow {\large \textcircled{\small 2}} &
e_2 = \exp \left[-(v_c+v_\sigma-4v_t + v_\tau +v_{\sigma\tau} -4v_{t\tau})\Delta \tau \right]
\nonumber\\
&& \langle 3 |  e^{-[
v_c + v_\sigma \vec \sigma_1 \cdot \vec \sigma_2
+ v_t S_{12}
+ v_\tau \vec \tau_1 \cdot \vec \tau_2
+ v_{\sigma \tau} \vec \tau_1 \cdot \vec \tau_2 \vec \sigma_1 \cdot \vec\sigma_2
+v_{t \tau} S_{12} \vec \tau_1 \cdot \vec \tau_2
] \Delta \tau}  | 3 \rangle \nonumber \\
&\Rightarrow {\large \textcircled{\small 3}} &
e_3 = \exp \left[-(v_c+v_\sigma+2v_t -3 v_\tau -3v_{\sigma\tau} -6v_{t\tau})\Delta \tau \right]
\nonumber\\
&& \langle 4 |  e^{-[
v_c + v_\sigma \vec \sigma_1 \cdot \vec \sigma_2
+ v_t S_{12}
+ v_\tau \vec \tau_1 \cdot \vec \tau_2
+ v_{\sigma \tau} \vec \tau_1 \cdot \vec \tau_2 \vec \sigma_1 \cdot \vec\sigma_2
+v_{t \tau} S_{12} \vec \tau_1 \cdot \vec \tau_2
] \Delta \tau}  | 4 \rangle \nonumber \\
&\Rightarrow {\large \textcircled{\small 4}} &
e_4 = \exp \left[-(v_c+v_\sigma-4v_t -3 v_\tau -3v_{\sigma\tau}+12v_{t\tau})\Delta \tau \right]
\nonumber\\
&& \langle 5 |  e^{-[
v_c + v_\sigma \vec \sigma_1 \cdot \vec \sigma_2
+ v_t S_{12}
+ v_\tau \vec \tau_1 \cdot \vec \tau_2
+ v_{\sigma \tau} \vec \tau_1 \cdot \vec \tau_2 \vec \sigma_1 \cdot \vec\sigma_2
+v_{t \tau} S_{12} \vec \tau_1 \cdot \vec \tau_2
] \Delta \tau}  | 5 \rangle \nonumber \\
&\Rightarrow {\large \textcircled{\small 5}} &
e_5 = \exp \left[-(v_c-3v_\sigma+v_\tau -3v_{\sigma\tau})\Delta \tau \right]
\nonumber\\
&& \langle 6 |  e^{-[
v_c + v_\sigma \vec \sigma_1 \cdot \vec \sigma_2
+ v_t S_{12}
+ v_\tau \vec \tau_1 \cdot \vec \tau_2
+ v_{\sigma \tau} \vec \tau_1 \cdot \vec \tau_2 \vec \sigma_1 \cdot \vec\sigma_2
+v_{t \tau} S_{12} \vec \tau_1 \cdot \vec \tau_2
] \Delta \tau}  | 6 \rangle \nonumber \\
&\Rightarrow {\large \textcircled{\small 6}} &
e_6 = \exp \left[-(v_c-3v_\sigma-3v_\tau +9v_{\sigma\tau})\Delta \tau \right].  \nonumber
\end{eqnarray}
From Eq.(\ref{av6linearizing}), we know that we need to solve for the coefficients $p_c, p_\sigma, p_t, p_\tau, p_{\sigma \tau}, p_{t \tau}$ from the already obtained values of $e_1, e_2, e_3, e_4, e_5, e_6$,
\begin{flalign}
e_1 &= p_c+p_\sigma+2 p_t + p_\tau +p_{\sigma\tau} +2p_{t\tau}
\nonumber\\
e_2 &= p_c+p_\sigma-4p_t + p_\tau +p_{\sigma\tau} -4p_{t\tau}
\nonumber\\
e_3 &= p_c+p_\sigma+2p_t -3 p_\tau -3p_{\sigma\tau} -6p_{t\tau}
\nonumber\\
e_4 &= p_c+p_\sigma-4p_t -3 p_\tau -3p_{\sigma\tau}+12p_{t\tau}
\nonumber\\
e_5 &= p_c-3p_\sigma+p_\tau -3p_{\sigma\tau}
\nonumber\\
e_6 &= p_c-3p_\sigma-3p_\tau + 9p_{\sigma\tau},  \nonumber
\end{flalign}
which can be written in the matrix form,
\begin{equation}
\left (
\begin{array}{c}
e_1 \\
e_2 \\
e_3 \\
e_4 \\
e_5 \\
e_6 \\
\end{array}
\right ) =
\left (
\begin{array}{rrrrrr}
1 & 1 & 2 & 1 & 1 & 2 \\
1 & 1 &-4 & 1 & 1 &-4 \\
1 & 1 & 2 &-3 &-3 &-6 \\
1 & 1 &-4 &-3 &-3 &12 \\
1 &-3 & 0 & 1 &-3 &0 \\
1 &-3 & 0 &-3 & 9 &0 \\
\end{array}
\right )
\left (
\begin{array}{c}
p_c \\
p_\sigma \\
p_t \\
p_\tau \\
p_{\sigma \tau} \\
p_{t \tau} \\
\end{array}
\right ) \,.
\end{equation}
The coefficients $p_c, p_\sigma, p_t, p_\tau, p_{\sigma \tau}, p_{t \tau}$ can be solved by matrix inversion,
\begin{equation}
\left (
\begin{array}{c}
p_c \\
p_\sigma \\
p_t \\
p_\tau \\
p_{\sigma \tau} \\
p_{t \tau} \\
\end{array}
\right ) = \frac{1}{48}
\left (
\begin{array}{rrrrrr}
18 & 9 & 6 & 3 & 9 & 3 \\
6 & 3 & 2 & 1 & -9 &-3 \\
6 & -6 & 2 &-2 & 0 & 0 \\
6 & 3 & -6 &-3 & 3 &-3 \\
2 & 1 & -2 & -1 &-3 &3 \\
2 &-2 & -2 & 2 & 0 &0 \\
\end{array}
\right )
\left (
\begin{array}{c}
e_1 \\
e_2 \\
e_3 \\
e_4 \\
e_5 \\
e_6 \\
\end{array}
\right ) \,.
\end{equation}
In other words, once we know $e_1, e_2, e_3, e_4, e_5, e_6$, the coefficients $p_c, p_\sigma, p_t, p_\tau, p_{\sigma \tau}, p_{t \tau}$ are written as,
\begin{flalign}
p_c &= (6 e_1+3e_2+2e_3+e_4+3e_5+e_6)/16,
\nonumber\\
p_\sigma &= (6 e_1+3e_2+2e_3+e_4-9e_5-3e_6)/48,
\nonumber\\
p_t &= (3 e_1-3e_2+e_3-e_4)/24,
\nonumber\\
p_\tau &= (2 e_1+e_2-2e_3-e_4+e_5-e_6)/16,
\nonumber\\
p_{\sigma \tau} &= (2 e_1+e_2-2e_3-e_4-3e_5+3e_6)/48,
\nonumber\\
p_{t \tau} &= (e_1-e_2-e_3+e_4)/24 \,. \nonumber
\end{flalign}
The six coefficients $p_c, p_\sigma, p_t, p_\tau, p_{\sigma \tau}, p_{t \tau}$ are the corresponding six $u^p(r_{ij})$ in Eq.(\ref{eVlinear}).

\chapter{Additional Calculations of the Response Functions}
\clearpage
\section{Angle Averaged Euclidean Response Functions}
\label{secAngleAveResponse}

For any operator $\hat{O}$ which is independent of the momentum $\bm{k}$, e.g., $\hat{O}$ can be $e^{-H\tau}$ or $e^{-V\tau}$ defined in Eq.(\ref{eVtaudef}), we show how to take the angle average of momentum $\bm{k}$ in calculating $\int \rho^\dagger(\bm{k}) \hat{O} \rho(\bm{k})  d \Omega /4\pi$ in the response functions.

In order to proceed, we first set up the coordinates as shown in Fig. \ref{angleave}.
The polar angle is $\theta$, the azimuthal angle is $\varphi$, and $d \Omega = \sin \theta d\theta d\varphi $. The length of $\bm{k}$ is $k$, its unit vector is $\hat{k}$, the $x$, $y$ and $z$ components of $\hat{k}$ are
\begin{align}
\hat{k}_x&=\sin \theta \cos \varphi , \\
\hat{k}_y&=\sin \theta \sin \varphi , \\
\hat{k}_z&= \cos \theta .
\label{xyzsetup}
\end{align}
The vector between particle $j$ and $i$ is defined as $\bm{r}_{ji}\equiv \bm{r}_j-\bm{r}_i$. Its length is $r_{ji}$, and its unit vector is $\hat{\bm{r}}_{ji}$ and its components are $\hat{r}_x$, $\hat{r}_y$ and $\hat{r}_z$.

\begin{figure}[tbh]
\centering
\includegraphics[scale=0.456]{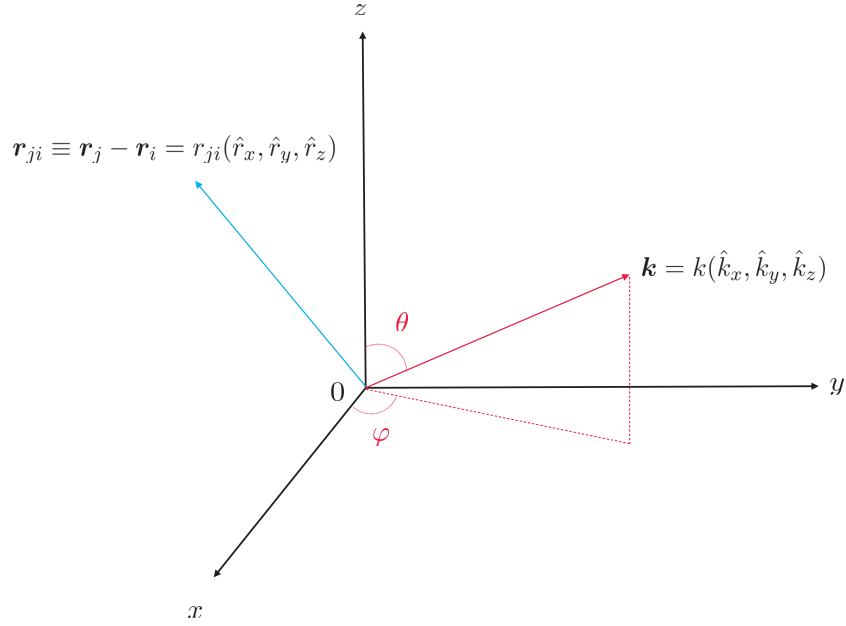}
\caption[The Coordinates for Calculating the Angle Averaged Euclidean Response Functions]{The coordinates for the angle averaged Euclidean response functions.}
\label{angleave}
\end{figure}

Note that the plane wave $e^{i\bm{k}\cdot\bm{r}_{ji}}$ can be expanded by spherical harmonics (\cite{jackson3rd}),
\begin{flalign}
e^{i\bm{k}\cdot\bm{r}_{ji}}
&= 4\pi \sum_{\ell=0}^\infty \sum_{m=-\ell}^{\ell} i^\ell j_\ell(kr_{ji}) Y_\ell^m(\hat{\bm{k}}) {Y_\ell^{m}}^*(\hat{\bm{r}}_{ji}) \nonumber \\
&= 4\pi \sum_{\ell=0}^\infty \sum_{m=-\ell}^{\ell} i^\ell j_\ell(kr_{ji}) {Y_\ell^{m}}^*(\hat{\bm{r}}_{ji}) Y_\ell^m(\theta,\varphi) ,
\label{planewaveEXP}
\end{flalign}
where $j_\ell$ are spherical Bessel functions (\cite{Handbook_abramowitz}), and the spherical harmonics $Y_\ell^m(\hat{\bm{k}})=Y_\ell^m(\theta,\varphi)$.
In this calculation we only need those functions with $\ell=0,2$.

For the spherical Bessel functions we need $j_0(kr_{ji})$ and $j_2(kr_{ji})$,
\begin{align}
j_0(kr_{ji}) &= \frac{\sin (kr_{ji})}{k r_{ji}},  \\
j_2(kr_{ji}) &= \left( \frac{3}{k^2 r^2_{ji}} -1  \right)\frac{\sin (kr_{ji})}{k r_{ji}} - \frac{3 \cos (kr_{ji})}{k^2 r_{ji}^2}  .
\end{align}

For $\ell = 0$, the spherical harmonics of $\hat{\bm{k}}$ and $\hat{\bm{r}}_{ji}$ are the same and they can be written as,
\begin{align}
Y_{0}^{0}(\theta,\varphi) = Y_{0}^{0}(\hat{\bm{k}})  = Y_{0}^{0}(\bm{r}_{ji}) = \sqrt{1\over 4\pi} .
\end{align}

For $\ell = 2$, the spherical harmonics of $\hat{\bm{k}}$ can be written in terms of $\theta$ and $\varphi$ as,
\begin{align}
Y_{2}^{-2}(\theta,\varphi)&= {1\over 4}\sqrt{15\over 2\pi}\cdot e^{-2i\varphi}\cdot\sin^{2}\theta ,\quad \\
Y_{2}^{-1}(\theta,\varphi)&= {1\over 2}\sqrt{15\over 2\pi}\cdot e^{- i\varphi}\cdot\sin    \theta\cdot  \cos\theta ,\quad \\
Y_{2}^{ 0}(\theta,\varphi)&= {1\over 4}\sqrt{ 5\over  \pi}\cdot (3\cos^{2}\theta-1), \quad \\
Y_{2}^{ 1}(\theta,\varphi)&=-{1\over 2}\sqrt{15\over 2\pi}\cdot e^{  i\varphi}\cdot\sin    \theta\cdot  \cos\theta, \quad \\
Y_{2}^{ 2}(\theta,\varphi)&= {1\over 4}\sqrt{15\over 2\pi}\cdot e^{ 2i\varphi}\cdot\sin^{2}\theta. \quad
\end{align}

The spherical harmonics of $\hat{\bm{r}}_{ji}$ for $\ell = 2$ can be written in terms of $\hat{r}_x$, $\hat{r}_y$ and $\hat{r}_z$ as,
\begin{align}
Y_{2}^{-2}(\hat{\bm{r}}_{ji})&= {1\over 4}\sqrt{15\over 2\pi}\cdot{(\hat{r}_x - i\hat{r}_y)^2}, \\
Y_{2}^{-1}(\hat{\bm{r}}_{ji})&= {1\over 2}\sqrt{15\over 2\pi}\cdot{(\hat{r}_x - i\hat{r}_y)\hat{r}_z}, \\
Y_{2}^{ 0}(\hat{\bm{r}}_{ji})&= {1\over 4}\sqrt{ 5\over  \pi}\cdot{(2\hat{r}_z^{2}-\hat{r}_x^{2}-\hat{r}_y^{2})}, \\
Y_{2}^{ 1}(\hat{\bm{r}}_{ji})&= -{1\over 2}\sqrt{15\over 2\pi}\cdot{(\hat{r}_x + i\hat{r}_y)\hat{r}_z},  \\
Y_{2}^{ 2}(\hat{\bm{r}}_{ji})&= {1\over 4}\sqrt{15\over 2\pi}\cdot{(\hat{r}_x + i\hat{r}_y)^2}.
\end{align}

The Spherical harmonics are orthogonal such that
\begin{eqnarray}
\int_{\theta=0}^\pi\int_{\varphi=0}^{2\pi}Y_\ell^m (\theta, \varphi) \, Y_{\ell'}^{m'}{}^* (\theta, \varphi) \, d\Omega=\delta_{\ell\ell'}\, \delta_{mm'},
\label{SHMortho}
\end{eqnarray}
and they satisfy,
\begin{eqnarray}
Y_\ell^{m}{}^* = (-1)^m Y_\ell^{-m}.
\label{SHMortho}
\end{eqnarray}

The following derived relations for spherical harmonics will be used,
\begin{flalign}
\cos^2\theta &=
\frac{4}{3}\sqrt{\frac{\pi}{5}}{Y_2^0}^*(\theta,\varphi)+\frac{1}{3}, \\
\sin\theta \cos\theta \cos\varphi &=
\sqrt{\frac{2\pi}{15}}
\left[{Y_2^{-1}}^*(\theta,\varphi) - {Y_{2}^{1}}^*(\theta,\varphi) \right], \\
\sin\theta \cos\theta \sin\varphi &=
\frac{1}{i}\sqrt{\frac{2\pi}{15}}
\left[{Y_2^{-1}}^*(\theta,\varphi) + {Y_{2}^{1}}^*(\theta,\varphi) \right], \\
\sin^2\theta \cos 2\varphi &=
2\sqrt{\frac{2\pi}{15}}
\left[{Y_2^2}^*(\theta,\varphi) + {Y_{2}^{-2}}^*(\theta,\varphi) \right], \\
\sin^2\theta \cos\varphi \sin\varphi &=
\frac{1}{i}\sqrt{\frac{2\pi}{15}}
\left[{Y_2^{-2}}^*(\theta,\varphi) - {Y_{2}^{2}}^*(\theta,\varphi) \right], \\
\sin^2\theta \sin^2\varphi &=
\frac{1}{3}
-\frac{2}{3}\sqrt{\frac{\pi}{5}}{Y_2^0}^*(\theta,\varphi)
-\sqrt{\frac{2\pi}{15}}
\left[{Y_2^{2}}^*(\theta,\varphi) + {Y_{2}^{-2}}^*(\theta,\varphi) \right] .
\end{flalign}

Before calculating the angle averaged response functions, we need to calculate several relevant angle averaged functions.

\begin{enumerate}
  \item Angle averaged function $\int   e^{i \bm{k} \cdot \bm{r}_{ji} } d \Omega  /4\pi$ can be calculated as,
\begin{flalign}
& \frac{1}{4\pi} \displaystyle \int   e^{i \bm{k} \cdot \bm{r}_{ji} } d \Omega \nonumber \\
&=
\sum_{\ell=0}^\infty \sum_{m=-\ell}^{\ell} i^\ell j_\ell(k{r}_{ji}) {Y_\ell^{m}}^*(\hat{\bm{r}}_{ji}) \int Y_\ell^m(\theta,\varphi) d\Omega  \nonumber \\
&= \sum_{\ell=0}^\infty \sum_{m=-\ell}^{\ell} i^\ell j_\ell(k{r}_{ji}) {Y_\ell^{m}}^*(\hat{\bm{r}}_{ji}) \int Y_\ell^m(\theta,\varphi) \sqrt{4\pi}
{Y_0^0}^* (\theta,\varphi) d\Omega \nonumber \\
&= \sum_{\ell=0}^\infty \sum_{m=-\ell}^{\ell} i^\ell j_\ell(k{r}_{ji}) {Y_\ell^{m}}^*(\hat{\bm{r}}_{ji})
\sqrt{4\pi} \delta_{\ell0}\, \delta_{m0}  \nonumber \\
&= j_0(k{r}_{ji}) .
\label{Ac}
\end{flalign}
  \item Angle averaged function $\int   e^{i \bm{k} \cdot \bm{r}_{ji} } \hat{k}_x \hat{k}_x d \Omega  /4\pi$ can be calculated as,
\begin{flalign}
& \frac{1}{4\pi}  \displaystyle  \int   e^{i \bm{k} \cdot \bm{r}_{ji} } \hat{k}_x \hat{k}_x d \Omega\nonumber \\
&=
\sum_{\ell=0}^\infty \sum_{m=-\ell}^{\ell} i^\ell j_\ell(k{r}_{ji}) {Y_\ell^{m}}^*(\hat{\bm{r}}_{ji}) \int Y_\ell^m(\theta,\varphi) \sin^2\theta \cos^2\varphi d\Omega  \nonumber \\
&=
\sum_{\ell=0}^\infty \sum_{m=-\ell}^{\ell} i^\ell j_\ell(k{r}_{ji}) {Y_\ell^{m}}^*(\hat{\bm{r}}_{ji}) \nonumber \\
& ~~~~ \times \int Y_\ell^m(\theta,\varphi)
\left\{\frac{1}{3}-\frac{2}{3}\sqrt{\frac{\pi}{5}}{Y_2^0}^*(\theta,\varphi)
+\sqrt{\frac{2\pi}{15}} \left[ {Y_2^{2}}^*(\theta,\varphi)
+{Y_2^{-2}}^*(\theta,\varphi) \right] \right\}
d\Omega \nonumber \\
&=
\sum_{\ell=0}^\infty \sum_{m=-\ell}^{\ell} i^\ell j_\ell(k{r}_{ji}) {Y_\ell^{m}}^*(\hat{\bm{r}}_{ji}) \nonumber \\
&~~~~\times  \left(\frac{\sqrt{4\pi}}{3}\delta_{\ell0}\, \delta_{m0}
-\frac{2}{3}\sqrt{\frac{\pi}{5}}\delta_{\ell2}\, \delta_{m0}
+\sqrt{\frac{2\pi}{15}}\delta_{\ell2}\, \delta_{m-2}
+\sqrt{\frac{2\pi}{15}}\delta_{\ell2}\, \delta_{m2}
 \right) \nonumber \\
&=
 \frac{\sqrt{4\pi}}{3}j_0(k{r}_{ji}){Y_0^0}^*(\hat{\bm{r}}_{ji})
+\frac{2}{3}\sqrt{\frac{\pi}{5}}j_2(k{r}_{ji}){Y_2^0}^*(\hat{\bm{r}}_{ji}) \nonumber \\
& ~~~~
-\sqrt{\frac{2\pi}{15}} \left[   j_0(k{r}_{ji}){Y_2^{-2}}^*(\hat{\bm{r}}_{ji})
- j_2(k{r}_{ji}){Y_2^2}^*(\hat{\bm{r}}_{ji}) \right] \nonumber \\
&=
\frac{1}{3} \left[
j_0(k{r}_{ji})+j_2(k{r}_{ji})(\hat{r}_z^2+\hat{r}_y^2-2\hat{r}_x^2)
\right].
\label{Akxkx}
\end{flalign}
  \item
Angle averaged function $\int   e^{i \bm{k} \cdot \bm{r}_{ji} } \hat{k}_x \hat{k}_y d \Omega  /4\pi$ can be calculated as,
\begin{flalign}
\label{Akxky}
&
\frac{1}{4\pi}  \displaystyle \int   e^{i \bm{k} \cdot \bm{r}_{ji} } \hat{k}_x \hat{k}_y d \Omega \nonumber \\
&=
\sum_{\ell=0}^\infty \sum_{m=-\ell}^{\ell} i^\ell j_\ell(k{r}_{ji}) {Y_\ell^{m}}^*(\hat{\bm{r}}_{ji}) \int Y_\ell^m(\theta,\varphi) \sin^2\theta \cos\varphi \sin\varphi d\Omega  \nonumber \\
&= \sum_{\ell=0}^\infty \sum_{m=-\ell}^{\ell} i^\ell j_\ell(k{r}_{ji}) {Y_\ell^{m}}^*(\hat{\bm{r}}_{ji})
\int Y_\ell^m(\theta,\varphi)
\frac{1}{i}\sqrt{\frac{2\pi}{15}}
\left[
Y_{2}^{-2}{}^*(\theta,\varphi)
-Y_2^{2}{}^*(\theta,\varphi)
\right]
d\Omega \nonumber \\
&= \sum_{\ell=0}^\infty \sum_{m=-\ell}^{\ell} i^\ell j_\ell(k{r}_{ji}) {Y_\ell^{m}}^*(\hat{\bm{r}}_{ji})
\frac{1}{i}\sqrt{\frac{2\pi}{15}}
\left(
\delta_{\ell2}\, \delta_{m-2} -\delta_{\ell2}\, \delta_{m2}
 \right) \nonumber \\
&= - j_2(k{r}_{ji})\left[
{Y_2^{-2}}^*(\hat{\bm{r}}_{ji}) \frac{1}{i}\sqrt{\frac{2\pi}{15}}
-{Y_2^{2}}^*(\hat{\bm{r}}_{ji}) \frac{1}{i}\sqrt{\frac{2\pi}{15}}
\right] \nonumber \\
&= -j_2(k{r}_{ji})\hat{r}_x\hat{r}_y.
\end{flalign}
  \item Angle averaged function $\int   e^{i \bm{k} \cdot \bm{r}_{ji} } \hat{k}_x \hat{k}_z d \Omega  /4\pi$ can be calculated as,
\begin{flalign}
&
\frac{1}{4\pi}  \displaystyle \int   e^{i \bm{k} \cdot \bm{r}_{ji} } \hat{k}_x \hat{k}_z d \Omega \nonumber \\
&=
\sum_{\ell=0}^\infty \sum_{m=-\ell}^{\ell} i^\ell j_\ell(k{r}_{ji}) {Y_\ell^{m}}^*(\hat{\bm{r}}_{ji}) \int Y_\ell^m(\theta,\varphi)
\sin\theta \cos\theta \cos\varphi d\Omega  \nonumber \\
&= \sum_{\ell=0}^\infty \sum_{m=-\ell}^{\ell} i^\ell j_\ell(k{r}_{ji}) {Y_\ell^{m}}^*(\hat{\bm{r}}_{ji}) \int Y_\ell^m(\theta,\varphi)
\sqrt{\frac{2\pi}{15}}
\left[
{Y_2^{-1}}^*(\theta,\varphi) - {Y_2^{1}}^*(\theta,\varphi)
\right]
d\Omega \nonumber \\
&= \sum_{\ell=0}^\infty \sum_{m=-\ell}^{\ell} i^\ell j_\ell(k{r}_{ji}) {Y_\ell^{m}}^*(\hat{\bm{r}}_{ji})
\left(
\sqrt{\frac{2\pi}{15}}\delta_{\ell2}\, \delta_{m-1}
-\sqrt{\frac{2\pi}{15}}\delta_{\ell2}\, \delta_{m1}
 \right) \nonumber \\
&= - j_2(k{r}_{ji})\left[
{Y_2^{-1}}^*(\hat{\bm{r}}_{ji})\sqrt{\frac{2\pi}{15}}
-{Y_2^{1}}^*(\hat{\bm{r}}_{ji})\sqrt{\frac{2\pi}{15}}
\right] \nonumber \\
&= -j_2(k{r}_{ji})\hat{r}_x\hat{r}_z.
\label{Akxkz}
\end{flalign}
  \item  Angle averaged function $\int   e^{i \bm{k} \cdot \bm{r}_{ji} } \hat{k}_y \hat{k}_y d \Omega  /4\pi$ can be calculated as,
\begin{flalign}
&
\frac{1}{4\pi}  \displaystyle \int   e^{i \bm{k} \cdot \bm{r}_{ji} } \hat{k}_y \hat{k}_y d \Omega \nonumber \\
&=
\sum_{\ell=0}^\infty \sum_{m=-\ell}^{\ell} i^\ell j_\ell(k{r}_{ji}) {Y_\ell^{m}}^*(\hat{\bm{r}}_{ji}) \int Y_\ell^m(\theta,\varphi) \sin^2\theta \cos^2\varphi d\Omega  \nonumber \\
&= \sum_{\ell=0}^\infty \sum_{m=-\ell}^{\ell} i^\ell j_\ell(k{r}_{ji}) {Y_\ell^{m}}^*(\hat{\bm{r}}_{ji}) \nonumber \\
&~~~  \times \int Y_\ell^m(\theta,\varphi)
\left\{\frac{1}{3}
-\frac{2}{3}\sqrt{\frac{\pi}{5}}{Y_2^0}^*(\theta,\varphi)
+\sqrt{\frac{2\pi}{15}} \left[ {Y_2^{2}}^*(\theta,\varphi)
+ {Y_2^{-2}}^*(\theta,\varphi) \right] \right\}
d\Omega \nonumber \\
&= \sum_{\ell=0}^\infty \sum_{m=-\ell}^{\ell} i^\ell j_\ell(k{r}_{ji}) {Y_\ell^{m}}^*(\hat{\bm{r}}_{ji}) \nonumber \\
&~~~  \times
\left[\frac{\sqrt{4\pi}}{3}\delta_{\ell0}\, \delta_{m0}
-\frac{2}{3}\sqrt{\frac{\pi}{5}}\delta_{\ell2}\, \delta_{m0}
+\sqrt{\frac{2\pi}{15}} \left(\delta_{\ell2}\, \delta_{m-2}
+\delta_{\ell2}\, \delta_{m2} \right)
 \right] \nonumber \\
&= \frac{\sqrt{4\pi}}{3}j_0(k{r}_{ji}){Y_0^0}^*(\hat{\bm{r}}_{ji})
+\frac{2}{3}\sqrt{\frac{\pi}{5}}j_2(k{r}_{ji}){Y_2^0}^*(\hat{\bm{r}}_{ji}) \nonumber \\
& ~~~
-\sqrt{\frac{2\pi}{15}}j_0(k{r}_{ji}){Y_2^{-2}}^*(\hat{\bm{r}}_{ji})
-\sqrt{\frac{2\pi}{15}}j_2(k{r}_{ji}){Y_2^2}^*(\hat{\bm{r}}_{ji}) \nonumber \\
&= \frac{1}{3} \left[
j_0(k{r}_{ji})+j_2(k{r}_{ji})(\hat{r}_z^2+\hat{r}_x^2-2\hat{r}_y^2)
\right].
\label{Akyky}
\end{flalign}
  \item Angle averaged function $\int   e^{i \bm{k} \cdot \bm{r}_{ji} } \hat{k}_y \hat{k}_z d \Omega  /4\pi$ can be calculated as,
\begin{flalign}
&
\frac{1}{4\pi}  \displaystyle \int   e^{i \bm{k} \cdot \bm{r}_{ji} } \hat{k}_y \hat{k}_z d \Omega \nonumber \\
&=
\sum_{\ell=0}^\infty \sum_{m=-\ell}^{\ell} i^\ell j_\ell(k{r}_{ji}) {Y_\ell^{m}}^*(\hat{\bm{r}}_{ji}) \int Y_\ell^m(\theta,\varphi)
\sin\theta \cos\theta \sin\varphi d\Omega  \nonumber \\
&= \sum_{\ell=0}^\infty \sum_{m=-\ell}^{\ell} i^\ell j_\ell(k{r}_{ji}) {Y_\ell^{m}}^*(\hat{\bm{r}}_{ji}) \int Y_\ell^m(\theta,\varphi)
\frac{1}{i}\sqrt{\frac{2\pi}{15}}
\left[
 {Y_2^{-1}}^*(\theta,\varphi) + {Y_2^{1}}^*(\theta,\varphi)
\right]
d\Omega \nonumber \\
&= \sum_{\ell=0}^\infty \sum_{m=-\ell}^{\ell} i^\ell j_\ell(k{r}_{ji}) {Y_\ell^{m}}^*(\hat{\bm{r}}_{ji})
\left(
\frac{1}{i}\sqrt{\frac{2\pi}{15}}\delta_{\ell2}\, \delta_{m-1}
-\frac{1}{i}\sqrt{\frac{2\pi}{15}}\delta_{\ell2}\, \delta_{m1}
 \right) \nonumber \\
&= - j_2(k{r}_{ji})\left[
{Y_2^{-1}}^*(\hat{\bm{r}}_{ji})\frac{1}{i}\sqrt{\frac{2\pi}{15}}
+{Y_2^{1}}^*(\hat{\bm{r}}_{ji})\frac{1}{i}\sqrt{\frac{2\pi}{15}}
\right] \nonumber \\
&= -j_2(k{r}_{ji})\hat{r}_x\hat{r}_z.
\label{Akykz}
\end{flalign}
  \item  Angle averaged function $\int   e^{i \bm{k} \cdot \bm{r}_{ji} } \hat{k}_z \hat{k}_z d \Omega  /4\pi$ can be calculated as,
\begin{flalign}
&
\frac{1}{4\pi}  \displaystyle  \int   e^{i \bm{k} \cdot \bm{r}_{ji} } \hat{k}_z \hat{k}_z d \Omega \nonumber \\
&=
\sum_{\ell=0}^\infty \sum_{m=-\ell}^{\ell} i^\ell j_\ell(k{r}_{ji}) {Y_\ell^{m}}^*(\hat{\bm{r}}_{ji}) \int Y_\ell^m(\theta,\varphi) \cos^2\theta d\Omega  \nonumber \\
&= \sum_{\ell=0}^\infty \sum_{m=-\ell}^{\ell} i^\ell j_\ell(k{r}_{ji}) {Y_\ell^{m}}^*(\hat{\bm{r}}_{ji}) \int Y_\ell^m(\theta,\varphi)
\left[\frac{1}{3}
+\frac{4}{3}\sqrt{\frac{\pi}{5}}{Y_2^0}^*(\theta,\varphi)
 \right]
d\Omega \nonumber \\
&= \sum_{\ell=0}^\infty \sum_{m=-\ell}^{\ell} i^\ell j_\ell(k{r}_{ji}) {Y_\ell^{m}}^*(\hat{\bm{r}}_{ji})
\left(\frac{\sqrt{4\pi}}{3}\delta_{\ell0}\, \delta_{m0}
+\frac{4}{3}\sqrt{\frac{\pi}{5}}\delta_{\ell2}\, \delta_{m0}
 \right) \nonumber \\
&= \frac{\sqrt{4\pi}}{3}j_0(k{r}_{ji}){Y_0^0}^*(\hat{\bm{r}}_{ji})
-\frac{4}{3}\sqrt{\frac{\pi}{5}}j_2(k{r}_{ji}){Y_2^0}^*(\hat{\bm{r}}_{ji}) \nonumber \\
&= \frac{1}{3} \left[
j_0(k{r}_{ji})+j_2(k{r}_{ji})(\hat{r}_x^2+\hat{r}_y^2-2\hat{r}_z^2)
\right].
\label{Akzkz}
\end{flalign}
\end{enumerate}

With the help of Eqs.(\ref{Ac}) to (\ref{Akzkz}), the 5 angle averaged response functions $\int \rho^\dagger(\bm{k}) \hat{O} \rho(\bm{k})  d \Omega /4\pi$  can be conveniently calculated.

\begin{enumerate}
  \item For the angle averaged nucleon coupling $\rho_N (\bm{k})$ we have,
\begin{eqnarray}
&&
\frac{1}{4\pi} \int \rho_N^\dagger(\bm{k}) \hat{O} \rho_N(\bm{k})  d \Omega \nonumber \\
&=&
\frac{1}{4\pi}  \int  \left(\sum_{i=1}^{A} e^{-i \bm{k} \cdot \bm{r}_i }\right)
\hat{O}
\left(\sum_{j=1}^{A} e^{i \bm{k} \cdot \bm{r}_j } \right)  d \Omega  \nonumber  \\
&=& \frac{1}{4\pi}
\sum_{i=1}^{A} \sum_{j=1}^{A} \left( \int   e^{i \bm{k} \cdot \bm{r}_{ji} }
 d \Omega \right) \hat{O} \nonumber \\
&=& \sum_{i=1}^{A} \sum_{j=1}^{A} j_0(k{r}_{ji}) \hat{O} .
\end{eqnarray}
  \item For the angle averaged proton coupling $\rho_p (\bm{k})$ we have,
\begin{eqnarray}
&&
\frac{1}{4\pi} \int \rho_p^\dagger(\bm{k}) \hat{O} \rho_p(\bm{k})  d \Omega \nonumber \\
 &=& \frac{1}{4\pi}
 \int  \left(\sum_{i=1}^{A} e^{-i \bm{k} \cdot \bm{r}_i } \frac{1+\tau_{iz}}{2} \right)
\hat{O}
\left(\sum_{j=1}^{A} e^{i \bm{k} \cdot \bm{r}_j } \frac{1+\tau_{jz}}{2}  \right)  d \Omega  \nonumber  \\
&=& \frac{1}{4\pi}
 \sum_{i=1}^{A} \sum_{j=1}^{A} \left( \int e^{i \bm{k} \cdot \bm{r}_{ji} }  d \Omega \right)
\frac{1+\tau_{iz}}{2} \hat{O} \frac{1+\tau_{jz}}{2} \nonumber \\
&=&
 \sum_{i=1}^{A} \sum_{j=1}^{A} j_0(k{r}_{ji})
\frac{1+\tau_{iz}}{2} \hat{O} \frac{1+\tau_{jz}}{2} .
\end{eqnarray}
  \item For the angle averaged isovector coupling $\rho_\tau (\bm{k})$ we have,
\begin{eqnarray}
&&
\frac{1}{4\pi} \int \rho_\tau^\dagger(\bm{k}) \hat{O} \rho_\tau(\bm{k})  d \Omega \nonumber \\
&=& \frac{1}{4\pi}
 \int  \left(\sum_{i=1}^{A} e^{-i \bm{k} \cdot \bm{r}_i } \tau_{iz} \right)
\hat{O}
\left(\sum_{j=1}^{A} e^{i \bm{k} \cdot \bm{r}_j } \tau_{jz} \right)  d \Omega \nonumber  \\
&=& \frac{1}{4\pi}
 \sum_{i=1}^{A} \sum_{j=1}^{A} \left( \int e^{i \bm{k} \cdot \bm{r}_{ji} }  d \Omega \right)
\tau_{iz} \hat{O}\tau_{jz} \nonumber \\
&=&
 \sum_{i=1}^{A} \sum_{j=1}^{A} j_0(k{r}_{ji})
\tau_{iz} \hat{O}\tau_{jz} .
\end{eqnarray}
  \item For the angle averaged spin-longitudinal coupling $\rho_{\sigma \tau L} (\bm{k})$ we have,
\begin{eqnarray}
&&
\frac{1}{4\pi} \int \rho_{\sigma \tau L}^\dagger(\bm{k}) \hat{O} \rho_{\sigma \tau L}(\bm{k})  d \Omega \nonumber \\
&=&
\frac{1}{4\pi} \int  \left[\sum_{i=1}^{A} e^{-i \bm{k} \cdot \bm{r}_i } (\bm{\sigma}_i \cdot \hat{\bm{k}}) \tau_{iz}  \right]
\hat{O}
\left[\sum_{j=1}^{A} e^{i \bm{k} \cdot \bm{r}_j } \frac{1+\tau_{jz}}{2} (\bm{\sigma}_j \cdot \hat{\bm{k}}) \tau_{jz}  \right]  d \Omega  \nonumber \\
&=&
\frac{1}{4\pi}
\sum_{i=1}^{A} \sum_{j=1}^{A} \left( \int e^{i \bm{k} \cdot \bm{r}_{ji} } \hat{k}_x \hat{k}_x d \Omega \right)
\sigma_{ix} \tau_{iz} \hat{O} \sigma_{jx} \tau_{jz} \nonumber \\
&~+& \frac{1}{4\pi} \sum_{i=1}^{A} \sum_{j=1}^{A} \left( \int e^{i \bm{k} \cdot \bm{r}_{ji} } \hat{k}_y \hat{k}_y d \Omega \right)
\sigma_{iy} \tau_{iz} \hat{O} \sigma_{jy} \tau_{jz} \nonumber \\
&~+& \frac{1}{4\pi} \sum_{i=1}^{A} \sum_{j=1}^{A} \left( \int e^{i \bm{k} \cdot \bm{r}_{ji} } \hat{k}_z \hat{k}_z d \Omega \right)
\sigma_{iz} \tau_{iz} \hat{O} \sigma_{jz} \tau_{jz} \nonumber \\
&~+& \frac{1}{4\pi} \sum_{i=1}^{A} \sum_{j=1}^{A} \left( \int e^{i \bm{k} \cdot \bm{r}_{ji} } \hat{k}_x \hat{k}_y d \Omega \right)
\left(\sigma_{iy} \tau_{iz} \hat{O} \sigma_{jx} \tau_{jz} + \sigma_{ix} \tau_{iz} \hat{O} \sigma_{jy} \tau_{jz}  \right)\nonumber \\
&~+& \frac{1}{4\pi} \sum_{i=1}^{A} \sum_{j=1}^{A} \left( \int e^{i \bm{k} \cdot \bm{r}_{ji} } \hat{k}_x \hat{k}_z d \Omega \right)
\left(\sigma_{iz} \tau_{iz} \hat{O} \sigma_{jx} \tau_{jz} + \sigma_{ix} \tau_{iz} \hat{O} \sigma_{jz} \tau_{jz} \right) \nonumber \\
&~+& \frac{1}{4\pi} \sum_{i=1}^{A} \sum_{j=1}^{A} \left( \int e^{i \bm{k} \cdot \bm{r}_{ji} } \hat{k}_y \hat{k}_z d \Omega \right)
\left( \sigma_{iz} \tau_{iz} \hat{O} \sigma_{jy} \tau_{jz} + \sigma_{iy} \tau_{iz} \hat{O} \sigma_{jz} \tau_{jz} \right) \nonumber \\
&=&
\frac{1}{3} \sum_{i=1}^{A} \sum_{j=1}^{A}
\left[
j_0(k{r}_{ji})+j_2(k{r}_{ji})(\hat{r}_z^2+\hat{r}_y^2-2\hat{r}_x^2)
\right]
\sigma_{ix} \tau_{iz} \hat{O} \sigma_{jx} \tau_{jz} \nonumber \\
&~+& \frac{1}{3} \sum_{i=1}^{A} \sum_{j=1}^{A}
\left[
j_0(k{r}_{ji})+j_2(k{r}_{ji})(\hat{r}_x^2+\hat{r}_z^2-2\hat{r}_y^2)
\right]
\sigma_{iy} \tau_{iz} \hat{O} \sigma_{jy} \tau_{jz} \nonumber \\
&~+& \frac{1}{3} \sum_{i=1}^{A} \sum_{j=1}^{A}
\left[
j_0(k{r}_{ji})+j_2(k{r}_{ji})(\hat{r}_x^2+\hat{r}_y^2-2\hat{r}_z^2)
\right]
\sigma_{iz} \tau_{iz} \hat{O} \sigma_{jz} \tau_{jz} \nonumber \\
&~-& \sum_{i=1}^{A} \sum_{j=1}^{A} j_2(k{r}_{ji})\hat{r}_x\hat{r}_y
\left(\sigma_{iy} \tau_{iz} \hat{O} \sigma_{jx} \tau_{jz} + \sigma_{ix} \tau_{iz} \hat{O} \sigma_{jy} \tau_{jz}  \right)\nonumber \\
&~-& \sum_{i=1}^{A} \sum_{j=1}^{A} j_2(k{r}_{ji})\hat{r}_x\hat{r}_z
\left(\sigma_{iz} \tau_{iz} \hat{O} \sigma_{jx} \tau_{jz} + \sigma_{ix} \tau_{iz} \hat{O} \sigma_{jz} \tau_{jz} \right) \nonumber \\
&~-& \sum_{i=1}^{A} \sum_{j=1}^{A} j_2(k{r}_{ji})\hat{r}_y\hat{r}_z
\left( \sigma_{iz} \tau_{iz} \hat{O} \sigma_{jy} \tau_{jz} + \sigma_{iy} \tau_{iz} \hat{O} \sigma_{jz} \tau_{jz} \right).
\end{eqnarray}
  \item For the angle averaged spin-transverse coupling $\rho_{\sigma \tau T} (\bm{k})$ we have,
\begin{eqnarray}
&&
\frac{1}{4\pi} \int \rho_{\sigma \tau T}^\dagger(\bm{k}) \hat{O} \cdot \rho_{\sigma \tau T}(\bm{k})  d \Omega \nonumber \\
&=&
\frac{1}{4\pi}
 \int  \left[\sum_{i=1}^{A} e^{-i \bm{k} \cdot \bm{r}_i } (\bm{\sigma}_i \times \hat{\bm{k}}) \tau_{iz} \right]
\hat{O} \cdot
\left[\sum_{j=1}^{A} e^{i \bm{k} \cdot \bm{r}_j } (\bm{\sigma}_j \times \hat{\bm{k}}) \tau_{jz}  \right]  d \Omega  \nonumber \\
&=& \frac{1}{4\pi} \sum_{i=1}^{A} \sum_{j=1}^{A}
\left( \int e^{i \bm{k} \cdot \bm{r}_{ji} } \hat{k}_x \hat{k}_x d \Omega
+\int e^{i \bm{k} \cdot \bm{r}_{ji} } \hat{k}_z \hat{k}_z d \Omega
\right)
\sigma_{iy} \tau_{iz} \hat{O} \sigma_{jy} \tau_{jz}\nonumber \\
&~+& \frac{1}{4\pi} \sum_{i=1}^{A} \sum_{j=1}^{A}
\left( \int e^{i \bm{k} \cdot \bm{r}_{ji} } \hat{k}_x \hat{k}_x d \Omega
+\int e^{i \bm{k} \cdot \bm{r}_{ji} } \hat{k}_y \hat{k}_y d \Omega
\right)
\sigma_{iz} \tau_{iz} \hat{O} \sigma_{jz} \tau_{jz}\nonumber \\
&~+& \frac{1}{4\pi} \sum_{i=1}^{A} \sum_{j=1}^{A}
\left( \int e^{i \bm{k} \cdot \bm{r}_{ji} } \hat{k}_y \hat{k}_y d \Omega
+\int e^{i \bm{k} \cdot \bm{r}_{ji} } \hat{k}_z \hat{k}_z d \Omega
\right)
\sigma_{ix} \tau_{iz} \hat{O} \sigma_{jx} \tau_{jz}\nonumber \\
&~-& \frac{1}{4\pi} \sum_{i=1}^{A} \sum_{j=1}^{A} \left( \int e^{i \bm{k} \cdot \bm{r}_{ji} } \hat{k}_x \hat{k}_y d \Omega \right)
\left(\sigma_{ix} \tau_{iz} \hat{O} \sigma_{jy} \tau_{jz} + \sigma_{iy} \tau_{iz} \hat{O} \sigma_{jx} \tau_{jz}  \right)\nonumber \\
&~-& \frac{1}{4\pi} \sum_{i=1}^{A} \sum_{j=1}^{A} \left( \int e^{i \bm{k} \cdot \bm{r}_{ji} } \hat{k}_x \hat{k}_z d \Omega \right)
\left(\sigma_{ix} \tau_{iz} \hat{O} \sigma_{jz} \tau_{jz} + \sigma_{iz} \tau_{iz} \hat{O} \sigma_{jx} \tau_{jz} \right) \nonumber \\
&~-& \frac{1}{4\pi} \sum_{i=1}^{A} \sum_{j=1}^{A} \left( \int e^{i \bm{k} \cdot \bm{r}_{ji} } \hat{k}_y \hat{k}_z d \Omega \right)
\left( \sigma_{iy} \tau_{iz} \hat{O} \sigma_{jz} \tau_{jz} + \sigma_{iz} \tau_{iz} \hat{O} \sigma_{jy} \tau_{jz} \right) \nonumber \\
&=&
\frac{1}{3} \sum_{i=1}^{A} \sum_{j=1}^{A}
\left[
2j_0(k{r}_{ji})+j_2(k{r}_{ji})(2\hat{r}_y^2-\hat{r}_x^2-\hat{r}_z^2)
\right]
\sigma_{iy} \tau_{iz} \hat{O} \sigma_{jy} \tau_{jz}\nonumber \\
&~+& \frac{1}{3} \sum_{i=1}^{A} \sum_{j=1}^{A}
\left[
2j_0(k{r}_{ji})+j_2(k{r}_{ji})(2\hat{r}_z^2-\hat{r}_x^2-\hat{r}_y^2)
\right]
\sigma_{iz} \tau_{iz} \hat{O} \sigma_{jz} \tau_{jz}\nonumber \\
&~+& \frac{1}{3} \sum_{i=1}^{A} \sum_{j=1}^{A}
\left[
2j_0(k{r}_{ji})+j_2(k{r}_{ji})(2\hat{r}_x^2-\hat{r}_y^2-\hat{r}_z^2)
\right]
\sigma_{ix} \tau_{iz} \hat{O} \sigma_{jx} \tau_{jz}\nonumber \\
&~+& \sum_{i=1}^{A} \sum_{j=1}^{A} j_2(k{r}_{ji})\hat{r}_x \hat{r}_y
\left(\sigma_{ix} \tau_{iz} \hat{O} \sigma_{jy} \tau_{jz} + \sigma_{iy} \tau_{iz} \hat{O} \sigma_{jx} \tau_{jz}  \right)\nonumber \\
&~+& \sum_{i=1}^{A} \sum_{j=1}^{A} j_2(k{r}_{ji})\hat{r}_x \hat{r}_z
\left(\sigma_{ix} \tau_{iz} \hat{O} \sigma_{jz} \tau_{jz} + \sigma_{iz} \tau_{iz} \hat{O} \sigma_{jx} \tau_{jz} \right) \nonumber \\
&~+& \sum_{i=1}^{A} \sum_{j=1}^{A} j_2(k{r}_{ji})\hat{r}_y \hat{r}_z
\left( \sigma_{iy} \tau_{iz} \hat{O} \sigma_{jz} \tau_{jz} + \sigma_{iz} \tau_{iz} \hat{O} \sigma_{jy} \tau_{jz} \right) .
\end{eqnarray}

\end{enumerate}

\section{Euclidean Response Functions in Different Coordinates}
\label{secResconvert}

In all the $\rho(\bm{k})$, we can choose $\bm{r}_i$ to be particle $i$'s absolute position, we can also choose $\bm{r}_i$ to be particle $i$'s relative position in the center of mass frame.
The response functions calculated using the center of mass $\bm{r}_i$ multiply by a factor of $e^{-k^2 \tau /2M}$ will be those calculated using the absolute coordinate $\bm{r}_i$, where $M=Am$ is the total mass of system. This can be explained by decomposing the separable Hamiltonian into the center of mass part $H_{CM}=\bm{P}^2/2M$, where $\bm{P}=\sum_{i=1}^{A}\bm{p}_i$ is the total momentum, and a relative part $H_{rel}$ first (\cite{messiah2014quantum}), then make use of the momentum translation operators.

The relevant part in Eq.(\ref{Ektauresponse}) in absolute coordinate is
\begin{equation}
\!\!\! \langle \Phi_0| \sum_j e^{-i \bm{k} \cdot \hat{\bm{r}}_j } e^{-H\tau} \sum_n e^{ i \bm{k} \cdot \hat{\bm{r}}_n } | \Phi_0 \rangle
= \langle \Phi_0| \sum_j e^{-i \bm{k} \cdot \hat{\bm{r}}_j } e^{- \frac{\bm{P}^2}{2M}\tau} e^{- H_{rel} \tau} \sum_n e^{ i \bm{k} \cdot \hat{\bm{r}}_n } | \Phi_0 \rangle.
\label{Ektauresponsecore}
\end{equation}
The total momentum $\bm{P}$ of the ground state $| \Phi_0 \rangle $  is 0, and the momentum translation operator $e^{ i \bm{k} \cdot \hat{\bm{r}}_n }$ will move particle $n$'s momentum by $\bm{k}$, therefore $\sum_{n}e^{ i \bm{k} \cdot \hat{\bm{r}}_n }$ will change the system's total momentum from 0 to $\bm{k}$ and we have
\begin{equation}
\sum_{n}e^{ i \bm{k} \cdot \hat{\bm{r}}_n } | \Phi_0 \rangle =| \bm{k} \rangle  .
\end{equation}
Also consider that the Hamiltonian is translational invariant, so $[H,\bm{P}]=0$ and therefore $[H_{rel},\bm{P}]=0$, so $e^{- H_{rel} \tau}$ and $e^{- \frac{\bm{P}^2}{2M}\tau}$ commute, then Eq.(\ref{Ektauresponsecore}) becomes,
\begin{align}
& \langle \Phi_0| \sum_j e^{-i \bm{k} \cdot \hat{\bm{r}}_j } e^{- \frac{\bm{P}^2}{2M}\tau} e^{- H_{rel} \tau} \sum_n e^{ i \bm{k} \cdot \hat{\bm{r}}_n } | \Phi_0 \rangle \nonumber \\
& = \langle \Phi_0| \sum_j e^{-i \bm{k} \cdot \hat{\bm{r}}_j } e^{- H_{rel} \tau} e^{- \frac{\bm{P}^2}{2M}\tau}  \sum_n e^{ i \bm{k} \cdot \hat{\bm{r}}_n } | \Phi_0 \rangle \nonumber \\
&= e^{- \frac{\bm{k}^2 \tau}{2M}}  \langle \Phi_0| \sum_j e^{-i \bm{k} \cdot \hat{\bm{r}}_j } e^{- H_{rel} \tau} \sum_n e^{ i \bm{k} \cdot \hat{\bm{r}}_n } | \Phi_0 \rangle  .
\label{responsecomp1}
\end{align}

The relevant part in Eq.(\ref{Ektauresponse}) in the center of mass coordinate is
\begin{equation}\label{EktauresponsecoreCM}
 \langle \Phi_0| \sum_j e^{-i \bm{k} \cdot (\hat{\bm{r}}_j - \hat{\bm{R}}_{CM} )   } e^{- \frac{\bm{P}^2}{2M}\tau} e^{- H_{rel} \tau} \sum_n e^{ i \bm{k} \cdot (\hat{\bm{r}}_n - \hat{\bm{R}}_{CM} )  } | \Phi_0 \rangle ,
\end{equation}
where $\hat{\bm{R}}_{CM}=\frac{ \sum_{i} \hat{\bm{r}}_i}{A}$ is the center of the mass operator. Note that $e^{ - i\bm{k} \cdot \hat{\bm{R}}_{CM} }$ is also a momentum translation operator which changes the system's total momentum by $-\bm{k}$,
\begin{equation}
e^{ - i\bm{k} \cdot \hat{\bm{R}}_{CM} } | \Phi_0 \rangle = e^{ - i\bm{k} \cdot \frac{ \sum_{i} \hat{\bm{r}}_i}{A} } | \Phi_0 \rangle =  | -\bm{k} \rangle  .
\end{equation}
So, the combined operator $\sum_n e^{ i \bm{k} \cdot (\hat{\bm{r}}_n - \hat{\bm{R}}_{CM} )  } = e^{ - i\bm{k} \cdot \hat{\bm{R}}_{CM} } \sum_{n}e^{ i \bm{k} \cdot \hat{\bm{r}}_n }  $ does not change the system's total momentum.
Therefore, the ground state's total momentum $\bm{P}$ remains zero after being operated by $\sum_n e^{ \pm i \bm{k} \cdot (\hat{\bm{r}}_n - \hat{\bm{R}}_{CM} )  }$.
Also, $\hat{\bm{R}}_{CM}$ commutes with $H_{rel}$ so $e^{ \pm i\bm{k} \cdot \hat{\bm{R}}_{CM} } $ and $e^{- H_{rel} \tau}$ commute with each other.
Taking all these into consideration, Eq.(\ref{EktauresponsecoreCM}) becomes,
\begin{align}
& \langle \Phi_0| \sum_j e^{-i \bm{k} \cdot (\hat{\bm{r}}_j - \hat{\bm{R}}_{CM} )   } e^{- \frac{\bm{P}^2}{2M}\tau} e^{- H_{rel} \tau} \sum_n e^{ i \bm{k} \cdot (\hat{\bm{r}}_n - \hat{\bm{R}}_{CM} )  } | \Phi_0 \rangle \nonumber \\
&= \langle \Phi_0| \sum_j e^{-i \bm{k} \cdot \hat{\bm{r}}_j   }
e^{ i\bm{k} \cdot \hat{\bm{R}}_{CM} } e^{- H_{rel} \tau}
e^{ - i\bm{k} \cdot \hat{\bm{R}}_{CM} } \sum_{n}e^{ i \bm{k} \cdot \hat{\bm{r}}_n }  | \Phi_0 \rangle \nonumber \\
& = \langle \Phi_0| \sum_j e^{-i \bm{k} \cdot \hat{\bm{r}}_j } e^{- H_{rel} \tau} \sum_n e^{ i \bm{k} \cdot \hat{\bm{r}}_n } | \Phi_0 \rangle
\label{responsecomp2}
\end{align}

Comparing Eq.(\ref{responsecomp1}) and Eq.(\ref{responsecomp2}), we see that the center of mass result Eq.(\ref{responsecomp2}) need to multiply by the factor of $e^{-\frac{\bm{k}^2 \tau}{2M}}$ to match the result in absolute coordinate Eq.(\ref{responsecomp1}). Take $^4$He as an example, in our plots, $k=350$ MeV/c, $m=939$ MeV/$\rm{c}^2$, $M=4m=8$, so the factor $e^{- \frac{\bm{k}^2 \tau}{2M}}$ becomes $e^{- 16.3\tau}$.
We calculate our response functions in the center of mass frame first because it will give lower variance compared with absolute coordinate (laboratory frame). We then multiply our results by $e^{- 16.3\tau}$ in order to compare with the results in \citeauthor{Carlson94a}'s 1994 paper which are marked as the red dots in Figs. \ref{response} and \ref{responseN2LO}.

\newpage
\phantomsection
\renewcommand{\chaptername}{BIOGRAPHICAL SKETCH}
\addcontentsline{toc}{part}{BIOGRAPHICAL SKETCH}
\biographicalpage{
Born in Shanghai, China on Feb 22, 1987, Rong Chen received his bachelor's degree from Shanghai University in June, 2009.
With postgraduate recommendation, he was directly admitted by Shanghai Jiao Tong University as a master degree student in August 2009,
where he worked with prof. Lie-Wen Chen on various topics of infinite nuclear matter with isospin-dependent phenomenological nucleon-nucleon effective interactions.
He received the Master of Science degree in June 2012 from Shanghai Jiao Tong University.
In August 2012, he entered Arizona State University to pursue his PhD degree, and he worked as teaching and research assistant thereafter.
He did research rotations with profs. Cecilia Lunardini during fall 2012 about neutrino physics and Tanmay Vachaspati during spring 2013 about cosmology.
Then he worked with prof. Lunardini again from fall 2013.
However the cooperation reached an end in early 2014,
due to some reason which may only be fully understood after many years.
Therefore his PhD career was once entwined with neutrinos, but never joined.
At that time the chair of the graduate program committee prof. Ralph V. Chamberlin introduced him to prof. Kevin E. Schmidt,
who kindly accept him as one of his graduate students without too much hesitations.
He started to worked with prof. Schmidt since mid 2014 on quantum Monte Carlo calculations.
The first project from mid 2014 to late 2016 about the inclusion of the full three-body interactions in the approximated propagator in auxiliary field diffusion Monte Carlo was not successful enough to lead to a PhD project.
During summer 2016, he attended the TALENT School on Nuclear Quantum Monte Carlo Methods held in North Carolina State University, which was mainly organized by prof. Dean Lee.
Late 2016 was the turning point, with prof. Schmidt's suggestion and patient guidance, he began to work on the first continuous space path integral quantum Monte Carlo calculations of light nuclei, which became his Ph.D project and publishable results were obtained.
He gave a \href{http://meetings.aps.org/Meeting/4CS19/Session/J05.2}{talk} about it at 2019 Annual Meeting of the APS Four Corners Section.
He successfully defended his PhD dissertation on July 9, 2020.
}
\end{document}